\documentclass[a4paper,twoside,BCOR10mm,amsbsy,openright]{scrreprt}

\usepackage[ngerman,english]{babel}
\usepackage[headsepline]{packages/scrpage2}
		\automark[section]{chapter}
		\ihead[\headmark]{\headmark}
		\ohead[\pagemark]{\pagemark}
	    \ofoot[]{}

\usepackage[utf8x]{inputenc}

\usepackage{lmodern}
\usepackage[T1]{fontenc}
\addtokomafont{caption}{\small}
\usepackage{packages/graphicx}

\usepackage{packages/geometry/geometry}
\usepackage{color}
\usepackage{packages/float/float}
\usepackage{packages/psfrag/psfrag}

\usepackage{amsmath}
\usepackage{amssymb}
\usepackage{packages/csquotes/csquotes}

\usepackage{textcomp}
\usepackage{packages/was/upgreek}
\usepackage{packages/mh/mathtools}
\mathtoolsset{showonlyrefs}

\usepackage{packages/dstroke/dsfont}
\usepackage{packages/units/nicefrac}
\usepackage{packages/multirow/multirow}
\usepackage{multicol}
\usepackage{packages/booktabs/booktabs}
\usepackage{packages/multirow/bigdelim}
\usepackage{ifthen}
\usepackage[section,above,below]{packages/placeins/placeins}
\usepackage{microtype}
\usepackage[Lenny]{packages/fncychap/fncychap} 
        \ChNameVar{\fontsize{14}{16}\usefont{T1}{lmss}{m}{n}\selectfont}
        \ChNumVar{\fontsize{60}{62}\usefont{T1}{lmss}{m}{n}\selectfont}
        \ChTitleVar{\Huge\bfseries\sffamily}

\usepackage{packages/siunitx/siunitx} 
\usepackage[numbers]{packages/natbib/natbib}
\usepackage[dvips]{packages/pdfpages/pdfpages}

\usepackage[
            pdftitle={Oblique-incidence excitation of surface plasmon polaritons on small metal wires},
            pdfauthor={Arian Kriesch},
            pdfsubject={Experimental examination of scattering excitation of SPP.},
            pdfkeywords={spp, surface plasmon polaritons, plasmonics, scattering, mie},
            breaklinks=true,
            hypertexnames=false,
            pdfpagelabels=true,
            pdfpagemode=UseOutlines,
            hyperindex=true,
            colorlinks=false,     
            bookmarksdepth=3
            ]{hyperref}

\usepackage{packages/breakurl/breakurl}
\usepackage[refpage]{packages/nomencl/nomencl}
            \makenomenclature

\newcommand{\e}{\text{e}}
\renewcommand{\i}{\text{i}}

\newcommand{\sfrac}[2]{\textstyle{\frac{#1}{#2}}}

\newcommand{\vc}[1]{\boldsymbol{#1}}            
\newcommand{\vcg}[1]{\boldsymbol{#1}}           
\newcommand{\bvc}[1]{\boldsymbol{\hat{#1}}}     

\newcommand{\degC}{\celsius}                          
\newcommand{\del}{\partial}
\newcommand{\den}{\mbox{d}}
\renewcommand{\pi}{\uppi}

\renewcommand{\Re}{\text{Re}}
\renewcommand{\Im}{\text{Im}}

\newenvironment{changemargin}[2]{
  \begin{list}{}{%
    \setlength{\topsep}{0pt}%
    \setlength{\leftmargin}{#1}%
    \setlength{\rightmargin}{#2}%
    \setlength{\listparindent}{\parindent}%
    \setlength{\itemindent}{\parindent}%
    \setlength{\parsep}{\parskip}%
  }%
  \item[]}{\end{list}}

\renewcommand{\eqdeclaration}[1]{~(eq.~#1)}
\renewcommand{\pagedeclaration}[1]{~(p.~#1)}

\RequirePackage{ifthen}
 \renewcommand{\nomgroup}[1]{%
 \ifthenelse{\equal{#1}{M}}{\item[\large{\textsf{\textbf{Nomenclature and constants}}}]}{%
 \ifthenelse{\equal{#1}{S}}{\item[\large{\textsf{\textbf{Abbreviations and setup components}}}]}{}}}

\makeatletter
  {%
  \endlist
  \nompostamble}
\makeatother

\newcounter{savedtocdepth}
\newcommand*{\SaveTocDepth}[1]{%
  \addtocontents{toc}{%
    \protect\setcounter{savedtocdepth}{\protect\value{tocdepth}}%
    \protect\setcounter{tocdepth}{#1}%
  }%
}

\newcommand*{\RestoreTocDepth}{%
  \addtocontents{toc}{%
    \protect\setcounter{tocdepth}{\protect\value{savedtocdepth}}%
  }%
} 

\begin{document}

\includepdf{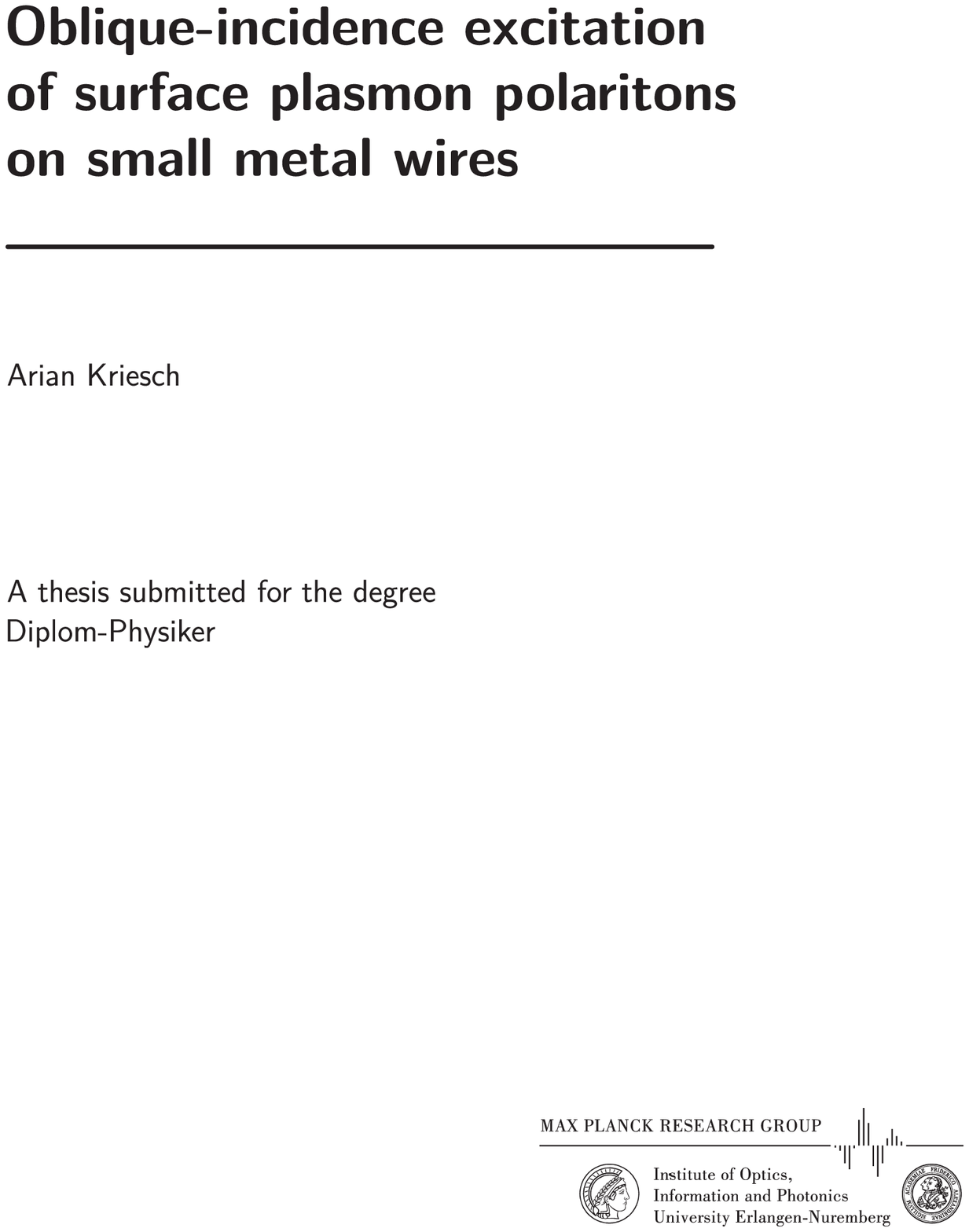}

\begin{titlepage}
\begin{changemargin}{-0.5cm}{-0.5cm}

    \begin{center}
    \begin{sf}

        \vspace{4em}
    \Huge{\textbf{Oblique-incidence excitation\\
    of surface plasmon polaritons\\
    on small metal wires}}\\ \rule{\linewidth}{0.5mm}\\
        \vspace{5em}
    \Large{ A thesis submitted for the degree\\
            Diplom-Physiker\\[1em]
            Fachbereich Physik\\
            der Friedrich-Alexander-Universit\"{a}t Erlangen-N\"{u}rnberg\\[1em]
            $\text{13}^\text{th}$ August 2008
    }
        \vspace{4em}

    \vfill
\begin{minipage}{0.5\textwidth}
\begin{flushleft} \Large
Author:\\
Arian Kriesch
\end{flushleft}
\end{minipage}
\begin{minipage}{0.5\textwidth}
\begin{flushright} \Large
Supervisor: \\
Prof. Dr. Philip St.J. Russell
\end{flushright}
\end{minipage}

    \vfill

    \begin{minipage}[t]{70mm}
    \includegraphics[width=70mm]{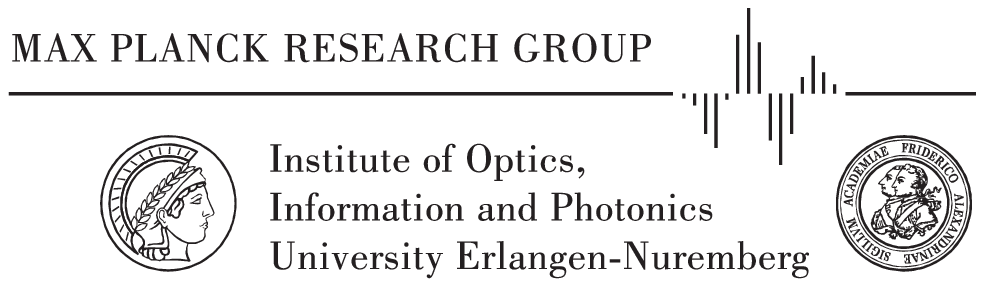}
    \end{minipage}


    \end{sf}
    \end{center}

\enlargethispage{3\baselineskip}

\end{changemargin}
\end{titlepage}


~
\vfill

{\centering\parbox[t]{0.6\textwidth}{\centering%
        \small{
        Kriesch, Arian\\
        \textit{Oblique-incidence excitation of surface plasmon polaritons on small metal wires} \vspace{0.5em}

        Thesis for the academic degree ``Diplom Physiker''

        Accomplished at the Max-Planck-Research-Group for Optics, Information and Photonics at the University of Erlangen-Nuremberg, division 3\\ \vspace{0.5em}

        Supervisors:

        Prof. Dr. Philip St.J. Russell

        Dr. Markus Schmidt

        $13^\text{th}$ August, 2008\\ \vspace{0.5em}

        Copyright \copyright  2008, Arian Kriesch


        }
        }}

\thispagestyle{empty} 
\clearpage{\pagestyle{empty}\cleardoublepage}

~
\vfill

\begin{flushright}\parbox[t]{\textwidth/2}{%
        \small{``Anybody who has been seriously engaged in scientific work of any kind realizes that over the entrance to the gates of the temple of science are written the words: Ye must have faith. It is a quality which the scientist cannot dispense with.''\\ \slshape{(Max Planck, 1923)}}}
\end{flushright}

\thispagestyle{empty} 
\clearpage{\pagestyle{empty}\cleardoublepage}


\begingroup\begin{changemargin}{+0.3cm}{+0.3cm}
\let\clearpage\relax
\let\cleardoublepage\relax
\let\cleardoublepage\relax
\enlargethispage{4\baselineskip}

\section*{Abstract}
    \small

    This work reports on the experimental investigation of surface plasmon polaritons (SPP) on cylindrical wires of small diameters. It has been accomplished within the scope of a diploma thesis at the Max-Planck Research Group for Optics, Information and Photonics of the Friedrich-Alexander University of Erlangen-Nuremberg.

    Applying a new technique that was developed by this group and recently reported \citep{schmidt-optl-numerical-07,schmidt-prb-arrays-08,schmidt-science,fio-2007}, single wire and wire array samples of gold (Au) and silver (Ag) with small diameters ($\SI{400}{nm}<\varnothing<\SI{3}{\micro m}$) and high aspect ratios ($\lesssim \SI{75000}{}$) in photonic crystal fibers and single hole capillaries are fabricated. Additionally, effective bulk metal wires of Au and Ag are created by a hybrid technique, including fiber tapering and magnetron sputter deposition for a large number of different diameters between \SI{13}{\micro m} to \SI{50}{\micro m}.

    First results of optical in-fiber excitation of SPP \citep{schmidt-prb-arrays-08} suggested an in depth study of the dispersion relation of surface plasmon polaritons. The measurement of the absorption, corresponding to the excitation of SPP modes was realized with a new goniometer-based experimental setup for oblique incidence laser beam scattering at a single wavelength. A device was developed that allows a simultaneous measurement of the scattering amplitude in s- and p-polarization under varied incident angle.

    For comparison, a model \citep{schmidt-optl-numerical-07} for the dispersion of discrete SPP modes and their angles of excitation is derived. The computed results are for the first time compared with a theoretical approach to electromagnetic scattering theory for cylinders, that is calculated analogue to Mie scattering \citep{mie-1908} and to an exact solution of Maxwell's equations for cylindrical SPP. The validity of the model is proved for higher incident angles. The first oblique incidence examination of SPP on metal wires for this size range is reported. The experimental results for low incident angles show good accordance to the numerical results from scattering theory.

\vfill

\pdfbookmark[1]{Abstract}{Abstract}
\section*{Zusammenfassung}

    Das Thema dieser Arbeit ist die experimentelle Untersuchung von Plasmon-Polaritonen auf der Oberfläche von Drähten geringen Durchmessers (SPP). Die Untersuchungen wurden im Rahmen einer Diplomarbeit an der Max-Planck-Forschungsgruppe für Optik, Information und Photonik der Friedrich-Alexander-Universtät Erlangen-Nürnberg durchgeführt.

    Auf Grundlage einer neuen Technik, die an dieser Gruppe entwickelt wurde  \citep{schmidt-optl-numerical-07,schmidt-prb-arrays-08,schmidt-science,fio-2007}, werden Einzeldrähte und Drahtarrays aus Gold (Au) und Silber (Ag) mit geringen Durchmessern ($\SI{400}{nm}<\varnothing<\SI{3}{\micro m}$) und großem Längenverhältnis ($\lesssim \SI{75000}{}$) mittels Matrizen gezielt gefertigter Photonischer Kristallfasern hergestellt. Außerdem wird eine kombinierte Technik aus Fasertapern und Magnetronsputtern zur Herstellung von Proben mit Durchmessern zwischen \SI{13}{\micro m} und \SI{50}{\micro m} angewandt.

    Erste Ergebnisse aus der Anregung von SPP durch glasfasergeführtes Licht \citep{schmidt-prb-arrays-08} legten eine detailliertere Untersuchung der Dispersionsrelationen solcher angeregten Oberflächenmoden nahe. Die hier angewandte Technik, diese anzuregen, basiert auf einem neuen, goniometerbasierten Aufbau zur Messung der Streuamplitude eines schräg zur Drahtachse einfallenden Laserstrahls bei konstanter Wellenlänge. Das optische System wurde so entwickelt, dass eine simultane Messung der Streuamplituden einfallender s- und p-Polarisation unter Winkelvariation möglich ist.

    Ergebnisse auf Basis eines Modells \citep{schmidt-prb-arrays-08} solcher SPP-Moden werden erstmals mit nu\-me\-ri\-schen Ergebnissen verglichen, deren Berechnung nach einer Herleitung von winkelabhängigen Streu\-am\-pli\-tu\-den und Feldverteilungen für die Streuung an einem Zylinder analog der Mie-Lösung \citep{mie-1908} der Maxwellgleichungen erfolgt. Außerdem erfolgt ein Vergleich mit der exakten Lösung der Maxwell-Gleichungen für SPP. Der Gültigkeitsbereich des Modells wird dabei auf größere Einfallswinkel bestimmt. Hiermit wird erstmals die Anregung von SPP auf Drähten dieser geringen Größe unter schräger Inzidenz untersucht. Die Messergebnisse für geringen Einfallswinkel zeigen gute Übereinstimmung mit den Berechnungen der Streu-Theorie.
\end{changemargin}
\endgroup			

\vfill
\thispagestyle{empty} 
\clearpage{\pagestyle{empty}\cleardoublepage}

\tableofcontents
\newpage

\chapter{Introduction \label{ch_introduction}}
\pagenumbering{arabic}
\section{The examination of surface plasmon modes on wires}

    Surface plasmons, collective vibrations of the electron gas propagating on the surface of metals, have experienced a renaissance during the last years since their first experimental examination and theoretical treatment in modern physics by \citet{sommerfeld} (1899), \citet{zenneck} (1907) and \citet{mie-1908} (1908).

    The optical effects of surface plasmon (SP) excitations, particularly the colour effects of metal colloids were phenomenologically known by craftsmen much earlier. One of the first and most impressive applications of dichroic glass is the `Lycurgus cup' from the late roman era (4th century AD), which shows a different colour for transmitted and reflected white light.

    Today, since the initial experimental studies on \citet{raether}, SP are finding an increasing number of applications \citep{barnes-2003}. Especially the advance in optical technology recently allows an in depth optical examination of surface plasmons that were before in a wide range of frequencies preferentially examined by electron diffraction and loss measurements.

    The advantages of surface plasmons can be attributed to its intrinsic nature of being confined in the near vicinity of metal-dielectric interfaces. This confinement leads to a strong enhancement of the electromagnetic fields, which makes SP extraordinarily sensitive to surface conditions. This sensitivity can be utilized e.g. for sensing applications~\citep{ozbay-2006}.

    Surface plasmons also offer new possibilities for nanoscale photonic circuits \citep{barnes-2003}, for the fabrication of subwavelength waveguides~\citep{bozhevolnyi} and may offer possibilities to realize a basis for optical switches and circuits below the diffraction limit of the applied wavelength~\citep{krasavin}.

    In addition they are shown to be well applicable to construct collectors for electromagnetic radiation near the surface of samples. These allow the measurement of near field distributions and reveal new insights into the radiative properties of samples, scanning near field optical microscopy~(SNOM), an application that has already entered commercial development.

    The progress lead finally to the concept of specifically `designed materials' to promote desired dispersions that are not achieved with natural materials~\citep{pendry-2004}.

    The common focus of the theoretical examination of surface plasmon polaritons (SPPs) always was on spherical particles as these are of special importance in many interdisciplinary applications from biology, the treatment of organic cells to atmospheric physics and aerosol science.

    Nevertheless, the theoretical examination of surface plasmon modes on cylinders over the last few years gathered increased attention throughout literature \citep{liaw-wu,lukyanchuk,schroeter-dereux,hasegawa,schider,zayats-04,novotny,schroeter-dereux}. Experimental studies for many of these calculations still have to be accomplished, although in some geometries first results are promising for further in depth examination~\citep{gonzalez,abushagur}.\\

    Recently by \citet{schmidt-prb-arrays-08} the fabrication and examination of metal wire arrays with an unsurpassed high length to diameter ratio and good surface smoothness was reported. These samples were fabricated by a new technique inside the hole structure of photonic crystal fibers (PCF)~\citep{russell-2006}.

    The resonance frequencies of SPPs on these wires were determined by measuring transmission spectra of the fibers. The results show good accordance with theoretical multipole calculations~\citep{schmidt-optl-numerical-07} and a proposed~\citep{miziumski-theory,schmidt-prb-arrays-08,schmidt-theory} model for the description of the dispersion of SPPs.\\

    The aim of this work is to answer a number of questions that appeared in the course of these studies and are not yet treated by any other experimental work:

    \begin{enumerate}
      \item What is the range of validity of the proposed model for the propagation of radiative SPPs on metal wires?
      \item Does a description of the scattering of light on metal wires comprise SPP resonances?
      \item Does an experimental scattering examination of exciting SPPs on small metal wires reveal resonances that match the predicted dispersion relation in a range of wave vectors that is not feasible by in-fiber coupling?
      \item Is it possible to relate the observations from scattering to a near field description?
    \end{enumerate}

\section{Overview and structure of the thesis}

    This thesis comprises the following parts:

    As a background for the further theory, in the first part of the thesis the theoretical description of electromagnetic waves is introduced in general and the linear optical properties of matter, especially for metals are discussed (sec.\;\ref{ch_theory}).

    Subsequently (sec.\;\ref{ch_theory-planar-spp}) an overview over the occurrence of planar surface plasmon polaritons is given. This theory is then (sec.\;\ref{chapter_theory-model}) applied to the special case of surface plasmon polaritons on a closed two dimensional surface, leading to the dispersion relations for discrete surface plasmon modes on cylindrically curved surfaces.

    From electromagnetic scattering theory (sec.\;\ref{chapter_theory-scattering}) an analytic approach, analogue to the Mie solution~\citep{mie-1908} for infinite right circular cylinders is derived.

    The range of validity of the model is evaluated by a comparison of computed results for the applied materials and experimental parameters (sec.\;\ref{ch_theoresults-model}) to the exact solution of Maxwell's equations for radiative SPP modes in the same case (sec.\;\ref{ch_theoresults-exact}).

    From the derived Mie theory, scattering field distributions and absorption patterns are computed and compared to the exact solution and the model (sec.\;\ref{chapter_theory-scatterincoeff}). An analysis of the near field distributions and an approach for their explanation and interconnection to the mode excitation is presented. \\

    \noindent To test the theory, two different types of samples are fabricated,

    \begin{enumerate}
      \item freestanding, metal coated silica wires for diameters $D > \lambda$ (sec. \ref{ch_sample-freestanding}) and
      \item nanowires embedded in silica fiber for diameters $D \lesssim \lambda$ (sec. \ref{ch_sample-silica}).
    \end{enumerate}

    The functional principle of the experimental measurement setup for probing these samples is explained~(sec.~\ref{ch_experimental-setup}). Results of the accomplished experiments are presented and discussed in section \ref{ch_results-discussion} with a comparison to the derived theories. The conclusion from the experimental and theoretical investigations closes the work (sec. \ref{ch_cnclusion}) and propositions for further studies of the observed phenomena are given.

\cleardoublepage

\chapter{Theoretical background \label{ch_theory}}

\begin{flushright}\parbox[t]{\textwidth/2}{%
        \small{``Whether you can observe a thing or not depends on the theory which you use. It is the theory which decides what can be observed.''\\ \slshape{(Albert Einstein, 1926)}}}
\end{flushright}

\section{From Maxwell's equations to surface plasmons}

    \subsection{Maxwell's equations \label{ch_theory-maxwell}}
    \footnotetext{A list of mathematical and physical symbols and abbreviations that are used in the following chapters can be found in appendix \ref{ch_appendix-conventions} with references to the page where they are first introduced.}Electromagnetic waves are in general described by equations, which were first collated from previously already known physical laws to a consistent system by James Clerk Maxwell in 1865 \citep{maxwell}. Usually they are written in the modern form of four linear\footnote{As the equations are linear, linear combinations of their solutions solve the equations again.} differential equations \footnote{In theoretical physics often the cgs system is applied. Here for a better insight into the experimental meaning of the results all derivations base on the SI notation. For an in depth discussion it is referred to \citep{jackson}}:

    \begin{subequations}
    \begin{align}
    \nabla \cdot \vc{D} &= \rho                                \label{eq-maxwell-divd}\\
    \nabla \times \vc{H} - \frac{\del}{\del t} \vc{D} &= \vc{j}  \label{eq-maxwell-roth}\\
    \nabla \times \vc{E} + \frac{\del}{\del t} \vc{B} &= 0 \label{eq-maxwell-rote}\\
    \nabla \cdot \vc{B} &= 0                                   \label{eq-maxwell-divb}
    \end{align}\label{eq-maxwell}
    \end{subequations}

    Coulomb's law \eqref{eq-maxwell-divd} expresses that the source for the electric displacement vector $\vc{D}$ is the free electric charge density $\rho$. The extended Amp\`{e}re's law \eqref{eq-maxwell-roth} states that the vortices of the magnetic vector field $\vc{H}$ as well as change in an electric vector field is caused by the electric current density vector $\vc{j}$. The vortices of the electric vector field $\vc{E}$ are produced by a change in the magnetic induction vector $\vc{B}$ and vice versa, following Faraday's law \eqref{eq-maxwell-rote}. Gau\ss' law states that the magnetic field is solenoidal; there are no magnetic monopoles \eqref{eq-maxwell-divb}\footnote{No magnetic monopoles have been verified up to now. Indeed, an elegant theoretical explanation of the origin of the quantization of electric charges by \citet{dirac} is based on the assumption that magnetic monopoles actually exist. An idea that is well described in \citep{jackson}}.

    In vacuum the relations of electric displacement and electric field to magnetic induction and magnetic field are connected by the dielectric permittivity of the vacuum $\varepsilon_0=\SI{8.8542e-12}{As/(Vm)}$ and the magnetic permeability of the vacuum $\mu_0 = 4 \pi \times \SI{e-7}{Vs/(Am)}$:

    \nomenclature[me ]{$\varepsilon$}{Complex relative dielectric permittivity $\varepsilon=\varepsilon_1+\i \varepsilon_2$, which is dimensionless. Expressed in terms of the refractive index $\varepsilon_1=n^2-\kappa^2$ and $\varepsilon_2=2 n \kappa$ (See also \ref{ch_appendix-conventions}).}
    \nomenclature[me ]{$\varepsilon_0$}{The dielectric permittivity (also called `diel. constant') of vacuum is $\varepsilon_0= \SI{8.8542e-12}{(A s) / (V m)}$}
    \nomenclature[mm ]{$\mu$}{Complex relative magnetic permeability, in dimensionless nomenclature. Wavelength dependent material constant. Defined consistent to $\varepsilon$ as $\mu=\mu_\text{abs}/\mu_0$.}
    \nomenclature[mm ]{$\mu_0$}{Complex magnetic permeability $\mu_0 =  4 \pi \times \SI{e-7}{(V s) / (A m)} $.}
    \begin{align}
    \vc{D} &= \varepsilon_0 \vc{E}    &     \vc{B} &= \mu_0 \vc{H} \label{eq_de-bh-vac}
    \end{align}

    The vacuum speed of light $c_0=\SI{2.99792458e8}{m/s}$ in the SI system defines a basic physical constant which can be determined by measurement. From a measurement of $\varepsilon_0$, $\mu_0$ can be calculated using the relation:
    \nomenclature[mc]{$c_0$}{The vacuum speed of light is a constant, defined by the special theory of relativity. $c= \sqrt{\mu_0 \varepsilon_0}^{-1}=\SI{2.99792458e8}{m/s}$.}

    \begin{equation}
    c_0 = \frac{1}{\sqrt{\varepsilon_0\mu_0}}
    \end{equation}

%

    \subsection{Light-matter interaction \label{ch_theory-susceptibility}}

    It is possible to introduce the response of matter to electromagnetic radiation by introducing additional terms into equations \refeq{eq_de-bh-vac}, called the electric polarization $\vc{P}$ and the magnetization $\vc{M}$:

    \begin{align}
        \vc{D} &= \varepsilon_0 \vc{E} + \vc{P}       &       \vc{B} &= \mu_0 ( \vc{H} + \vc{M} ) \label{eq_polarization-magnetization}
    \end{align}

    Since the wavelength of visible light is large compared to atomic distances, it is often sufficient to treat the response of media to light macroscopically. The electric polarization $\vc P$ and the magnetization $\vc{M}$ are defined as functions of the electric field $\vc{P}(\vc{E})$ and the magnetic field $\vc{M}(\vc{H})$, representing the material response. This dependency can be expressed in good approximation, applying an expansion in power series in terms of sums over vector entries:

    \begin{subequations}
    \begin{align}
        P_i &= P_0 +  \sum\limits_{j=1}^{3} \eta_{ij}^{(1)} E_j +  \sum\limits_{j=1}^{3} \sum\limits_{k=1}^{3} \eta_{ijk}^{(2)} E_j E_k +  \sum\limits_{j=1}^{3} \sum\limits_{k=1}^{3} \sum\limits_{l=1}^{3} \eta_{ijkl}^{(2)} E_j E_k E_l + \ldots   \label{eq_polarization-sum-p} \\
        M_i &= M_0 +  \sum\limits_{j=1}^{3} \chi_{ij}^{(1)} H_j +  \sum\limits_{j=1}^{3} \sum\limits_{k=1}^{3} \chi_{ijk}^{(2)} H_j H_k +  \sum\limits_{j=1}^{3} \sum\limits_{k=1}^{3} \sum\limits_{l=1}^{3} \chi_{ijkl}^{(2)} H_j H_k H_l + \ldots  \label{eq_polarization-sum-m}
    \end{align}\label{eq_polarization-sum}
    \end{subequations}
    \nomenclature[mp]{$\vc P$}{Macroscopic electric polarization of a medium.}
    \nomenclature[mm]{$\vc M$}{Macroscopic magnetization of a medium.}

    Thus the dependency is written as a successive sum over dielectric susceptibility tensors $\eta_{j,k,l,\ldots}^{(n)}$ and magnetic susceptibility tensors $\chi_{j,k,l,\ldots}^{(n)}$ plus a bias of polarization $P_0$ and magnetization $M_0$. The terms of $n=1$ are responsible for a linear dependency on the electric and magnetic field. Higher order terms ($n>1$) describe nonlinear effects, which can be neglected as well as the bias terms for most optical materials under the assumption of field strengths which are reasonably lower than inner atomic Coulomb fields.

    Due to the isotropic nature of the materials that are treated throughout the course of this work\footnote{Otherwise optical effects like birefringence result (for further details see e.g. \citep[ch. xi]{landau-lifschitz} or \citep{born-wolf}).}, the two dimensional $3 \times 3$ tensors $\eta$ and $\chi$ can be written as scalar quantities.

     \begin{align}
     \vc P &= \eta \vc E \label{eq_lin-polarization} \\
     \vc M &= \chi \vc H \label{eq_lin-magnetization}
    \end{align}

    Assuming a homogeneous material, both are additionally independent of the spatial coordinates. With those material constants the dependency of $\vc{D}(\vc{E})$ and $\vc{B}(\vc{H})$ for electromagnetic waves in linear, homogeneous and isotropic media can be defined\footnote{It shall be pointed out that throughout this work $\varepsilon$ and $\mu$ are defined as relative, dimensionless values. For details it is referred for appendix \ref{ch_appendix-conventions}.}.

    From equations (\refeq{eq_lin-polarization}, \refeq{eq_lin-magnetization}) the material dependent dielectric permittivity $\varepsilon$ and the magnetic permeability tensor respectively scalar $\mu$ are defined:

    \begin{align}
    \varepsilon \coloneqq& \, \eta / \varepsilon_0 +1      &       \mu \coloneqq& \, \chi +1 \\
    \vc{D} =& \, \varepsilon_0 \varepsilon \vc{E}    &       \vc{B} = & \, \mu_0\mu \vc{H}
    \end{align}

    Different to the dielectric permittivity $\varepsilon > 1$, the magnetic permeability in physically real cases is only limited to be positive $0 < \mu$ as the magnetic susceptibility complies $-1 <\chi \gtreqqless 0$. \footnote{See \citep[\S31]{landau-lifschitz} for details.} This is the mathematical expression of para- and diamagnetism respectively. Only by so called synthetic meta-materials, this rule is broken, which also allow classically forbidden negative refractive indices. Quantitatively for all non ferromagnetic substances the magnetic permeability can be assumed to be unity in most cases in optics as $|\mu-1|\lesssim 10^{-6}$.

    The material properties also influence the current density which is generally not zero but $\vc{j}=\sigma \vc{E}\label{text_constituent-j}$. This relation is also known as the differential form of Ohm's law, where $\sigma$ represents the specific conductivity of a material. The value of this material constant distinguishes conductors $\sigma \neq 0$, among which metals show extraordinary high $\sigma$, from insulators $\sigma \approx 0$.
    \nomenclature[ms]{$\sigma$}{The specific conductivity of a material.}

    By the so called `Maxwell relation' the complex refractive index can be defined from $\varepsilon$ and $\mu$ as an alternative quantity for the material response to electromagnetic waves:

    \begin{align}
    n = \sqrt{  \mu \epsilon } = n_\text{r} + \i \kappa \label{eq_maxwell-relation}
    \end{align}%
    \nomenclature[mn]{$n$}{Complex index of refraction, defined by $n=n_r+\i \kappa$ $n=\sqrt{\varepsilon \mu}$.}

    Its real part $n_\text{r}= c_0/c$ represents the quotient of the vacuum speed of light over the phase velocity in the specific material. Its imaginary part $\kappa$, which is also called `extinction coefficient' represents the absorption in the medium, as described by Beer's law (see p. \pageref{eq_k-dispersionrel}).

    \subsection{Dispersion \label{ch_theory-dispersion} }

    For the introduction of the dielectric susceptibility the material properties were described macroscopically. This approach leads to correct results as it equals a spatial summation over the response to the incident field of a large number ($ N > \SI{1e6}{}$)\footnote{An approximate number of the atoms inside a macroscopical volume with the size of optical wavelengths. A concise microscopic derivation can be found in \citep[Ch.~6.6]{jackson}.} of electrons and atomic nuclei. This number is sufficient to smooth microscopic fluctuations which allows a macroscopic treatment in the nonretarded limit for static fields or field fluctuations that are slow enough for the material to respond instantaneously.

    Indeed, propagating electromagnetic fields intrinsically cannot be described electrostatically. All media aside from vacuum show a dependency of the optical material properties on the frequency of the electromagnetic field. This dependency is called dispersion. Nevertheless, with this effect under consideration all derived relations remain valid, only the dielectric permittivity and magnetic permeability have to be described as frequency dependent quantities $\varepsilon(\omega)$, $\mu(\omega)$\footnote{To be precise, this approach is only valid as long as monochromatic waves are treated, otherwise it is an approximation.}.

    The reason for dispersion is obvious from the point of view of a microscopic description of the material. For an e.g. harmonic time dependency an additional factor enters the electromagnetic fields $E=E_0 \e^{-\i \omega t}$. Including this temporal dependency, the material response can no longer be assumed to be instantaneous, a phase discrepancy between driving field and radiated field enters. For a quantitative derivation a microscopic description of the material is necessary\footnote{The dispersion is derived classically, omitting quantum mechanical effects, as a thoroughly quantum mechanical approach follows the same derivation with a replacement of the classical values by quantum mechanical expectation values and leads to the same results. For a derivation that accounts for quantum mechanics it is referred to \citet{becker} and \citet{adler}.}.

    The polarization $\vc P$ (eq. \refeq{eq_lin-polarization}) of a volume element of a medium, can therefore be written as a superposition of molecular polarizations, expressed as the spatial mean molecular dipole moment $\langle \vc{p} \rangle$ of $N$ molecules with a molecular polarizability $\gamma_\text{mol}$ in that volume:

    \begin{align}
        \vc{P} &= N \langle \vc p \rangle \\
        \langle \vc p \rangle &= \varepsilon_0 \gamma_\text{mol} (\vc E + \vc E_\text{i})
    \end{align}

    Generally not only the macroscopic field $\vc E$ causes the molecular polarization $\vc{p}$ but intermolecular interactions add a second term $\vc E_\text{i}$. Applying the local field approximation\footnote{See \citep[p. 186]{jackson} for details.}, a relation between the polarization and the dielectric susceptibility is obtained:

    \begin{align}
        \vc{P} &= N \gamma_\text{mol} \left( \varepsilon_0 \vc E + \nicefrac{1}{3} \vc P \right) \\
        \eta & \overset{\text{\eqref{eq_lin-polarization}}}{=} \frac{N \gamma_\text{mol}}{1-\nicefrac{1}{3} N \gamma_\text{mol}}
    \end{align}

    The molecular polarizability $\gamma_\text{mol}$ in this expression represents the same information about the material response as the dielectric permittivity. The equation that connects $\gamma_\text{mol}$ and $\varepsilon$ is called Clausius-Mosotti relation\footnote{For the assumption of optical frequencies, therefore $\mu \approx 1$ and $n^2=\varepsilon/\varepsilon_0$ this equation is also called Lorenz-Lorentz relation.} and includes the density of the material by the number $N$ of molecules inside the considered volume:

    \begin{equation}
        \gamma_\text{mol} = \frac{3}{N} \left( \frac{\varepsilon-1}{\varepsilon +2} \right)
    \end{equation}

    \noindent For media of low density however the neighbour interaction is negligible ($\vc E_\text{i} \approx 0$).\\

    A simple oscillator model is a sufficient approach for many applications to derive the frequency dependency of the dielectric permittivity. The electrons of all molecules are then assumed to be bound, each in a Lorentz oscillator potential. This adiabatic approximation bases on the assumption that the motion of the nuclei can be neglected. In fact, this is true up to very long wavelengths (infrared) due to the very different masses of electrons and atomic cores\footnote{Detailed treatment of the Lorentz oscillator and the free electron gas approximations can be found in the classical solid state literature, e.g. \citep{ashcroft,kittel}.}.

    The equation of motion of an electron with the charge $-e$, mass $m_\text{e}$ and displacement $\vc r$ from its equilibrium position is an inhomogeneous linear differential equation of second order:

    \begin{align}
        -e \vc E (\vc r,t)  = m_\text{e} (\ddot{\vc r} + \gamma \dot{ \vc r} + \omega_0^2 \vc r ) \label{eq_dispersion-motion-diel}
    \end{align}

    \noindent with the resonance frequency $\omega_0$, considering a damping coefficient $\gamma$

    Solving this equation for the assumed time harmonic field $\vc E= \vc E_0 \e^{-\i \omega t}$ results in a molecular dipole moment $\vc p$ and via a summation to the macroscopic polarization $\vc P$ of:

    \begin{equation}
        \vc P = N \vc p = N \frac{e^2}{m_\text{e}} \frac{ \vc E}{\omega_0^2-\omega^2-\i \omega \gamma} .
    \end{equation}

    Applying this equation to equation \eqref{eq_lin-polarization} the explicit dependency of the dielectric permittivity $\varepsilon (\omega) = \eta / \varepsilon_0+1$ on frequency is obtained:

    \begin{align}
        \varepsilon(\omega) = 1 + \frac{N}{\varepsilon_0} \frac{e^2}{m_\text{e}} \frac{1}{ \omega_0^2- \omega^2 -\i \omega \gamma} \label{eq_epsilon-dispersion}
    \end{align}

    \begin{figure}[tbp]
        \centering

            \psfrag{epsilon}[c][c][1.0][0]{$\Re(\varepsilon)$, $\Im(\varepsilon)$}
            \psfrag{Driving frequency}[c][c][1.0][0]{Driving frequency $\omega$}
            \psfrag{reo}[c][c][1.0][0]{$\Re(\varepsilon)$}
            \psfrag{imo}[c][c][1.0][0]{$\Im(\varepsilon)$}
            \psfrag{ND}[c][c][1.0][0]{ND}
            \psfrag{AD}[c][c][1.0][0]{AD}
            \psfrag{o_0}[B][B][1.0][0]{$\omega_0$}
            \psfrag{o-g}[Bc][Bl][1.0][0]{$\omega_0-\nicefrac{\gamma}{2}$}
            \psfrag{o+g}[Bc][Br][1.0][0]{$\omega_0+\nicefrac{\gamma}{2}$}
            \psfrag{val}[c][c][1.0][0]{$\frac{N e^2}{\varepsilon_0 m_\text{e} \omega_0^2}$}
            \psfrag{0}[c][c][1.0][0]{$0$}
            \psfrag{1}[c][c][1.0][0]{$1$}

            \includegraphics[width=0.8\textwidth]{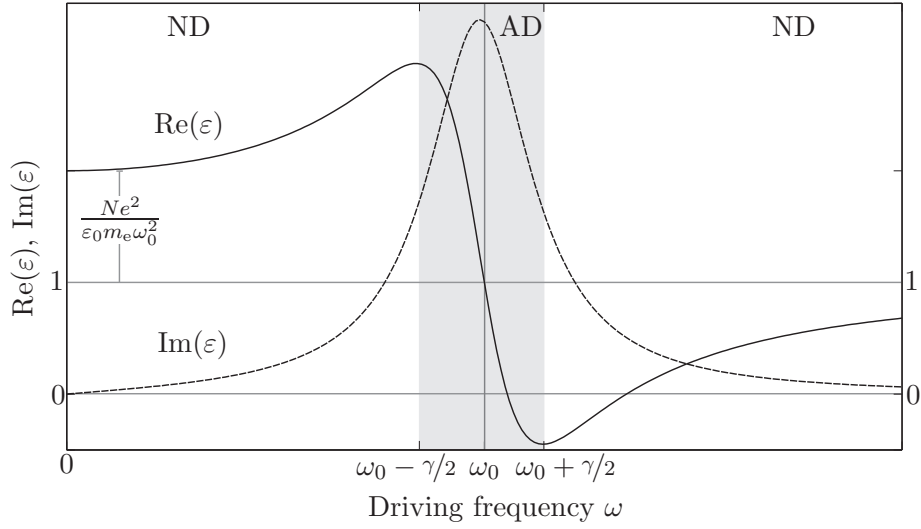}

        \caption[Plot of the real and imaginary part of the dispersion of $\varepsilon(\omega)$ for a single Lorentz resonance.]{Real $\Re(\varepsilon)$ and imaginary part $\Im(\varepsilon)$ of the dielectric permittivity at a resonance frequency $\omega_0$ in the Lorentz oscillator model, including damping $\gamma \neq 0$.\footnotemark \label{graphic_dispersion}}
\end{figure} 
\footnotetext{The dispersion curve is calculated with numeric values that are not representative for real materials but that point out the resonance behaviour. A plot of the real dispersion for silica glass from the Sellmeier equation is depicted in figure \ref{graphic_sio2-sellmeier}.} 

    Neglecting damping ($\gamma=0$) the dielectric permittivity becomes infinite at the resonance frequency ($\omega=\omega_0$). In fact, damping cannot be omitted as it accounts for all dissipative processes (e.g. due to collisions between molecules) occur. Including these damping effects ($\gamma \neq 0$) the polarization becomes complex.

    The imaginary part $\Im(\varepsilon (\omega))$, as displayed in fig.\;\ref{graphic_dispersion}, is strongly peaked at a value of $\omega$ that is slightly smaller than the resonance frequency, indicating maximum excitation, while the characteristic of the real part $\Re(\varepsilon (\omega))$ is slightly more complicated. For low frequency $\omega \ll \omega_0$ the DC property of the medium prevails.

    Approaching the resonance frequency $\omega<\omega_0$ the $\Re(\varepsilon(\omega))$ increases and reaches a maximum at $\omega=\omega_0-\gamma/2$. This positive gradient behaviour is called normal dispersion (ND) and is the characteristic of most transparent substances in the visible wavelength regime. In this regime $\varepsilon_\text{red}<\varepsilon_\text{blue}$, therefore different spectral components exhibit different phase velocities $c(\omega)$ and separate as soon as they pass material interfaces that are not parallel. For further increasing frequency $\omega_0-\gamma/2 < \omega$ a negative gradient follows, a region of so called anomalous dispersion until the frequency approaches $\omega=\omega_0+\gamma/2$ and a second normal dispersion region begins.

    For frequencies $\omega \gg \omega_0$ (ac, i.e. high frequency limit) damping becomes negligible, $\varepsilon(\omega)$ becomes predominantly real and equation \eqref{eq_epsilon-dispersion} can be significantly simplified to:

    \begin{align}
        \varepsilon(\omega) & \approx 1- \frac{\omega_\text{p}^2}{\omega^2} , && \omega_\text{p}^2 = \frac{N e^2}{\varepsilon_0 m_\text{e}}, \label{eq_plasma-freq}
    \end{align} 

    \noindent where $\omega_\text{p}$ represents the plasma frequency of the material.
    \nomenclature[mop]{$\omega_\text{p}$}{Plasma frequency of a material.}\\

    For real materials the previous assumption of a Lorentz resonator at resonance frequency $\omega_0$ must be extended. It is a good approach to consider a model in which $N$ molecules, each with $Z$ electrons are situated in a volume element. Of these $Z$ electrons $f_j$ are distributed over $j$ different discrete oscillator frequencies $\omega_j$. Thus the oscillator strengths $f_j$ must fulfil the sum rule $\sum_j f_j = Z$. This represents a superposition of more than one of the previously discussed resonances. The obtained dispersion relation is:

    \begin{align}
        \varepsilon(\omega) = 1 + \frac{N}{\varepsilon_0} \frac{e^2}{m_\text{e}} \sum\limits_j \frac{f_j}{\omega_j^2-\omega^2-\i\omega\gamma}
    \end{align}
    \nomenclature[mnu]{$\nu$}{Frequency $[\nu]=\SI{}{s^{-1}}=\SI{}{Hz}$.}
    \nomenclature[mom]{$\omega$}{The angular frequency $\omega= 2 \pi \nu$ $[\omega]=\SI{}{s^{-1}}=\SI{}{Hz}$.}

    \subsubsection*{Dielectrics \label{ch_theory-sellmeier} }

    Especially for the dispersion of optically transparent substances, as explained, the resonances do not lie in the spectrum of the electromagnetic waves. \citet{cauchy} replaced the derived formulae were for this case by an expansion of the resonance behaviour into power series with respect to $\lambda= 2 \pi c_0/ \omega$. The resulting formula is called Cauchy's formula\footnote{A detailed derivation of Cauchy's and Sellmeier's formula, aside from the mentioned original publications, can be found in \citet{born-wolf}.}:

    \begin{equation}
        \sqrt{\varepsilon(\lambda)} = 1+ A + \frac{B}{\lambda^2}+ \frac{C}{\lambda^4} + \ldots \; ,
    \end{equation}

    where $A$, $B$, $C,\ldots$ are determined from the oscillator frequencies and $f_j$. Usually in this formula only $A$ and $B$ are taken into account. The elegance of this simple model lies in its mathematical simplicity, nevertheless, its weak point is that it only describes the dispersion relation for regions of normal dispersion.

    To circumvent this deficiency, \citet{sellmeier} proposed a description that builds upon this approach and includes anomalous dispersion. He expressed the dispersion of dielectrics phenomenologically in terms of experimentally determined coefficients $B_j$ for the resonance frequencies $\omega_j$:

    \begin{equation}
        \varepsilon(\omega) = \sum\limits_{j=1}^m{ \frac{B_j \omega_j^2}{\omega_j^2-\omega^2} } \label{eq_sellmeier}
    \end{equation}

    Usually the approximation is truncated with $m=4$, which is sufficient for a good modelling of the dispersion properties of e.g. silica in a wavelength range from the ultra-violet, throughout the visible wavelength to the infra-red range ($\lambda\lesssim \SI{2.3}{\micro m}$). A plot of the dispersion for real silica glass (amorphous $\text{SiO}_2$), calculated from the Sellmeier equation \ref{eq_sellmeier} is presented in fig.\;\ref{graphic_sio2-sellmeier}.

    \subsubsection*{Metals \label{ch_theory-drude} }

    For a large number of metals, most notably alkaline metals but also for example aluminium and silver, the kinetic energy of the electrons is significantly larger than the potential energy that is caused by the lattice atoms. Thus the Drude model \citep{drude} of the free electron gas is expected and experienced experimentally to be a very appropriate model\footnote{For a well explained and in depth study of the optical properties of metals it is referred to \citet[ch. 5, 12]{dressel}.}: Stationary atoms, surrounded by a quasi-free plasma of electrons.

    Taking this model into account, the total complex dielectric permittivity of metals can be expressed in terms of two separable constituents, the contribution of bound electrons $\varepsilon_\text{bound}$ and the contribution of the free electron gas $\varepsilon_\text{free}$ as follows:

    \begin{equation}
        \varepsilon (\omega) = \varepsilon_\text{bound} (\omega) + \varepsilon_\text{free} (\omega) \label{eq_metal-epsilon}
    \end{equation}

    The contribution from the free electron gas can be calculated analogous to the derivation of the dispersion of the bound electron oscillators. The equation of motion of free electrons is comparable to that for bound electrons (equation \ref{eq_dispersion-motion-diel}), solely without a harmonic potential:

        \begin{equation}
            -e \vc E(\vc r,t) = m_\text{e} (\ddot{\vc r} + \gamma \dot{\vc r})       \; , \label{eq_dispersion-motion-metal}
        \end{equation}

    \noindent where the damping factor has usually the order of $\gamma \approx \SI{1e14}{s^{-1}}$, which can be intuitively interpreted as a relaxation time of the free electron gas ($\tau = 1/ \gamma \approx \SI{1e-14}{s}$). For the homogeneous solution of this differential equation the following expression is simply obtained:

        \begin{equation}
            \vc r = \vc r_0 - \frac{1}{\gamma} v_0 \e^{-\gamma t} \; ,
        \end{equation}

    \noindent which is the description of an exponentially decelerated motion with the decay time $\tau$. An inhomogeneous solution for a harmonic wave $\vc E = \vc E_0 \e^{-\i \omega t}$ represents a periodic motion of the electron gas. For $N$ electrons this gives rise for a current density $\vc j = N e  \dot{ \vc r}$. Applying the constituent relation $\vc j = \sigma \vc E$ (see p. \pageref{text_constituent-j}) to $\vc j (\omega)$, also the conductivity $\sigma (\omega)$ can be obtained:

    \begin{eqnarray}
        \vc r (\omega) =& - &\frac{e}{m_\text{e}(\omega^2 + \i \gamma \omega )} \vc E \\
        \vc j (\omega) =&   &\frac{N e^2}{m_\text{e}(\gamma - \i \omega )} \vc E \\
        \sigma (\omega) =&  &\frac{N e^2}{m_\text{e} ( \gamma -\i \omega)}
    \end{eqnarray}

    For low frequencies $\omega \ll \gamma$, $\sigma$ can be approximated by its static value of $ \sigma_0 = N e^2 / (m_\text{e} \gamma) $ and becomes real. In this case the following equation for the relation of the imaginary part of the dielectric permittivity and the conductivity is valid:

     \begin{equation}
     \Im(\varepsilon(\omega)) = \sigma \varepsilon_0 / \omega
     \end{equation}

    For high frequencies $\omega \gg \gamma$, which is the case for optical frequencies, $\sigma$ becomes largely imaginary (compared to its real part) and the mentioned relation becomes invalid as the real and imaginary contribution to $\varepsilon$ are not separable any more.

    For low enough frequencies the major contribution to the total dielectric permittivity $\varepsilon$ in equation \eqref{eq_metal-epsilon} is caused by the free electron gas in metals. The contribution of the bound electrons can then be neglected. From the dispersion relation of $\sigma$ via equation \eqref{eq_metal-epsilon} and the relation $ \Im(\varepsilon(\omega)) = \sigma \varepsilon_0 / \omega $ the total dielectric permittivity of metals in the Drude model can be derived:

    \begin{align}
        \varepsilon_\text{metal} (\omega) = 1- \frac{\omega_\text{p}^2}{\omega^2-\i \omega \gamma } \label{eq_drude-dispersion} \; ,\\
        \omega_\text{p} = \frac{N e^2}{ \varepsilon_0 m_\text{e} } \; , \label{eq_plasma-freq-metal}
    \end{align}

    \noindent where, analogue to equation \eqref{eq_plasma-freq} for the high frequency limit of the Lorentz oscillator model, $ \omega_\text{p}$ is the plasma frequency of the free electron gas. For the frequency range below that plasma frequency $\omega \ll \omega_\text{p}$, metals retain their metallic, opaque and reflective character. For high enough frequencies however, the Drude free electron model's dielectric permittivity $\varepsilon_\text{free}$ is described sufficiently by \eqref{eq_plasma-freq}, applying the metal plasma frequency. In the very high frequency limit eventually $\lim_{\omega\rightarrow \infty} \varepsilon_\text{free}(\omega) = 1$.

    For many metals, also those under consideration in the regarded experiment, not only the response of the free electrons, described by the Drude model, has to be taken into account but also electron excitations between different bands, the so called `interband transitions'\footnote{analogue to the `interband transitions' for metals, for semiconductors `intraband transitions' occur which are not discussed within the scope of this work.}:

    \begin{equation}
        \varepsilon_\text{metal}(\omega) = \varepsilon (\omega) + \varepsilon_\text{inter}(\omega)
    \end{equation}

    Due to their specific electron configuration (the d-band, close to the Fermi surface is filled), this effect in the visible range arises especially for noble metals for frequencies $\omega > \omega_\text{p}$. For frequency ranges where these interband-transitions occur the Drude-model is not valid anymore.

    Values for the plasma frequencies of typical metals, derived from experimentally determined electron densities $N$ \footnote{A comparison to theoretical and electron energy loss (EEL) studies can be found in\citep{dressel}.} are shown in table \ref{tab_plasmafreq}. A theoretical treatment of the effect is besides the scope of this work\footnote{Concise derivations can be found in \citet{ashcroft}.}. Usually in these cases experimental data is applied (see ch. \ref{ch_sample-fabrication}).

    An analysis of the band structure of gold reveals two interband absorption edges in the visible regime, which explain the yellowish colour of the metal. In contrast, the band structure of silver does not allow interband absorption in the visible, causing the typical neutral colour, which together with the high reflectivity of silver is the reason for its widespread application as material for mirrors.

    \begin{table}[H]
    \centering
    \begin{tabular}{ccccc}\toprule
        Metal & Valency    &   N in $10^{28} \,\text{m}^{-3}$    &   $\omega_\text{p}$ in \SI{1e16}{Hz} &$\lambda_\text{p}$ in nm  \\\midrule
        Ag & 1    &   5.86    &   1.36 &    138\\
        Au & 1    &   5.90    &   1.37 &    138\\
        Al & 3    &   18.1    &   2.40 &    79 \\\bottomrule
    \end{tabular}
    \caption{Plasma frequencies of silver, gold and aluminum from \citep{fox}. The plasma frequencies $\omega_\text{p}$ are calculated with experimental data for the electron density $N$ \citep{wyckoff} from equation \ref{eq_plasma-freq-metal}.\label{tab_plasmafreq}}
    \end{table}

    \subsubsection*{Kramers-Kronig relation}

    The real and imaginary parts of the dielectric permittivity $\varepsilon(\omega)$ are in fact not independent. The relation between both is described by the Kramers-Kronig relation\footnote{The Kramers-Kronig-relation was first derived by H. A. Kramers (1927) \citep{kramers} and R. L. Kronig (1926) \citep{kronig} independently. An in depth explanation can be found in \citet{dressel}} which shall be briefly explained.

    From the frequency dependency of the dielectric permittivity $\varepsilon(\omega)$ the frequency, therefore time dependency enters the correlation of $\vc{E}$ to $\vc{D}$. By a Fourier decomposition of $\vc{D}(\vc{r},\omega)$ into its monochromatic components and a Fourier transformation into the time domain, the following relation can be derived \citep[pp.~381-388]{jackson}:

    \begin{align}
     \vc{D}(\vc{r},t) &= \varepsilon_0 \left( \vc{E}(\vc{r},t) + \int\limits_{0}^\infty \den\tau G(\tau) \vc{E}(\vc{r},t-\tau)\right) \label{eq_kk-d-e}\\
     G(\tau)          &= \frac{1}{2\pi} \int\limits_{0}^\infty \den\omega \left( \varepsilon(\omega)-1 \right) \e^{-\i \omega \tau} \label{eq_kk-g}
     \end{align}

    This equation describes a retarded dependency between $\vc{D}$ and $\vc{E}$ in space, $G$ being the Fourier transformed of $\eta = \varepsilon(\omega)-1$. The causality principle imposes that the response of a medium at a time $t$ at a spacial coordinate $\vc{r}$ can only result from a field that existed before $t$ at $\vc{r}$. With this premise the lower boundaries of the integrals in equations (\refeq{eq_kk-d-e},\label{eq_kk-d-g}) are chosen to be $0$ and not $-\infty$.

    A corollary of equation \eqref{eq_kk-d-e} is that the dielectric permittivity is an analytical function in the complex upper half plane ($\Im(\omega) \geq 0$), $\varepsilon(-\omega) = \varepsilon^\ast (\omega^\ast)$. By partial integration of equation \eqref{eq_kk-g} an asymptotic series is achieved:

    \begin{align}
        \varepsilon(\omega) - 1 &= \frac{\i G(0)}{\omega} - \frac{G'(0)}{\omega^2} + \ldots
    \end{align}

    Physically in this series the first summand must be $0$. Thus, for the high frequency $\omega$ limit $\Re(\varepsilon(\omega)/\varepsilon_0 - 1)$ decreases with the order of $\mathcal{O}(\omega^{-2})$ and $\Im(\varepsilon(\omega)/\varepsilon_0 - 1)$ decreases with the order of $\mathcal{O}(\omega^{-3})$.
    \nomenclature[mo]{$\mathcal{O}(x^n)$}{Landau symbol. A function $f \in \mathcal{O}(x^n)$ has a dependency on variable $x$ of order $n$. Following the definition precisely the Landau symbol $\mathcal{O}$ defines an asymptotic upper bound for the function $f$.}

    As the previous corollary is valid, the application Cauchy's integral theorem is allowed for relating the real and imaginary parts of $\varepsilon(\omega)$ along the real axis\footnote{
        Where $P$ represents the Cauchy principal part that is defined for a function $f(x)$ with a singularity at $x_0$ by:

        \begin{equation}
        \int\limits_{-\infty}^{\infty} P(f(x)) \den x = P \int\limits_{-\infty}^{\infty} f(x) \den x = \lim_{\delta \rightarrow +0} \left( \int\limits_{-\infty}^{x_0-\delta} \den x f(x) - \int\limits_{x_0+\delta}^{\infty} \den x f(x)   \right)
        \end{equation}
    }:


    \begin{alignat}{3}
        \Re ( \varepsilon(\omega)) &= 1+ && \pi^{-1} P \int\limits_{-\infty}^\infty \den \omega' \frac{\Im (\varepsilon(\omega'))}{\omega'-\omega}\\
        \Im ( \varepsilon(\omega)) &= -  && \pi^{-1} P \int\limits_{-\infty}^\infty \den \omega' \frac{\Re (\varepsilon(\omega'))-1}{\omega'-\omega}
    \end{alignat}

    Experimentally the Kramers-Kronig relation predicates that for the determination of the dispersion of the complex permittivity $\varepsilon(\omega)$ it is only necessary to measure the absorption of the medium, which is connected to the imaginary part ($\Im(\varepsilon(\omega))$). The real part ($\Re(\varepsilon(\omega))$), can then be calculated. However it is important to point out, that for that objective the measured data has to cover a broad spectral range and the sample must not be transparent.\\

    \subsection{The vector and scalar wave equations}

    \subsubsection*{The homogeneous Maxwell's equations}

    In most optical applications the Maxwell equations can be significantly simplified by the assumption of isotropy of a dielectric material or a metal. In both cases the free charge density $\rho=0$ \footnote{Also for a metal with no external field as well as for a metal with a time oscillating field $\rho=0$.} becomes zero. The electric current density in a metal $\vc{j} = \sigma \vc{E}$ depends on the electric conductivity $\sigma$ of the material and the electric field $\vc{E}$, in the case of a dielectric medium $\vc{j} = 0.$

    For the following applications additionally a description of homogeneous media is sufficient. In this case, the dielectric function $\varepsilon$ as well as the magnetic permeability $\mu$ are translationally invariant in space. Under these assumptions Maxwell's equations \eqref{eq-maxwell} can be written with divergence free $\vc{D}$ and $\vc{B}$ homogeneously:

    \begin{subequations}
    \begin{align}
    \nabla \cdot \vc{D} &= 0                                \label{eq-maxwell-divd-hom}\\
    \nabla \times \vc{H} - \varepsilon_0 \varepsilon \frac{\del}{\del t} \vc{E} &= 0  \label{eq-maxwell-roth-hom}\\
    \nabla \times \vc{E} + \mu_0 \mu  \frac{\del}{\del t} \vc{H} &= 0 \label{eq-maxwell-rote-hom}\\
    \nabla \cdot \vc{B} &= 0                                   \label{eq-maxwell-divb-hom}
    \end{align}\label{eq-maxwell-hom}
    \end{subequations}

    \subsubsection*{The vector wave equations}

    From equations \refeq{eq-maxwell-roth-hom} and \refeq{eq-maxwell-rote-hom} the vector wave equations, second-order linear\footnote{Therefore solutions can be linearly combined to obtain new solutions.} partial differential equations, for $\vc{E}$ respectively $\vc{H}$ are derived, which are also called Helmholtz equations\footnote{With the standard notation of the Laplace differential operator $\triangle \vc{A}=\nabla\cdot(\nabla \cdot \vc{A})$ and the dispersion relation \ref{eq_k-dispersionrel}.}:

    \begin{align}
        \nabla \times \left( \nabla \times \vc{E} \right) = \i \omega\mu \nabla \times \vc{H} &= \omega^2 \varepsilon \mu \vc{E} & \nabla \times \left( \nabla \times \vc{H}\right) = -\i \omega\varepsilon \nabla \times \vc{E} &= \omega^2 \varepsilon \mu \vc{H}\\
        \triangle \vc{E} + k^2 \vc{E} &= 0   &   \triangle \vc{H} + k^2 \vc{H} &= 0   \label{eq-vectorwaveequations}
    \end{align}

    \noindent where $k$ is the complex wave number, being the absolute value of the complex wave vector $k=|\vc{k}|$ of a propagating electromagnetic wave.

    Its imaginary part represents the absorption of the wave in the medium $\Im(k(\omega)) = \alpha/2$ where the intensity of the plane wave decreases as $I(l) = I_0 \e^{-\alpha l}$ (Beer-Lambert's absorption law).

    In transparent media ($\Im \left(\varepsilon(\omega) \right) \approx 0$) the wave number can be assumed to be real. From equations \eqref{eq-vectorwaveequations} it is possible to express the dependency of $k$ on the angular frequency $\omega$ by the dispersion relation:

    \begin{align}
    k (\omega) &= \frac{\omega}{c}n (\omega) \overset{\text{\eqref{eq_maxwell-relation}}}{=} \frac{\omega}{c}\sqrt{\mu(\omega) \epsilon(\omega)} \label{eq_k-dispersionrel}
    \end{align}
    \nomenclature[mk]{$\vc k (\omega)$}{Propagation vector of an electromagnetic wave. The absolute value is called propagation constant $k(\omega)$.}

    The phase velocity of light and its dispersion can then be expressed as:

    \begin{align}
        v = \frac{\omega}{k} = \frac{c_0}{\mu(\omega) \epsilon(\omega)} \overset{\text{\eqref{eq_maxwell-relation}}}{=} \frac{c_0}{n(\omega)} \label{eq_phase-velocity}
    \end{align}
    \nomenclature[mv]{$v$}{Velocity, particularly the phase velocity of an electromagnetic wave, $v = \frac{\omega}{k}$.}

    \subsubsection*{Plane waves \label{ch_plane-waves-sol}}

    The simplest, nontrivial solution of equation \eqref{eq-vectorwaveequations} describes propagating `plane waves'.

    From solving \eqref{eq-vectorwaveequations} as well as Maxwell's equations \eqref{eq-maxwell-hom} solutions for the electric and magnetic field are obtained. They contain a harmonic (sinusoidal) time dependency ($t$) and a dependency on the spatial coordinate $\vc r$:

    \begin{align}
        \vc E (\vc r,t) &= \vc E_0 \e^{\i( \vc k \vc r - \omega t ) } \\
        \vc H (\vc r,t) &= \vc H_0 \e^{\i( \vc k \vc r - \omega t )   }
    \end{align}

    From the divergence equations (\refeq{eq-maxwell-divd-hom}, \refeq{eq-maxwell-divb-hom}) it can be seen that $\vc E \cdot \vc k = 0$ and $\vc H \cdot \vc k = 0$. Therefore, plane waves, have an orthogonal trihedron of the field vectors and the wave vector $\vc k \perp \vc E \perp \vc H \perp \vc k$, which is the reason to call them also transverse waves.\footnote{Due to their importance and as it will be used for the numerical calculations in section \ref{ch_scattering-s}, in appendix \ref{ch_plane-waves} the Poynting vector $\vc S$ is explained that describes the energy transport and direction of electromagnetic waves, particularly for plane waves.}.


    Particularly for a complex wave vector $\vc k = \vc k_\text{re} + \i \vc k_\text{im}$ the imaginary part $\vc k_\text{im}$ determines the amplitude of the field vectors $\vc E_0 \e^{-\vc k_\text{im} \cdot \vc r}$ respectively $\vc H_0 \e^{-\vc k_\text{im} \vc r}$. The real part $\vc k_\text{re}$ determines the phase of the wave $\varphi = \vc k_\text{re} \cdot \vc r - \omega t$ at a certain spatial and temporal coordinate.

    Therefore $\vc k_\text{re}$ is perpendicular to surfaces of constant phase and $\vc k_\text{im}$ is per\-pen\-di\-cu\-lar to surfaces of constant amplitude of the wave. In vacuum, $\vc k_\text{re} \parallel \vc k_\text{im}$. The plane wave is then called homogeneous, otherwise inhomogeneous.

    The surfaces of constant phase ($\vc k_\text{im}=0$) propagate through the space with phase velocity $v$ \eqref{eq_phase-velocity}.

    \subsubsection*{The scalar wave equations}

    The problem of solving the vector wave equations generally, particularly for different coordinate systems, can be reduced to solving a scalar wave function, which is an easier task. Therefore (following \citep{bohren-huffman}) a vector function $\vc{M}$ \footnote{Those vector functions subsequently are called vector harmonics, harmonic solutions being the term for solutions of the Laplace equation.} is constructed from a scalar function $\psi$ (subsequently called the generating function), and a constant vector $\vc{c}$, called pilot vector, that can be written as\footnote{applying the identity $\nabla \cdot ( \nabla \times \vc{A}) = 0$}:

    \begin{align}
        \vc{M} &= \nabla \times (\vc{c} \psi)    &    \nabla \cdot \vc{M} &= 0
    \end{align}\vspace{-28pt}
    \begin{align}
        \triangle \vc{M} + k^2 \vc{M}  =  \nabla \times  \left( \vc{c} ( \triangle \psi + k^2 \psi ) \right) \label{eq-vector-harmonic-M}
    \end{align}

    $\vc{M}$ satisfies the vector wave equation if $\psi$ solves the scalar wave equation:

    \begin{equation}
        \triangle \psi + k^2 \psi = 0  \label{eq_scalar-wave-equation}
    \end{equation}

    The choice of $\psi$ depends on the coordinate system in which the problem is described as it has to satisfy the wave equation in that particular case. The vector $\vc c$ then points into a particularly exposed direction in the geometric symmetry of the coordinate system. For the simplest nontrivial case, for plane waves, the solution is represented by:

    \begin{align}
    \psi &=  \psi_0 \e^{\i k z - i \omega t} \; ,
    \end{align}

    \noindent where the wave number, the absolute value of the wave vector $k=|\vc{k}|$ is defined for an angular frequency $\omega$ by its dispersion relation in equation \eqref{eq_k-dispersionrel}.

    $\vc{M}$ is perpendicular to $\vc{c}$. A second vector harmonic $\vc{N}$ can be constructed, which also satisfies the vector wave equation, both vector harmonics fulfil the Maxwell requirements (eq.s \refeq{eq-maxwell-roth-hom}, \refeq{eq-maxwell-rote-hom}) for an electromagnetic field:

    \begin{align}
        \nabla \times \vc{N} &= k \vc{M}      &       \nabla \times \vc{M} = k \vc{N}  \label{eq-vector-harmonic-N}
    \end{align}

\section{Planar surface plasmon polaritons \label{ch_theory-planar-spp}}

    This work aims for an examination of exciting surface plasmon polaritons on cylindrical samples. For a theoretical study of the SPP dispersion relation on the surface of wires, first the dispersion relation in the planar case is derived which is then applied to cylindrical geometry.

    \subsection{Background}

    The nature of long-range Coulomb interaction in the free electron gas of metals allows the excitation of collective vibrations of the electron plasma with respect to the crystal lattice by incident electromagnetic waves. These oscillations are called plasmons. The frequency of such plasmons in the bulk volume is the plasma frequency of the metal according to the Drude model \citep{drude} $\omega_\text{p}$, as previously derived in \eqref{eq_plasma-freq-metal}. The direction of vibration of the electron gas for volume plasmons is longitudinal; the plasma frequency for metals usually lies in the range of $\omega_\text{p} \approx \SI{1e16}{Hz}$ (see table \ref{tab_plasmafreq}).

    In presence of a material interface, thus a boundary of the metal, the symmetry is broken and new modes arise. This effect was in principle already described mathematically in the early studies of Sommerfeld \citep{sommerfeld} and Zenneck \citep{zenneck}\footnote{These works were already referred to by \citet{debye-1909} in his theoretical work on scattering on cylinders that are mentioned in section \ref{ch_scattering}.}. First experimental observation of dips in the spectra of metal gratings, caused by those `Sommerfeld's surface waves' had already been reported as `anomalous intensity dips' without explanation by \citet{wood} in 1902. These theoretical and experimental works were not combined until \citet{fano} in 1941.

    Shortly after Fano's conclusion, \citet{ritchie} observed unexpectedly low loss in perpendicular incidence electron diffraction experiments on thin metal foils, followed by several extensive analogous experimental studies \citep{powell1,powell2}. The results could be explained completely by a theoretical derivation of \citet{stern}, who called the observed quantized excitations `surface plasmons' (SP)\footnote{Several later reviews and books resume the latest theoretical and experimental developments and the basic principles \citep{pitarke,zayats-04,raether,sambles,welford-1991,nisoli,maier-book,maier-2005,barnes-2003,liaw-wu} of surface plasmons (SP) and surface plasmon polaritons (SPP) in planar geometry, spheres and cylinders as well as different optical coupling, excitation and measurement methods.}.

    After a long time of extended experiments in the visible wavelength regime especially with different prism coupling setups \citep{otto-1968,ulrich} and extended experiments on the excitation of surface plasmons with electron guns, current research on planar surface plasmon polaritons focusses on the excitation via surface structuring. This approach is supported by the technical progress in microstructuring techniques. Particularly the field of surface plasmonics has attracted special interest after \citet{ebbesen} reported on extraordinarily low loss in sub-wavelength hole arrays in a metal substrate. Thus also the theoretical description of SPP on structured surfaces recently attracted particular interest \citep{zandi}.\\


    There is more than one possible type of surface polaritons. An electromagnetic wave that passes a medium induces a polarization (as described in sections \ref{ch_theory-susceptibility}, \ref{ch_theory-dispersion}) which couples back and therefore by a superposition of incident and emitted field modifies the total field. This coupled excitation is called polariton. Surface polaritons (SP) are those with the characteristic that the wave is bound to a material surface.

    The SP that this work's aim is to examine are excitations of the electron gas, thus called surface plasmon-polaritons (SPP). In general also coupling to phonons could occur, in that case leading to surface phonon-polaritons.
    \nomenclature[ssp]{SP}{Surface plasmon, surface bound electromagnetic excitation (non-retarded) of the electron plasma.}
    \nomenclature[sspp]{SPP}{Surface plasmon polariton, surface bound electromagnetic excitation (retarded) of the electron plasma.}\\


    \subsection{Dispersion of planar surface plasmon polaritons \label{ch_planar-spp-dispersion}}

    In the following, the the dispersion relation of surface plasmon polaritons (SPP) on a single, planar interface is derived. An interface is observed between two nonmagnetic media 1 and 2 with the wavelength dependent complex dielectric permittivities $\varepsilon_1 (\omega)$ and $\varepsilon_2 (\omega)$. The origin of the orthogonal coordinate system (as depicted in fig\;\ref{graphic_spp-dispersion}) is chosen that the coordinate $z$ is perpendicular to the material interface at $z=0$ ($z<0$: medium 1, $z>0$: medium 2).

    A harmonic plane wave in this system can be described in terms of an electric field vector $\vc E$ and a magnetic field vector $\vc H$. As both materials are taken as free of external charges $\rho=0$ and currents $\vc j =0$, $\vc E$ and $\vc H$ are defined as solutions of the homogeneous Maxwell equations \eqref{eq-maxwell-hom}.

    \begin{figure}[]
        \centering

            \psfrag{dielectric}[Bl][Bl][0.9][0]{dielectric}
            \psfrag{eps_1}[Bl][Bl][0.9][0]{$\varepsilon_1(\omega)$}
            \psfrag{metal}[Bl][Bl][0.9][0]{metal}
            \psfrag{eps_2}[Bl][Bl][0.9][0]{$\varepsilon_2(\omega)$}
            \psfrag{0}[c][c][0.9][0]{$0$}
            \psfrag{x}[c][c][0.9][0]{$x$}
            \psfrag{y}[c][c][0.9][0]{$y$}
            \psfrag{z}[c][c][0.9][0]{$z$}
            \psfrag{E}[c][c][0.9][0]{$|E_x|$}
            \psfrag{k}[c][c][0.9][0]{$\vc \beta$}

            \includegraphics[width=0.7\textwidth]{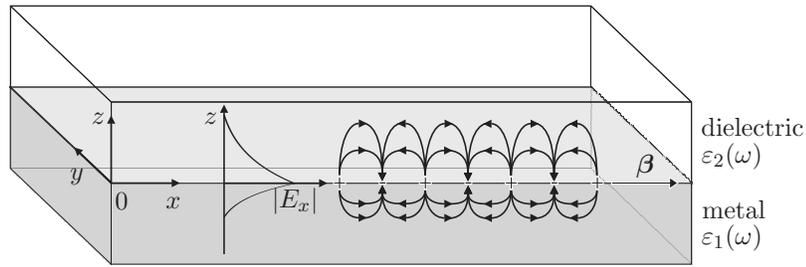}

        \caption[Surface plasmon polariton, propagating along a single, planar, metal-dielectric interface.]{Schematic of a surface plasmon polariton with propagation vector $\vc \beta$ on a single planar metal to dielectric interface. $z$ is the normal vector. The TM field with amplitude $|E_x|$ penetrates both adjacent media, decaying exponentially. Schematically the alternating SPP charge distribution with correlated fields is indicated. \label{graphic_spp-dispersion}}
\end{figure} 
    \begin{figure}[tb]
        \centering

            \psfrag{normalized frequency}[c][c][0.9][0]{normalized frequency $\omega/\omega_\text{p}$}
            \psfrag{normalized wave vector}[c][c][0.9][0]{normalized wave vector $\beta c_0 /\omega_\text{p}$}
            \psfrag{vacuum line}[c][c][0.9][55]{air light line $\omega=c_0 k$}
            \psfrag{silica line}[c][c][0.90][43.5]{silica light line $\omega=c_0 k \varepsilon_{\text{SiO}_2}^{-\nicefrac{1}{2}}$}
            \psfrag{o_sps}[Bl][Bl][0.9][0]{$\omega_{\text{SPP-SiO}_2}$ }
            \psfrag{o_spv}[Bl][Bl][0.9][0]{$\omega_\text{SPP-air}$ }
            \psfrag{o_sv}[Bl][Bl][0.9][0]{$\omega_\text{SP-air}$}
            \psfrag{o_ss}[Bl][Bl][0.90][0]{$\omega_{\text{SP-SiO}_2}$}
            \psfrag{retarded}[cl][cl][0.90][0]{retarded limit}
            \psfrag{nonretarded}[cr][cr][0.90][0]{nonretarded limit}

\psfrag{x01}[t][t][0.9]{0}%
\psfrag{x02}[t][t][0.9]{0.2}%
\psfrag{x03}[t][t][0.9]{0.4}%
\psfrag{x04}[t][t][0.9]{0.6}%
\psfrag{x05}[t][t][0.9]{0.8}%
\psfrag{x06}[t][t][0.9]{1}%
\psfrag{x07}[t][t][0.9]{0}%
\psfrag{x08}[t][t][0.9]{0.5}%
\psfrag{x09}[t][t][0.9]{1}%
\psfrag{x10}[t][t][0.9]{1.5}%
\psfrag{q}[t][t][0.9]{2}%
\psfrag{x12}[t][t][0.9]{2.5}%
\psfrag{x13}[t][t][0.9]{3}%
%
\psfrag{v01}[r][r][0.9]{0}%
\psfrag{v02}[r][r][0.9]{0.1}%
\psfrag{v03}[r][r][0.9]{0.2}%
\psfrag{v04}[r][r][0.9]{0.3}%
\psfrag{v05}[r][r][0.9]{0.4}%
\psfrag{v06}[r][r][0.9]{0.5}%
\psfrag{v07}[r][r][0.9]{0.6}%
\psfrag{v08}[r][r][0.9]{0.7}%
\psfrag{v09}[r][r][0.9]{0.8}%
\psfrag{v10}[r][r][0.9]{0.9}%
\psfrag{v11}[r][r][0.9]{1}%
\psfrag{v12}[r][r][0.9]{0}%
\psfrag{v13}[r][r][0.9]{0.5}%
\psfrag{v14}[r][r][0.9]{1}%
\psfrag{v15}[r][r][0.9]{1.4}%

            \includegraphics[width=0.8\textwidth]{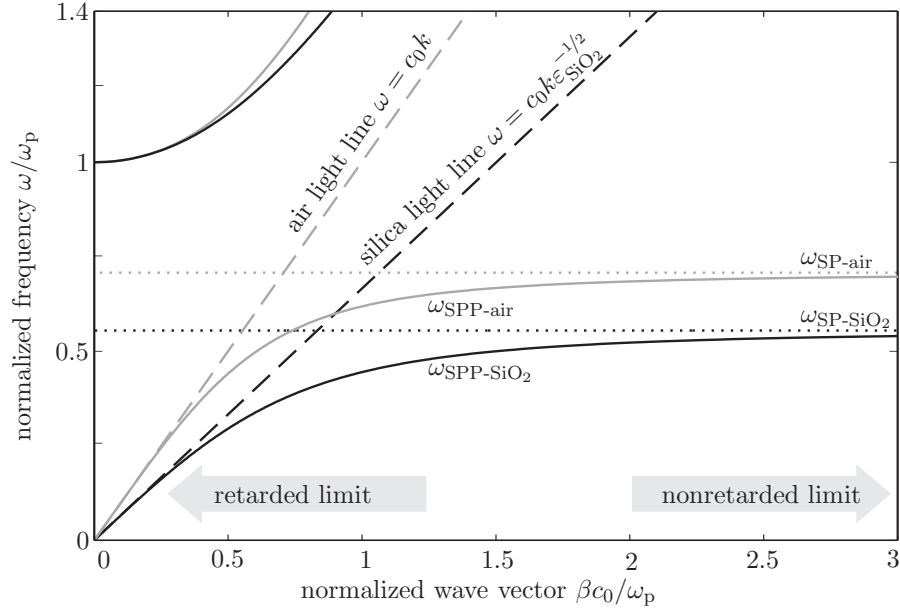}

        \caption[Dispersion curves for planar surface plasmon polaritons on a single metal interface to air and to silica.]{Dispersion curves for planar surface plasmon polaritons on a single metal dielectric interface, the metal dispersion being described by the Drude model \eqref{eq_planar-sp-drude}. The black, solid line below the light line indicates the real part of the SPP dispersion $\Re (\beta(\omega))$ for a metal to silica glass interface, whose high frequency asymptote $\omega_\text{SP}$ \eqref{eq_spp-disp-omega_s} is plotted as a dotted line and whose low frequency asymptote is the silica light line for free wave propagation in the dielectric. Analogue, in grey the dispersion and limits for a SPP on a metal to air ($\varepsilon_2=1$) interface is shown. For both cases the branches above the light lines represent radiative modes.} \label{graphic_spp-dispersion}
\end{figure} 

    Usually the solution vectors $\vc E$ and $\vc H$ are classified with respect to the surface direction into p-polarized light ($\vc E \parallel x-z-\text{plane}$) and s-polarized light ($\vc E \perp x-z-\text{plane}$)\footnote{
        p-polarized light generally is light, whose $\vc{E} \parallel \text{plane of incidence}$ (p=parallel), which
        is also meant by the term TM-polarized (transversal magnetic). For s-polarized light
        $\vc E \perp \text{plane of incidence}$, which is also known as TE-polarized light.
        The latter etymologically derives from the German word 'senkrecht' for perpendicular.
        The plane of incidence is always oriented perpendicular to the surface in which the $\vc k$ vector of the light lies.
    }. %
    For waves that travel along the interface, a component of $\vc E$ is necessary that is perpendicular to the $x-y-\text{plane}$. Therefore no s-polarized (TE) surface plasmon polaritons exist. Only those field vectors are regarded with the additional constraint to be bound to the surface, thus the fields must be evanescent for $z>0$ and $z<0$. The remaining solutions for the arbitrarily chosen propagation direction $x$, can be written for both media $i = 1,2$ as:

    \begin{align}
        \vc E_i &= \left( \begin{array}{c}E_{i,x} \\  0 \\ E_{i,z}\end{array} \right) \e^{-\kappa_i |z|} \e^{- \i (\beta x -\omega t)} \label{eq_spp-e-field} \; ,\\
        \vc H_i &= \left( \begin{array}{c}0  \; ,\\  H_{i,y} \\ 0 \end{array} \right) \e^{-\kappa_i |z|} \e^{- \i (\beta x -\omega t) } \label{eq_spp-h-field}
    \end{align}

    \noindent where $\beta$ is introduced as the propagation constant, which is the wave number $k$ of the surface plasmon, propagating along the surface. As $\beta$ is defined on the surface, describing the SPP on the surface, this wave number must be the same value in both media. From the Maxwell equations \eqref{eq-maxwell-hom} with \eqref{eq_spp-e-field},\eqref{eq_spp-h-field} the following relations are obtained:

    \begin{align}
        \i \kappa_1 H_{1,y} &\overset{\text{\eqref{eq-maxwell-roth-hom}}}{=} + \omega \varepsilon_0 \varepsilon_1 E_{1,x}  &
        \i \kappa_2 H_{2,y} &\overset{\text{\eqref{eq-maxwell-roth-hom}}}{=} - \omega \varepsilon_0 \varepsilon_2 E_{2,x} \label{eq_spp-kappa}
    \end{align}

    \noindent and the following relation for the propagation constant $\beta$ can be written from \eqref{eq_spp-e-field},\eqref{eq_spp-h-field}, interpreting $\kappa$ as a $z$ component of the propagation vector:

    \begin{align}
        \kappa_i = \sqrt{\beta^2 - \varepsilon_i \frac{\omega^2}{c^2} } \label{eq_spp-kappasqrt}
    \end{align}

    Until this point the fields in both media 1 and 2 have been treated separately. However on the interface at $z=0$
    the electromagnetic waves occur a discontinuity in the material properties. Due to the law of the conservation of energy\footnote{The proof results in a closed integral perpendicular to the material interface plane as shown e.g. in \citep[ch. 1.3]{stratton}.} it can be derived that on material boundaries the tangential ($x$ and $y$) components of $\vc{E}$ and $\vc{H}$ must be continuous. Applying this condition to equations \eqref{eq_spp-kappa}, the result is the following linear system of equations:

    \begin{align}
        H_{1,y} \left( \begin{array}{c} \frac{\kappa_1}{\varepsilon_1} \\ 1 \end{array} \right) + H_{2,y}
        \left( \begin{array}{c} \frac{\kappa_2}{\varepsilon_2} \\ -1 \end{array} \right) = 0 \label{eq_spp-boundarycondition} \;,
    \end{align}

    \noindent which is solved, if the determinant is zero which is the resulting retarded SPP condition:

    \begin{equation}
        \frac{\varepsilon_1}{\kappa_1} + \frac{\varepsilon_2}{\kappa_2} = 0 \label{eq_spp-condition}
    \end{equation}

    Firstly for a physical interpretation, $\kappa_1,\kappa_2$ are assumed to be real ($\in \mathbb{R}$). This equation is only fulfilled , as $\varepsilon_2(\omega)$, being the dielectric permittivity of a dielectric is positive and real, if $\varepsilon_1(\omega)$ of the other material is real and negative. This is true for metals in a frequency range where no volume electromagnetic waves can propagate.

    From equation \eqref{eq_spp-condition} the dispersion relation for the propagation constant of planar surface plasmon polaritons \citep{ritchie} is obtained:


    \begin{equation}
        k_{\text{SPP}}(\omega) =\beta(\omega) = \frac{\omega}{ \text{c}_0 } \sqrt{ \frac{\varepsilon_1(\omega)\varepsilon_2(\omega)}{\varepsilon_1(\omega) + \varepsilon_2(\omega) } } \label{eq-planar-sp}
    \end{equation}

    This dispersion relation can also be written with respect to the free space wave number $k_0 = n_0 \omega / c_0$ ($n_0=n_\text{vac}=1$), in terms of an effective refractive index $n_\text{eff}$ of the medium for SPP:

    \begin{align}
         \beta &= k_0 n_\text{eff} & n_\text{eff} &= \sqrt{ \frac{\varepsilon_1(\omega)\varepsilon_2(\omega)}{\varepsilon_1(\omega) + \varepsilon_2(\omega) } } \label{eq_spp-n_eff}
    \end{align}

    For the previously derived dispersion $\varepsilon_\text{metal}(\omega)$ for metals in the Drude model \eqref{eq_drude-dispersion}, the SPP dispersion relation \eqref{eq-planar-sp} can be rewritten in terms of the plasma frequency $\omega_\text{p}$ \eqref{eq_plasma-freq-metal} of the material:

    \begin{align}
        \beta(\omega) = \frac{\omega}{c_0} \sqrt{ \frac{\omega^2-\omega_\text{p}^2 }{2 \omega^2 - \omega_\text{p}^2 } } \label{eq_planar-sp-drude}
    \end{align}

    As depicted in fig\;\ref{graphic_spp-dispersion}, for a metal ($\varepsilon_1(\omega)$) to dielectric ($\varepsilon_2(\omega)$) interface, $\omega (\beta)$ from equation \eqref{eq-planar-sp}, in the limit of $\omega (\beta \approx 0)$ the SPP dispersion approaches the dielectric light line $c_0 k \varepsilon_2^{-\nicefrac{1}{2} }$, which describes the dispersion of an electromagnetic wave, freely propagating in medium 2. For higher wave numbers $\beta$, the SPP dispersion function increases monotonically but never crosses the light line.

    Therefore planar SPP can neither radiate light into the dielectric medium nor can they be excited by light, incident through the dielectric.

    This conclusion is at least true for perfectly plane interfaces. The excitation of SPP can nevertheless be achieved by either surface roughness or gratings or attenuated total reflection (ATR) \citep{otto-1968,kretschmann} from inside e.g. a prism whose internally reflecting surface is situated near enough to the surface of the metal that light with tunable $k$ component parallel to the interface, can be coupled into SPP.\\

    Although throughout recent literature the terms surface plasmon and surface plasmon polariton are used almost interchangeable, in its precise definition a difference shall be pointed out\footnote{This is emphasized and presented in detail by \citet{welford-1991}.}:

\begin{enumerate}
  \item For low wave numbers $\beta$ strong coupling between the electromagnetic wave and the polarization of the medium occurs, retardation plays an important role, which means mathematically that the finiteness of the speed of light $c_0$ is taken into account. The derived electrodynamical solution \eqref{eq-planar-sp} then describes a surface plasmon-polariton (SPP), that travels along the material interface.
  \item If the propagation constant $\beta$ is much larger than the corresponding plasma frequency of the metal (see eq. \eqref{eq_plasma-freq-metal}) $k\gg\omega_\text{p}/c$, the electromagnetic wave is sufficiently mismatched from the induced polarization that limited coupling occurs. In this non-retarded limit, the SPP condition \eqref{eq_spp-condition} reduces to $\varepsilon_1 + \varepsilon_2 = 0$. Non-propagating collective vibrations of the electron plasma near the metal interface occur. The phenomenon is then called a surface plasmon (SP).
\end{enumerate}
    Strictly speaking, the SP approximation is valid as long as the wave number lies in the range of $\omega_\text{s}/c \ll k \ll k_\text{F}$ where $k_\text{F}$ represents the absolute value of the Fermi wave vector \citep{pitarke}. In fig.\;\ref{graphic_spp-dispersion} it can be seen that in the nonretarded limit for high $k$ the SPP approaches the classical, nondispersive SP frequency that with the Drude model from equation \eqref{eq_planar-sp-drude} can be written, as \citet{ritchie} showed in his pioneering work as:

    \begin{align}
        \omega_\text{SP} = \omega_\text{p} / \sqrt{1+\varepsilon_2} \label{eq_spp-disp-omega_s}
    \end{align}

    The upper branches in fig.\;\ref{graphic_spp-dispersion} represent radiative modes for both material combinations, describing electromagnetic waves with frequencies larger than the plasma frequency of the metal $\omega > \omega_\text{p}$, propagating through the metal. It becomes transparent then. As it can be also seen from fig.\;\ref{graphic_spp-dispersion}, there exists a forbidden range of wave vectors $\beta$ which reaches from $\omega_\text{SSP}$ to the plasma frequency $\omega_\text{p}$.

    In the ansatz for the derivation of the dispersion relation for SPP (\refeq{eq_spp-e-field},\refeq{eq_spp-h-field}) the value $\kappa$ was introduced to insert the assumption of exponentially decaying field strengths with $|z|$ in both media. By applying the dispersion relation \eqref{eq-planar-sp} to equation \eqref{eq_spp-kappasqrt}, this parameter can be expressed as:

    \begin{align}
        \kappa_i = \frac{\omega}{c_0} \sqrt{ \frac{-\varepsilon_i^2}{\varepsilon_1+\varepsilon_2} }
    \end{align}

    This relation allows to determine the so called skin-depth $\delta_i = 1 / \kappa_i$ which describes the $\e^{-1}$ depth of penetration of the electromagnetic fields into the adjacent media (see fig.\;\ref{graphic_spp-dispersion}). It is simple to show that the field penetrates deeper into the dielectric than into the metal by applying the dielectric permittivities, thus $\delta_1 < \delta_2$. For this reason for one metal and frequency, the field penetrates deeper into air than into silica glass.

    For a comprehensive description of the propagation of SPP on real metal to dielectric surfaces, another effect cannot be omitted. As the dielectric permittivity of real metals has a complex part $\Im(\varepsilon) \neq 0$, the propagating SPP undergoes attenuation. Regarding equation \eqref{eq_spp-condition} this enters the description in terms of an imaginary part of the wave vector. The  $\e^{-1}$ decay length $l_\text{d}$ of the SPP is then:

    \begin{align}
        l_\text{d}^{-1} = 2 \Im \left(\beta(\omega) \right) \overset{\text{\eqref{eq-planar-sp}}}{=} \frac{\omega}{c_0} \sqrt{ \frac{ \varepsilon_1^3(\omega) \varepsilon_2^2(\omega) }{ |\varepsilon_1(\omega)| \left(|\varepsilon_1(\omega)|-\varepsilon_1(\omega) \right)^3 } }
    \end{align}

\section{Surface plasmon polaritons on a wire}

\subsection{Background and validity}

    \citet{schmidt-prb-arrays-08} have recently presented and discussed an approach to convert the dispersion of SPP from the planar case to cylindrical geometry. This model was first introduced by Miziumski \citep{miziumski-theory,miziumski-proceeding} (based upon \citet{pfeiffer-theory}) without verification of its validity for very high mode orders ($m>40$). Also \citet[pp.\;524-537]{slater} already proposed the approach.

    \citet{liaw-wu} have studied in depth SPP on bent metal dielectric interfaces and numerically demonstrated that the dispersion of SPP modes on cylinder like structures generally for high enough radii converges to the solution of planar SPP, which in principle confirms the validity of the approach. It shall be mentioned that the dispersion relation of SPP on lossless circular cylinders is the subject of discussion of several publications \citep{novotny,khosravi}. Also the case of metal coated dielectric cylinders is the subject of discussion of several recent publications \citep{schroeter-dereux}.

    In addition to the planar SPP solutions, in other geometries e.g. on very small particles, so called localized surface plasmons (LSP) are well known and experimentally proved \citep[p.142]{zayats-04}. They are solutions of the wave equation for the non-retarded SP approximation for appropriate boundary conditions and have resonance frequencies that are therefore determined by the size and shape of the particles and the dielectric permittivity of the particle material. These non-retarded solutions are valid as long as the size of the particle is small compared to the wavelength of the electromagnetic wave ($D \ll \lambda$).

    Particularly on the surface of cylinders with very small diameters $D \ll \lambda$ and very high length to diameter aspect ratios $l \gg \lambda$ SP excitations arise that share properties of planar SPP and LSP. $D \ll \lambda $ allows non-retarded SP to be excited, $l \gg \lambda$ gives rise to retarded modes \citep{zayats-04}. The result are discrete SP modes on the surface that approach the solution of planar SPP (eq. \refeq{eq-planar-sp}) for high orders. The solutions have both, radiative and non-radiative branches of modes for different frequencies, except for the lowest mode which is purely nonradiative.

    Thus in the examined case of relatively large radii (compared to $\lambda$), the assumption to apply the proposed model seems valid, at least for frequencies below the plasma frequency $\omega_\text{p}$ of the metal, as \citet{novotny} reasons, a case that is separately discussed in detail by \citet{martinos-economou}.

    However, in this work the derivation of this model for the dispersion and the incident angles of excitation for the SPP modes on the surface of a metal cylinder is shown. The result is compared to an exact solution of Maxwell's equations for the mode dispersion in sec.\;\ref{ch_theoresults-exact}.

\subsection{Model for the dispersion relation \label{chapter_theory-model}}

    For a model of the propagation of surface plasmons on a cylinder, the cylinder is treated as a closed two dimensional surface on which the SPP propagates as described in section \ref{ch_planar-spp-dispersion}.\\

\begin{figure}
        \centering

            \psfrag{D}[c][c][1.0][0]{$D$}
            \psfrag{k}[l][l][1.0][0]{$\vc{k_\text{SP} }$}
            \psfrag{b}[l][l][1.0][0]{$\vc{\beta}$}
            \psfrag{Kp}[l][l][1.0][0]{$\vc{k_\varphi }$}
            \psfrag{k_l}[l][l][1.0][0]{$\vc{k_\text{inc}}$}
            \psfrag{a}[l][l][1.0][0]{\textcolor{white}{$\alpha$} }
            \psfrag{k_lbeta}[l][l][1.0][0]{$\vc{\beta_{\text{inc}}}$}
            \psfrag{Incident plane waves}[l][l][1.0][0]{Incident plane waves}
            \psfrag{Propagating SP}[l][l][1.0][0]{Propagating SPP}
            \includegraphics[width=\textwidth]{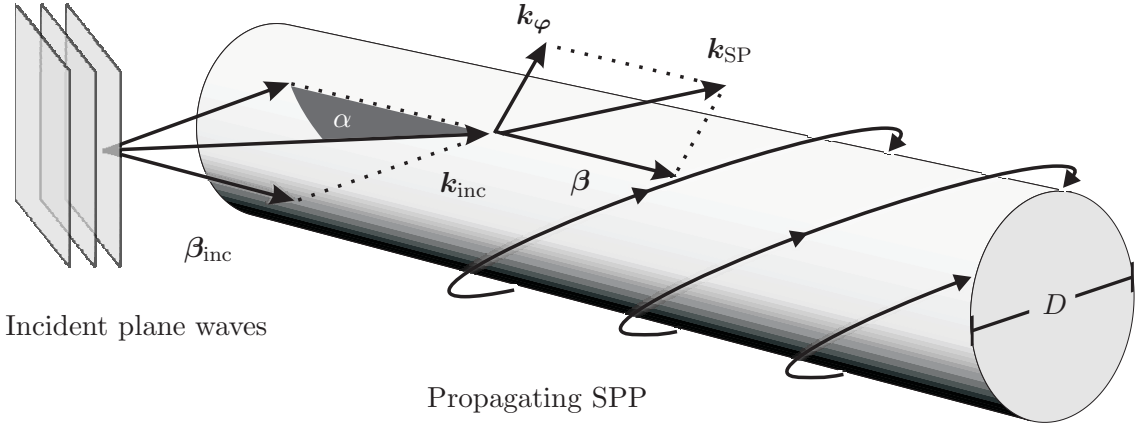}

        \caption[Schematic of spiralling surface plasmons on a wire referring to the derivation of a model for the resonant modes in section \ref{chapter_theory-model}.]{Schematic of plane waves with the wave vector $\vc{k_\text{inc}}$ and the angle of incidence $\alpha$ hitting a metal cylinder with a diameter $D$. Resonances occur at angles where $\vc{\beta_{\text{inc}}}$ fits the parallel propagation vector $\vc{\beta}$ which is the axial component of the wave vector of the surface plasmon polariton $\vc{k_\text{SPP} }$. The perpendicular component $\vc{k_\varphi }$ lies in a tangential plane to the surface of the cylinder on which the surface plasmons form discrete spiralling modes.\label{graphic_model-drawing}}


\end{figure}

    Following \citet{schmidt-prb-arrays-08}, the wave vector $\vc{k_{\text{SP}} }$ of a surface plasmon, propagating on the surface of a cylinder can be geometrically decomposed (see fig.\;\ref{graphic_model-drawing}) into a tangential component $\vc k_{\varphi}$ and a longitudinal component $\vcg{\beta}$. $\beta = |\vc \beta|$
    respectively is the propagation constant of the surface plasmon along the axis of the cylinder:

    \nomenclature[mb]{$\beta$}{Propagation constant of surface plasmons. $\beta_m$ represents discrete propagation constants, $\vc{\beta}$ represents in this geometry the propagation vector.}

    \begin{align}
        | \vc{k_{\text{SPP}}} |^2 &= | \vc{k_{\varphi}} |^2 + | \vcg{\beta} |^2 \\
         \beta               &= \sqrt{  k_{\text{SPP}} ^2 -  k_{\varphi} ^2 }  \label{model_beta}
    \end{align}

    The surface of the cylinder is approximated by a plane tangential to the cylinder surface in each point $\vc{r}$. Thus the dispersion relation of a planar SPP along an interface of material 1 inside the cylinder ($\varepsilon_i=\varepsilon_1$) and the surrounding material 2 ($\varepsilon_o=\varepsilon_2$) from eq.~\eqref{eq-planar-sp} is applied to eq.~\eqref{model_beta}:

    \begin{align}
    \beta(\omega)  &\overset{\text{\eqref{eq-planar-sp}}}{=} \sqrt{ k_0^2   \frac{\varepsilon_i(\omega)\varepsilon_o(\omega)}{\varepsilon_i(\omega) + \varepsilon_o(\omega)}   - k_{\varphi}^2  }
    \end{align}

    The model that is derived bases on the assumption that only integral numbers of field nodes can form. This means that $2 \pi R = m \lambda$ where $R=D/2$ is the radius of the cylinder and $m$ is the mode order, which can be shortened to $k=m/R$. An additional geometrical phase change of $\pi$ per \ang{180} around the perimeter of the wire is added, which results from comparison to an exact derivation of the mode dispersion, thus shifting $m$ to $(m-1)$:

    \nomenclature[ml]{$\lambda$}{Wavelength in m, usually indicated as $[\lambda]=\SI{1}{nm}$. The relation between wavelength and frequency is $\lambda=c/\nu$.}


    \begin{align}
        k_\varphi &= \frac{(m-1)}{R} \\
    \beta_m(\omega)  &= k_0 \sqrt{ \frac{\varepsilon_i(\omega)\varepsilon_o(\omega)}{\varepsilon_i(\omega) + \varepsilon_o(\omega)}  - \left( \frac{c_0}{ \omega } \right)^2 \left( \frac{(m-1)}{R} \right)^2 } \label{model_beta_m}
    \end{align}

    Application of the dispersion relation from equation \eqref{eq_k-dispersionrel}  for a wave vector to the propagation constant $ \beta = \omega n / c_0$ for each mode $\beta_m$ to equation \eqref{model_beta_m} constitutes the following relation for the effective refractive index $n_m$ of the SPP of mode $m$, as it is defined in eq.~\eqref{eq_spp-n_eff}:

    \begin{align}
        n_m(\omega) &= \sqrt{ \frac{\varepsilon_i(\omega)\varepsilon_o(\omega)}{\varepsilon_i(\omega) + \varepsilon_o(\omega)}  - \left( \frac{(m-1)\lambda}{D \pi} \right)^2 } &
                    &= \sqrt{ \frac{\varepsilon_i(\omega)\varepsilon_o(\omega)}{\varepsilon_i(\omega) + \varepsilon_o(\omega)}  - \left( \frac{c_0}{\omega} \right)^2 \left( \frac{2(m-1)}{D } \right)^2 } \label{eq_model-n-eff}
    \end{align}

    For reasons of convenience the radius $R$ is replaced by the diameter of the cylinder $D$. To examine the surface plasmons experimentally by side scattering it is necessary to derive the incident angles for a matching of the wave vectors of incident light $\vc{k_\text{inc}}$ in the surrounding material 2 to the $\vc{k_{\text{SPP}}}$ vectors of the surface plasmon polariton modes. The light occurs with an incident angle $\alpha$ with respect to the cylinder axis. Therefore $\vc{k}$ is decomposed (as shown in fig.\;\ref{graphic_model-drawing}):
    \nomenclature[ma]{$\alpha$}{ Incident angle of light with respect to cylinder respectively wire axis.}

    \begin{align}
        \cos(\alpha_m(\omega)) &= \frac{\beta_m(\omega)}{k_\text{inc}(\omega)}  = \frac{n_m(\omega)}{|\sqrt{\varepsilon_o(\omega)}|} \\
        \alpha_m(\omega)       &= \arccos\left(  \frac{1}{|\sqrt{\varepsilon_o(\omega)}|}\sqrt{  \frac{\varepsilon_i(\omega)\varepsilon_o(\omega)}{\varepsilon_i(\omega) + \varepsilon_o(\omega) }   - \left( \frac{c}{\omega} \right)^2 \left( \frac{2(m-1)}{D} \right)^2 }  \right) \label{eq-model-plasmon-incangles}
    \end{align}

    It is important to mention the translation of the polarization condition for the excitation of SPP on a cylinder.

    Making the step from the planar case to the model of cylinder-SPPs, ``rolling up'' the planar surface, in principle the basic condition does not change as the model's assumptions are not relaxed but tightened by additional boundary conditions.

\begin{figure}[hp]
        \centering

            \psfrag{x}[c][c][0.9][0]{$x$}
            \psfrag{y}[l][l][0.9][0]{$y$}
            \psfrag{z}[l][l][0.9][0]{$z$}
            \psfrag{phi}[l][l][0.9][0]{$\varphi$}
            \psfrag{a}[l][l][0.9][0]{$\alpha$}
            \psfrag{e_i}[l][l][0.9][0]{$\varepsilon_\text{i}$}
            \psfrag{e_o}[l][l][0.9][0]{$\varepsilon_\text{o}$}
            \psfrag{ei}[l][l][0.9][0]{$\hat{\vc{e_\text{inc}}}$}
            \psfrag{es}[l][l][0.9][0]{$\hat{\vc{e_\text{sca}}}$}
            \psfrag{R}[l][l][0.9][0]{$R$}
            \psfrag{E}[l][l][0.9][0]{$\vc{E}$}
            \psfrag{H}[l][l][0.9][0]{$\vc{H}$}
            \psfrag{case p}[l][l][0.9][0]{case s}
            \psfrag{case s}[l][l][0.9][0]{case p}
            \psfrag{incident plane wave}[c][c][0.9][0]{incident plane wave}
            \psfrag{(E_i,H_i)}[c][c][0.9][0]{$(\vc{E_\text{inc}},\vc{H_\text{inc}})$}
            \psfrag{internal field}[c][c][0.9][0]{internal field}
            \psfrag{(E_1,H_1)}[c][c][0.9][0]{$(\vc{E_\text{i}},\vc{H_\text{i}})$}
            \psfrag{scattered field}[c][c][0.9][0]{scattered field}
            \psfrag{(E_s,H_s)}[c][c][0.9][0]{$(\vc{E_\text{sca}},\vc{H_\text{sca}})$}
            \includegraphics[width=\textwidth]{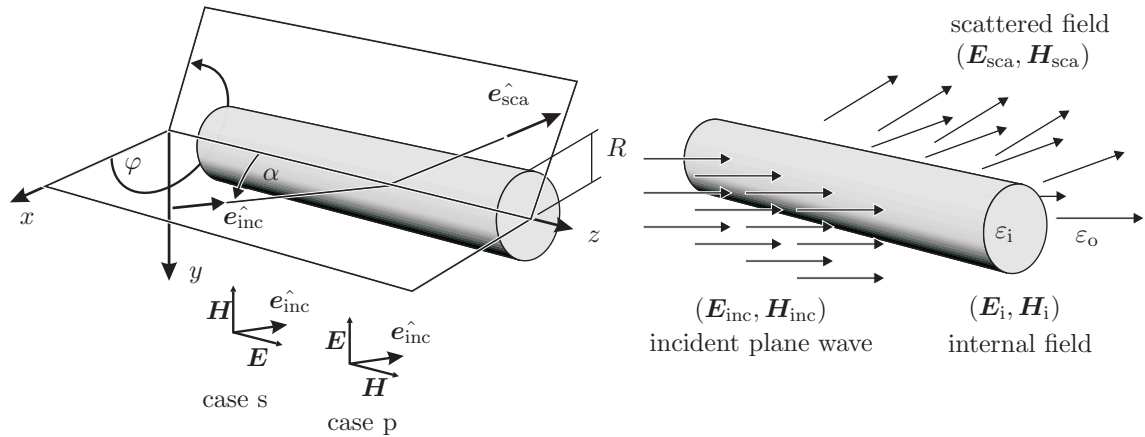}

        \caption[Cylinder and cylindrical coordinate system with annotations as used for the derivation of the scattering theory in section \ref{chapter_theory-scattering}]{Schematic drawing of the coordinate system for the derivation of the scattering theory. The incident field is described by $(\vc{E_\text{inc}},\vc{H_\text{inc}})$, the internal field in the cylinder by $(\vc{E_\text{i}},\vc{H_\text{i}})$, the scattered field by $(\vc{E_\text{sca}},\vc{H_\text{sca}})$. The total fields outside the cylinder are a superposition of the first and the third.\label{graphic_scattering-drawing}}


\end{figure}

    Consider a plane wave that approaches with perpendicular incidence ($\alpha = \ang{90}$) that hits a point of the wire, never exactly in the xz-plane (fig.\;\ref{graphic_scattering-drawing}) but at least slightly higher or lower ($y\neq0$).

    The model bases on the assumption that a local approximation of the cylinder surface as planar is valid. Thus, by definition a plane wave polarized perpendicular to the xz-plane for such a tangential surface is p-polarized (TM). Respectively, if it is polarized parallel to the xz-plane, for any tangential plane it is s-polarized (TE).

    In section \ref{ch_planar-spp-dispersion} it is explained, that SPP on a planar metal-dielectric interface can only be TM-modes (p-polarized). These can only be excited by incident light with a matching field component in p-polarization.

    This condition for SPP excitation is therefore also valid for the cylinder geometry; p-polarized ($\perp$ to the xz-plane) incident waves can excite the described spiralling SPP modes, s-polarized incident waves ($\parallel$ to the xz-plane) cannot.

    As soon as the assumption of $\alpha = \ang{90}$ abandoned, the principle does not change. Only in the case of polarization $\parallel$ to the xz-plane, the incident electromagnetic field vector $\vc E_\text{inc}$ can be decomposed into a radial field component and one that remains s-polarized.

\newpage

\section{Scattering of light by a small cylinder\label{chapter_theory-scattering}}

The process used to probe the surface plasmons on metallic wires is that of scattering of a plane, monochromatic, electromagnetic wave with predefined linear polarization. This process can be well described in the formalism of scattering theory.\\

\subsection{Background}

An exact solution of the Maxwell equations for scattering of plane waves on spheres of the size of the wavelength was first derived by Gustav Mie and Ludvig Lorenz in 1908, eponyms of this solution method, which is also called ``Lorenz-Mie scattering''. Mie particularly obtained a theoretical explanation for the colour effects that were observed on light, scattering on wavelength sized spherical particles of gold in a colloidal solution \citep{mie-1908}. The modern formulation of this theory in terms of Bessel functions was first published by \citet{debye-1909} along with a consistent interpretation.

Unlike the scattering solution for the Maxwell equations that was derived by Lord Rayleigh (NP 1904) \citep{rayleigh-theory} being an approximation for much smaller particle diameters than the wavelength, the Mie solution is generally also valid for scattering obstacles whose dimensions are larger than $\lambda$. The two limits of the Mie description are the Rayleigh theory for particle diameters of $\varnothing \ll \lambda$ and classical beam optics for $\varnothing \gg \lambda$.

The Mie ansatz for the solution of the spherical problem can be transferred to particles of other shape. Of special interest in this work are amongst those cylinders, having one spatial degree of symmetry less than spheres, which means a slightly more complicated mathematical description.

In the case of an incident plane wave, travelling perpendicular to the cylinder axis (see section \ref{ch_theory-scattering-perpendicular}), the problem reduces again and the solution is quite comparable to the Mie theory for spheres. The first complete solution of this case for the assumption of a dielectric cylinder was already derived by \citet{rayleigh-cylinder}\footnote{Which is well presented in \citep{stratton}. A complete historical review about the founders of the modern scattering theory can be found in \citep{kerker}.}, a comparable treatment was presented even earlier by \citet{hondros-debye}.

This theoretical description not only was cross checked by a large number of experimental and theoretical \citep{sun,rother} applications but it was also experimentally proved explicitly e.g. by \citet{abushagur}. Recently \citet{martinos-economou} have theoretically demonstrated the optical absorption spectrum, applying for this special case of perpendicular light incidence on small wires for s- and p-polarization.

\citet{she-2008} have extended the description of normal incidence to small ($\varnothing \ll \lambda$) cylinders that are coated. They assumed the cylinder consisting of dielectric material and the coating consisting of a metal (or a metamaterial) and vice versa, neglecting dissipative effects. The theory is validated by a comparison to an analytic description accounting for retardation of the surface plasmon modes, derived by \citet{schroeter-dereux} for dielectric cylinders ($\varnothing = \SI{1}{\micro m}$) that are surface coated by a thin layer of metal.\\

The first analytical solution of scattering on a dielectric cylinder for oblique incidence following the Mie ansatz was in some abstract formulation presented by \citet{wait} who also compared this plain wave solution to incident spherical waves \citep{wait-spherical}. \citet{kerker-paper,kerker-negative} subsequently treated the problem more extensively. In the case of oblique incidence the scattering problem cannot be simplified significantly, thus the theoretical description becomes more extensive.\\

\label{text_scattering-constraints}\citet{kuik} lately have shown by applying the T-matrix method\footnote{The T-matrix method (T: transmission matrix) is an alternative numerical method, also called extended boundary condition method', that allows for exact scattering computations for particles with arbitrary shapes and optical properties. For details about this method it is referred to \citep{waterman-t}.} that the scattering solution for cylinders of finite longitudinal extension for perpendicular and oblique incidence does not converge to the exact solution for infinite cylinders as the aspect ratio (ratio of length over diameter) increases, an examination that was with exact solutions also done by \citet{wang-hulst}. \citet{lock} has extended the Mie type solutions for incident plane waves on infinitely long cylinders to non plane wave real beams.

Nevertheless the derivation for infinitely long cylinders and plane waves that is presented is assumed to show good agreement with the conditions of the experiment, as the extension of the experimental beam, that can be well treated as a plane wave, is large compared to the lateral extension of the wires but small compared to the longitudinal extension of the wires. Therefore particularly effects of scattering on the wire's edges that were considered in the mentioned publication can be omitted in the examined case.\\

The following analytical derivation of the scattering amplitudes and field distributions from scattering by an infinite right circular cylinder follow the derivations of the theory of scattering on particles of different shapes by \citet{bohren-huffman} and the earlier treatise of \citet{hulst}. Both concentrate mainly on the more common case of scattering on spheres, thus their in depth discussion of the spherical geometry is in the following transferred to the cylindrical geometry.

The results of a subsequent computation (presented in sec.\;\ref{chapter_theory-scatterincoeff}) of the theoretical scattering amplitudes should be substantially comparable to the experimental results of this work as well to the scattering by metallic nanowires embedded in a dielectric matrix as to the scattering on coated cylinders. The latter with the assumption of a high ratio of the coating thickness to the skin depth so that it can be treated as bulk material for the plasmonic effects.

\subsection{Cylindrical coordinate system}

The scattering problem is solved in the intrinsic coordinate system of a cylinder,  $\vc{r}=(r,\varphi,z)$, the system's intrinsic symmetry is exploited, thus the complexity of the problem reduces and an analytical solution becomes possible. As drawn in fig.\;\ref{graphic_scattering-drawing} the cylinder axis defines the coordinate axis $z$, the extension of the cylinder is defined by its radius $R$ and its length is taken as infinite.

\nomenclature[mr]{$\vc{r}$}{Spacial coordinate vector, usually three dimensional. May be defined different in different coordinate systems, $(x,y,z)$ in cartesian coordinates, $(r,\varphi,z)$ in spherical cylindrical coordinates.}

The incident light is assumed to be a plane wave whose $\vc k_\text{inc} \| \bvc e_\text{inc}$ vector lies, in cartesian coordinates, in the $x$-$y$ plane, enclosing an incident angle of $\alpha$ \footnote{
    The definition of the incident angle $\alpha$ follows the convention of the experimental chapter \ref{ch_experimental-setup}. Note that this is in accordance with the convention of \citet{bohren-huffman} and most later publications referring to this. Different to that \citet{hulst} along with nearly all publications earlier than \citet{bohren-huffman} defines $-\pi/2<\alpha<+\pi/2$.}%
with $z$. The observed portion of the scattered field's propagation direction ($\vc k_\text{sca} \| \bvc e_\text{sca}$) lies in a plane with the $z$ axis that encloses the azimuthal angle $\varphi$ with the incident $x$-$y$ plane.\\

\subsection{Field expansion in vector cylindrical harmonics}

The starting point of this derivation is the scalar wave equation \eqref{eq_scalar-wave-equation} that has to be written in cylindrical coordinates:

\begin{align}
    \frac{1}{r}\frac{\partial}{\partial r}\left(r \frac{\partial \psi}{\partial r}\right)
   + \frac{1}{r^2}\frac{\partial^2 \psi}{\partial \varphi^2}
   + \frac{\partial^2 \psi}{\partial z^2} + k^2\psi = 0 \label{eq_scalar-wave-equation-cyli}
\end{align}

In these coordinates the generating function $\psi$ is calculated using a separation ansatz to be:

\begin{align}
\psi_n(r,\varphi,z) &= Z_n(\rho)\e^{\i n \varphi}\e^{\i hz}& \rho&=r\sqrt{k^2-h^2} \quad (n=0,\pm1,\ldots) \label{eq-cylindrical-psi}
\end{align}

Where $h$ and $n$ are separation constants, $h$ is chosen to fulfil the boundary conditions. In this case for a plane wave with an incident angle $\alpha$, $h=-k_o \cos{\alpha}$. $Z_n(\rho)$ are Bessel functions of integral order $n$. Their arguments are scaled, dimensionless radii $\rho$.\\

In the case of cylindrical coordinates $\psi$ generates the cylindrical vector harmonics $\vc{M}$ and $\vc{N}$ as defined in equations \eqref{eq-vector-harmonic-M} respectively \eqref{eq-vector-harmonic-N} to be\footnote{If indicated as in this equation, vectors are referring to the cylindrical coordinates, their entries corresponding to $r,\varphi,z$.}:

\begin{align}
\vc M_n &= \sqrt{k^2-h^2}\e^{\i(n\varphi+hz)}\left(
  \begin{array}{c}
    \i n\frac{1}{\rho}Z_n(\rho)
    \\ -Z_n'(\rho)
    \\ 0
  \end{array}\right)_{\bvc{r},\bvc{\varphi},\bvc{z}}
\\
\vc N_n &= \sqrt{k^2-h^2}\frac{1}{k}\e^{\i(n\varphi+hz)}\left(
  \begin{array}{c}
    \i hZ_n'(\rho)
    \\ -hn\frac{1}{\rho}Z_n(\rho)
    \\ \sqrt{k^2-h^2}Z_n(\rho)
  \end{array}\right)_{\bvc{r},\bvc{\varphi},\bvc{z}}
\end{align}

These vector cylindrical harmonics are orthogonal, forming a basis. $M_n, N_n$ are not normalized pair wise but over the summation of all orders $n$.\\

\begin{figure}[tb]
        \centering

            \psfrag{M_0(x,z)}[l][l][0.9][0]{\textcolor{white}{$\vc{M_0(x,z)}$}}
            \psfrag{M_0(x,y)}[l][l][0.9][0]{\textcolor{white}{$\vc{M_0(x,y)}$}}
            \psfrag{N_0(x,z)}[l][l][0.9][0]{\textcolor{white}{$\vc{N_0(x,z)}$}}
            \psfrag{N_0(x,y)}[l][l][0.9][0]{\textcolor{white}{$\vc{N_0(x,y)}$}}
            \psfrag{M_1(x,z)}[l][l][0.9][0]{\textcolor{white}{$\vc{M_1(x,z)}$}}
            \psfrag{M_1(x,y)}[l][l][0.9][0]{\textcolor{white}{$\vc{M_1(x,y)}$}}
            \psfrag{N_1(x,z)}[l][l][0.9][0]{\textcolor{white}{$\vc{N_1(x,z)}$}}
            \psfrag{N_1(x,y)}[l][l][0.9][0]{\textcolor{white}{$\vc{N_1(x,y)}$}}
            \psfrag{M_2(x,z)}[l][l][0.9][0]{\textcolor{white}{$\vc{M_2(x,z)}$}}
            \psfrag{M_2(x,y)}[l][l][0.9][0]{\textcolor{white}{$\vc{M_2(x,y)}$}}
            \psfrag{N_2(x,z)}[l][l][0.9][0]{\textcolor{white}{$\vc{N_2(x,z)}$}}
            \psfrag{N_2(x,y)}[l][l][0.9][0]{\textcolor{white}{$\vc{N_2(x,y)}$}}
            \psfrag{-1.5}[c][c][0.5][0]{-1.5}
            \psfrag{-1}[c][c][0.5][0]{-1}
            \psfrag{-0.5}[c][c][0.5][0]{-0.5}
            \psfrag{0}[c][c][0.5][0]{0}
            \psfrag{0.5}[c][c][0.5][0]{0.5}
            \psfrag{1}[c][c][0.5][0]{1}
            \psfrag{1.5}[c][c][0.5][0]{1.5}
            \psfrag{2}[c][c][0.5][0]{2}
            \psfrag{2.5}[c][c][0.5][0]{2.5}
            \psfrag{3}[c][c][0.5][0]{3}
            \psfrag{3.5}[c][c][0.5][0]{3.5}
            \psfrag{4}[c][c][0.5][0]{4}
            \psfrag{4.5}[c][c][0.5][0]{4.5}
            \psfrag{5}[c][c][0.5][0]{6}
            \psfrag{5.5}[c][c][0.5][0]{5.5}
            \psfrag{6}[c][c][0.5][0]{6}
            \psfrag{6.5}[c][c][0.5][0]{6.5}
            \psfrag{7}[c][c][0.5][0]{7}
            \psfrag{7.5}[c][c][0.5][0]{7.5}
            \psfrag{8}[c][c][0.5][0]{8}
            \psfrag{8.5}[c][c][0.5][0]{8.5}
            \psfrag{x}[c][c][0.6][0]{$x$}
            \psfrag{y}[c][c][0.6][0]{$y$}
            \psfrag{z}[c][c][0.6][0]{$z$}
            \includegraphics[width=\textwidth]{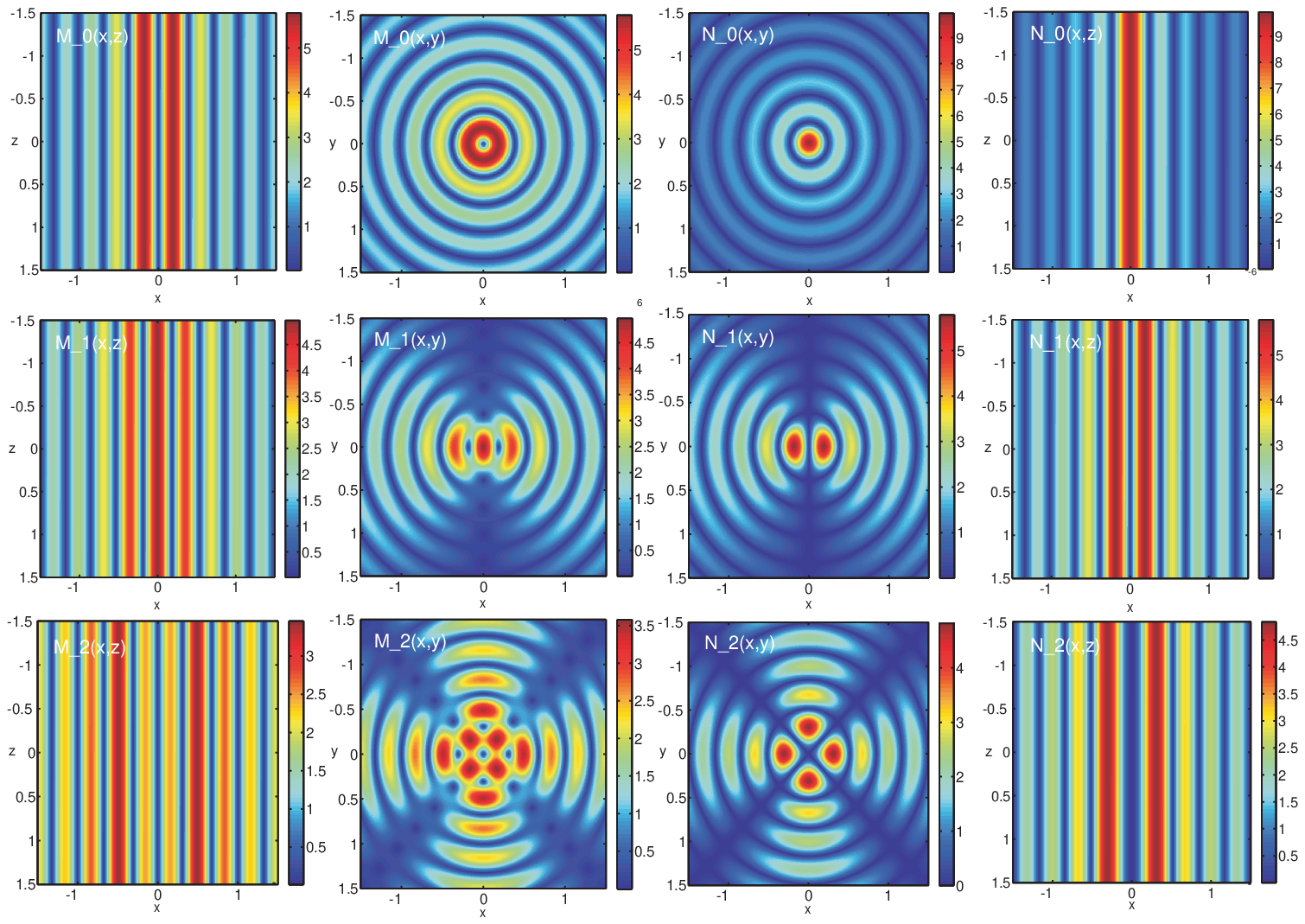}

        \caption[Spatial plots of vector cylindrical harmonics for the applied wavelength of $\lambda=\SI{632.8}{nm}$]{Spatial plots of the first orders of the vector cylindrical harmonics. The first row displays the amplitudes of the functions $\Re(M_0)^2$ and $\Re(N_0)^2$ in two of the perpendicular main planes, the second row for $\Re(M_1)^2$ and $\Re(N_1)^2$, respectively the third row for $\Re(M_2)^2$ and $\Re(N_2)^2$. All calculations are performed for a wavelength of $\lambda=\SI{632.8}{nm}$ and free space i.e. $\varepsilon=1$. The spatial coordinates $x,y,z$ are displayed in \SI{}{\micro m}, the color scales arbitrarily normalized.}

\end{figure} 

The electric and magnetic fields involved in this scattering process are divided into three domains (see the right part of fig.\;\ref{graphic_scattering-drawing}). Without any scattering object the incident electromagnetic field would simply pass the space undisturbed. The fields describing this wave throughout the whole space with $\varepsilon_\text{o}$ are called $\vc{E_\text{inc}}$ and $\vc{H_\text{inc}}$.

As soon as the material of a cylinder, placed in that space is described by an $\varepsilon_\text{i} \neq \varepsilon_\text{o}$ the electromagnetic field inside the cylinder changes and is described by only $\vc E_\text{i}$ and $\vc H_\text{i}$. From the surface of the cylinder a third electromagnetic wave is emitted, which represents the scattered fields $\vc{E_\text{sca}}$ and $\vc H_\text{sca}$.

To calculate the total field components in the space outside the cylinder, a superposition of $\vc E_\text{o}=\vc E_\text{inc}+\vc E_\text{sca}$ respectively $\vc H_\text{o}=\vc H_\text{inc}+\vc H_\text{sca}$ is necessary.
\nomenclature[me]{$\vc E_\text{o},\vc{E_\text{i}}$}{Electric field outside (o) respectively inside (i) the cylinder.}
\nomenclature[me]{$\vc H_\text{o},\vc{H_\text{i}}$}{Magnetic field outside (o) respectively inside (i) the cylinder.}
\nomenclature[me]{$\vc E_\text{inc}$}{The incident electric field vector.}
\nomenclature[me]{$\vc E_\text{sca}$}{The scattered electric field vector.}
\nomenclature[mh]{$\vc H_\text{inc}$}{The incident magnetic field vector.}
\nomenclature[mh]{$\vc H_\text{sca}$}{The scattered magnetic field vector.}

All electric and magnetic fields are to be expanded in the vector cylindrical harmonics as in this representation the boundary conditions are easiest to apply.
Two different cases for perpendicular polarizations are solved separately, linear combinations of which represent all possible polarizations of the incident plane wave. The time dependency of the fields can be expressed as $\e^{\i\omega t}$ and will hereafter be omitted as the whole problem can be solved statically.
For s-polarized incident light (\textbf{case s}), thus $\vc E_\text{inc} \parallel x-z-\text{plane}$ the expansion results in:

\begin{align}
  \vc E_\text{inc} &= \vc E_0 \e^{\i \vc k_\text{o}\vc r} = E_0 \left( \begin{array}{c}-\cos{\alpha} \\ 0 \\ \sin{\alpha} \end{array}\right) \e^{\i\vc k_\text{o}\vc{r}}
  = \sum\limits_{n=-\infty}^{\infty}E_n\vc{N}_n^{(1)}  & ; \;
  E_\text{n} &= E_0 \frac{(-\i)^n}{k_0 \, \sin{\alpha}} \label{eq-expansion-e_i1}
\end{align}

Expressing the expansions for each of the fields in terms of $E_0$ the following relations are defined:

\begin{subequations}
\begin{alignat}{10}
  &\vc{E}_{\text{inc}}
  &&=
  &&\sum\limits_{n=-\infty}^{\infty}E_n\vc{N}_n^{(1)}
  && \quad
  && \vc{H}_{\text{inc}}
  &&=
  \frac{-\i k}{\omega\mu}
  &&\sum\limits_{n=-\infty}^{\infty}E_n \vc{M}_n^{(1)} \label{eq-expansion-inc}
  \\
  &\vc{E}_{i}
  &&=
  &&\sum\limits_{n=-\infty}^{\infty}E_n\left(f_n\vc{N}_n^{(1)}+g_n\vc{M}_n^{(1)}\right)
  &&
  &&\vc{H}_{i}
  && = \frac{-\i k}{\omega\mu}
  &&\sum\limits_{n=-\infty}^{\infty} E_n \left(f_n\vc{M}_n^{(1)}+g_n\vc{N}_n^{(1)}\right) \label{eq-expansion-i}
  \\
  &\vc{E}_{\text{sca}}
  &&= -
  &&\sum\limits_{n=-\infty}^{\infty}E_n\left(b_n\vc{N}_n^{(3)}+\i a_n\vc{M}_n^{(3)}\right)
  &&
  &&\vc{H}_{\text{sca}} &&= \frac{\i k}{\omega\mu}
  &&\sum\limits_{n=-\infty}^{\infty} E_n \left(b_n\vc{M}_n^{(3)}+\i a_n\vc{N}_n^{(3)}\right) \\ \vphantom{H}\label{eq-expansion-sca}
\end{alignat}
\label{eq-expansion-all}
\end{subequations}

For the perpendicular, the so called p-polarization of the incident light (\textbf{case p}, $\vc{E_\text{i}} \perp x-z-\text{plane}$) the expansion for the incident field changes to:

\begin{align}
  \vc{E}_\text{inc} &= \vc E_0 \e^{\i\vc{k_\text{o}}\vc{r}} = E_0 \left(\begin{array}{c} 0\\1\\0 \end{array}\right)  \e^{\i\vc{k_\text{o}}\vc{r}}
  = -\i\sum\limits_{n=-\infty}^{\infty}E_n\vc{M}_n^{(1)} \label{eq-expansion-e_i2}
\end{align}

\subsection{Boundary conditions and analytical solution \label{ch_scattering-solution-fields}}

From a macroscopic point of view on the boundary between the two materials $\varepsilon_1=\varepsilon_\text{i}$ (in this case inside the cylinder) to $\varepsilon_2=\varepsilon_\text{o}$ (in this case outside the cylinder) the electromagnetic waves occur a discontinuity in the material properties. For the further discussion an abbreviation is defined as $\varepsilon_\text{i} / \varepsilon_\text{o} =:m^2$.

\nomenclature[mko]{$\vc k_\text{o}$, $\vc k_\text{i}$}{Wave vector of an electromagnetic wave in the medium outside the cylinder (o) or inside the cylinder (i).}

This occurs on the surface of the cylinder at $r=R$. Due to the law of the conversation of energy%
\footnote{analogue to the derivation of the dispersion of planar SPP on page \pageref{eq_spp-boundarycondition}. Equations \refeq{eq-e-h-boundary} base upon the assumptions of finite current density and finite conductivity which are both true in physically real cases.}%
it can be derived that on material boundaries the tangential ($\varphi$ and $z$) components of $\vc{E}$ and $\vc{H}$ must be continuous.

\begin{align}
    \bvc{e_\text{r}} \times (\vc{E_\text{o}}-\vc{E_\text{i}}) &= 0  &   \bvc{e_\text{r}} \times (\vc{H_\text{o}}-\vc{H_\text{i}}) &= 0  \label{eq-e-h-boundary}
\end{align}

This leads to a linear system of 4 equations that can for \textbf{case p} be written in matrix notation as:

\begin{flalign}
\begin{array}{rcccccccccrl}
      \ldelim({4}{0pt}[]
      & hn\frac{1}{k_\text{i}}J_n(\eta)
      & \eta J'_n(\eta)
      & hn\frac{1}{k_\text{o}}H_n^{(1)}(\xi)
      & \xi {H^{(1)}}'_n(\xi)
      	\rdelim){4}{0pt}[]
      & \ldelim({4}{0pt}[]
      & f_n
        \rdelim){4}{0pt}[]
      & \multirow{4}{*}{=}
      \ldelim({4}{0pt}[]
      & hn\frac{1}{k_\text{o}}J_n(\xi)
        \rdelim){4}{0pt}[]
      &
      & \}
      & {\scriptstyle E_\phi}
\\
      & \eta k_\text{i} J'_n(\eta)
      & hnJ_n(\eta)
      & \xi  k_\text{o} {H^{(1)}}'_n(\xi)
      & hn H^{(1)}_n(\xi)
      &
      & g_n
      &
      & k_\text{o}\xi J'_n(\xi)
      &
      & \}
      & {\scriptstyle H_\phi}
\\
      & \frac{1}{k_\text{i}}\eta^2J_n(\eta)
      & 0
      & \frac{1}{k_\text{o}}\xi^2H^{(1)}_n(\xi)
      & 0
      &
      & b_n
      &
      & \frac{1}{k_\text{o}}\xi^2J_n(\xi)
      &
      & \}
      & {\scriptstyle E_z}
\\
      & 0
      & \eta^2J_n(\eta)
      & 0
      & \xi^2H^{(1)}_n(\xi)
      &
      & \i a_n
      &
      & 0
      &
      & \}
      & {\scriptstyle H_z}
\\
      & \multicolumn{2}{c}{\underbrace{
      		\hphantom{ hn\frac{1}{k_\text{i}}J_n(\eta)
      			\hspace{\arraycolsep} \eta J'_n(\eta) }
      		}_{(\vc{E_\text{i}},\vc{H_\text{i}})}}
      & \multicolumn{2}{c}{\underbrace{
      		\hphantom{ hn\frac{1}{k_\text{i}}J_n(\eta)
      			\hspace{\arraycolsep} \eta J'_n(\eta) }
      		}_{(\vc{E_\text{sca}},\vc{H_\text{sca}})}}
      &
      &
      &
      & \underbrace{
      		\hphantom{\frac{1}{k_\text{o}}\xi^2J_n(\xi)}
      		}_{(\vc{E_\text{inc}},\vc{H_\text{inc}})}
      &
      &
  \end{array}\\
  \vphantom{x}\label{eq-scattering-coefficientmatrix}
\end{flalign}

    For \textbf{case s}, as $\vc{E_\text{inc}}$ is defined different (see equation \ref{eq-expansion-e_i2}) the right side of equation \ref{eq-scattering-coefficientmatrix} changes to:

\begin{align}
\begin{array}{rcccccccccrl}
      \ldelim({4}{0pt}[]
      & -\i \xi J_n'(\xi)
        \rdelim){4}{0pt}[]
      &
      & \}
      & {\scriptstyle E_\phi}\\
      & -\i h n J_n(\xi)
      &
      & \}
      & {\scriptstyle H_\phi}\\
      & 0
      &
      & \}
      & {\scriptstyle E_z}\\
      & -\i \xi^2 J_n(\xi)
      &
      & \}
      & {\scriptstyle H_z}\\
      & \underbrace{
      		\hphantom{-\i h n J_n(\xi)}
      		}_{(\vc{E_\text{inc}},\vc{H_\text{inc}})}
      &
      &
      &
  \end{array}
  \label{eq-scattering-case-s}
\end{align}

    In these equations $J_n$ are Bessel functions of first kind and integer order $n$, $J^\prime_n = \frac{1}{2} (J_{n-1}(x) - J_{n+1}(x) )$ are their first derivatives\footnote{Useful information about the Bessel functions, the derivatives and the Bessel differential equation are gathered in appendix \ref{ch_bessel-functions}} \citep{abramowitz}. Respectively $H^{(1)}_n=J_n+\i Y_n$ are the Bessel functions of third kind (also called Hankel functions of first kind) whose first derivatives are ${H^{(1)}}^\prime_n = \frac{1}{2} (H^{(1)}_{n-1}(x) - H^{(1)}_{n+1}(x) )$. It can be shown that Hankel functions of first type $H^{(1)}_n$ describe outgoing and Hankel functions of second type $H^{(2)}_n=J_n-\i Y_n$ describe incoming waves. Thus, it is according that in this case $H^{(1)}_n$ is the choice for the description of the scattered wave.\\

    For reasons of clarity the following two abbreviations are introduced for the limits of $\rho$ (see equation \refeq{eq-cylindrical-psi}) on the cylinder surface from inside the cylinder $\rho_\text{i}$ and from outside the cylinder $\rho_\text{o}$:

    \begin{alignat}{5}
      \eta & \overset{\text{\ref{eq-cylindrical-psi}}}{=} \sqrt{k_\text{i}^2-h^2}R &&= k_\text{o}R
         \sqrt{m^2-\cos^2\alpha}
         \quad && &&= \quad \rho_\text{i}(r=R-\delta) \label{eq_scattering-eta}
    \\
      \xi & \overset{\text{\ref{eq-cylindrical-psi}}}{=} \sqrt{k_\text{o}^2-h^2}R &&=  k_\text{o}R
         \sqrt{1-\cos^2\alpha} &&= k_\text{o}R \sin\alpha  \quad &&= \quad \rho_\text{o}(r=R+\delta) \label{eq_scattering-xi}
    \end{alignat}

    By analytically solving the equation system \eqref{eq-scattering-coefficientmatrix} relations for the coefficients $a_n$, $b_n$, $f_n$ and $g_n$ are achieved. From these with equation \eqref{eq-expansion-all} the electric and magnetic field vectors inside the cylinder $(\vc{E_i},\vc{H_i})$ and the scattered field vectors outside the cylinder $(\vc{E_\text{sca}},\vc{H_\text{sca}})$ can be evaluated.

    The solution vector $(a_n,b_n,f_n,g_n)$ includes rather extensive expressions. After some simplifications and introducing abbreviation expressions $A_n$,$B_n$,$C_n$,$D_n$,$V_n$,$W_n$ the following standard notation\footnote{This notation can be found in \citet{bohren-huffman}, slightly modified in \citet{hulst} and certain recent papers (e.g. \citep{martinos-economou,barabas}).} can be achieved for the two perpendicular polarizations in \textbf{case p} and \textbf{case s} for the scattered light outside the cylinder:

    \begin{subequations}
    \begin{align}
    a_{n\text{s}} &= \frac{C_nV_n - B_nD_n}{W_nV_n + \i D_n^2}      &      b_{n\text{s}}  &= \frac{W_nB_n + \i D_n C_n}{W_nV_n + \i D_n^2} \label{eq_scattering-an-bn-s} \\
    a_{n\text{p}} &= - \frac{A_n V_n - \i C_n D_n }{W_n V_n + \i D_n^2}    &    b_{n\text{p}}  &= -\i \frac{C_n W_n + A_n D_n}{W_n V_n + \i D_n^2} \label{eq_scattering-an-bn-p}\\
    \end{align}\label{eq_scattering-an-bn}
    \end{subequations}

    In these equations against a number of useful abbreviations is introduced, shortening the expressions significantly:

    \begin{align}
        A_n &= \i \xi \left( \xi J_n'(\eta) J_n(\xi) - \eta J_n(\eta) J_n' (\xi)  \right)   \\
        B_n &= \xi \left( m^2 \xi J_n'(\eta) J_n(\xi) - \eta J_n(\eta) J_n'(\xi)   \right)  \\
        C_n &= n \cos(\alpha) \eta J_n(\eta) J_n (\xi) (\xi^2/\eta^2 -1) \\
        D_n &= n \cos(\alpha) \eta J_n(\eta) H_n^{(1)}(\xi) (\xi^2/\eta^2 -1) \\
        V_n &= \xi \left( m^2 \xi J_n'(\eta) H_n^{(1)}(\xi) - \eta J_n (\eta) H_n'^{(1)}(\xi)  \right) \\
        W_n &= \i \xi \left( \eta J_n(\eta)  H_n'^{(1)}(\xi) - \xi J_n'(\eta) H_n^{(1)} (\xi)   \right)
    \end{align}

    The scattering coefficients show symmetries that are caused by the symmetries of the Bessel functions $J_{-n}=(-1)^nJ_n$ and $Y_{-n}=(-1)^nY_n$, which was already demonstrated by \citet{kerker-paper}:

    \begin{alignat}{8}
    a_{-n\text{s}} &= -a_{n \text{s}}
    && \quad\quad\quad\quad
    && b_{-n\text{s}} &&= b_{n\text{s}}
    && \quad\quad\quad\quad
    &&  a_{0\text{s}} &&= 0
    \\
    a_{-n\text{s}} &= a_{n \text{p}}
    &&
    && b_{-n\text{s}} &&= -b_{n\text{p}}
    &&
    &&  b_{0\text{p}} &&= 0
    \\
     &
    &&
    && a_{n\text{s}} && = - b_{n\text{p}}
    &&
    && &&
    \end{alignat}

\subsection{Perpendicular incidence \label{ch_theory-scattering-perpendicular}}

    The special case of perpendicular incidence ($\alpha=\pi/2$) is of particular interest as equations (\refeq{eq_scattering-an-bn-s}, \refeq{eq_scattering-an-bn-p}) can be simplified significantly as $\xi = k_\text{o}R\sin(\alpha) = k_\text{o}R$ and $\eta = k_\text{o}R \sqrt{m^2 - \cos^2(\alpha)} = k_\text{o}Rm$. Thus, the scattering coefficients can be expressed explicitely as:

    \begin{align}
    a_{n \text{p} \bot} &= 0\\
    b_{n \text{p} \bot} &= \frac{J_n(mkR) J_n' (kR) - kR J_n'(mkR) J_n(kR)}{J_n(mkR) {H_n^{(1)}}'(kR) - m J_n'(mkR) H_n^{(1)} (kR)}\\
    a_{n \text{s} \bot} &= \frac{m J_n'(kR) J_n(mkR) - J_n(kR) J_n'(mkR)}{mJ_n(mkR) {H_n^{(1)}}'(kR) - J_n'(mkR) H_n^{(1)}(kR)}\\
    b_{n \text{s} \bot} &= 0
    \end{align}

    As the complexity of these simplified equations is much lower than that for arbitrary incident angle $\alpha$ the case of perpendicular incidence has attracted special attention in literature and was discussed e.g. by \citet{martinos-economou} \footnote{In this publication the coordinate system thus also the scattering coefficients are defined different from the demonstrated derivation.}.

\subsection{Scattering efficiencies \label{ch_theory-scatteringefficiency} }

    The scattering process cannot only be described in terms of the scattered field distribution. The natural measurable scalar quantity of electromagnetic field distributions is the intensity $I(\vc{r})$ defining the total electromagnetic energy flux per area (and is sometimes also called irradiance). The intensity mathematically is derived from the nonobservable Poynting vector $\vc S$ of an electromagnetic wave.

    \begin{align}
     I(\vc r) &= |\vc{S} (\vc r)|      &        \vc S (\vc r)= \nicefrac{1}{2} \, \Re \left( \vc E (\vc r)\times \vc H (\vc r)^{\ast} \right)
    \end{align}

    The Poynting vector includes additional information of the direction of energy flux\footnote{For details it is referred to appendix \ref{ch_plane-waves}.}. In general $\vc S$ is a complex vector, whose real and imaginary parts represent the physical energy flux respectively an analogon to the idle power. In this thesis the Poynting vector is defined as the real part only. From $\vc S$ it is possible to derive a more general description of the scattering efficiency.


    In precise terminology the scattering process is divided into two contributing effects, together forming the total extinction, which is also called attenuation:

    \begin{equation}
        \text{extinction} = \text{scattering} + \text{absorption}
    \end{equation}

    Both processes, scattering and absorption withdraw energy from the incident electromagnetic wave thus reducing the intensity of the transmitted wave. Scattering describes a deflexion of electromagnetic waves, thus a change in the energy flux distribution. Absorption comprises the dissipative processes involved in the scattering, including also the absorption that is connected to plasmon excitation and the predominantly ohmic losses on the SPP.

    The electromagnetic energy flux through a surface $A$ is defined by the surface integral over the vector component of $\vc{S}$ perpendicular to the surface ($\vc{S}\cdot \bvc{e}_\perp$). The definition of a surrounding, closed sphere with surface $A'$ and radius $R'>R$ surrounding the scattering cylinder leads to an integral equation for the net energy flux to ($W>0$) respectively from the sphere ($W<0$), the first being omitted in our application of scattering of an electromagnetic wave from a source outside the sphere:

    \begin{equation}
        W = - \int\limits_{A'} \den A \vc{S} \cdot \bvc{e}_\text{r} \label{eq_scattering-w1}
    \end{equation}

    For the described system the law of the conservation of energy must be fulfilled. Thus the sum of energy flux through the surface $A'$ for all processes participating in the total scattering process must balance out to $0$, generally including absorption.

    Then the energy that is absorbed by the scattering obstacle can be expressed as $W_\text{abs}=W_\text{inc}-W_\text{sca}+W_\text{ext}$. For a non\-ab\-sor\-bing surrounding me\-dium (always valid as $\Im(\varepsilon(\text{SiO}_2))=0$ and $\Im(\varepsilon(\text{vac.}))=0$), without any scattering obstacle the incident waves passed the volume element that is re\-stric\-ted by the surface $A'$ without loss of energy.

    \noindent As this assumption is true,~eq.\;\eqref{eq_scattering-w1} can be simplified to the following expression for $A'$:

    \begin{equation}
        W_\text{ext} = W_\text{abs} + W_\text{sca}
    \end{equation}

    These energy flux values do not yet only depend on the scattering configuration but also on the incident irradiation $I$. Therefore a new variable is defined by dividing this linear dependency out, leading to the so called cross sections $[C_n]=\SI{1}{m^2}$:

    \begin{align}
        C_\text{ext} & = \frac{W_\text{ext}}{I_\text{inc}} \\
        C_\text{ext} & = C_\text{abs} + C_\text{sca}
    \end{align}

    From these scattering cross sections of the unit of area a unitless value: The scattering efficiency $Q_n$ which is defined by dividing the projection of the geometrical cross section $G$ (that can be calculated for each particle shape) on a plane perpendicular to the incident beam  $C_n$ as $Q_n = C_n / G$. The total scattering absorption efficiency is then given by:

    \begin{equation}
        Q_\text{abs}=Q_\text{ext}-Q_\text{sca} \label{eq_scattering-abs} \; .
    \end{equation}

    From a strict point of view the transfer of this theory to the case of an infinitely extended cylinder is not as obvious to be valid. It is well defined for spatially confined particles that can be simply encased by a closed plane. But as the treated cylinder is infinitely long, its scattering $C_\text{sca}$ and absorption cross section $C_\text{abs}$ are of course infinite.

    However, the unitless definition of $Q_n$~(eq.\;\refeq{eq_scattering-abs}) allows a differential description of the scattering and absorption efficiency even for infinitely extended cylinders and infinitely extended plane waves.

    Expressions for the absorption efficiency of spherical and cylindrical particles were also calculated from the Rayleigh theory. \citet{lukyanchuk} presents the Rayleigh scattering efficiency of an infinitely long cylinder as:

    \begin{align}
        Q_\text{sca} &= \frac{\pi^2}{4} \left| \frac{m^2 -1 }{m^2 + 1} \right|^2 2 \pi R \; \frac{ \varepsilon_\text{i}^{\nicefrac{1}{2} } }{\lambda} \; .
    \end{align}

    However this description does only take into account dissipative effects ($\Im(\varepsilon_\text{i}) \neq 0$) and not radiative damping that is inherently included in the Mie scattering description \citep{lukyanchuk}. Additionally the Rayleigh solution, as mentioned, is only applicable for small scattering samples ($\lambda \gg D$).

    The derived scattering theory can also be used to calculate the scattering efficiencies $Q_n$, which do not show the weak points of the Rayleigh approximation. Therefore usually an amplitude scattering matrix $T$ is derived that includes all effects and connects the incident field components ($\vc E_\text{inc}, \vc H_\text{inc}$) in cylindrical coordinates to corresponding scattered field components ($\vc E_\text{inc}, \vc H_\text{inc}$).

    In the case of a cylinder, the surface of integration of the Poynting vector is defined as a concentric cylinder with length $L'$, surface $A'$ and radius $R'$ \citep{bohren-huffman}. Integrated are the components of the Poynting vector perpendicular to this surface, the radial components $(\vc S)_r$:

    \begin{align}
        W_\text{sca} &\overset{\text{\eqref{eq_scattering-w1}}}{=} R' L' \int_0^{2 \pi} \den \varphi (\vc S_\text{sca})_r   &
        W_\text{ext} &\overset{\text{\eqref{eq_scattering-w1}}}{=} R' L' \int_0^{2 \pi} \den \varphi (\vc S_\text{ext})_r   \\
        \vc S_\text{sca} & = \nicefrac{1}{2} \Re \left( \vc E_\text{sca} \times \vc H_\text{sca}^\ast \right) &
        \vc S_\text{ext} & = \nicefrac{1}{2} \Re \left( \vc E_\text{inc} \times \vc H_\text{sca}^\ast  + \vc E_\text{sca} \times \vc H_\text{inc}^\ast\right)
    \end{align}

    Introducing the expansions of $E_\text{sca}$ and $E_\text{inc}$~(eq.\;\refeq{eq-expansion-inc}, \refeq{eq-expansion-sca}) into these equations and integrating, the scattering and extinction efficiencies of the infinite circular cylinder are obtained:

    \begin{subequations}
    \begin{alignat}{4}
        Q_\text{sca, s}  & = \frac{ W_\text{sca} }{ 2 R L' I_\text{inc} } = \; && \frac{2}{k_0 R} &&\left( |b_{\text{s}0}|^2 + 2 \sum_{n=1}^\infty \left( |b_{\text{s}n}|^2 +  |a_{\text{s}n}|^2  \right)   \right)   \\
        Q_\text{ext, s} &= \frac{ W_\text{ext} }{ 2 R L' I_\text{inc} } = \; && \frac{2}{k_0 R} \Re &&\left(  b_{\text{s}0} +  2 \sum_{n=1}^\infty b_{\text{s}n} \right) \\
        Q_\text{sca, p}  & = && \frac{2}{k_0 R} &&\left( |a_{\text{p}0}|^2 + 2 \sum_{n=1}^\infty \left( |a_{\text{p}n}|^2 +  |b_{\text{p}n}|^2  \right)   \right)   \\
        Q_\text{ext, p} &= && \frac{2}{k_0 R} \Re &&\left(  a_{\text{p}0} +  2 \sum_{n=1}^\infty a_{\text{p}n} \right)
    \end{alignat} \label{eq_scattering-efficiencies}
    \end{subequations}

    \section{Asymptotic scattered field and intensity pattern \label{ch_scattering-pattern}}

    In the experiment the observation point is comparably far away from the scattering centre. Introducing far field approximations, also the form of the observed intensity pattern can be derived\footnote{Details about the derivation can be found in \citet{bohren-huffman}.}.

    For large distances from the sample, as $k_0 R \sin(\alpha) \gg 1$, the scattered field can be asymptotically written as:

    \begin{align}
        \vc E_\text{sca,s} =& - \vc E_0 \e^{-\i \pi/4} \sqrt{\frac{2}{\pi K_0 R \sin(\alpha)}} \e^{\i k_0 \left(r \sin(\alpha) - z \cos(\alpha) \right)} \; \cdot \\
        & \sum_n (-1)^n \e^{\i n \varphi} \left(  a_{\text{s}n} \bvc e_\varphi + a_{\text{s}n} ( \cos(\alpha) \bvc e_r + \sin(\alpha) \bvc e_z ) \right)  \\
        \vc E_\text{sca,p} =& - \vc E_0 \e^{-\i \pi/4} \sqrt{\frac{2}{\pi K_0 R \sin(\alpha)}} \e^{\i k_0 \left(r \sin(\alpha) - z \cos(\alpha) \right)} \; \cdot \\
        & \sum_n (-1)^n \e^{\i n \varphi} \left(  - a_{\text{p}n} \bvc e_\varphi - b_{\text{p}n} ( \cos(\alpha) \bvc e_r + \sin(\alpha) \bvc e_z ) \right)  \\
    \end{align}

    The wave fronts of the scattered wave (planes of constant phase) in cartesian coordinates are found to satisfy the relation:

    \begin{align}
        f(x,y,z) = r \sin(\alpha) - z \cos(\alpha)
    \end{align}

    This relation describes cones of half angle $\alpha$. Therefore the scattering pattern that is expected from the experiment is the projection of cones. Concisely for $\alpha = \ang{90}$, the cone is expected to reduce to a cylinder. In a plane perpendicular to $z$ at $\alpha=0$, for angles $\alpha< \ang{45}$, the projection of the calculated intensity forms ellipses with decreasing eccentricity, approaching $\alpha = \ang{0}$. For $\alpha = \ang{45}$ the pattern is a parabola and for further increased incident angles, hyperbolas are observed. 

\chapter{Sample fabrication \label{ch_sample-fabrication}}

    \vspace{-\baselineskip}
    The following experiments were performed by the using three different materials: Silica (fused silica glass, $\text{SiO}_2$), gold (Au) and silver (Ag). For the computation of the dispersion of the surface plasmons (sec.\,\ref{chapter_theory-model}) as well as for the computation of scattering field distributions and excitation angles (sec.\,\ref{chapter_theory-scattering}) the knowledge of the material dispersion $\varepsilon(\omega)$ of these materials is essential.

    Of additional interest are the physical properties regarding the fabrication process of the samples. Thus, an overview over the relevant information of all three materials is presented. With this information background, the techniques that are applied for the fabrication of two different types of samples are summarized:
    \begin{enumerate}
      \item Free standing wires, consisting of a silica fiber substrate with a diameter that is varied between $\SI{10}{\micro m} <D< \SI{40}{\micro m}$, coated with a comparably thick layer of sputtered metal.
      \item Thin wires with a diameter $\SI{0.4}{\micro m} <D< \SI{1}{\micro m}$ which are fabricated by drawing a specifically designed fiber, analogue to the fabrication of photonic crystal fibers (PCF) with a single hole or a hole array. The metal that is subsequently filled into its holes forms arrays of wires inside the remaining silica glass matrix.
    \end{enumerate}

    \begin{table}[bt]
        \begin{small}
        \centering
        \begin{tabular}{ccccccc}\toprule
          Material   &   \multicolumn{3}{l}{Processing properties}   &   \multicolumn{3}{l}{Optical properties} \\\midrule
             &  $T_\text{m}$ (\textcelsius) &  $T_\text{b}$(\textcelsius)  &  $\alpha_\text{exp}$  &  Type  &  Model  &  $\varepsilon(\SI{632.8}{nm})$ \\\cmidrule(r){2-4}\cmidrule(l){5-7}
          Ag                    & \SI{961}{}   &   \SI{2162}{}&  \SI{18e-6}{}    &  n.m. fcc & Drude &  $-16.1396+1.0814 \i$\\
          Au                    & \SI{1063}{}   &   \SI{2856}{}&  \SI{14e-6}{}    &  n.m. fcc & experimental & $-9.56+0.81\i$ \\
          $\text{SiO}_2$        & \SI{1730}{}%
          \footnotemark
                                                    &  \SI{2230}{}              &   \SI{5.5e-7}{}& amorphous & Sellmeier & 2.123\\\bottomrule
        \end{tabular}
        \end{small}
        \caption[Overview of processing and optical properties of silver, gold and silica glass.]{Overview of the processing (melting $T_\text{m}$ (in \textcelsius) and boiling point $T_\text{b}$ (in \textcelsius), thermal expansion coefficient $\alpha_\text{exp}$ in \SI{1e-6}{K^{-1}} at room temperature) and optical properties: Material type, dispersion model, dielectric permittivity $\varepsilon(\lambda=\SI{632.8}{nm})$ of the noble metals silver (Ag) and gold (Au) and of the applied type of fused silica glass ($\text{SiO}_2$) with a pureness of \SI{99.99}{\%}.}\label{tab_material-properties}
    \end{table}
    \footnotetext{As silica is an amorphous solid, no discrete melting point exists. Provided is therefore the softening point at \SI{1730}{\celsius}, additional parameters which are important for tapering is the annealing point at $T=\SI{1075}{\celsius}$ and the strain point at $T=\SI{1180}{\celsius}$, which are all defined by different grades of material strength (see for more details \citep{hilgenberg-glass}).}

    \section{Material properties \label{ch_material-properties}}
    \subsection*{Silica}

    For the sample fabrication pure silica glass is used\footnote{Hilgenberg 0620 with a purity of \SI{99.99}{\%}. Its physical properties from \citep{hilgenberg-glass} are included in table \ref{tab_material-properties}.}. The amorphous material is fabricated by fusing crystalline silica ($\text{SiO}_2$). Silica glass is an insulator ($\sigma \approx 0$) and transparent in the visible wavelength regime (see section \ref{ch_theory-sellmeier}, p.~\pageref{ch_theory-sellmeier}). As the $\text{SiO}_2$ molecules exhibit electronic vibration resonances in the ultraviolet, the absorption in the UV increases for $\lambda<\SI{200}{nm}$. Vibrational absorption makes silica intransparent in the infrared for $\lambda>\SI{2000}{nm}$.

    The dispersion of silica glass is well described by the Sellmeier equation \eqref{eq_sellmeier}, which practically is often written in terms of dimensionless coefficients $A_j$ and coefficients $\lambda_j$ of the dimension of the wavelength as follows:

    \begin{align}
        \varepsilon(\lambda) = 1 + \frac{A_1 \lambda^2}{\lambda^2-\lambda_1^2} + \frac{A_2 \lambda^2}{\lambda^2-\lambda_3^2} + \frac{A_3 \lambda^2}{\lambda^2-\lambda_3^2} \label{eq_sellmeier-comp}
    \end{align}

    The Sellmeier coefficients always have to be experimentally determined. In this case data for bulk fused silica from \citet{agrawal} is taken, who fitted the Sellmeier equation \ref{eq_sellmeier-comp} into a measured dispersion curve from \citet{sotobayashi}:

        \begin{table}[H]
        \centering
        \begin{tabular}{cccc}\toprule
          $j$   &   1   &   2   &   3\\\midrule
          $A_j$ &   \SI{0.6961663}{} & \SI{0.407426}{} & \SI{0.8974794}{}\\
          $\omega_j$ in $\text{m}^{-1}$ & \SI{2.753703e16}{} & \SI{1.620465e16}{} & \SI{1.903416e14}{} \\
          $\lambda_j$ in m & \SI{0.068043e{-6}}{} & \SI{0.1162414e{-6}}{} & \SI{9.896161e{-6}}{} \\\bottomrule
        \end{tabular}
        \end{table}

\begin{figure}[p]
        \centering

            \psfrag{eps(633nm)=2.123}[c][c][0.9][0]{$\varepsilon(\SI{633}{nm})=2.123$}
            \psfrag{eps}[c][c][0.9][0]{$\varepsilon(\lambda)$}
            \psfrag{reeps-SiO2}[lc][lc][0.9][0]{$\Re( \varepsilon_{\text{SiO}_2}(\lambda) )$}
            \psfrag{wavelength}[c][c][0.9][0]{wavelength $\lambda$}
            \psfrag{nm}[c][c][0.9][0]{nm}
            \includegraphics[width=0.8\textwidth]{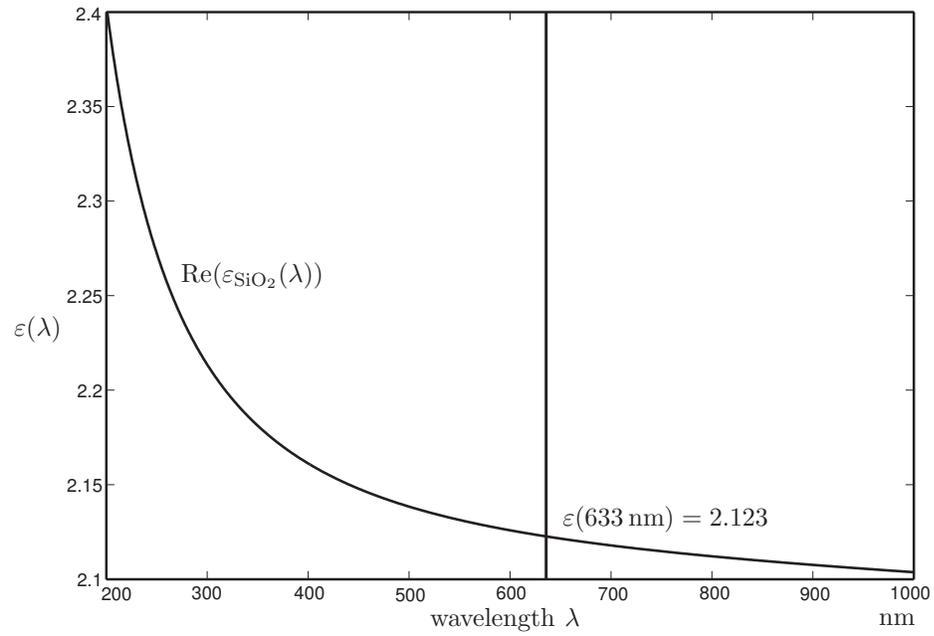}

        \caption[Material dispersion of fused silica glass (amorphous $\text{SiO}_2$), calculated from the Sellmeier equation.]{Material dispersion of silica (amorphous $\text{SiO}_2$), calculated from the Sellmeier equation \eqref{eq_sellmeier-comp}. The imaginary part of the dielectric permittivity of silica is negligible, its real part is displayed.\label{graphic_sio2-sellmeier}}

\end{figure}

    As depicted in fig.\;\ref{graphic_sio2-sellmeier} the dielectric permittivity of silica glass in the visible range is approximately $\varepsilon \approx 2.123$ ($n \approx 1.457$), which decreases by $1 \%$ in towards the region of $\lambda \approx \SI{1.5}{\micro m}$.

    \subsection*{Gold}
    As mentioned in section \ref{ch_theory-drude} (p.\;\pageref{ch_theory-drude}), the Drude model is not applicable for gold in the visible spectrum, as this metal exhibits strong interband-transitions in this wavelength range. Therefore for the computations commonly used experimental data from \citet[pp.~286-295]{palik} is interpolated for a wavelength range of $\SI{200}{nm}<\lambda<\SI{1000}{nm}$. The experimental data in the wavelength range of interest is obtained by \citet{theye}, measuring the reflectance and transmittance of thin ($\SI{10}{}-\SI{25}{nm}$), semi-transparent evaporated Au films on superpolished fused silica substrates.

    \begin{figure}[tbp]
        \centering

            \psfrag{eps(633nm)= -9.56 + 0.81i}[c][c][0.8][0]{\colorbox{white}{$\varepsilon (\SI{633}{nm})= -9.56 + 0.81 \i$} }
            \psfrag{imeps-Au}[c][c][0.8][0]{$\Im(\varepsilon (\lambda))$}
            \psfrag{reeps-Au}[lc][lc][0.8][0]{$\Re( \varepsilon (\lambda) )$}
            \psfrag{wavelength}[c][c][0.8][0]{wavelength $\lambda$}
            \psfrag{nm}[c][c][0.8][0]{nm}
            \includegraphics[width=0.8\textwidth]{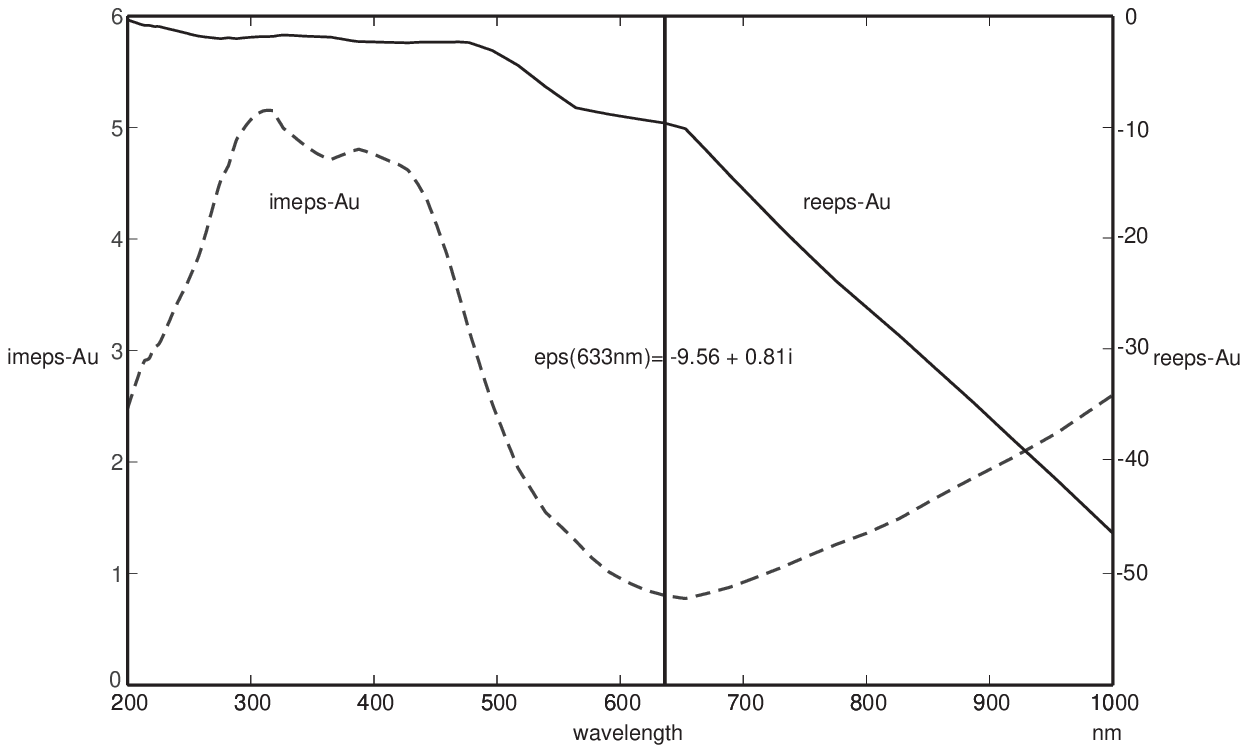}

        \caption[Material dispersion of gold for a wavelength range of $\SI{200}{nm}<\lambda<\SI{1000}{nm}$, based on experimental data.]{Material dispersion of gold, interpolated from experimental data \citep{palik}. For Au in this wavelength region interband transitions are not negligible and the Drude model (see section \ref{ch_theory-drude}) is not valid, as is clearly noticeable by the increased absorption  ($\Im(\varepsilon(\lambda))$) for $\lambda<\SI{600}{nm}$, compared to the dispersion of silver as calculated by the Drude model and depicted in figure \ref{graphic_ag-drude}.\label{graphic_au-palik}}

\end{figure}

    Recent results of density functional calculations and REELS\footnote{REELS: Reflection electron energy-loss spectroscopy} measurements on the dielectric properties of silver (Ag) and gold (Au) in a range from infrared to ultraviolet by \citet{werner} indicate that the widely used experimental dispersion dataset from \citet{palik} deviate from new values. This deviation increases for the high energy range. For the applied lower energy range ($\lambda=\SI{633}{nm}:\; \SI{1.96}{eV}$) the results from \citep{palik} do not deviate significantly, at least nor new, better experimental neither numeric results have been reported. Thus the used experimental values are still the best available for the desired wavelength range.

    \subsection*{Silver}

    The optical properties of silver are well described by the Drude model of the free electron gas as described in section \ref{ch_theory-drude} (p. \pageref{ch_theory-drude}).


    Different to gold, for silver interband-transitions play a negligible role in the visible frequency range. The interband absorption edge of silver is in the UV at around $\lambda = \SI{310}{nm}$ ($\omega = \SI{6.08e15}{Hz}$).

    The complex dielectric permittivity $\varepsilon_\text{Ag}(\omega)$ can therefore be written using the dispersion relation \eqref{eq_drude-dispersion}, separated into its real and imaginary part and converted to a convenient form:

    \begin{figure}[tbp]
        \centering

            \psfrag{eps(633nm)= -16.1396 + 1.0814i}[c][c][0.8][0]{\colorbox{white}{$\varepsilon (\SI{633}{nm})= -16.1396 + 1.0814 \, \i$} }
            \psfrag{imeps-Ag}[c][c][0.8][0]{$\Im(\varepsilon(\lambda))$}
            \psfrag{reeps-Ag}[lc][lc][0.8][0]{$\Re( \varepsilon(\lambda) )$}
            \psfrag{wavelength}[c][c][0.8][0]{wavelength $\lambda$}
            \psfrag{nm}[c][c][0.8][0]{nm}
            \includegraphics[width=0.8\textwidth]{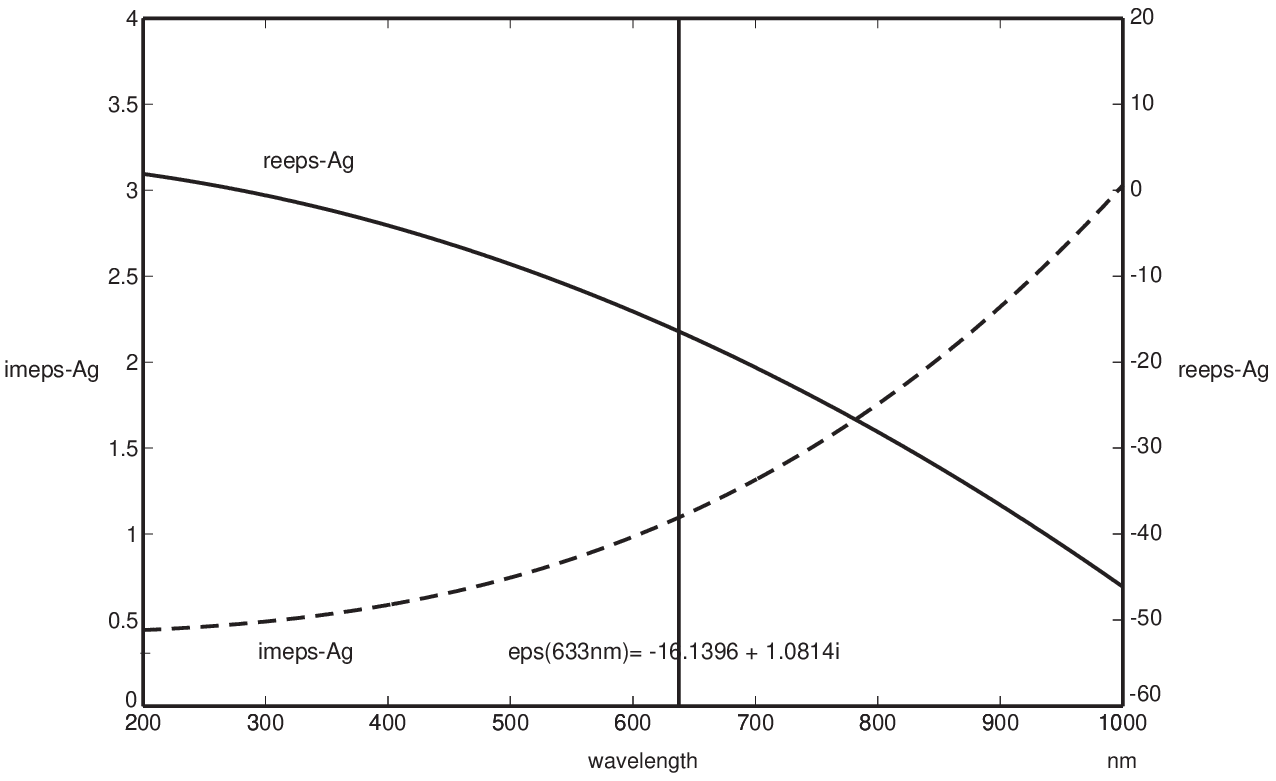}

        \caption[Material dispersion of silver (Ag) for a wavelength range of $\SI{350}{nm}<\lambda<\SI{1000}{nm}$, based the Drude model.]{Material dispersion of silver (Ag), computed from equation \eqref{eq_drude-ag-re},\eqref{eq_drude-ag-im} for the Drude free electron gas model (sec.\;\ref{ch_theory-drude}).\label{graphic_ag-drude}}

\end{figure} 

    \begin{align}
        \Re( \varepsilon_\text{Ag} (\omega )) &= A_1 - \frac{A_2^2}{\omega^2 + A_3^2} \label{eq_drude-ag-re} \; , \\
        \Im( \varepsilon_\text{Ag} (\omega) ) &= B_3 + \frac{B_1^2 B_2}{ \omega^3 + \omega B_2^2c }  \; , \label{eq_drude-ag-im}
    \end{align}

    \noindent where the coefficients for silver, calculated from the plasma frequency of silver $\omega_\text{p}$ (as listed in table \ref{tab_plasmafreq}, p. \pageref{tab_plasmafreq}) are $A_1 = \SI{3.9000}{}$, $A_2 = \SI{1.3325e16}{}$, $A_3 = \SI{6.992407e6}{}$, $B_1 = \SI{1.0907e17}{}$, $B_2 = \SI{1.46511e12}{}$, $B_3 = \SI{0.42062}{}$. A plot of the dispersion, computed from these values is shown in fig.\;\ref{graphic_ag-drude}.

    \section{Free-standing wires \label{ch_sample-freestanding}}

    For the investigation of SPP on wires with a diameter $\SI{10}{\micro m} <D< \SI{40}{\micro m}$ a coating technique was applied to fabricate samples. This method involves tapering of circular silica fiber and sputtering with a layer of metal with a thickness large enough ($d>\SI{0.5}{\micro m}$) to treat them as virtually bulk metal wires.

    The advantage of this method lies in its flexibility to access a wide range of diameters precisely with a high uniformity of the diameter and shape along the longitudinal direction. The technique of heat tapering of silica glass fibers makes it is possible to fabricate wires with very high aspect ratios. At least sample lengths of \SI{0.4}{cm} are necessary as the wire diameter on the one hand must be much smaller than the probe beam diameter to fulfil the condition of plane waves (p.\;\pageref{text_scattering-constraints}). On the other hand for a variation of the incident angle, the effective projected beam diameter into the longitudinal direction of the wire increases with low incident angles (compare to fig.\;\ref{graphic_setup-scheme}), making a sufficient length of the sample necessary.

    The film preparation of these fibers is performed using the magnetron sputter deposition technique, allowing well reproducible results with a high surface uniformity and low surface roughness.

    \subsection{Tapering \label{ch_tapering}}

    As starting material for the core substrate silica glass fibers with diameters of $D''=\SI{125}{\micro m}$ or $D''=\SI{200}{\micro m}$ are taken. These fibers are stripped of their protective coating and cleaned accurately with a cascade of acetone ($\text{CH}_3\text{COCH}_3$), isopropanol ($\text{C}_3\text{H}_8\text{O}$) and ethanol ($\text{C}_2\text{H}_6\text{O}$) to achieve a clean surface without any residuals.

\begin{figure}[width=\textwidth]
        \centering

            \psfrag{a}[c][c][1.5][0]{\textbf{\textsf{a}}}
            \psfrag{b}[c][c][1.5][0]{\textbf{\textsf{b}}}
            \psfrag{c}[c][c][1.5][0]{\textbf{\textsf{c}}}
            \psfrag{Ddd}[c][c][0.9][0]{$D''$}
            \psfrag{Dd}[c][c][0.9][0]{$D'$}
            \psfrag{D}[c][c][0.9][0]{$D$}
            \psfrag{l}[c][c][0.9][0]{\colorbox{white}{$l$}}
            \psfrag{A}[c][c][0.9][0]{A}
            \psfrag{B}[c][c][0.9][0]{B}
            \psfrag{C}[c][c][0.9][0]{C}
            \psfrag{Dg}[c][c][0.9][0]{D}
            \psfrag{TS}[c][c][0.9][0]{TS}
            \psfrag{flame}[c][c][0.9][0]{flame}

            \psfrag{T/T_m}[c][c][0.9][0]{$T/T_\text{m}$}
            \psfrag{0.2}[c][c][0.9][0]{$0.2$}
            \psfrag{0.3}[c][c][0.9][0]{$0.3$}
            \psfrag{0.4}[c][c][0.9][0]{$0.4$}
            \psfrag{0.5}[c][c][0.9][0]{$0.5$}
            \psfrag{0.6}[c][c][0.9][0]{$0.6$}

            \psfrag{magnetrons}[c][c][0.9][0]{magnetrons}
            \psfrag{turntable}[c][c][0.9][0]{turntable}
            \psfrag{plasma}[c][c][0.9][0]{\colorbox{white}{Ar-plasma}}
            \psfrag{sample}[c][c][0.9][0]{sample}
            \psfrag{vacuum chamber}[c][c][0.9][0]{vacuum chamber}
            \psfrag{Cr}[c][c][0.9][0]{Cr}
            \psfrag{Ag/Au}[c][c][0.9][0]{Ag/Au}
            \psfrag{N}[c][c][0.5][0]{\colorbox{white}{N}}
            \psfrag{S}[c][c][0.5][0]{\colorbox{white}{S}}

            \includegraphics[width=\textwidth]{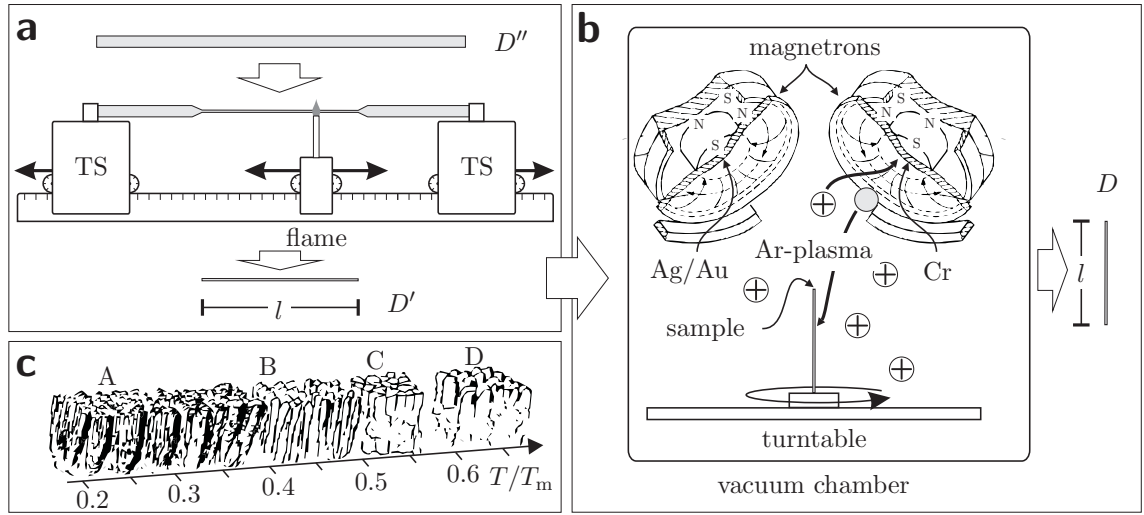}

        \caption[Schematic drawing of the tapering and sputtering process for sample fabrication.]{Schematic drawing of the tapering and sputtering process for sample fabrication. Part (a) displays the flame tapering process, reducing the diameter of a fiber sample from $D''$ down to $D'$. This sample is then sputtered with a virtually monoatomic contact layer of Cr and a thick layer of Ag or Au by magnetron sputter coating. The final, virtually bulk metal sample has a diameter of $D$. Inset (c) illustrates the dependency of the corning size and shape on the sputtering temperature $T$ with respect to the melting temperature $T_m$ of the metal.\footnotemark  \label{graphic_tapering-sputtering}}

\end{figure}

    In the next step (fig.\,\ref{graphic_tapering-sputtering}\,a) the diameter of the fiber is tapered down by the application of a butane-oxygen flame that is continuously moved along the fiber with a computer controlled, well defined speed and increasing movement distance while a controlled tension is applied between both ends of the fiber. The temperature of the fiber for this process is kept below the temperature where the silica's viscosity decreases to far, thus its circular cylindrical shape is well maintained. This technique is commonly used \citep{cleo-tapering-2007} and well described e.g. by \citet{birks}.

\footnotetext{Drawing c and the magnetron sketches are reproduced following \citep{bergauer}.}

    The fiber diameter $D''$ for a certain length $l$ is uniformly tapered down to a smaller diameter $D'$ with a fast transition region to the original fiber diameter on both sides. The fluctuations of $D'$ over the tapered length are measured with an optical microscope to be $\Delta D' < \SI{0.01}{\micro m}$. For the sample fabrication usually a length of $l \approx \SI{2}{cm}$ from the center region of a uniformly tapered length of $\SI{4}{cm}$ is made at a final fiber diameter of $\SI{10}{\micro m}<D'< \SI{40}{\micro m}$.

    \subsection{Sputtering}

    The tapered fibers with diameters $D'$ are vertically mounted on the rotating turntable inside the vacuum chamber of a sputtering system\footnote{AJA OTC Orion series uhv, also used was a smaller single magnetron sputtering device  Emtech K575X.}, which is equipped with more than three magnetron sputtering devices.

    Between the silica glass substrate and the deposited silver or gold layer always a very thin contact layer of Cr is sputtered to achieve a better adhesion between both materials\footnote{This work shall only give a short summary of the applied magnetron technology. For a more detailed treatement the reader is referred to the very good introductions \citep{martin-1986} and the recent lecture script \citep{bergauer}.}.\\

    Inside the chamber a vacuum of (\SI{1e-6}{{bar}}) is established with a turbomolecular pump system. A low pressure argon atmosphere (\SI{1.3e-6}{bar}) is then conditioned in which a gas discharge ignites a continuous plasma. Argon ions from this plasma are accelerated by the magnetic field of the magnetron devices and impact into the sputtering target (Cr, Ag, Au) where neutral atoms of the sputtering target are knocked out. These single atoms travel with the starting energy towards the sample surface where they assemble to form a metallic layer on the glass substrate.

    The lower the pressure inside the chamber, the higher is the purity of the sputtered material and the higher is the energy of the impinging sputtering molecules on the sample. On the other hand, as the gas pressure decreases below a limit, the plasma extinguishes and interrupts the sputtering process. Sputtering deposition offers generally the advantage of comparable high surface smoothness.

    \begin{figure}
        \centering

            \psfrag{1}[c][c][1.5][0]{\textbf{\textsf{\textcolor{white}{1}}}}
            \psfrag{2}[c][c][1.5][0]{\textbf{\textsf{2}}}
            \psfrag{3}[c][c][1.5][0]{\textbf{\textsf{\textcolor{white}{3}}}}
            \psfrag{4}[c][c][1.5][0]{\textbf{\textsf{\textcolor{white}{4}}}}
            \psfrag{5}[c][c][1.5][0]{\textbf{\textsf{\textcolor{white}{5}}}}

            \psfrag{22.64mum}[c][c][0.9][0]{\SI{22.64}{\micro m}}
            \psfrag{1mum}[c][c][0.9][0]{\SI{1}{\micro m}}
            \psfrag{100nm}[c][c][0.9][0]{\textcolor{white}{\SI{100}{nm}}}
            \psfrag{670nm}[c][c][0.9][0]{\textcolor{white}{\SI{670}{nm}}}
            \psfrag{10mum}[c][c][0.9][0]{\textcolor{white}{\SI{10}{\micro m}}}
            \psfrag{C}[c][c][0.9][0]{\textcolor{white}{C}}
            \psfrag{Au}[c][c][0.9][0]{\textcolor{white}{Au}}
            \psfrag{Silica}[c][c][0.9][0]{\textcolor{white}{Silica}}
            \psfrag{SEM 3.74k x}[c][c][0.9][0]{\textcolor{white}{SEM \SI{3.74}{k}$\times$}}
            \psfrag{SEM 1.5k x}[c][c][0.9][0]{\textcolor{white}{SEM \SI{1.5}{k}$\times$}}
            \psfrag{SEM 104k x}[c][c][0.9][0]{\textcolor{white}{SEM \SI{104}{k}$\times$}}
            \psfrag{FIB-SEM 1.5k x}[c][c][0.9][0]{\textcolor{white}{FIB-SEM \SI{1.5}{k}$\times$}}
            \psfrag{FIB-SEM 72k x}[c][c][0.9][0]{\textcolor{white}{FIB-SEM \SI{72}{k}$\times$}}
            \psfrag{SEM 44.9k x}[c][c][0.9][0]{SEM \SI{44.9}{k}$\times$}
            \includegraphics[width=1\textwidth]{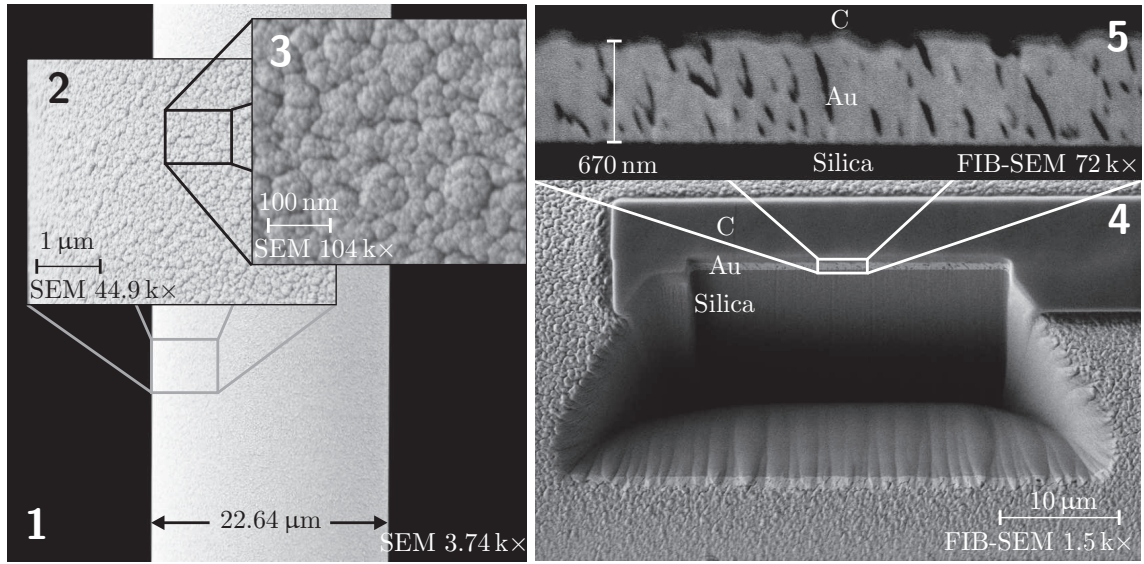}

        \caption[Scanning electron micrograph and FIB cut into the surface of a Au sputtered sample.]{Scanning electron micrograph and FIB cut into the surface of a sample with diameter \SI{22.64}{\micro m} and a thickness of the sputtered gold layer of \SI{670}{nm}. Insets 1 to 3 show SEM micrographs of the surface with different magnifications. Insets 4 and 5 depict the result of a focused ion beam cut perpendicularly into that surface. During the cutting process a thin layer of carbon (C) was deposited on top of the gold (Au). Clearly the structure of the gold layer is visible. The nm range thin contact layer of Cr below the Au layer is not visible.  \label{graphic_fib-sem}}

\end{figure} 

    To achieve a minimum surface roughness (A), thus grain size, a precise adjustment of the sample temperature with respect to the melting point of the sputtered material ($T/T_m$) is essential as is illustrated by fig.\;\ref{graphic_tapering-sputtering}, inset (c). Too high temperatures support the formation of large grained material spheres on the substrate surface (D). A compromise was chosen, also taking into account that the very thin silica substrate fibers are very sensitive to high temperatures in the geometry as they are perpendicularly mounted in the sputtering chamber.\\

    To monitor the surface quality, usually examination with an optical microscope\footnote{Nikon Eclipse LV100 microscope with a maximum magnification of $100 \times$.} was sufficient. For adjusting the sputter deposition paramers, using a focussed ion beam (FIB) and SEM system\footnote{Zeiss Gemini Nvision 40 CrossBeam. The measurements were conducted with the kind help of Helga Hussy and Daniel Plo{\ss}}, a series of micrographs of the obtained samples was taken as shown in fig.\;\ref{graphic_fib-sem}. In pictures (1-3) micrographs of the surface of a sample with an outer diameter of \SI{22.64}{\micro m} are shown. The sample exhibits a rather small corning size of $<\SI{100}{nm}$.

    For the examination of a cross section of the sputtered layer a thin layer of carbon is first deposited on the sputtered layer of gold (Au), allowing a perpendicular fib cut with accelerated gallium ions into the layer, whilst inevitably the removed matter deposits around the cutting place. Insets (4,5) show the resulting view of a cross section of the sputtered Au layer. The layer thickness is, as calculated from the assumed sputtering rates, \SI{670}{nm}. A distinct structure inside the layer is visible with thin longitudinal hole sizes up to half the layer thickness. Successive cutting and SEM observation of the layer disclosed that these structures do not show a large lateral extension.

    Nevertheless it is assumed that the observed structure degrades the SPP excitation properties of the sample. A further reduction of the hole structure of the sample layer could therefore decrease the impurity level of the metal, hence further increase the conductivity and support the SPP excitation and guidance properties.

    \section{Wires in fiber silica matrix \label{ch_sample-silica}}

    To fabricate metal wires of very low diameters and very high aspect ratios a technique is applied that was recently published by \citet{schmidt-prb-arrays-08}. Silica glass fibers are drawn that provide a hole structure with longitudinal extension in a specifically defined order and size. These holes are then filled with metals, which form wires of very small diameters ($D\gtrsim \SI{400}{nm}$) and extraordinary high aspect ratios.

    \subsection{Fiber drawing \label{ch_fiber-drawing}}

    Special fibers with a single, centered hole with diameters down to $d=\SI{400}{nm}$ are drawn by the method of PCF drawing. These fibers are used as matrix for nanowire arrays. The drawing technique shall be briefly explained\footnote{An in depth review of the applications, properties and fabrication of photonic crystal fibers can be found in the reviews of \citet{russell-2006}\citep{russell-science}.}.

    \begin{figure}[width=\textwidth]
        \centering

            \psfrag{A}[c][c][1.5][0]{\textbf{\textsf{A}}}
            \psfrag{B}[c][c][1.5][0]{\textbf{\textsf{B}}}
            \psfrag{fiber}[c][c][0.9][0]{fiber}
            \psfrag{induction coil}[cr][cr][0.9][0]{induction coil}
            \psfrag{graphite susceptor}[cr][cr][0.9][0]{graphite susceptor}
            \psfrag{Au}[c][c][0.9][0]{Au}
            \psfrag{1000C}[c][c][0.9][0]{\SI{1000}{\degC}}
            \psfrag{pr cell}[tr][tr][0.9][0]{\parbox{20mm}{\flushright pressure\\cell}}
            \psfrag{N2pr}[tl][tl][0.9][0]{\parbox{20mm}{\flushleft $\text{N}_2$\\high pressure}}
            \psfrag{preform}[c][c][0.9][0]{preform}
            \psfrag{oven}[c][c][0.9][0]{oven}

            \includegraphics[width=0.75\textwidth]{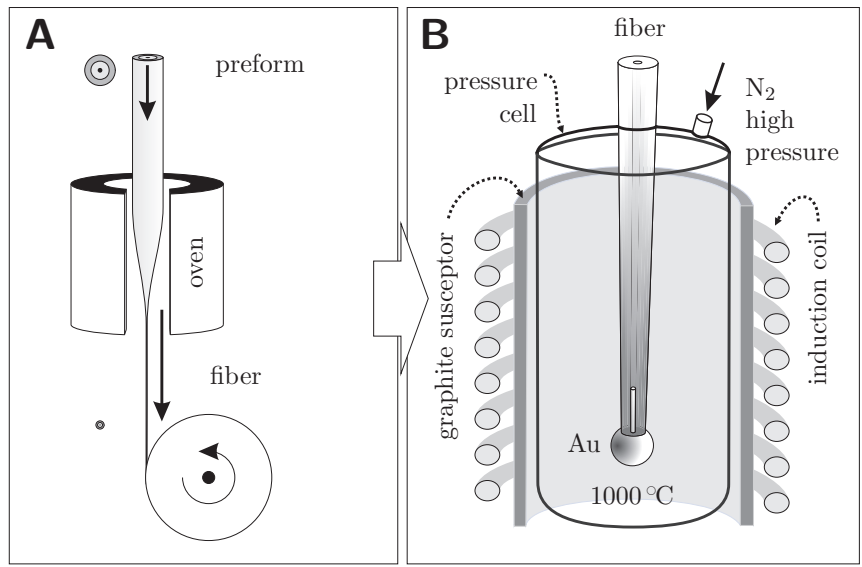}

        \caption[Schematic picture of the fiber drawing and the hollow fiber filling process for sample fabrication.]{Schematic drawing of the fiber drawing and the hollow fiber filling process for sample fabrication. The left drawing shows a principle sketch of the drawing process for the fabrication of PCF or single hole capillary fiber. Inset B shows the high pressure filling process inside an induction over to fabricate the wires inside the silica fiber matrix.  \label{graphic_drawing-filling}}

\end{figure} 

    The PCFs that are used for the samples are made using a stacking and drawing technique. The first stage is to produce a preform. This macroscopic array of capillaries with the same hole geometry that the final fiber is stacked by silica glass capillaries that are previously drawn from larger glass tubes to the desired diameters. These capillaries are stacked horizontally in a suitably shaped jig. The finished stack is inserted into a jacketing glass tube. The whole arrangement is mounted in the preform feed of a drawing tower (fig.\;\ref{graphic_drawing-filling}\,(a)).

    \begin{figure}
        \centering

            \psfrag{1}[c][c][1.5][0]{\textbf{\textsf{\textcolor{white}{1}}}}
            \psfrag{2}[c][c][1.5][0]{\textbf{\textsf{\textcolor{white}{2}}}}
            \psfrag{3}[c][c][1.5][0]{\textbf{\textsf{\textcolor{white}{3}}}}
            \psfrag{4}[c][c][1.5][0]{\textbf{\textsf{\textcolor{white}{4}}}}

            \psfrag{100mu}[c][c][0.9][0]{\textcolor{white}{\SI{100}{\micro m}} }
            \psfrag{200 nm}[c][c][0.9][0]{\textcolor{white}{\SI{200}{n m}} }
            \psfrag{4mu}[c][c][0.9][0]{\textcolor{white}{\SI{4}{\micro m}}}

            \includegraphics[width=1\textwidth]{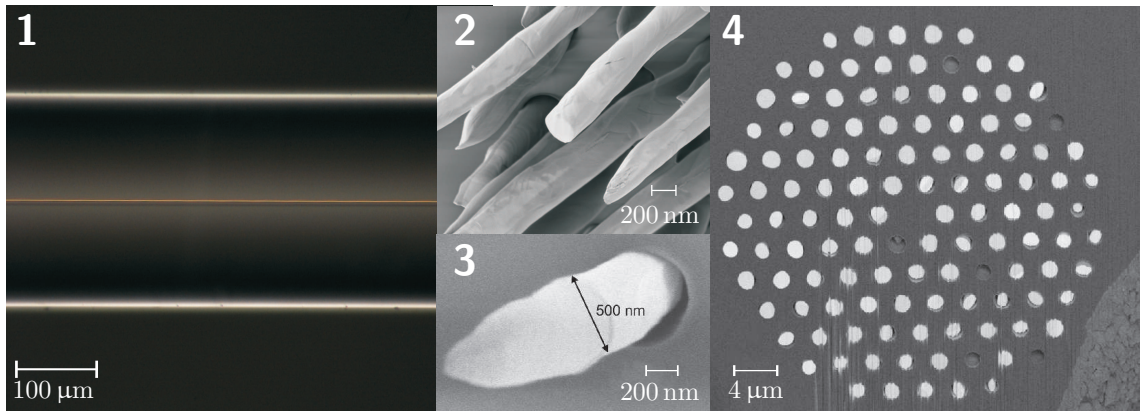}

        \caption[Optical microscope picture and scanning electron micrograph of wire gold filled single and array fibers.]{Lateral optical microscope picture of a gold filled single hole capillary fiber (1) and scanning electron micrographs of gold array filled photonic crystal fibers (2-4). (4) shows a FIB polished cleaved fiber end face. The wires extend the fiber end face after cleaving the fiber due to the high ductility of gold.  \label{graphic_sem-wires}}

\end{figure} 

    The preform feed slowly moves the preform into a resistance furnace which heats the silica glass up to temperatures near the softening temperature (tab.\;\ref{tab_material-properties}). Therefore the silica glass preform softens. Due to the drawing force from a cane pulling capstan the sizes of the preform decrease whilst intercapillary holes close, which is supported by an applied vacuum. The resulting canes once again are inserted into a jacket tube.

    The same process is applied again to this new preform, drawing it down to outer diameters of $\SI{125}{\micro m}$ or in this case to $\SI{200}{\micro m}$. The resulting hole size depends sensitively on all drawing parameters and can therefore be varied over a wide range from \SI{300}{nm} to several $\micro\text{m}$.\\

    The same procedure is applied for fabricating single hole capillary fibers where stacking dispenses and a silica glass tube is instead drawn in two steps and inserted into jacket tubes to achieve the very low diameters.

    The final silica glass fibers are coated with a UV hardened polymer coating to prevent it from mechanically harmful influences and to support their mechanical stability.

    \subsection{Fabrication of wires in fibers}

    As recently success on fabricating wire arrays inside silica glass photonic crystal fibers was reported \citep{schmidt-optl-numerical-07,schmidt-prb-arrays-08,schmidt-science,fio-2007}, for the sample fabrication in the course of this work  the same technique was applied.

    It can be seen from the values in table \ref{tab_material-properties} that the melting point of the applied metals is much lower than the softening point of silica glass. Therefore it is possible to pump molten silver and gold into the holes of PCF or single hole capillary. \\

    The fabricated fiber is first stripped of its protective coating which would start burning when heated to the appropriate temperatures. The fiber is then inserted into a pressure cell made of thick silica glass. The insertion port is well isolated to remain high pressure, thus the hole structure in the fiber provides the only way out of the cell. The fiber's inner end is heated up above the melting point of Au, respectively Ag. At the same time its end face dips completely into the liquid metal. When applying a high pressure inside the pressure cell, the metal intrudes the fiber. After cooling down the whole arrangement to room temperature, thin metal wires have formed inside the fiber matrix.

    Thin wires with a diameter of \SI{400}{nm} are achieved to produce successfully with this method. It is shown from electrical conductivity measurements, that the wires inside the fiber are continuous for gold samples. \\

    It is reasonable that due to the different thermal expansion coefficients of the metals and the surrounding silica matrix ($\alpha_\text{exp}$, tab.\;\ref{tab_material-properties}), tiny gaps at the material interface form under certain cooldown configurations. This is proved by an optical microscope examination from a side view on the fiber as well as from a comparison of previous measured experimental results, achieved by in-fiber optical excitation of SPP on the wire arrays to computed theoretical results \citep{schmidt-prb-arrays-08}.

    The fabricated fibers containing wire arrays and single wires are cleaned from remains from the filling process that are found on the outer surface of the fiber using a ultrasonic cleaning device with a cascade of extrane\footnote{A laboratory detergent to remove especially organic contaminations that is particularly used for the first step of sample preparation for SEM and other high vaccum applications.}, acetone ($\text{CH}_3\text{COCH}_3$), isopropanol ($\text{C}_3\text{H}_8\text{O}$) and ethanol ($\text{C}_2\text{H}_6\text{O}$).

    During the whole filling process they have to be handled with special care. For common fiber applications contaminations of the outer surface of the silica fibers do usually not play a significant role. In the case of side scattering, the light has to pass the outer surface without any measurable distortions. Especially contaminations that occur before or during the heating process are very difficult to impossible to remove after the cooldown. However, it is possible to reduce those to a sufficient level. Additionally recent further enhancement of the filling technique supports to avoid these effects. 

\chapter{Experimental setup \label{ch_experimental-setup}}

\vspace{-3em}
\begin{flushright}\parbox[t]{\textwidth/2}{%
        \small{``In every branch of knowledge the progress is proportional to the amount of facts on which to build, and therefore to the facility of obtaining data.''\\ \slshape{(James Clerk Maxwell, 1851
        )}}}
\end{flushright}

\section{General setup}

\begin{figure}[width=\textwidth, tbp]
        \centering

            \psfrag{A}[c][c][1.5][0]{\textbf{\textsf{A}}}
            \psfrag{Beam preparation}[c][c][0.9][0]{Beam optics}
            \psfrag{B}[c][c][1.5][0]{\textbf{\textsf{B}}}
            \psfrag{Scattering}[c][c][0.9][0]{Scattering}
            \psfrag{C}[c][c][1.5][0]{\textbf{\textsf{C}}}
            \psfrag{Measuring}[c][c][0.9][0]{Measuring}
            \psfrag{D}[c][c][1.5][0]{\textbf{\textsf{D}}}
            \psfrag{Control and data acquisition}[c][c][0.9][0]{Control and data acquisition}
            \psfrag{incident beam}[c][c][0.8][0]{\textcolor{white}{incident beam}}
            \psfrag{al}[c][c][0.8][0]{$\alpha$}
            \psfrag{g}[c][c][0.8][0]{$\gamma$}
            \psfrag{g_}[c][c][0.8][0]{$\gamma'$}
            \psfrag{detector}[c][c][0.8][0]{detector}
            \psfrag{wire}[c][c][0.8][0]{wire}

            \includegraphics[width=\textwidth]{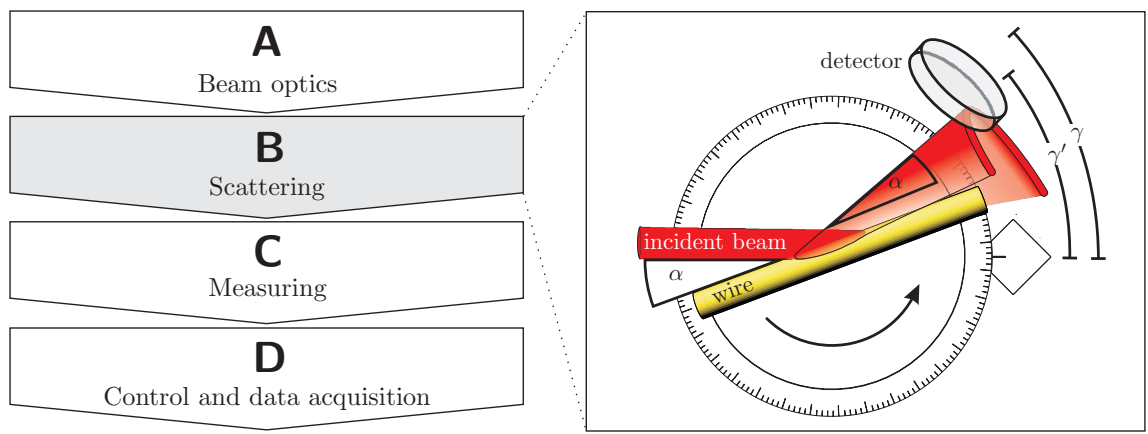}

        \caption[Schematic of the experimental setup.]{Principle scheme of the experimental setup. Each part of the experimental setup A, B, C and D is explained separately. The diagram on the right depicts the scattering process with incident angle $\alpha$. The scattering maximum cone is drawn with the scattering maximum angle indicated as $\gamma'=2 \alpha$ and the angle of observation as $\gamma$.\label{graphic_setup-scheme}}

\end{figure} 

\noindent The experimental setup\footnotetext{All abbreviations that are used during the explanation of the setup can be found in appendix \ref{ch_appendix-conventions} with a reference to the page where the abbreviation is first introduced and explained.} is designed for measuring optically the scattering amplitude of a sample, either small freestanding wire or embedded micro- to nano-wire in silica. The scattering amplitude (fig. \ref{graphic_setup-scheme}) is acquired by varying the incident angle $\alpha$ of a collimated laser beam at a wavelength of \SI{632.8}{nm}. The adjustable angle of observation $\gamma$ is kept constant for each measurement. The maximum intensity of the reflected light is observed at twice the incident angle $\gamma^\prime = 2 \alpha$, as derived in sec.\;\ref{ch_scattering-pattern}.
\nomenclature[mg]{$\gamma$}{Angle of observation, angle of the measurement photo detector with respect to the incident beam.}

For a constant wavelength $\lambda$ surface plasmon modes on the cylindrical surface are excitable only for discrete incident angles $\alpha_m$, as shown in sec.\;\ref{chapter_theory-model}. For these angles an abrupt coupling to the SPP and therefore a corresponding decrease of the reflectivity of the cylinder is anticipated. Of special interest are lower order SPP modes as these are predicted (see sec.\;\ref{ch_scattering}) to cause more distinct dips in the reflectivity. Particularly low incident angles are probed, as the mode number increases with larger incident angle $\alpha$. The setup also allows measuring larger incident angles $\alpha$ as long as the range of the scattering angle $\gamma'$ does not exceed the size of the measurement sensor (PD2) which is situated close to the scattering centre.

The measurements are controlled by means of a personal computer with a program that acquires amplitude values for each angle step by step. With the developed setup it is possible to measure the scattered amplitude  for two different, perpendicular polarizations simultaneously. Thus the setup provides access to the polarization dependent scattering amplitudes for s- and p-polarized light (sec.\;\ref{chapter_theory-scattering}). This not only makes a faster measurement possible but also minimizes artifacts from remaining vibrations or small movements of the probe.

\begin{figure}[width=\textwidth, bt]
        \centering

            \psfrag{A}[c][c][1.5][0]{\textbf{\textsf{A}}}
            \psfrag{B}[c][c][1.5][0]{\textbf{\textsf{B}}}
            \psfrag{C}[c][c][0.9][0]{C}
            \psfrag{D}[c][c][0.9][0]{D}
            \psfrag{Laser}[c][c][0.8][0]{Laser}
            \psfrag{OI}[l][l][0.9][0]{OI}
            \psfrag{M1}[l][l][0.9][0]{M1}
            \psfrag{M2}[l][l][0.9][0]{M2}
            \psfrag{HWP1}[r][r][0.9][0]{HWP1}
            \psfrag{PBS1}[l][l][0.9][0]{PBS1}
            \psfrag{A1}[l][l][0.9][0]{A1}
            \psfrag{chopper}[l][l][0.9][0]{chopper}
            \psfrag{M4}[l][l][0.9][0]{M4}
            \psfrag{M3}[l][l][0.9][0]{M3}
            \psfrag{A2}[l][l][0.9][0]{A2}
            \psfrag{PBS2}[l][l][0.9][0]{PBS2}
            \psfrag{M5}[l][l][0.9][0]{M5}
            \psfrag{M6}[l][l][0.9][0]{M6}
            \psfrag{HWP2}[l][l][0.9][0]{HWP2}
            \psfrag{L1}[l][l][0.9][0]{L1}
            \psfrag{PM-SMF}[l][l][0.9][0]{PM-SMF}
            \psfrag{path 1}[l][l][0.9][0]{path 1}
            \psfrag{path 2}[l][l][0.9][0]{path 2}
            \psfrag{stage 1}[l][l][0.9][0]{stage 1}
            \psfrag{stage 2}[l][l][0.9][0]{stage 2}
            \psfrag{FRM}[l][l][0.9][0]{FRM}
            \psfrag{BS3}[l][l][0.9][0]{BS3}
            \psfrag{PF}[l][l][0.9][0]{PF}
            \psfrag{PD1}[l][l][0.9][0]{PD1}
            \psfrag{PD2}[l][l][0.9][0]{PD2}
            \psfrag{L2}[l][l][0.9][0]{L2}
            \psfrag{Manual RS}[l][l][0.9][0]{manual RS}
            \psfrag{sample}[l][l][0.9][0]{sample}
            \psfrag{Motorized RS}[l][l][0.9][0]{motorized RS}
            \includegraphics[width=\textwidth]{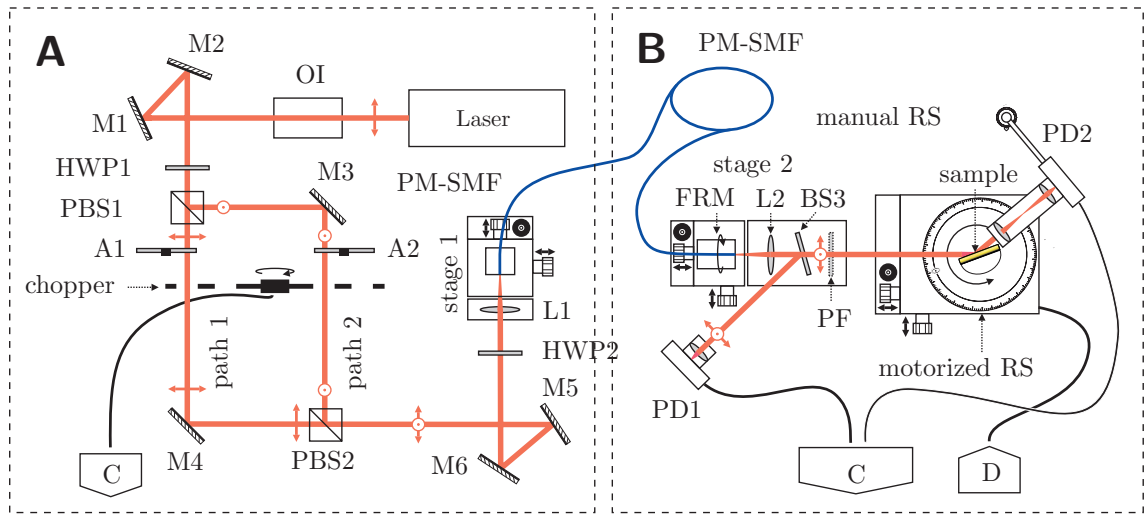}

        \caption[Schematic general diagram of the optical setup showing part A for polarization dividing beam chopping and B for the scattering process.]{The schematic diagram of part A of the setup shows the light source and the polarization dividing beam chopping as described in sec.\;\ref{chapter_beam-optics}. The resulting beam is coupled into polarization maintaining fiber (PM-SMF). Part B shows the main part of the scattering setup as described in section \ref{ch_scattering} with the reference photo detector (PD1), the rotated sample and the main photo detector (PD2).\label{graphic_setup-general}}

\end{figure} 


For the following explanations the entire setup is divided into parts A and B for the optical experiment, C for the electronics and measurement devices and D for the software, control and data acquisition system.\\

In \textbf{part A} the laser beam is prepared for the measurement as described in sec.\;\ref{chapter_beam-optics} (left inset in fig.\;\ref{graphic_setup-a}). For a simultaneous measurement of both polarizations the beam is divided into orthogonal polarizations, each of them being chopped by a different frequency and thus labelled by a frequency envelope on the signal. Hence the two polarization dependent signals can be separated again after probing. It is possible to measure the scattering response for each polarization as long as the two polarizations are not mixed before scattering and independent of possible mixing of the polarizations after the scattering process. The two modulated beams are superimposed again and launched into a polarization maintaining fiber (PM-SMF) that guides the light to part B of the setup.\\

\textbf{Part B} covers the scattering probing itself as depicted in fig.\;\ref{graphic_setup-b} and is described thoroughly in sec.\;\ref{ch_scattering}. The superimposed orthogonal polarizations are collimated and directed towards the sample. The scattering wire is rotated around the vertical axis perpendicular to the cylinder axis (fig.\;\ref{graphic_setup-scheme}) in small steps by a motorized rotation stage (motorized RS) that is controlled by the software as explained in part D. A reference signal from the beam before scattering is acquired by a photo detector (PD1), which is connected to part C of the setup by BNC cables. The scattering amplitude is measured by a {second}, amplified photo detector (PD2) at a constant but adjustable angle of observation $\gamma$. \\

\textbf{Part C} comprises the electronic processing of the acquired signals, the signal separation for the different polarizations by means of three lock-in amplifiers, the analog-to-digital conversion and noise suppression. The setup is described in detail in section \ref{chapter_measurement}.\\

In \textbf{part D} the computer program is presented which was developed in the proprietary visual programming environment National Instruments LabView. It provides control of the step motor, lock-in amplifiers and measurement devices, acquiring the measurement signals, processing them, cancelling noise. Additionally it acquires monitoring pictures via a digital adjustment microscope that looks on the rotated sample from top continuously allowing an error analysis of the acquired measurement data.

\section{Beam optics\label{chapter_beam-optics}}
\subsection{The light source\label{chapter_light-source}}

        \begin{figure}[width=\textwidth, btp]
        \centering

            \psfrag{A}[c][c][1.5][0]{\textbf{\textsf{A}}}
            \psfrag{B}[c][c][0.9][0]{B}
            \psfrag{C}[c][c][0.9][0]{C}
            \psfrag{Laser}[c][c][0.9][0]{Laser}
            \psfrag{OI}[l][l][0.9][0]{OI}
            \psfrag{M1}[l][l][0.9][0]{M1}
            \psfrag{M2}[l][l][0.9][0]{M2}
            \psfrag{HWP1}[r][r][0.9][0]{HWP1}
            \psfrag{PBS1}[l][l][0.9][0]{PBS1}
            \psfrag{A1}[l][l][0.9][0]{A1}
            \psfrag{chopper}[l][l][0.9][0]{chopper}
            \psfrag{stage 1}[l][l][0.9][0]{stage 1}
            \psfrag{M4}[l][l][0.9][0]{M4}
            \psfrag{M3}[l][l][0.9][0]{M3}
            \psfrag{A2}[l][l][0.9][0]{A2}
            \psfrag{PBS2}[l][l][0.9][0]{PBS2}
            \psfrag{M5}[l][l][0.9][0]{M5}
            \psfrag{M6}[l][l][0.9][0]{M6}
            \psfrag{HWP2}[l][l][0.9][0]{HWP2}
            \psfrag{L1}[l][l][0.9][0]{L1}
            \psfrag{PM-SMF}[l][l][0.9][0]{PM-SMF}
            \psfrag{path 1}[l][l][0.9][0]{path 1}
            \psfrag{path 2}[l][l][0.9][0]{path 2}
            \psfrag{f_outer}[l][l][0.9][0]{$f_\text{outer}$}
            \psfrag{f_inner}[l][l][0.9][0]{$f_\text{inner}$}
            \psfrag{f_innerformula}[l][l][0.9][0]{$f_\text{outer} = \frac{6}{5}f_\text{inner}$}
            \psfrag{amplitude}[l][l][0.9][0]{amplitude}
            \psfrag{t}[l][l][0.9][0]{$t$}

            \includegraphics[width=\textwidth]{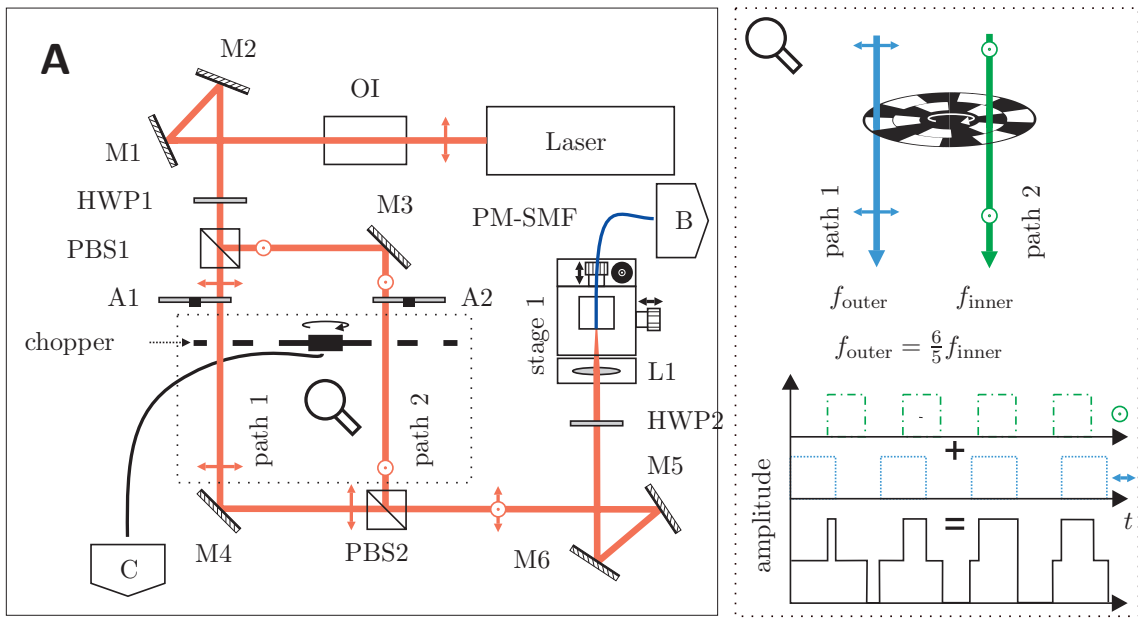}

        \caption[Detailed schematic diagram of part A of the setup and explanation of the polarization dividing beam chopping with two different frequencies in detail.]{Detailed schematic diagram of part A of the setup. The beams of perpendicular polarizations in the two paths are chopped and thus labelled with different frequencies $f_1=f_\text{outer}$ and $f_2=f_\text{inner}$ by one chopper before they are superimposed. The resulting waveform of the total amplitude is sketched as a function of time.\label{graphic_setup-a} }

        \end{figure} 

A Helium-Neon laser\footnote{Thorlabs HRP 120} with a power of \SI{12}{mW} operating at $\lambda=\SI{632.8}{nm}$ was utilized as a source of monochromatic, collimated light. Its beam is linearly polarized ($>500:1$), single moded ($\text{TEM}_{00}>99\%$) with a $1/\e^2$ diameter of \SI{0.88}{mm} and a divergence of \SI{0.92}{mrad}.

\nomenclature[s]{Laser}{HeNe laser}

As first test measurements showed, it is necessary to stabilize the laser against backscattered light from the scattering process and further optical components. Therefore an optical Faraday isolator\footnote{OFR/Thorlabs IO-2D-633-VLP Free-Space Isolator, 633 nm, 2 mm Aperture, isolation 35-40 dB} (OI) is introduced in the optical beam path after the laser.

\nomenclature[s]{OI}{Optical isolator}

The optical isolator passes light into the intended direction just with a low intrinsic loss of \SI{30}{\%}. In fact, a power of \SI{8}{mW} of the initial \SI{12}{mW} is measured to be transmitted. At the same time the polarization of light travelling the optical path backwards is attenuated by $35-\SI{40}{dB}$.

After the application of the optical isolator, fluctuations of the laser are well suppressed to a level where referencing the measured results leads to a sufficient signal-to-noise-ratio. The laser is operated at least some hours after switch-on to ensure that it is stabilized.

\subsection{Polarization labelling by beam chopping\label{chapter_chopping}}

With the task of measuring scattering effects for the two polarizations synchronously, a system was set up to deliver two perpendicular polarizations in precisely one beam path, each of them chopped with a different frequency. For that purpose a polarization dividing setup in the geometry of a Mach-Zehnder interferometer is employed as depicted in fig.\;\ref{graphic_setup-a}.\\

The linear polarization of the laser beam is tilted by a half-wave plate (HWP1), optimized for $\lambda=\SI{632.8}{nm}$ and is split by a polarizing beam splitter cube (PBS1)\footnote{Linos G335592000, \SI{5}{mm}, design wavelength \SI{633}{nm}}.
\nomenclature[s]{HWP1,HWP2}{Half wave plates with a design wavelength of \SI{633}{nm}}
By adjusting the angle of HWP1, thus it is possible to change the fraction of horizontally polarized light passing straight through PBS1 (path 1) and vertically polarized light that is by alignment reflected to an angle of \ang{90} with respect to path 1 (path 2). In order to be able to shut path 1 or path 2 separately without unbalancing HWP1 an adjustable optical attenuator (A1, A2) is passed by each path.\\

\nomenclature[s]{A1,A2}{Adjustable optical attenuators}
\nomenclature[s]{PBS1,PBS2}{Polarizing beam splitters}


Subsequently an optical chopper\footnote{Stanford Research Systems SR540 optical chopper system: Frequency stability \SI{250}{ppm/ \degC}, (typ.) frequency drift $< 2 \%$, $\SI{100}{Hz} < f < \SI{3700}{Hz}$, phase jitter (rms) \ang{0.2} (\SI{50}{Hz} to \SI{400}{Hz}) respectively \ang{0.5} (\SI{400}{Hz} to \SI{3.7}{kHz}) } is situated, chopping of both beams at once with two different frequencies. Therefore the chopper wheel consists of two rings, an outer ring formed by 6 holes for path 1 and an inner ring with 5 holes for path 2.

\nomenclature[s]{SR540}{Optical chopper system}

So beam 1 is chopped by  $f_\text{outer}$ and beam 2 by $f_\text{inner}=\frac{5}{6}f_\text{outer}$. The chopper itself is controlled by a stabilized controller (\SI{250}{ppm/ \degC} with a typical frequency drift $< 2 \%$). The chopper also measures the two frequencies via slotted optical switches on the chopper wheel and two optical barriers and delivers them as reference signal.

\nomenclature[sb]{BNC}{Bayonet Neill Concelman - coaxial connectors and cables with an impedance of \SI{50}{\Omega}.}
\nomenclature[sf]{$f$}{Technical frequency, particularly the signal envelope frequencies, unit $[f] = s^{-1}=\text{Hz}$.}

The two chopper hole rings are arranged in such a way that the duty cycles (the fraction of light time to $T=1/f$), of the inner and of the outer ring are similar, both \SI{50}{\%}. This implicates that there are phases (\SI{30}{\%} of the time) where both beams are open at once. Even if the polarization of both beams is well adjusted to be perpendicular, a decrease in the duty cycle minimizes the time that both beams pass the chopper simultaneously.

Such a reduction of the duty cycle is possible by applying two similar chopper wheel blades stacked and slightly tilted with respect to each other. This precisely reduces the opening aperture of each chopper hole opening in the direction of rotation. The chopper controller frequency measurement for one of the two rings is obfuscated due to technical reasons of the frequency measurement. This problem is solved by a  slight modification of the chopper wheels was applied to enable a correct frequency measurement again.

At the output of the Mach-Zehnder geometry the two beams from both paths are superimposed by a second polarizing beam splitter (PBS2) with low loss (typical PBS extinction ratio $>500:1$). Again the horizontally polarized beam path (1) passes the beam splitter straight through, the vertically polarized beam (path 2) is reflected. For a resulting parallel and precise superposition of the two beam paths an adjustment by tilting PBS1 and PBS2 as well as the mirrors M3 and M4 is required.

\nomenclature[s]{M1-M6}{Mirrors}

The two superimposed beams pass alignment mirrors (M5, M6) that allow tilting and translation of the beam and a second half-wave plate (HWP2). For the later scattering part a flexible adjustment of the launching position of the beam is necessary. Therefore the beam is coupled into three metres of bow tie polarization maintaining \SI{633}{nm} single mode fiber (PM)\footnote{Fibercore Limited HB 600} (effective mode diameter of \SI{3.2}{\micro m} and a numerical aperture of $\SI{0.14}{}-\SI{0.18}{}$).\\

\nomenclature[s]{PM}{Polarization maintaining single mode fiber.}

The second half-wave plate (HWP2) provides rotation of the two perpendicular polarizations to align them parallel to the two symmetry axes of the polarization maintaining fiber (PM-SMF) into which the beam is coupled by means of a fiber coupling stage\footnote{Elliot Scientific Gold Series XYZ Flexure Stage with High Precision Manual Adjusters}.

\nomenclature[s]{stage1,stage2}{Fiber coupling stage with micrometer screws.}

For efficient coupling, the beam is focused by an aspheric lens (L1) with a numeric aperture of \SI{0.5}{} on a straight cleaved fiber end face by a typical efficiency of $40$ to \SI{50}{\%}. Cladding modes are sufficiently removed by coiling the PM-fiber with a bending radius of \SI{15}{cm}.\\

\nomenclature[s]{L1,L2}{Aspheric biconvex lenses}

\section{Scattering \label{ch_scattering}}

    \noindent To perform the actual scattering process, two different physical configurations of the setup are applied:

    \begin{enumerate}
  \item A `bottom up' approach, explained in sec.\;\ref{ch_setup-freestanding}, is the most obvious way for probing scattering wires that are freestanding in air. It is also the most stable, as standing optomechanical standard laboratory equipment is designed for minimizing mechanical vibrational effects.

    Also silica embedded nanowire arrays can be probed in this configuration showing a superposition of the scattering effects on the surface of the supporting silica fiber, surrounding the wires, and the effects of scattering on the wires and wire arrays themselves.

  \item The latter measurement can be performed without the superposed effects, thus significantly more advanced, with the second configuration as described in section \ref{ch_setup-liquid}: Surrounding the wire supporting fibers with index matching liquid, which matches the index of refraction of silica whilst still enabling the rotation of the sample. This makes a more sophisticated `top down' setup necessary with the sample hanging down from the motor into the liquid. In this configuration the supporting fiber becomes invisible as the refractive index difference of its surface to the surrounding medium becomes zero.

    \end{enumerate}

            \begin{figure}[width=\textwidth]
        \centering

            \psfrag{B}[c][c][1.5][0]{\textbf{\textsf{B}}}
            \psfrag{C}[c][c][0.9][0]{C}
            \psfrag{D}[c][c][0.9][0]{D}
            \psfrag{L1}[l][l][0.9][0]{L1}
            \psfrag{PM-SMF}[l][l][0.9][0]{PM-SMF}
            \psfrag{path 1}[l][l][0.9][0]{path 1}
            \psfrag{path 2}[l][l][0.9][0]{path 2}
            \psfrag{stage 1}[l][l][0.9][0]{stage 1}
            \psfrag{stage 2}[l][l][0.9][0]{stage 2}
            \psfrag{stage 3}[l][l][0.9][0]{stage 3}
            \psfrag{stage 4}[l][l][0.9][0]{stage 4}
            \psfrag{RS}[l][l][0.9][0]{RS}
            \psfrag{FRM}[l][l][0.9][0]{FRM}
            \psfrag{BS3}[l][l][0.9][0]{BS3}
            \psfrag{PF}[l][l][0.9][0]{PF}
            \psfrag{PD1}[l][l][0.9][0]{PD1}
            \psfrag{PD2}[l][l][0.9][0]{PD2}
            \psfrag{L2}[l][l][0.9][0]{L2}
            \psfrag{Manual RS}[l][l][0.9][0]{manual RS}
            \psfrag{sample}[l][l][0.9][0]{sample}
            \psfrag{CCD}[l][l][0.9][0]{CCD}
            \psfrag{zoom}[l][l][0.9][0]{}
            \psfrag{20x}[l][l][0.9][0]{$20 \times$}

            \psfrag{100mum}[c][c][0.7][0]{\textcolor{white}{\SI{100}{\micro m}}}

            \psfrag{Motorized RS}[l][l][0.9][0]{motorized RS}
            \psfrag{a}[l][l][0.9][0]{\textcolor{white}{$\alpha$}}
            \psfrag{g}[l][l][0.9][0]{\textcolor{white}{$\gamma$}}
            \psfrag{50mum}[c][c][0.9][0]{\textcolor{white}{\SI{50}{\micro m}}}
            \includegraphics[width=\textwidth]{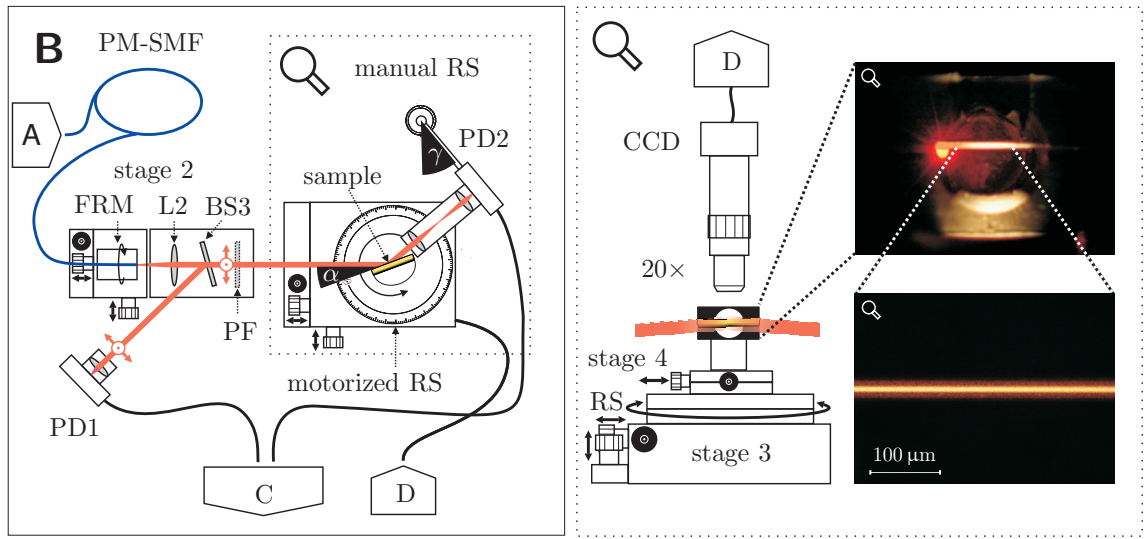}

        \caption[Detailed schematic diagram of part B of the experimental setup, showing the optical setup for the scattering probing.]{The schematic diagram of part B shows the main part of the scattering setup. As described in sec.\;\ref{ch_scattering} the light is collimated and directed onto the sample with the incident angle $\alpha$ which is stepwise successively varied by a motorized rotation stage. The angle of observation $\gamma$ is kept constant for each measurement.\label{graphic_setup-b}}

        \end{figure} 

    The setup in either configuration is particularly sensitive to vibrations as a tiny movement (of the order of wire's size) of the scattering target can cause a significant influence on the scattering measurement. To reduce setup inherent noise sources it was decided not to incorporate active vibration damping but to use other techniques.

    A method to minimize the influence of unavoidable mechanical vibrations and to increase the overall effective Pointing stability is using a larger beam diameter. For larger beam diameters the wire is situated in the centre of a Gaussian beam intensity profile with increased total width. With the same Gaussian slope thus a small movement from the centre does not change the incident intensity as much as for a small beam diameter. However, the beam diameter can not be extended infinitely in reality and an increase in the diameter always reduces the portion of scattered light, reducing also the signal amplitude that is detected.

    A suitable compromise of a beam waist of the collimated Gaussian laser beam of \SI{800}{\micro m} was chosen. Further noise reduction is realized by a sophisticated measurement and acquisition scheme as described in sections \ref{ch_setup-lockindetection} and \ref{ch_setup-software}.

\subsection{Optics}

    Outcoupling from the polarization maintaining fiber (PM) is provided by a second fiber coupling stage in front of the scattering part of the setup. The stage is situated on the height level of the scattering setup and is equipped with a fiber rotation mount (FRM), by means of which the two perpendicular superimposed polarizations are adjusted to be vertically and horizontally oriented. This results in scattering with incident s-, and p polarization with respect to the sample and its vertical rotation axis (as defined accodring to sec.\;\ref{chapter_theory-scattering}).

    \nomenclature[s]{FRM}{Fiber rotation mount}

    It is necessary to couple in both polarizations precisely according to the symmetry axes of the fiber, otherwise the polarizations will start mixing over the length of the fiber. Measurements show that a stable separation of the two polarizations of $I_\text{s}:I_\text{p}=200:1$ is achieved with a precise adjustment of HWP2.

    In order to prevent both fiber end faces from collecting dust and being effected by ambient influences they are covered and continuously purged with nitrogen.\\

    The beam is collimated by an aspheric lens (L2). The $\e^{-2}$ diameter of the collimated beam was measured by beam profiling to be $D \approx \SI{800}{\micro m}$. A small fraction of the beam is then reflected by a beam splitter (BS3) (reflectivity \SI{8}{\%}) on a photo detector (PD1) for the measurement of a reference signal. This coated pellicle beam splitter (BS3) is not polarization sensitive, minimizes ghosting and internal reflection and thus virtually eliminates the amount of backscattered light from the scattering process potentially hitting the reference sensor. Additionally, the choice of a pellicle beam splitter minimizes the lateral beam offset.

    \nomenclature[s]{BS3}{Coated pellicle beam splitter}

    As the angle of BS3 can be adjusted, is mechanically stabilized and is chosen to be near to perpendicular incidence, the portion of the reflected reference beam is stable and always of sufficient amplitude whilst the amplitude of the main beam is not reduced significantly.\\

    The smaller the wire diameters, the smaller are the geometrical and effective cross sections (compare to the theoretical treatment in sec.\;\ref{chapter_theory-scattering}) of the scattering target, leading to a decrease in the total scattered intensity.

    This can in principle be compensated by focussing the light to a smaller Gaussian beam waist. On the other hand this modification leads inevitably to a broadening of the real incident angle $\alpha \pm \delta$. The incident angle $\alpha$ being best peaked ($\delta \approx 0$) for plane waves with a collimated but not focussed beam.

    The latter case of perfect plane waves was treated in chapter \ref{ch_theory} theoretically. Any broadening of the incident angle spectrum obviously leads to a broadening of the expected resonance dips, thus a scattering geometry without a focussing objective for the scattering beam was finally preferred.\\

\subsection{Free-standing scattering setup \label{ch_setup-freestanding}}

\subsubsection*{Sample stages and beam centering}

    The scattering sample is placed on a higher level than the other beam optical components: it is mounted on top of a number of stages allowing a stepwise rotation for the measurement itself as well as a centering of the sample in all spatial directions (fig.\;\ref{graphic_setup-b}).

    As the beam direction, height, and the focal point are predetermined by the, once coupled and roughly adjusted, invariable position of stage 2, the whole rotation setup is placed on a xyz-translation stage (stage 3) with micrometer precision. By means of this, the sample is moved to the beam position. As this stage carries a light load but a rather high superstructure, it is necessary to balance it out by small weight loads to suppress mechanical vibrations.

\nomenclature[s]{stage3}{xyz-translation stage with micrometer screws and long traverse path.}

    On top of stage 3 the motorized precision rotation stage is placed providing variation of the incident angle $\alpha$ for the measurement process\footnote{Physik Instrumente (PI) M-037.DG servo motorized, referenced precision rotation stage with PI mercury servo motor controller, USB programmable.}. The rotation stage offers a rotation range of $>2\pi$ with a smallest specified step width of \SI{3.5}{\micro {rad}}, a maximum speed of \SI{6}{^\circ/s} with a reverse backlash of \SI{200}{\micro {rad}}. This motor is programmed to operate at a maximum speed of \SI{0.25}{^\circ/s}, the actual speed being calculated from the step width by the control program to avoid the backlash error at very small step widths.
\nomenclature[s]{motorized RS}{Motorized precision rotation stage, programmable.}

    The rotation stage provides a magnet-and-Hall-sensor referencing instrument to ensure a reproducible starting position with very high accuracy. The system is controlled by a matched dc servo controller with feedback.

    On top of the rotation stage an xy-stage with micrometer screws is mounted (stage 4), used for centering the scattering sample to the rotation axis of the rotation stage. This manual centering whilst observing the incident beam and the wire from top by means of a digital zoom microscope camera (see section \ref{ch_setup-visual}), is the first step for adjusting the sample and selecting the center of rotation to be situated on the sample surface with micrometre precision. It is subsequently followed by the adjustment by means of stage 3 to the position of the incident beam.

\nomenclature[s]{stage4}{xy translation stage.}

\subsubsection*{Sample mounting}

    The thin wire samples as well as the wires supported by silica fiber are permanently attached to black PVC holders (picture inset in fig.\;\ref{graphic_setup-b}). The synthetic material offers the intrinsic advantages of high absorption and not causing any plasmonic effects.

    The holder is designed as a V-groove with a centred circular hole of a diameter that is large enough, to absorb the whole extension of the incident beam also for low incident angles $\alpha$. Thus any possible reflection that is not caused by the sample is suppressed. The sample is hold straight and without movement which otherwise might be caused by air turbulences or resonant vibrations.

    At the same time the central hole is utilized for mounting the holder as it is threaded inside. By attaching the exchangeable sample holder to its mount with a transparent PVC screw it is possible to adjust the height of the sample to the incident beam in a position of perpendicular incidence. The beam causes a backward scattering pattern that can be well observed by means of a ground glass plate. The transparent screw lights up if the beam passes the sample when not exactly hitting the wire's centre.

    A side effect of this sample mounting is that half of the probing beam for an incident angle of $\alpha \approx 0$ is blocked, absorbed and diffusely backscattered by the black holder.

\subsection{In-liquid setup\label{ch_setup-liquid}}

    For the second type of the measurement setup, the sample is placed inside a box which is filled with index matching liquid\footnote{Cargille Labs: Immersion liquid 19571 $n=\SI{1.4587}{} \pm \SI{0.0005}{} at \lambda=\SI{589.3}{nm} and \SI{25}{\degC}$} matching the refractive index of silica. Hence, probing is possible without additional optical effects from the surface of the supporting silica fibers.

    Very fine surcface contaminations, relicts of the filling process (described in section \ref{ch_sample-silica}), cannot be removed completely and still potentially cause scattering on interface of glass to index matching liquid.\\

    The box is made of acrylic glass with laser beam windows of silica that are slightly tilted against each other to suppress internal reflections at the silica to air boundary as well as any potential etalon and cavity effects. To maintain maximum and equal intensity of both, s- and p-polarization nearly perpendicular incidence instead of Brewster angle incidence is chosen.

    The sample holder setup as described for the `bottom up' setup in the last paragraph is effectively turned upside down to a `top down' setup as gravitation shall be utilized to maintain the index matching liquid confined in the box. Thus, a rotation and mounting of the sample from the bottom is excluded. The motor and stages are operated from the top, the sample with the already described holders hangs into the liquid box.

\subsection{Visual process observation \label{ch_setup-visual}}

To allow a visual imaging observation of the scattering measurement as well as a visual support for the required precise adjustment of the sample relative to the incident beam, a camera equipped zoom microscope\footnote{Navitar machine vision 12x optical zoom microscope with CCD camera.} system is focussed on the sample from top (right part of fig.\;\ref{graphic_setup-b}).

\nomenclature[s]{microscope}{CCD microscope with adjustable zoom tubus.}

    The microscope tubus is provided with a 20x magnifying long working distance microscope objective with a working distance of \SI{20}{mm} allowing operation whilst scattering measurements are running.

    The microscope is equipped with a high resolution CCD video camera (UEye) that is programmed and read via a USB interface with proprietary low level machine commands. From the zoom microscope's aperture the sample is illuminated sufficiently to acquire pictures of the wire with magnifications that are during the measurements usually adjusted to $40 \times$, depending on the wire diameter and sample type.

    The illumination white light does not interfere with the measurement laser beam as its frequency, the power line frequency of \SI{50}{Hz} is filtered away by lock-in amplifiers.

\section{Measurement and referencing\label{chapter_measurement}}

\subsection{Optical detection \label{ch_exp-optical-detection}}

\begin{figure}[width=\textwidth]
        \centering

            \psfrag{C}[c][c][1.5][0]{\textbf{\textsf{C}}}
            \psfrag{D}[c][c][0.9][0]{D}
            \psfrag{aux}[c][c][0.9][0]{aux}
            \psfrag{ref}[c][c][0.9][0]{ref}
            \psfrag{in}[c][c][0.9][0]{in}
            \psfrag{reference signal}[l][l][0.9][0]{reference signal}
            \psfrag{scattering signal}[r][r][0.9][0]{scattering signal}
            \psfrag{Frequency generator}[c][c][0.9][0]{Frequency generator}
            \psfrag{and chopper control}[c][c][0.9][0]{and chopper control}
            \psfrag{f_inner}[c][c][0.9][0]{$f_\text{inner}$}
            \psfrag{f_outer}[c][c][0.9][0]{$f_\text{outer}$}
            \psfrag{R_L}[c][c][0.9][0]{$R_\text{load}$}
            \psfrag{PD1}[c][c][0.9][0]{PD1}
            \psfrag{PD2}[c][c][0.9][0]{PD2}
            \psfrag{s-pol}[c][c][0.9][0]{s-pol-ref}
            \psfrag{p-pol}[c][c][0.9][0]{p-pol-ref}
            \psfrag{Lockin amplifier Sr830}[c][c][0.9][0]{lock-in amplifier SR830}
            \psfrag{GPIB interface}[c][c][0.9][0]{GPIB}
            \psfrag{Module 1}[c][c][0.9][0]{lock-in module 1}
            \psfrag{Module 2}[c][c][0.9][0]{lock-in module 2}
            \psfrag{a/d}[c][c][0.9][0]{a/d}
            \psfrag{Dig. multimeter}[c][c][0.9][0]{dig. multimeter}
            \psfrag{R}[c][c][0.9][0]{R}
            \psfrag{PD1-s}[c][c][0.9][0]{$\text{PD1}_\text{s}$}
            \psfrag{PD2-s}[c][c][0.9][0]{$\text{PD2}_\text{s}$}
            \psfrag{PD2-p}[c][c][0.9][0]{$\text{PD2}_\text{p}$}
            \psfrag{X}[c][c][0.9][0]{X}

            \includegraphics[width=\textwidth]{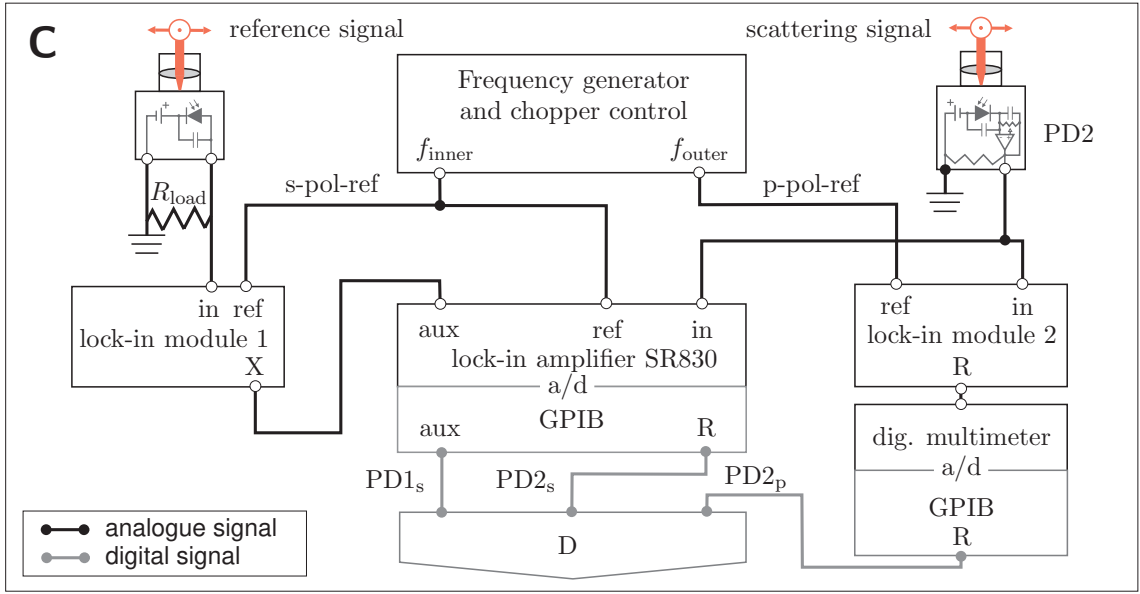}

        \caption[Detailed schematic diagram of the electronics and measurement devices, part C of the setup.]{Detailed schematic diagram of the electronics and measurement devices, part C of the setup. Analogue electronic signal connections are indicated in black, digital signal connections in grey. The signal waveforms of PD1 (reference signal) and PD2 (measured signal) are fed into 3 lock-in amplifiers. Two of them are digital lock-in modules. The resulting DC waveforms are acquired digitally and connected to a control PC via GPIB connections.  \label{graphic_setup-c}}

\end{figure}

    Two photo detectors are used (fig.\;\ref{graphic_setup-b} and fig.\;\ref{graphic_setup-c}) to detect the scattering amplitude (PD2) and to acquire a reference signal (PD1). The latter includes all amplitude fluctuations that are caused before the scattering process for a normalization of the scattering signal that is acquired by PD2.

\subsubsection*{Acquisition of the reference signal (PD1)}

    To measure the reference signal just before the scattering process a battery reverse biased Si-pin photodetector PD1\footnote{Thorlabs DET36A/M - High-Speed Si Detector, 350-1100 nm, 14 ns Rise Time} is operated at a wavelength and at a power level (typically $P \approx \SI{400}{\micro W}$) where it provides a good linear response with a spectral responsitivity of $\mathcal{R}(\SI{633}{nm})=\SI{0.4}{A/W}$.

\nomenclature[sp]{PD1}{Photo detector 1}

    The collimated beam is focussed by a lens (L2). Its focusses size hits the sensor area (\SI{3.6}{mm} x \SI{3.6}{mm}) centered. The reference sensor hence provides a reliable signal proportional to the chopped beam and not affected by possible vibrations. It effectively detects the sum of the intensities of both polarizations  a step before the scattering.

\nomenclature[s]{L2}{Focussing lens}

    PD1 provides the photocurrent directly out of the photo diode anode. This current is proportional to the incident optical power $I_\text{ref} = P \mathcal{R}$. It is converted into a proportional voltage $V_\text{ref} = P \mathcal{R} R_\text{load}$ by an adjustable resistance $R_\text{load}=\SI{50}{\Omega} \ldots \SI{50}{k \Omega}$.

    Depending on the fiber coupling efficiency and the desired amplitude of the beam, the voltage can by changing $R_\text{load}$  be easily adjusted not to exceed the input voltage level of the lock-in amplifier whilst maximizing the used voltage range to reduce electrical noise and to optimally utilize the available sampling width.

    In this case it is helpful that the modulation range of the reference signal is, apart from the intended chopping square modulation, low as it is only caused by the fluctuations in the beam intensity. Typically a load resistance of $R_\text{load}=\SI{50}{k \Omega}$ was applied to achieve a peak voltage of as near as possible to \SI{1}{V}, usually approximately \SI{0.7}{V}.

\nomenclature[sr]{$R_\text{load}$}{Adjustable load resistance for PD1.}

    At the same time the second important quantity for the precision of a measurement with the photo detector is the bandwidth \cite{hobbs} $f_\text{bw} = (2 \pi R_\text{load} C_\text{d})^{-1}$ and the rise time response $t_\text{r} = 0.35 / f_\text{bw}$ to the incident signal, $C_\text{d}$ being the photo diode capacitance.

    The electric circuit formed by the BNC cable, the photo detector, the load resistance and the measurement device is nothing else than an RC-circuit with a certain inertia, smoothing the edges of the chopped signal waveforms. An increase of $R_\text{load}$ thus decreases the bandwidth which results in smoothing waveform edges, observable on an oscilloscope.

    For the applied chopping frequencies of $f \approx \SI{320}{Hz}$ a sufficient response is achieved with the applied load resistance.

\subsubsection*{Acquisition of the scattered signal (PD2)}

    To measure the scattered amplitude a Si-pin photodetector PD2 \footnote{Thorlabs PDA 100A switchable gain, amplified silicon detector, $400-\SI{1100}{nm}$} is applied. It includes the same reverse biased photo diode as PD1 but with a switchable gain transimpedance amplifier with an operational amplifier (in fig.\;\ref{graphic_setup-c} schematically depicted as $\triangle$).

\nomenclature[sp]{PD2}{Photo detector 2}

    The transimpedance amplifier has the advantage that, in connection with the photo diode, it shows only a very low resistance. The photo diode therefore shows a highly linear response to the incident optical power. Also for high gain factors a response is achieved that is sufficient for the applied measurement frequencies. The BNC connection of the photo detector is connected in parallel to an internal \SI{50}{\Omega} resistance to match the \SI{50}{\Omega} impedance of the BNC cables.

    PD2 is usually operated with an amplification adjusted to $20-\SI{30}{dB}$. The sensor then shows a total responsitivity depending on the sensor responsitivity $\mathcal{R}(\SI{633}{nm})=\SI{0.35}{A/W}$ and the transimpedance gain $[\mathcal{G}]$:

\begin{equation}
\mathcal{R}_\text{tot} = \mathcal{R} \mathcal{G}
\end{equation}

    For a transimpedance amplification of \SI{20}{dB} a theoretical total responsitivity of $\mathcal{R}_\text{tot} \approx \SI{525}{} \pm 2 \%$ is calculated, for \SI{30}{dB} a total responsitivity of $\mathcal{R}_\text{tot} \approx \SI{16625}{} \pm 2 \%$. For some samples, especially those with very low diameter $\varnothing \lesssim \SI{20}{\micro m}$ the amplification is increased further. The noise of the amplifier then increases also but still remains bearable.

\subsection{Principles of lock-in amplification \label{ch_principle-lockin}}

    Lock-in amplifiers\footnote{Only the most important features of lock-in amplification are mentioned here. For an in depth discussion it is referred to \citep{hobbs,scofield,temple}} are used in a broad spectrum of applications due to their ability to measure signals with a high dynamic range (the ratio of noise contributions to the measurement signal) and a very low signal-to-noise ratio. They offer phase sensitive measurement and the separation of frequencies with a very narrow bandwidth.

    \begin{figure}[width=\textwidth]
        \centering

            \psfrag{V_in}[c][c][0.9][0]{$V_\text{in}$}
            \psfrag{V_ref}[c][c][0.9][0]{$V_\text{ref}$}
            \psfrag{V_PSD}[c][c][0.9][0]{$V_\text{PSD}$}
            \psfrag{V_VCO}[c][c][0.9][0]{$V_\text{VCO}$}
            \psfrag{X_out}[c][c][0.9][0]{$X_\text{out}$}
            \psfrag{Y_out}[c][c][0.9][0]{$Y_\text{out}$}
            \psfrag{V_out}[c][c][0.9][0]{$V_\text{out}$}
            \psfrag{AC signal amp}[c][c][0.9][0]{\parbox{20mm}{ \center{ac signal\\ amp.}} }
            \psfrag{Notch filters}[c][c][0.9][0]{notch filters}
            \psfrag{VCO}[c][c][0.9][0]{VCO}
            \psfrag{PSD_X}[c][c][0.9][0]{$\text{PSD}_\text{X}$}
            \psfrag{PSD_Y}[c][c][0.9][0]{$\text{PSD}_\text{Y}$}
            \psfrag{Low pass}[c][c][0.9][0]{low pass}
            \psfrag{DC amp}[c][c][0.9][0]{dc amp}
            \psfrag{+90}[c][c][0.6][0]{$+\ang{90}$}

            \includegraphics[width=\textwidth]{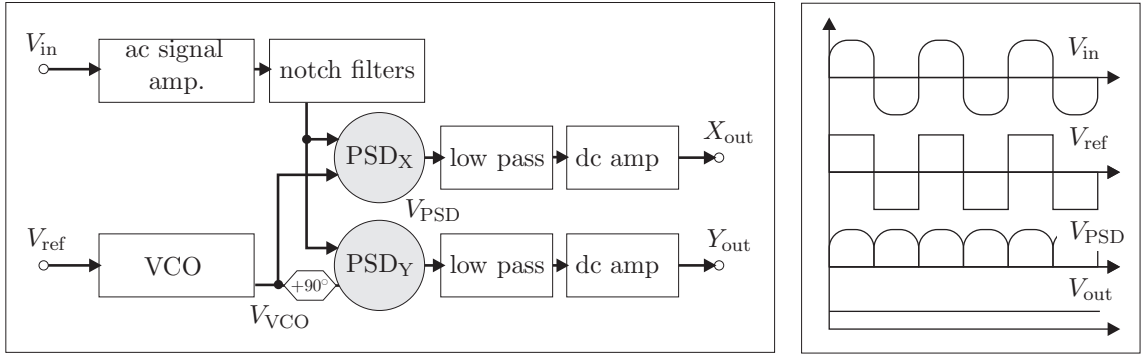}

        \caption[Schematic diagram of the signal processing of the lock-in amplifier.]{Schematic diagram of the signal processing of the lock-in amplifier. Two phase sensitive detectors (PSD) are displayed for measurement of amplitude ($X$) and quadrature ($Y$) as realized in dual channel lock-in amplifiers. The right inset displays schematically the processing of an edge smoothed rectangular input signal as it is typical for optical chopping measurements.  \label{graphic_lock-in}}

\end{figure} 

    In the setup for this work the lock-in amplifier's ability to separate signal contributions with different frequency modulation and the cancelling abilities for noise of other frequencies are particularly utilized. The second advantage cancels also contributions from disturbing ambient influences caused by e.g. other experiments in the same laboratory. Particularly all influences from laboratory illumination or power line interferences are cancelled as the power line frequency of \SI{50}{Hz} is filtered away very effectively.\\

    The operating principle of lock-in amplifiers with special attention on the features that are practically utilized in the setup is explained in the following (fig.\;\ref{graphic_lock-in}:\\

    The \textbf{voltage controlled oscillator} (VCO) is a sinusoidal (RS830) or rectangular (lock-in modules 1 and 2) waveform generator that is triggered by the frequency $f_\text{s,p}$ of the square shape reference signal $V_\text{ref}$ (frequency locking). The signal $V_\text{VCO}$ that is generated by this internal oscillator can be adjusted in phase to fit the measurement signal $V_\text{in}$ which is important for the next step (phase locking). The RS830 can automatically determine the phase mismatch, thus it is also applied (preliminarily triggered with the other frequency $f_\text{p}$) for the determination of the phase for p polarization that is then manually configured into lock-in module 2.

    \nomenclature[s]{VCO}{Voltage controlled oscillator. Part of the lock-in amplifiers}

    In this experiment no substantial phase shift over the measurement time is expected, hence for any phase shift far enough from $\pi/2$ a constant signal would be expected. Thus a correct phase locking to $\ang{0}$ simply provides a maximum output signal which is useful to minimize later electronic noise contributions.\\

    The \textbf{phase sensitive detectors} (PSD, also called demodulators or mixers) are the core parts of any lock-in-amplifier. Their function is in principle the multiplication of the reference oscillator signal $V_\text{VCO}$ with the measurement signal $V_\text{in}$ that already is roughly cleaned from other frequency components by a cascade of adjustable notch filters.
    \nomenclature[s]{PSD}{Phase sensitive detector. Part of the lock-in amplifiers.}

    A short mathematical explanation may clear the functional principle of the PSD. Assuming a noise-free sinusoidal signal voltage with angular frequency $\omega$ at the input $V_\text{in} = A \cos(\omega t)$. The reference signal $V_\text{ref}$ with the same frequency  and a respective phase shift of $-\varphi$ triggers the VCO to generate a signal with zero phase shift $V_\text{VCO} = B \cos (\omega t + \varphi)$.

    The output of $\text{PSD}_\text{X}$ is then obtained by a multiplication of both signals:

    \begin{align}
        V_\text{PSD} &= A \cos (\omega t) \cdot B \cos (\omega t + \varphi)\\
                    &= AB \left(\cos(2 \omega t) \cos (\varphi) - \cos (\omega t) \sin (\omega t) \sin (\varphi) \right)\\
                    &= AB \left( \left( \nicefrac{1}{2} + \nicefrac{1}{2} \cos(2 \omega t) \right) \cos(\varphi) -  \nicefrac{1}{2} \sin(2 \omega t) \sin (\varphi) \right)\\
                    &= \nicefrac{1}{2} AB \cos(\varphi) + \nicefrac{1}{2} AB \cos(2\omega t + \varphi)
    \end{align}

    For a constant amplitude $B$ (VCO) and a constant frequency $\omega$, thus the output is a DC signal, that is proportional to $A$ (signal), the cosine of the phase difference $\varphi$ and modulated with $2\omega t$, which physically means that the first harmonic component remains.

    Technically the PSD consists of signal splitting, inversion and triggered gating elements. It is followed by a low pass filter providing time averaging for an adjustable time scale, called integration time (which is usually chosen to be \SI{300}{ms}) and whose purpose is to eliminate the $2\omega t$ modulation. For a lock-in amplifier this averaging is thus intrinsically necessary, revealing the major disadvantage of the instrument: Lock-in amplifiers are comparably slow, becoming slower for higher measurement accuracy.

    In all applied lock-in amplifiers the PSDs are realized as digital electronic circuits with the main advantage of real linear multiplication with a high dynamic reserve and good harmonic rejection.\\

    The \textbf{dynamic reserve} of a lock-in amplifier is the adjustable ratio of the largest tolerable noise signal to the full scale signal, quantitatively expressed in dB. At the lock-in amplifier the dynamic reserve at which it may be operated without an input overload can be selected. Higher dynamic reserve allows measuring of signals with a large noise contribution at the expense of a low gain and a/d conversion noise. Thus when applying a lock-in amplifier always a compromise has to be found.\\

    The \textbf{signal outputs} of the lock-in amplifiers deliver dc signals, proportional to the incoming ac signal amplitude at the correct frequency and phase. $X_\text{out}$ represents the amplitude component with $\ang{0}$ phase shift. This is passed for the reference signal at the output of the lock-in module 1 only as this module has a single PSD design. RS830 and lock-in module 2 deliver also the quadrature amplitude $Y_\text{out}$ ($\ang{90}$ phase shift) and the radius R of the vector in a $X$-$Y$- plane. The latter value, R is acquired for both scattering signals for further data processing as this has the advantage to be intrinsically independent of any phase shifts.

\subsection{Polarization dependent lock-in detection and noise handling \label{ch_setup-lockindetection}}

    The two polarizations (s and p) being scattered by the sample are detected by only one photo detector (PD2). This 'one photo detector' principle makes it necessary to separate the two signals after detection. The separation is realized using of lock-in amplifiers that are locked to the chopping 'frequency labels' of the two different polarizations (sec.\;\ref{chapter_chopping}).

\subsubsection*{Electronic separation of the measurement signals}

    The application of `one photo detector to detect both polarizations' has the major advantage that both polarizations can be observed in exactly one angle of observation synchronously. A little spatial offset would already cause a large offset in the angle of observation $\gamma$ as the sensor PD2 is placed very near to the sample to capture a maximum scattered light.

    On the other hand the conversion from optical power to electric current by the photo diode at this point already has reduced one information dimension, only detecting the intensity $I=|\vc{E}_\text{s}+\vc{E}_\text{p}|^2$ and not the two dimensional (in the plane of the sensor) electric (or magnetic) field vectors $\vc{E}_{\text{s},\text{p} }$.

    The clearer the polarizations of the two beams, chopped with $f_\text{s}$ and $f_\text{p}$ are separated (i.e. $\vc{E}_\text{s} \perp \vc{E}_\text{p}$), the less interference between both polarizations occurs. In the case of a non zero parallel component of the beam portions from path 1 and path 2, interference is unavoidably caused by slight optical path length differences, already at the order of the wavelength of light, in the Mach-Zehnder like chopping setup.

    These interference effects would affect the measurement accuracy thus reducing the acquired information of the whole measurement irreversibly. As a separation of the two polarizations of $200:1$ is practically measured to be achieved with the chopping setup for precise PM-fiber coupling (as mentioned in section \ref{ch_scattering}) the interference effects between the two polarizations are low enough to allow the desired measurements with high accuracy. This interference reduction is supported by a reduction of the chopping duty cycle (as described in sec.\;\ref{chapter_chopping})

\begin{figure}[tb]
        \centering

            \psfrag{Normalized total scattering amplitude}[c][c][1][0]{Normalized total scattering amplitude}
            \psfrag{t}[l][l][1][0]{$t$}
            \psfrag{10e-4s}[l][l][1][0]{\SI{1e-4}{s}}

            \includegraphics[width=\textwidth]{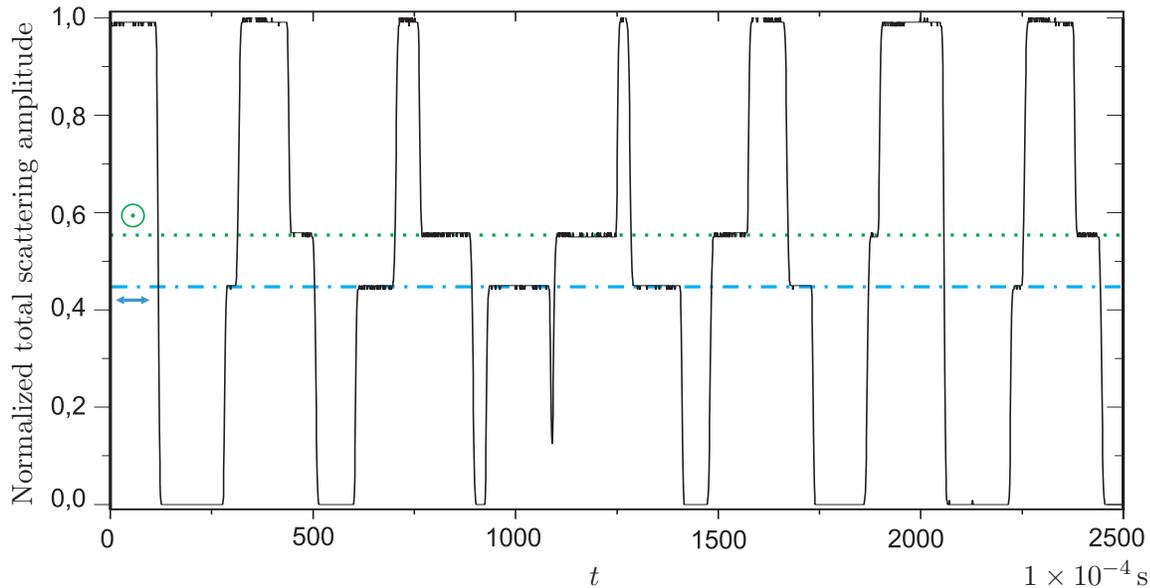}

        \caption[Graph of the typical signal waveform, experimentally acquired.]{By means of a digital oscilloscope acquired typical signal waveform (PD2) during a measurement (compare to figure \ref{graphic_setup-a}). The two amplitudes of the s-polarized (green line, $\odot$) and the p-polarized (blue line, $\leftrightarrow$) beam portion are indicated. The noise level on each single amplitude are of the same order as the noise amplitude with both beam portions at once, indicating that interbeam-interference is very low. The remaining noise is mainly due to laser noise and is mostly cancelled by reference division.\label{graphic_setup-chopping-waveform}}

\end{figure}

    In principle, optical effects that mixed the polarization of both `beams' at the scattering process whilst not destroying the coherence of the light could lead to the same interference noise. This is not expected to happen from theoretical assumptions. It is shown by oscilloscope measurement of the signal waveform (fig.\;\ref{graphic_setup-chopping-waveform}) that with a precise adjustment of the beams and optical components interference effects in fact do not play a significant role.\\

\subsubsection*{Scattering signal $\text{PD2}_\text{s}$}

    PD2 is connected to the single voltage signal input of a dual channel digital DSP lock-in-amplifier SR830 \footnote{Stanford Research Systems SR830, for detail it is referred to appendix \ref{ch_appendix-instrumentation}} (fig.\;\ref{graphic_setup-c}). The lock-in amplifier's internal sinus oscillator is externally synchronized to the (reference) frequency of the s-polarized beam  ($f_\text{inner}=f_\text{s}$) which is delivered from the frequency generator and chopper control.

\nomenclature[spd2s]{$\text{PD2}_\text{s}$}{Experimentally acquired signal portion for the s-polarization from photo detector PD2}

    The two superposed signals of the s- and the p-polarized beam portions are of the same amplitude. Additional amplitude noise caused by interference, detection or optomechanical parts of the setup on the PD2 signal (fig.\;\ref{graphic_setup-chopping-waveform}) cause a total signal-to-noise-ratio (SNR) of the order of $\SI{1e3}{} - \SI{1e4}{} $. The laser noise contribution to the total noise is higher but still much lower than the signal amplitude.

    Therefore the lock-in amplifier can be operated with low dynamic reserve (sec.\;\ref{ch_principle-lockin}), thus allowing a higher measurement accuracy as the a/d noise and the dc gain noise decrease\footnote{The SR830's low noise adjustment still allows a relatively high dynamic range even in low dynamic range mode of operation, compared to analogue lock-in amplifiers as it is constructed to minimize the total noise level and a further decrease in the dynamic range would not lead to a significantly further decreased noise level due to the digital detection process.}. The low dynamic reserve setting of the SR830 changes for different gain settings. Not discussing this dependency in detail it shall be mentioned that the sensitivities (maximum full scale signal detection amplitude) of the SR830 are usually, depending on a good combination of the PD2 gain factor and the SR830 and lock-in module 2 gain factors, chosen to be $\SI{200}{mV},\SI{500}{mV},\SI{1}{V}$. With an adjustment of `low dynamic reserve' this quantitatively results in a dynamic reserve of $\SI{4}{dB},\SI{6}{dB},\SI{0}{dB}$.

    The SR830 is programmed and the measurement data is acquired via its GPIB PC interface.
\nomenclature[s]{GPIB}{General Purpose Interface Bus/IEEE-488. Digital communication interface system, applied to control the measurement and motor systems.}

\subsubsection*{Scattering signal $\text{PD2}_\text{p}$}

    The photo detector PD2 for the scattering signal is not only connected to the SR830 but also to the signal input of the lock-in module 2\footnote{Femto LIA-MVD-200L - Digital dual channel lock-in module.} (fig\;\ref{graphic_setup-c}). This instrument is an integrated lock-in module that in principle operates equal to the SR830 with two (amplitude and quadrature) phase sensitive detectors (PSDs).

    The main difference is that it lacks the ability to automatically lock the phase which has to be adjusted manually. The vector radius output R is obtained as for the RS830. Thus referencing the phase might seem unnecessary which is in fact not true as for a wrong phase adjustment the simpler (compared to the SR830) square signal VCO (not a sine generator) could cause a/d signal mismatch artifacts.
\nomenclature[spd2p]{$\text{PD2}_\text{p}$}{Experimentally acquired signal portion for the p-polarization from photo detector PD2}

    The lock-in module 2 is on its reference signal input connected to the frequency generator and chopper control's output for the $f_\text{outer}$ frequency that labels the p-polarization beam portion.

    As the lock-in module lacks an integrated a/d converter it is connected to a programmable digital precision multimeter\footnote{Agilent 34410A, 6.5 digits digital multimeter} which is programmed via its GPIB interface. The full range R-output amplitude of the lock-in module 1 is maximized to \SI{10}{V}. The detection precision of the multimeter is adjusted respectively to achieve a maximum a/d conversion precision, minimum a/d discrimination artifacts and to minimize electronic noise.

\subsubsection*{Reference signal $\text{PD1}_\text{s}$}

    The output of the photo detector PD1 is connected to the signal input of the lock-in amplifier module 1 (fig.\;\ref{graphic_setup-c}). This instrument is predominantly identical to the lock-in amplifier module 2 but has only one PSD and provides only the phase $\ang{0}$ amplitude $X_\text{out}$ as an output, not the quadrature $Y_\text{out}$ or the radius $R_\text{out}$.
\nomenclature[spd2s]{$\text{PD1}_\text{s}$}{Experimentally acquired signal portion for the s-polarization from photo detector PD1, the reference signal.}

    The reference signal input of the lock-in amplifier module 1 is connected to the frequency generator and chopper control frequency output $f_\text{inner}$ together with the SR830, for frequency locking on the s-polarized beam portion. The lock-in module 1 in the beginning of the measurement cycle is manually adjusted to the phase of the frequency reference for s-polarization. This phase does in fact not change over the measurement and not significantly over different measurements, also for this signal always the maximum $\ang{0}$ amplitude is achieved to acquire with the same advantages as mentioned before.

    It was chosen only to acquire the reference signal for one of the two chopping frequencies thus beam polarizations (s) as preliminary measurements showed that for an optimized coupling the noise of the beam signals of both polarizations does not differ significantly.

    The SR830 is in addition to its main purpose also utilized as an a/d convertor with the ability to intrinsically acquire samples exactly synchronized to the SR830 lock-in sampling, when correctly programmed. Therefore it supplies external DC signal input channels which though provide only a a/d conversion resolution of \SI{16}{bit}.

    After some tests it was decided not to use these input channels for the measurement of the scattering signal $\text{PD2}_\text{p}$ as the electronic noise contribution distorted the measurement signals too severely. Nevertheless, for the scattering signal a very high dynamic range over the whole measurement, from 0 signal to maximum range amplitude has to be detected, which is not necessary for the reference signal $\text{PD1}_\text{s}$.

    The reference signal is expected to contain only the noise fluctuations that it is intended to provide for referencing, no additional amplitude drift over the measurement as it is modulated on the scattered signals. Therefore the a/d convertors of the SR830 with the reference signal are always driven at maximum input amplitude, there providing the full \SI{16}{bit} resolution for the whole measurement process which is sufficient for the reference signal.

    The SR830 provides the a/d convertor signal for $\text{PD1}_\text{s}$ on the second of two measurement channels via the GPIB programming interface.

\section{Data acquisition and control software \label{ch_setup-software} }

\begin{figure}[width=\textwidth]
        \centering

            \psfrag{D}[c][c][1.5][0]{\textbf{\textsf{D}}}
            \psfrag{C}[c][c][0.9][0]{C}
            \psfrag{B}[c][c][0.9][0]{B}

            \psfrag{GPIB}[c][c][0.9][0]{GPIB}
            \psfrag{USB}[c][c][0.9][0]{USB}
            \psfrag{save}[c][c][0.9][0]{user input}
            \psfrag{save}[c][c][0.9][0]{configuration}
            \psfrag{measurement loop}[c][c][0.9][0]{measurement loop}
            \psfrag{s. waveform acq.}[c][c][0.9][0]{synchronized waveform acq.}
            \psfrag{motor drive}[c][c][0.9][0]{  \parbox{20mm}{ \center{motor drive\\control\\error corr.}}  }
            \psfrag{exact real}[c][c][0.9][0]{exact real}
            \psfrag{angle alpha}[c][c][0.9][0]{angle $\alpha$}
            \psfrag{init system ch motor}[c][c][0.9][0]{\parbox{52mm}{ \center{initialization - system checks\\motor referencing} }}
            \psfrag{user input}[c][c][0.9][0]{user input}
            \psfrag{configuration}[c][c][0.9][0]{configuration}
            \psfrag{Lock-in Sr830}[c][c][0.9][0]{lock-in SR830}
            \psfrag{(PD2-s)}[c][c][0.9][0]{($\text{PD2}_\text{s}$)}
            \psfrag{Lock-in 2}[c][c][0.9][0]{lock-in 2}
            \psfrag{(PD2-p, via multimeter a/d)}[c][c][0.9][0]{($\text{PD2}_\text{p}$, via multimeter a/d)}
            \psfrag{Lock-in 1}[c][c][0.9][0]{lock-in 1}
            \psfrag{(PD1-s, via SR a/d)}[c][c][0.9][0]{($\text{PD1}_\text{s}$, via SR830 a/d)}
            \psfrag{al++}[c][c][0.9][0]{$\alpha=\alpha+\delta\alpha$}
            \psfrag{vis. control}[c][c][0.9][0]{ \parbox{20mm}{ \center{vis. control\\microscope\\picture acq.} }}
            \psfrag{save}[c][c][0.9][0]{save}
            \psfrag{to meas file}[c][c][0.9][0]{to file}
            \psfrag{reset systems close errors}[c][c][0.9][0]{\parbox{55mm}{ \center{reset systems - close connections\\error handling} }}
            \psfrag{PD2-s/PD1-s}[c][c][0.8][0]{$\text{PD2}_\text{s}/\text{PD1}_\text{s}$}
            \psfrag{PD1-s}[c][c][0.8][0]{$\text{PD1}_\text{s}$}
            \psfrag{PD2-p/PD1-s}[c][c][0.8][0]{$\text{PD2}_\text{p}/\text{PD1}_\text{s}$}
            \psfrag{t}[c][c][0.8][0]{$t$}

            \psfrag{mean-s}[c][c][0.8][0]{$\bar{s}$}
            \psfrag{sigma-s}[c][c][0.8][0]{$\sigma(s)$}
            \psfrag{mean-ref}[c][c][0.8][0]{$\overline{\text{ref}}$}
            \psfrag{sigma-ref}[c][c][0.8][0]{$\sigma(\text{ref})$}
            \psfrag{mean-p}[c][c][0.8][0]{$\bar{p}$}
            \psfrag{sigma-p}[c][c][0.8][0]{$\sigma(p)$}

            \psfrag{s}[c][c][0.8][0]{$s$}
            \psfrag{p}[c][c][0.8][0]{$p$}
            \psfrag{al}[c][c][0.8][0]{$\alpha$}
            \psfrag{ref}[c][c][0.8][0]{ref}

            \includegraphics[width=\textwidth]{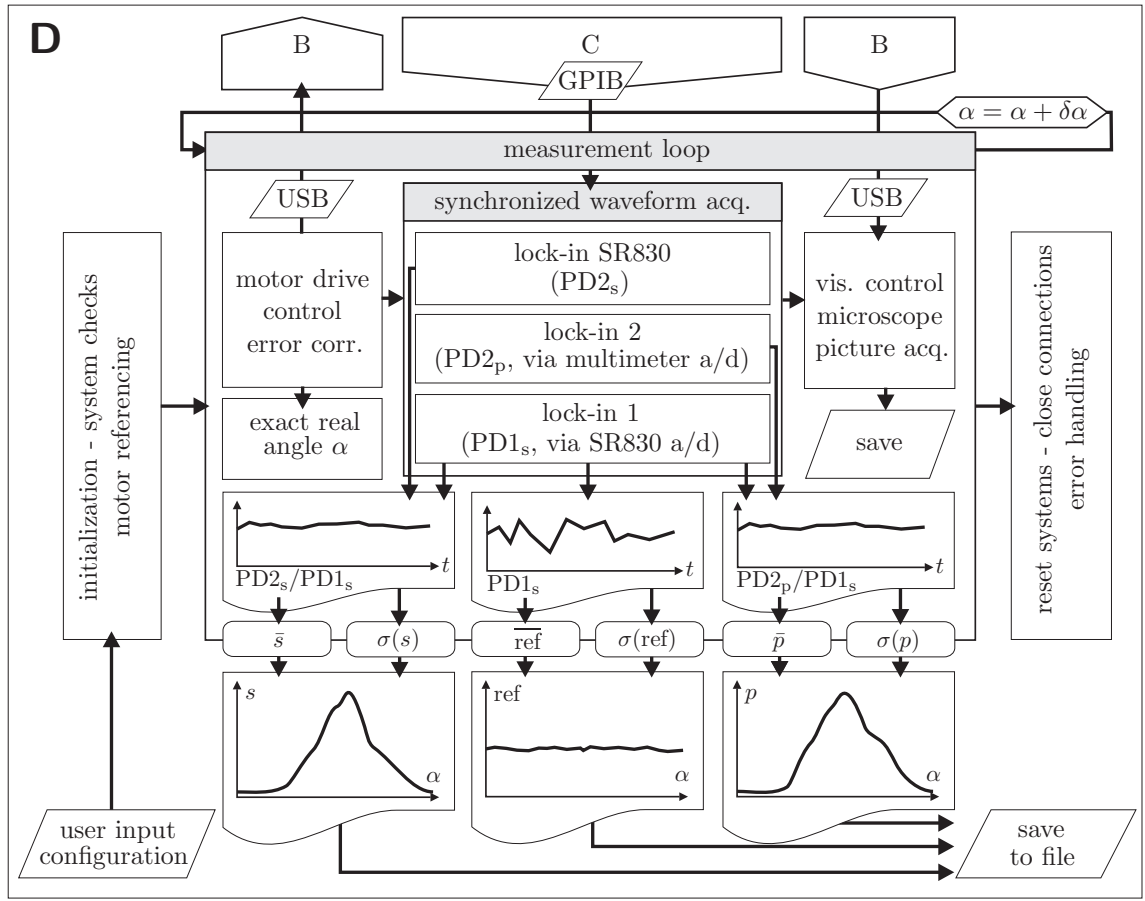}

        \caption[Simplified schematic diagram of the measurement and control software, part D of the experimental setup.]{Schematic diagram of the measurement and control software, part D of the experimental setup. \label{graphic_setup-d}}

\end{figure} 

    The great advantage of lock-in amplifier detection is its low measurement noise level and its capability to separate electric signals of different frequency (sec.\;\ref{ch_setup-lockindetection}). On the other hand the operating principle of a lock-in amplifier makes measurements comparably slow. Additionally a requirement for a large $\alpha$ and amplitude resolution requires a high number of measurement steps. A complete measurement over an incident angle range $\alpha= \ang{0} \ldots \ang{10}$ with an angle resolution of $\ang{0.01}$ takes more than 3 to 5 hours.

    At least these reasons motivate a complete automation of the experimental measurement including a computer assisted processing of the acquired measurement values and several error corrections. The software to control and automate the experimental measurements is written in the visual programming language National Instruments LabView 8.2. For precise synchronization the communication with the measurement instruments was programmed as low level command functions, embedded in the visual programming language. A simplified schematic of the software elements is depicted in fig.\;\ref{graphic_setup-d}.

    The interface of the software part \textbf{D} to the electronic measurement instruments of the setup \textbf{C} (see section \ref{ch_setup-lockindetection}) is realized by the digital data communication protocol GPIB. Via this protocol the RS830 digital lock-in amplifier and the digital multimeter are controlled and their data is read and transferred for processing into the PC and the control software. The motorized precision rotation stage (setup part B, fig.\;\ref{graphic_setup-b}) is controlled via a USB connection. It is programmed with a proprietary low level command language as well as the control camera of the zoom microscope that observes the scattering process from top (see also fig.\;\ref{graphic_setup-b}).

    The whole software is modularized and includes precise error handling and diagnostics. The measurement results are saved in reduced, human readable format on hard disk.

\subsection*{Initialization and referencing}

    To start the program, the measurement parameters are input by the user. It is necessary to drive the motorized rotation stage to its reproducible reference start point, to partly manually adjust the sample in the beam and to drive the stage to its starting position\footnote{The rotation stage is equipped with a Hall sensor for referencing and counts its motor position from that point of reference.}. After a change of the sample slight differences in the holders can change the $\alpha=\ang{0}$ position. This has to be adjusted by observation of the scattering pattern and the measured amplitude as well as the controlled zoom microscope camera.

    The measurement parameters are the lock-in sensitivity, the sampling rate for the synchronous acquisition of the measurement data, the angle $\alpha$ range of measurement and the number of steps to measure, respectively the angle $\alpha$ resolution.

    Finally all instruments are set to the measurement parameters and prepared for acquisition. All further communication with the instruments is software triggered.

\subsection*{Measurement loop}

    The measurement loop itself consists of three subsequent steps. First the sample is driven to the desired incident angle $\alpha$. In the next step synchronized scattering amplitude detection takes place. In the third step the observation microscope camera is triggered to acquire a control image.

    The motor for each position reports back the exact real position that will be used for further steps and for the data acquisition including backlash correction. The position is measured and calculated by the servo motor's parameters and counter.

    For the acquisition of the scattering data it is necessary to achieve a precise synchronization of the different measurement devices (SR830 and digital multimeter). These commands are implemented on a basic hardware command level with correction factors that are the result of test measurements.

    With a lock-in hardware integration time of usually \SI{300}{ms} and a sampling rate of usually \SI{128}{Hz}, adjusted to fit to the measured low frequency noise contributions, up to \SI{10}{s} data for the scattered s-polarization via the SR830 ($\text{PD2}_\text{s}$) is acquired. For the scattered p-polarization the multimeter is read synchronously to achieve a $\text{PD2}_\text{p}$ waveform. The reference signal $\text{PD1}_\text{s}$ waveform (as described in section \ref{ch_setup-lockindetection}) is acquired via the internal a/d converter unit of the SR830.

\subsection*{Signal processing}

    All three waveforms show the noise modulation of the scattered beam. The main portion of the signal noise is cancelled or at least reduced significantly by the lock-in amplifiers. Remaining noise on the lock-in output signals is dominated by low frequency contributors that modulate the square chopped signal amplitude and pass the frequency filters. The so called laser residual intensity noise (RIN) \citep{hobbs} is the dominant contributor to the total noise level of the scattering beam and is far above the shot noise level. HeNe lasers usually show a very low phase noise but this does neither way influence the measurement in this experimental setup.

    As the total noise modulation is not assumed to simply disappear by a time integration it is an increase in real measurement precision to divide the signals of $s(t):= \text{PD2}_\text{s}(t) / \text{PD1}_\text{s}(t)$ and $p(t):=\text{PD2}_\text{p}(t)/\text{PD1}_\text{s}(t)$ for the waveforms pointwise and after that calculating the time averaged mean values $\overline{s}$ and $\overline{p}$.

    By displaying the acquired waveforms from the different photo detectors the coincidence of the low frequency ($1-\SI{100}{Hz}$) common noise can clearly be identified. Before the reference division signal-to-noise-ratios of \SI{1e2}{} to \SI{1e3}{} are observed, after the pointwise reference division this increases to \SI{1e3}{} to \SI{2e5}{} always depending on the conditions, amplitudes and gain factors of the current measurement.

    To simplify a later data analysis also the mean values of the original scattered amplitudes and the mean value of the reference signal are processed. To achieve a measure of the waveform fluctuations over time, the standard deviations $\sigma$ of all values are calculated and bundled simultaneously. As the light source fluctuations are divided out, the standard deviation of the divided values should be lower than the original ones (usually $\sigma_\text{s,p} \approx \SI{1e-4}{}$).

    After the measurement of the scattering amplitudes a reference picture is automatically taken from top view by the microscope to allow later manual crosschecking of observed peaks to possible not intended reasons.

    The calculated scattering amplitudes ($\overline{s},\overline{p}$) are displayed over the incident angle $\alpha$ and stored. The loop continues its functions with increasing $\alpha$ until the measurement is finished.

\subsection*{Closing the measurement}

After finishing the measurement loop the systems are driven back to their initial configuration. All acquired data along with possible error messages and the measurement parameters are saved in a measurement file.

\chapter{Theoretical results \label{ch_theory-results}}

\section{Numerical computation of the dispersion model \label{ch_theoresults-model}}

    \begin{figure}
        \centering

            \psfrag{m1}[c][c][0.7][0]{$m=1$}
            \psfrag{m101}[r][r][0.7][0]{$m=101$}
            \psfrag{633}[c][c][0.7][0]{\SI{633}{nm}}
            \psfrag{light line}[c][c][0.7][50]{air light line}

            \psfrag{normalized wave vector}[c][c][0.9][0]{normalized wave vector $\beta \, c_0/\omega_\text{p}$}
            \psfrag{x12}[t][t][0.7]{0}%
            \psfrag{x13}[t][t][0.7]{0.1}%
            \psfrag{x14}[t][t][0.7]{0.2}%
            \psfrag{x15}[t][t][0.7]{0.3}%
            \psfrag{x16}[t][t][0.7]{0.4}%
            \psfrag{x17}[t][t][0.7]{0.5}%
            \psfrag{x18}[t][t][0.7]{0.6}%

            \psfrag{normalized frequency}[c][c][0.9][0]{normalized frequency $\omega/\omega_\text{p}$}
            \psfrag{v12}[r][r][0.7]{0.1}%
            \psfrag{v13}[r][r][0.7]{0.15}%
            \psfrag{v14}[r][r][0.7]{0.2}%
            \psfrag{v15}[r][r][0.7]{0.25}%
            \psfrag{v16}[r][r][0.7]{0.3}%
            \psfrag{v17}[r][r][0.7]{0.35}%
            \psfrag{v18}[r][r][0.7]{0.4}%
            \psfrag{v19}[r][r][0.7]{0.45}%
            \psfrag{v20}[r][r][0.7]{0.5}%
            \psfrag{v21}[r][r][0.7]{0.55}%
            \psfrag{v22}[r][r][0.7]{0.6}%

            \psfrag{va12}[l][l][0.7]{0.1}%
            \psfrag{va13}[l][l][0.7]{0.15}%
            \psfrag{va14}[l][l][0.7]{0.2}%
            \psfrag{va15}[l][l][0.7]{0.25}%
            \psfrag{va16}[l][l][0.7]{0.3}%
            \psfrag{va17}[l][l][0.7]{0.35}%
            \psfrag{va18}[l][l][0.7]{0.4}%
            \psfrag{va19}[l][l][0.7]{0.45}%
            \psfrag{va20}[l][l][0.7]{0.5}%
            \psfrag{va21}[l][l][0.7]{0.55}%
            \psfrag{va22}[l][l][0.7]{0.6}%

            \psfrag{wavelength}[c][c][0.9][0]{wavelength $\lambda$}

            \psfrag{vr12}[c][c][0.7]{1385}%
            \psfrag{vr13}[c][c][0.7]{923}%
            \psfrag{vr14}[c][c][0.7]{693}%
            \psfrag{vr15}[c][c][0.7]{554}%
            \psfrag{vr16}[c][c][0.7]{462}%
            \psfrag{vr17}[c][c][0.7]{396}%
            \psfrag{vr18}[c][c][0.7]{346}%
            \psfrag{vr19}[c][c][0.7]{308}%
            \psfrag{vr20}[c][c][0.7]{277}%
            \psfrag{vr21}[c][c][0.7]{252}%
            \psfrag{vr22}[c][c][0.7]{231}%

            \psfrag{angle of incidence}[c][c][0.9][0]{angle of incidence $\alpha$ in \textdegree}
            \psfrag{x0}[t][t][0.7]{0}%
            \psfrag{x10}[t][t][0.7]{10}%
            \psfrag{x20}[t][t][0.7]{20}%
            \psfrag{x30}[t][t][0.7]{30}%
            \psfrag{x40}[t][t][0.7]{40}%
            \psfrag{x50}[t][t][0.7]{50}%
            \psfrag{x60}[t][t][0.7]{60}%
            \psfrag{x70}[t][t][0.7]{70}%
            \psfrag{x80}[t][t][0.7]{80}%
            \psfrag{x90}[t][t][0.7]{90}%

            \includegraphics[width=\textwidth]{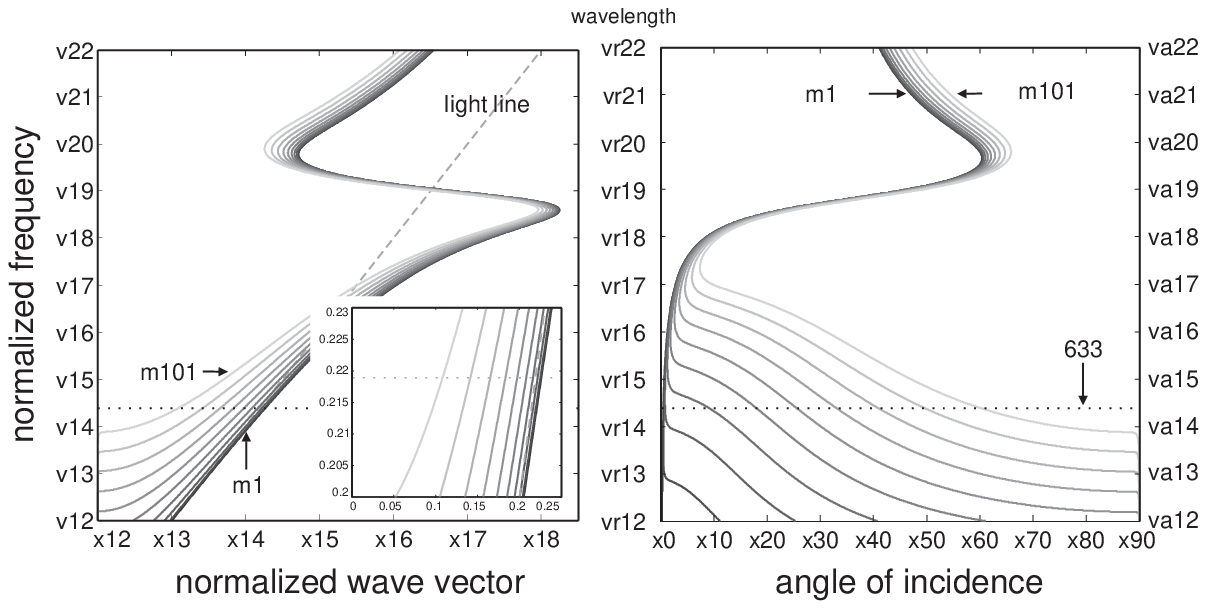}

        \caption[Dispersion of SPP on a silver cylinder of $D=\SI{20}{\micro m}$ in air.]{Dispersion of SPP on a silver cylinder of $D=\SI{20}{\micro m}$, surrounded by air. Displayed are the dispersions of modes $m=1$ (in black) to $m= 101$ (in grey) in steps of $10$. The left figure shows the dispersion in terms of the normalized SPP wave vector $\beta \, c_0/ \omega_\text{p}$ (eq. \refeq{model_beta_m}). The inset is a magnification for the wavelength range around the laser wavelength ($\lambda=\SI{632.8}{nm}$). The right figure depicts the incident angles for excitation of the same modes (eq.~\refeq{eq-model-plasmon-incangles}).\label{graphic_spp-ag-cylinder}}
\end{figure} 

    The model for SPP modes on cylindrical samples, derived in section \ref{chapter_theory-model} is used to compute\footnote{These computations are implemented in computer programs, written in the proprietary programming language Matlab.} the dispersion of SPP (eq.\;\refeq{model_beta_m}) and the angles of incidence ($\alpha$) for exciting modes ($m$) (eq.\;\refeq{eq-model-plasmon-incangles}). The computations are accomplished for the experimentally relevant material configurations of silver (Ag) or gold (Au) in silica (SiO2 glass) or air.

    For the dispersion of the dielectric permittivity of silica glass, experimental values (see tab.~\refeq{eq_sellmeier-comp}, fig.~\ref{graphic_sio2-sellmeier}) and the Sellmeier equation are applied. For the optical properties of gold, experimental data is used (fig.~\ref{graphic_au-palik}). Silver is modelled by the Drude model (fig.~\ref{graphic_ag-drude})\footnote{For details about the material properties see sec. \ref{ch_material-properties}.}.
    
    \begin{figure}
        \centering

            \psfrag{m1}[c][c][0.7][0]{$m=1$}
            \psfrag{m150}[r][r][0.7][0]{$m=151$}
            \psfrag{mr150}[l][l][0.7][0]{$m=151$}
            \psfrag{633}[c][c][0.7][0]{\SI{633}{nm}}
            \psfrag{silica light line}[c][c][0.7][50]{silica light line}

            \psfrag{normalized wave vector}[c][c][0.9][0]{normalized wave vector $\beta \, c_0/\omega_\text{p}$}
            \psfrag{x12}[t][t][0.7]{0}%
            \psfrag{x13}[t][t][0.7]{0.1}%
            \psfrag{x14}[t][t][0.7]{0.2}%
            \psfrag{x15}[t][t][0.7]{0.3}%
            \psfrag{x16}[t][t][0.7]{0.4}%
            \psfrag{x17}[t][t][0.7]{0.5}%
            \psfrag{x18}[t][t][0.7]{0.6}%
            \psfrag{x19}[t][t][0.7]{0.7}%
            \psfrag{x20}[t][t][0.7]{0.8}%
            \psfrag{x21}[t][t][0.7]{0.9}%

            \psfrag{normalized frequency}[c][c][0.9][0]{normalized frequency $\omega/\omega_\text{p}$}
            \psfrag{v12}[r][r][0.7]{0.1}%
            \psfrag{v13}[r][r][0.7]{0.15}%
            \psfrag{v14}[r][r][0.7]{0.2}%
            \psfrag{v15}[r][r][0.7]{0.25}%
            \psfrag{v16}[r][r][0.7]{0.3}%
            \psfrag{v17}[r][r][0.7]{0.35}%
            \psfrag{v18}[r][r][0.7]{0.4}%
            \psfrag{v19}[r][r][0.7]{0.45}%
            \psfrag{v20}[r][r][0.7]{0.5}%
            \psfrag{v21}[r][r][0.7]{0.55}%
            \psfrag{v22}[r][r][0.7]{0.6}%

            \psfrag{va12}[l][l][0.7]{0.1}%
            \psfrag{va13}[l][l][0.7]{0.15}%
            \psfrag{va14}[l][l][0.7]{0.2}%
            \psfrag{va15}[l][l][0.7]{0.25}%
            \psfrag{va16}[l][l][0.7]{0.3}%
            \psfrag{va17}[l][l][0.7]{0.35}%
            \psfrag{va18}[l][l][0.7]{0.4}%
            \psfrag{va19}[l][l][0.7]{0.45}%
            \psfrag{va20}[l][l][0.7]{0.5}%
            \psfrag{va21}[l][l][0.7]{0.55}%
            \psfrag{va22}[l][l][0.7]{0.6}%

            \psfrag{wavelength}[c][c][0.9][0]{wavelength $\lambda$}

            \psfrag{vr12}[c][c][0.7]{1385}%
            \psfrag{vr13}[c][c][0.7]{923}%
            \psfrag{vr14}[c][c][0.7]{693}%
            \psfrag{vr15}[c][c][0.7]{554}%
            \psfrag{vr16}[c][c][0.7]{462}%
            \psfrag{vr17}[c][c][0.7]{396}%
            \psfrag{vr18}[c][c][0.7]{346}%
            \psfrag{vr19}[c][c][0.7]{308}%
            \psfrag{vr20}[c][c][0.7]{277}%
            \psfrag{vr21}[c][c][0.7]{252}%
            \psfrag{vr22}[c][c][0.7]{231}%

            \psfrag{angle of incidence}[c][c][0.9][0]{angle of incidence $\alpha$ in \textdegree}
            \psfrag{x0}[t][t][0.7]{0}%
            \psfrag{x10}[t][t][0.7]{10}%
            \psfrag{xa20}[t][t][0.7]{20}%
            \psfrag{x30}[t][t][0.7]{30}%
            \psfrag{x40}[t][t][0.7]{40}%
            \psfrag{x50}[t][t][0.7]{50}%
            \psfrag{x60}[t][t][0.7]{60}%
            \psfrag{x70}[t][t][0.7]{70}%
            \psfrag{x80}[t][t][0.7]{80}%
            \psfrag{x90}[t][t][0.7]{90}%

            \includegraphics[width=\textwidth]{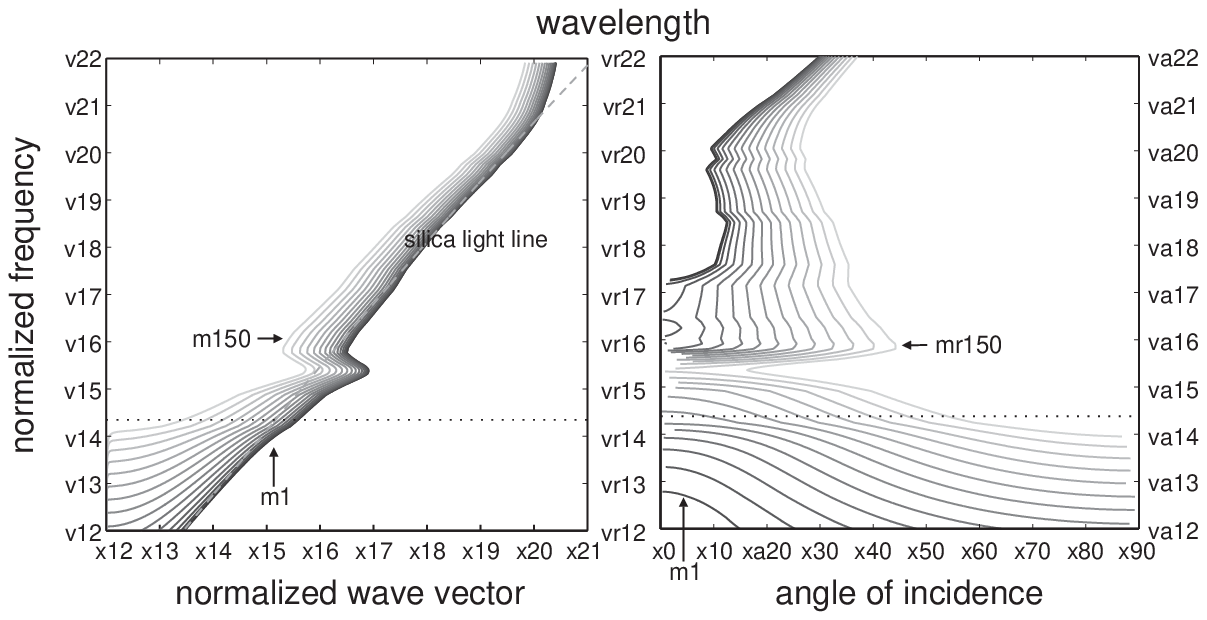}

        \caption[Dispersion of SPP on a gold cylinder of $D=\SI{20}{\micro m}$ in silica.]{Dispersion of SPP on a gold cylinder (fig.\;\ref{graphic_au-palik}) of $D=\SI{20}{\micro m}$, embedded in silica (fig.\;\ref{graphic_sio2-sellmeier}). Displayed are the dispersions of modes $m=1$ (in black) to $m= 151$ (in grey) in steps of $10$. In both figures the wavelength of the applied laser $\lambda=\SI{632.8}{nm}$ is indicated by a dotted line. The left figure shows the dispersion in terms of the normalized SPP wave vector $\beta \, c_0/ \omega_\text{p}$ (eq. \refeq{model_beta_m}). The right figure depicts the incident angles for excitation of the same modes (eq.~\refeq{eq-model-plasmon-incangles}).\label{graphic_spp-au-cylinder}}
\end{figure} 
    \begin{figure}
        \centering

            \psfrag{ml1}[r][r][0.7][0]{$m=1$}
            \psfrag{ml2}[r][r][0.7][0]{$m=2$}
            \psfrag{m1}[c][c][0.7][0]{$m=1$}
            \psfrag{m2}[c][c][0.7][0]{$m=2$}
            \psfrag{m342}[r][r][0.7][0]{$m=342$}
            \psfrag{m202}[r][r][0.7][0]{$m=202$}

            \psfrag{angle of incidence}[c][c][0.9][0]{angle of incidence $\alpha$ in \textdegree}
            \psfrag{x12}[t][t][0.7]{0}%
            \psfrag{x13}[t][t][0.7]{10}%
            \psfrag{x14}[t][t][0.7]{20}%
            \psfrag{x15}[t][t][0.7]{30}%
            \psfrag{x16}[t][t][0.7]{40}%
            \psfrag{x17}[t][t][0.7]{50}%
            \psfrag{x18}[t][t][0.7]{60}%
            \psfrag{x19}[t][t][0.7]{70}%
            \psfrag{x20}[t][t][0.7]{80}%
            \psfrag{x21}[t][t][0.7]{90}%

            \psfrag{cylinder radius in mum}[c][c][0.9][0]{cylinder radius in $\micro \text{m}$.}

            \psfrag{v12}[c][c][0.7]{0}%
            \psfrag{v13}[c][c][0.7]{2}%
            \psfrag{v14}[c][c][0.7]{4}%
            \psfrag{v15}[c][c][0.7]{6}%
            \psfrag{v16}[c][c][0.7]{8}%
            \psfrag{v17}[c][c][0.7]{10}%
            \psfrag{v18}[c][c][0.7]{12}%
            \psfrag{v19}[c][c][0.7]{14}%
            \psfrag{v20}[c][c][0.7]{16}%
            \psfrag{v21}[c][c][0.7]{18}%
            \psfrag{v22}[c][c][0.7]{20}%

            \includegraphics[width=\textwidth]{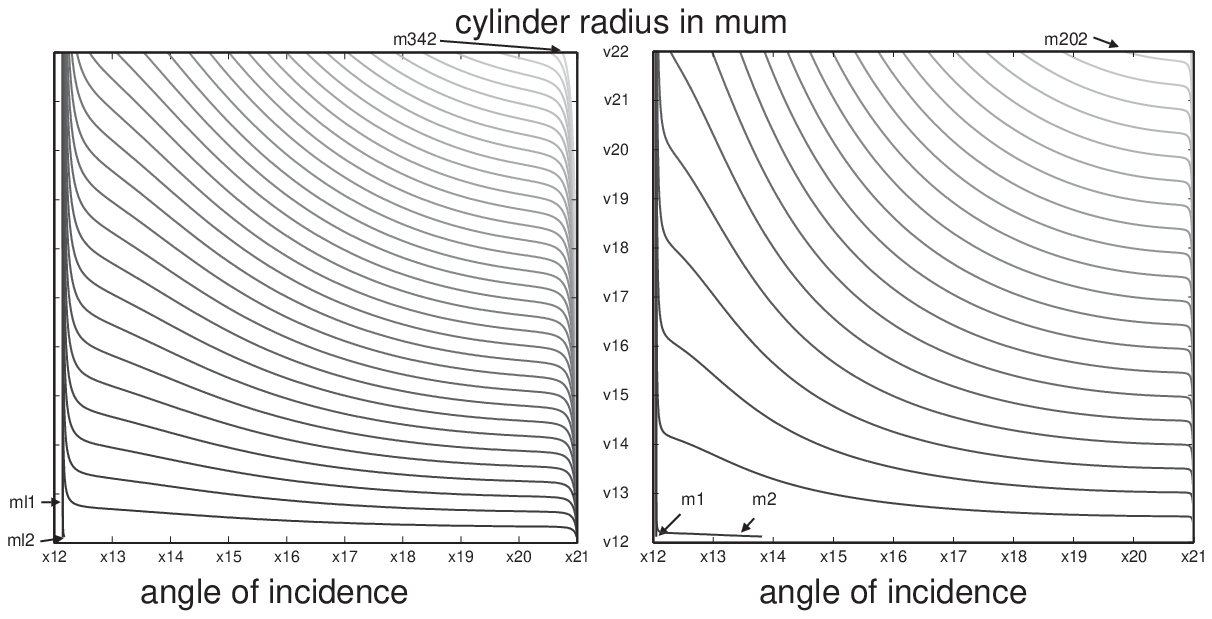}

        \caption[Dispersion of SPP modes on cylinders of gold in silica and silver in air over the radius and incident angle.]{Dispersion of SPP modes on cylinders of gold in silica (left) and silver in air (right) over the radius and incident angle. The excitation angles are calculated (eq. \refeq{model_beta_m}) for a wavelength of $\lambda=\SI{632.8}{nm}$. Displayed are the modes $m = 1,2,12,22, \ldots$ \label{graphic_spp-r-alpha}}
\end{figure} 

    In figure \ref{graphic_spp-ag-cylinder} the dispersion of SPP modes on the surface of a silver cylinder with diameter \SI{20}{\micro m} is shown, calculated from equation \eqref{model_beta_m}. For a better visibility only modes $m= 1,\ldots,101$ in steps of 10 are shown.

    The plot can be directly compared to the case of planar SPP (fig.\ref{graphic_spp-dispersion}). It shall be pointed out that, in comparison to the undamped dispersion (fig.\,\ref{graphic_spp-dispersion}), also for planar SPP a finite maximum wave number $\beta$ is approached at the surface plasmon frequency $\omega_\text{SP}$ if real materials are considered.

    In this case, including damping, the quasibound, leaky part of the dispersion relation in the region $\omega_\text{SP} <\omega < \omega_\text{p}$ is not forbidden any more. This ``backbending effect'' can be observed in all dispersion curves which consider losses\footnote{To avoid redundancy, for the case of planar SPP it is here not depicted explicitely. It can be easily found in common textbooks, e.g. \citet[p.~29]{maier-2005}.}. The effect can also be observed in figures \ref{graphic_spp-ag-cylinder} and \ref{graphic_spp-au-cylinder}.

    The depicted curves correspond well to the theoretical dispersion relations that were calculated by \citet{pfeiffer-theory}, as long as an undamped Drude model is assumed instead of the complex dielectric permittivity. Pfeiffer et al. presented an exact solution of radiative SPP modes below $\omega_\text{p}$ on the surfaces of metal cylinders for this simplified case.

    The inset (fig.~\ref{graphic_spp-ag-cylinder}, left) displays the dispersion in the wavelength region of experimental interest. The laser wavelength at $\lambda=\SI{632.8}{nm}$ is indicated as a horizontal dashed line.
    For the experimentally relevant wavelength regime it can be clearly seen that the dispersion curves of the modes lie above the light line. Thus an excitation of these radiative SPP modes is possible.\\

    For a gold wire in silica, as depicted in figure \ref{graphic_spp-au-cylinder}, the dispersion is slightly more complicated than for a silver wire due to the interband absorption in the visible regime (compare fig.~\ref{graphic_au-palik}). It is qualitatively the same as for a freestanding gold wire, only the remaining edge from the forbidden region in the undamped planar case smoothes out even further, which also smoothes the gap in the plot over the angles of excitation for further modes of lower order. Therefore a second plot for the air-gold configuration is not presented here.

    In the limit of very large diameters the dispersion of the cylindrical SPP approaches the dispersion of planar SPP as presented in section \ref{ch_theory-planar-spp}; for an infinite diameter, a cylinder is a single planar interface. This behaviour is obviously confirmed analytically, referring to eq. \eqref{model_beta_m}.\\

    For both configurations depicted in this section, the angles for excitation of the modes in a wide range of radii ($\SI{0.1}{\micro m}<R<\SI{20}{\micro m}$) are calculated, applying eq. \eqref{eq-model-plasmon-incangles}. The range of radii as well as the wavelength ($\lambda = \SI{632.8}{nm}$) are chosen according to the experiment. The resulting angles  $\alpha$ are depicted in figure \ref{graphic_spp-r-alpha}.

    It is apparent from this chart that the SPP modes for gold cylinders in silica spread further than those for silver cylinders in air. The modes of low order can be seen to converge for increasing radius at low $\alpha$ in all cases. Higher order modes converge with decreasing radius towards $\alpha = \ang{90}$. From the model it can be therefore expected that for low $\alpha$, thus in the case of striking incidence, for sufficiently large radius a number of distinct low order SPP excitations is observed whose spacing is very small, followed for larger $\alpha$ by an increasing mode spacing. This prediction is done under the assumption that the model is valid in the region of interest.

    \section{Comparison to an exact analytic solution \label{ch_theoresults-exact}}

    \citet{schmidt-theory} have reported a different approach to determine the dispersion relation of long-range spiralling SPP-modes on metallic nanowires. Applying the theory usually used to calculate guided modes in optical wave guides \citep{snyder-love} to the structure of metallic wires in dielectrics, an exact solution for the dispersion of different kinds of modes on the surface is obtained.

    In detail, the approach starts from solving Maxwell's equations in cylindrical coordinates, applying boundary conditions on the metal-to-dielectric interfaces. The resulting dispersion relation can be written as \citep{schmidt-theory}:

    \begin{align}
        (q_\text{o}^2 \psi_\text{i} + q_\text{i}^2 \psi_\text{o})(\varepsilon_\text{i}q_\text{o}^2 \psi_\text{i} + \varepsilon_\text{o} q_\text{i}^2 \psi_\text{o}) - m^2 n_m^2 (\varepsilon_\text{o} - \varepsilon_\text{i})^2 = 0
    \end{align}

    \noindent where $m$ is the azimuthal mode order. The radial wavevectors outside $q_\text{o}$ and inside the cylinder $q_\text{i}$ and the abbreviations $\psi_\text{o}$ and $\psi_\text{i}$ are defined as follows:

    \begin{align}
        \psi_\text{o} &= m - k_0 q_\text{o} R  \frac{K_{m+1}(k_0 q_\text{o} R)}{K_m (k_0 q_\text{o} R)} &
        \psi_\text{i} &= m - k_0 q_\text{i} R  \frac{J_{m+1}(k_0 q_\text{i} R)}{K_m (k_0 q_\text{i} R)} \\
        q_\text{o} &= \pm \sqrt{n_m^2-\varepsilon_\text{o}} & q_\text{i} &= \sqrt{\varepsilon_\text{i} - n_m^2}
    \end{align}

    \noindent $J_m(x)$ being the Bessel function of first kind, $K_m(x)$ being the modified Bessel function of kind (both of order $m$). $n_m$ represents the effective refractive index of mode $m$ as also described in the model (eq. \eqref{eq_model-n-eff}). In the mentioned publication the validity of the model whose results are discussed in sec.\;\ref{ch_theoresults-model} is cross checked with computed dispersion results from this exact approach. It is found to agree increasingly well for higher order modes.

    For a determination of the range of validity of the model in the relevant radius and wavelength ranges for radiative modes (above the silica respectively vacuum light line), the solutions for the appropriate modes are searched and the corresponding dispersions are calculated\footnote{These calculations were kindly contributed by Dr. Markus Schmidt.}.

    The obtained effective refractive indices $n_m$ are then converted to excitation angles analogue to eq.~\eqref{eq-model-plasmon-incangles}.

\begin{figure}
        \centering

            \psfrag{m5}[c][c][0.7][0]{$m=5$}
            \psfrag{m12}[c][c][0.7][0]{$m=12$}
            \psfrag{model}[l][l][0.7][0]{dispersion model}
            \psfrag{exact}[l][l][0.7][0]{exact solution}
            \psfrag{r}[r][r][0.9][90]{radius $R$ in \SI{}{\micro m}}

            \psfrag{v01}[r][r][0.7]{0}%
            \psfrag{v02}[r][r][0.7]{0.1}%
            \psfrag{v03}[r][r][0.7]{0.2}%
            \psfrag{v04}[r][r][0.7]{0.3}%
            \psfrag{v05}[r][r][0.7]{0.4}%
            \psfrag{v06}[r][r][0.7]{0.5}%
            \psfrag{v07}[r][r][0.7]{0.6}%
            \psfrag{v08}[r][r][0.7]{0.7}%
            \psfrag{v09}[r][r][0.7]{0.8}%
            \psfrag{v10}[r][r][0.7]{0.9}%
            \psfrag{v11}[r][r][0.7]{1}%
            \psfrag{v12}[r][r][0.7]{1}%
            \psfrag{v13}[r][r][0.7]{1.5}%
            \psfrag{v14}[r][r][0.7]{2}%
            \psfrag{v15}[r][r][0.7]{2.5}%
            \psfrag{v16}[r][r][0.7]{3}%
            \psfrag{v17}[r][r][0.7]{3.5}%
            \psfrag{v18}[r][r][0.7]{4}%
            \psfrag{v19}[r][r][0.7]{4.5}%
            \psfrag{v20}[r][r][0.7]{5}%

            \psfrag{angle}[c][c][0.9][0]{angle of incidence $\alpha$ in \textdegree}
            \psfrag{x01}[t][t][0.7]{0}%
            \psfrag{x02}[t][t][0.7]{0.1}%
            \psfrag{x03}[t][t][0.7]{0.2}%
            \psfrag{x04}[t][t][0.7]{0.3}%
            \psfrag{x05}[t][t][0.7]{0.4}%
            \psfrag{x06}[t][t][0.7]{0.5}%
            \psfrag{x07}[t][t][0.7]{0.6}%
            \psfrag{x08}[t][t][0.7]{0.7}%
            \psfrag{x09}[t][t][0.7]{0.8}%
            \psfrag{x10}[t][t][0.7]{0.9}%
            \psfrag{x11}[t][t][0.7]{1}%
            \psfrag{x12}[t][t][0.7]{0}%
            \psfrag{x13}[t][t][0.7]{10}%
            \psfrag{x14}[t][t][0.7]{20}%
            \psfrag{x15}[t][t][0.7]{30}%
            \psfrag{x16}[t][t][0.7]{40}%
            \psfrag{x17}[t][t][0.7]{50}%
            \psfrag{x18}[t][t][0.7]{60}%
            \psfrag{x19}[t][t][0.7]{70}%
            \psfrag{x20}[t][t][0.7]{80}%
            \psfrag{x21}[t][t][0.7]{90}%

            \includegraphics[width=0.75\textwidth]{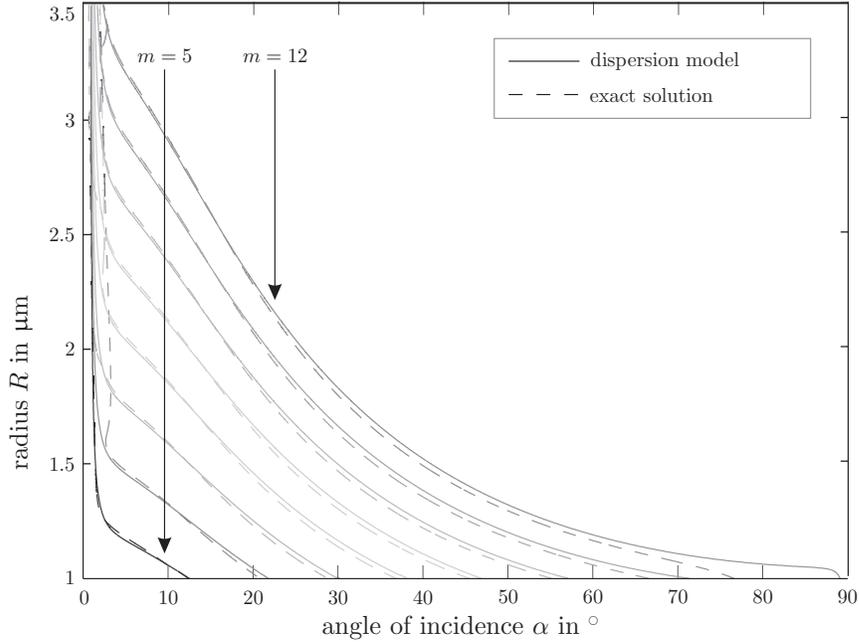}

        \caption[Comparison of the dispersion of cylindrical surface modes from the derived model to the dispersion from an exact solution of the guided modes.]{Comparison of the dispersion of cylindrical surface modes from the derived model (lines) to the dispersion from an exact solution of the guided modes (dashed). Calculated are modes $m = 5 \ldots 12$ on a gold cylinder ($R = 2 \ldots \SI{10}{\micro m}$) in air.\label{graphic_exact-vs-model}}
\end{figure}

    A comparison of the resulting angles $\alpha$ for radii $\SI{1}{\micro m}<R<\SI{3.5}{\micro m}$ from the model of spiralling SPP and the exact solution is depicted in figure \ref{graphic_exact-vs-model}.

    The model shows good agreement with the exact solutions for lower radii per mode, respectively for higher $\alpha$ for all modes. The deviation of both solutions increases with lower incident angles. Particularly for small angles from the graphs in fig.\;\ref{graphic_exact-vs-model} it is apparent that the dispersions do not coincide.

    \section{Numerical computation of the scattering coefficients \label{chapter_theory-scatterincoeff}}

    \begin{figure}
        \centering

        \psfrag{1}[c][c][1][0]{\textcolor{white}{\textbf{\textsf{1}}}}
        \psfrag{2}[c][c][1][0]{\textcolor{white}{\textbf{\textsf{2}}}}
        \psfrag{3}[c][c][1][0]{\textcolor{white}{\textbf{\textsf{3}}}}
        \psfrag{4}[c][c][1][0]{\textcolor{white}{\textbf{\textsf{4}}}}
        \psfrag{5}[c][c][1][0]{\textcolor{white}{\textbf{\textsf{5}}}}
        \psfrag{6}[c][c][1][0]{\textcolor{white}{\textbf{\textsf{6}}}}
        \psfrag{Au,air}[c][c][1][0]{\textcolor{white}{\textbf{\textsf{Au-air}}}}
        \psfrag{Ag,air}[c][c][1][0]{\textcolor{white}{\textbf{\textsf{Ag-air}}}}

        \psfrag{angle}[c][c][0.7]{incident angle $\alpha$ in $\degree$}%
        \psfrag{radius}[c][c][0.7]{radius in \SI{}{\micro m}}%

\psfrag{a00}[t][t][0.7]{$0$}%
\psfrag{a01}[t][t][0.7]{$1$}%
\psfrag{a02}[t][t][0.7]{$2$}%
\psfrag{a03}[t][t][0.7]{$3$}%
\psfrag{a04}[t][t][0.7]{$4$}%
\psfrag{a05}[t][t][0.7]{$5$}%
\psfrag{a06}[t][t][0.7]{$6$}%
\psfrag{a07}[t][t][0.7]{$7$}%
\psfrag{a08}[t][t][0.7]{$8$}%
\psfrag{a09}[t][t][0.7]{$9$}%
\psfrag{a10}[t][t][0.7]{$10$}%
\psfrag{a11}[t][t][0.7]{$11$}%
\psfrag{a12}[t][t][0.7]{$12$}%
\psfrag{a13}[t][t][0.7]{$13$}%
\psfrag{a14}[t][t][0.7]{$14$}%
\psfrag{a15}[t][t][0.7]{$15$}%
\psfrag{a16}[t][t][0.7]{$16$}%
\psfrag{a17}[t][t][0.7]{$17$}%
\psfrag{a18}[t][t][0.7]{$18$}%
\psfrag{a19}[t][t][0.7]{$19$}%
\psfrag{a20}[t][t][0.7]{$20$}%
\psfrag{a21}[t][t][0.7]{$21$}%
\psfrag{a22}[t][t][0.7]{$22$}%
\psfrag{a30}[t][t][0.7]{$30$}%
\psfrag{a40}[t][t][0.7]{$40$}%
\psfrag{a50}[t][t][0.7]{$50$}%
\psfrag{a60}[t][t][0.7]{$60$}%
\psfrag{a70}[t][t][0.7]{$70$}%
\psfrag{a80}[t][t][0.7]{$80$}%

\psfrag{r01}[c][c][0.7]{$1$}%
\psfrag{r02}[c][c][0.7]{$2$}%
\psfrag{r03}[c][c][0.7]{$3$}%
\psfrag{r04}[c][c][0.7]{$4$}%
\psfrag{r05}[c][c][0.7]{$5$}%
\psfrag{r06}[c][c][0.7]{$6$}%
\psfrag{r07}[c][c][0.7]{$7$}%
\psfrag{r08}[c][c][0.7]{$8$}%
\psfrag{r09}[c][c][0.7]{$9$}%
\psfrag{r10}[c][c][0.7]{$10$}%
\psfrag{r11}[c][c][0.7]{$11$}%
\psfrag{r12}[c][c][0.7]{$12$}%
\psfrag{r13}[c][c][0.7]{$13$}%
\psfrag{r14}[c][c][0.7]{$14$}%
\psfrag{r15}[c][c][0.7]{$15$}%
\psfrag{r16}[c][c][0.7]{$16$}%
\psfrag{r17}[c][c][0.7]{$17$}%
\psfrag{r18}[c][c][0.7]{$18$}%
\psfrag{r19}[c][c][0.7]{$19$}%
\psfrag{r20}[c][c][0.7]{$20$}%
\psfrag{r21}[c][c][0.7]{$21$}%
\psfrag{r22}[c][c][0.7]{$22$}%
\psfrag{r30}[c][c][0.7]{$30$}%
\psfrag{r40}[c][c][0.7]{$40$}%
\psfrag{r50}[c][c][0.7]{$50$}%
\psfrag{r60}[c][c][0.7]{$60$}%
\psfrag{r70}[c][c][0.7]{$70$}%
\psfrag{r80}[c][c][0.7]{$80$}%

\psfrag{sv00}[l][l][0.7]{ $ 0$}%
\psfrag{sv0.1}[l][l][0.7]{$ 0.1$}%
\psfrag{sv0.2}[l][l][0.7]{$ 0.2$}%
\psfrag{sv0.3}[l][l][0.7]{$ 0.3$}%
\psfrag{sv0.4}[l][l][0.7]{$ 0.4$}%
\psfrag{sv0.5}[l][l][0.7]{$ 0.5$}%
\psfrag{sv0.6}[l][l][0.7]{$ 0.6$}%
\psfrag{sv0.8}[l][l][0.7]{$ 0.8$}%
\psfrag{sv01}[l][l][0.7]{$ 1$}%
\psfrag{sv1.2}[l][l][0.7]{$ 1.2$}%
\psfrag{sv1.4}[l][l][0.7]{$ 1.4$}%
\psfrag{sc1.6}[l][l][0.7]{$ 1.6$}%
\psfrag{sv1.8}[l][l][0.7]{$ 1.8$}%
\psfrag{sv02}[l][l][0.7]{$ 2$}%

\psfrag{sh00}[t][t][0.7]{$0$}%
\psfrag{sh0.1}[t][t][0.7]{$0.1$}%
\psfrag{sh0.2}[t][t][0.7]{$0.2$}%
\psfrag{sh0.3}[t][t][0.7]{$0.3$}%
\psfrag{sh0.4}[t][t][0.7]{$0.4$}%
\psfrag{sh0.5}[t][t][0.7]{$0.5$}%
\psfrag{sh0.6}[t][t][0.7]{$0.6$}%
\psfrag{sh0.8}[t][t][0.7]{$0.8$}%
\psfrag{sh01}[t][t][0.7]{$1$}%
\psfrag{sh1.2}[t][t][0.7]{$1.2$}%
\psfrag{sh1.4}[t][t][0.7]{$1.4$}%
\psfrag{sc1.6}[t][t][0.7]{$1.6$}%
\psfrag{sh1.8}[t][t][0.7]{$1.8$}%
\psfrag{sh02}[t][t][0.7]{$2$}%

        \includegraphics[width=0.95\textwidth]{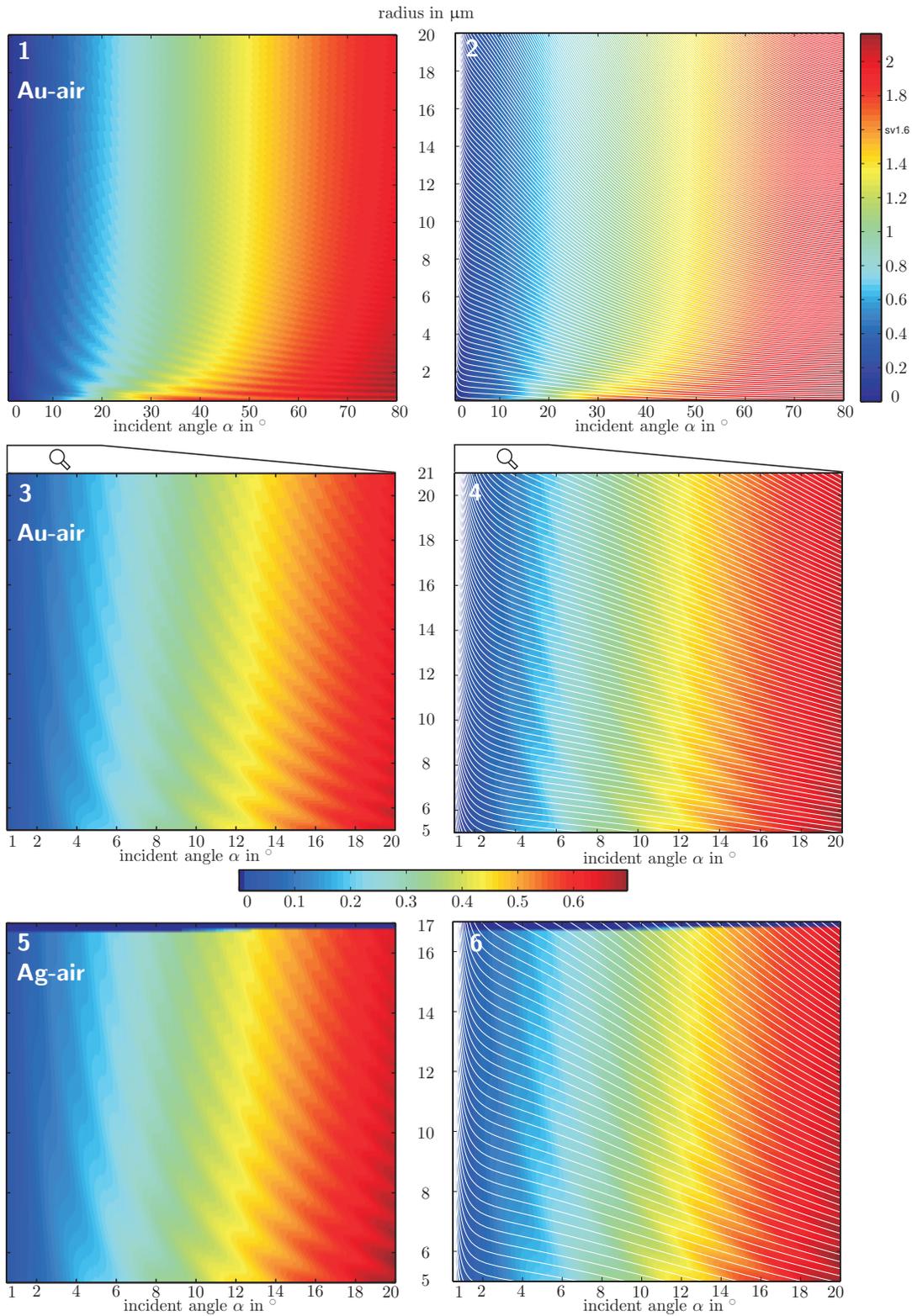}

        \caption[Absorption efficiency for different angles and diameters on a single of gold and silver in air for $\lambda = \SI{632.8}{nm}$ and p-polarized light.]{Absorption efficiency for different angles and diameters on a single of gold and silver in air. The left graphs (1,3,5) show the amplitude of the scattering efficiency $Q_\text{abs}$ as a colourplot over a variation of incident angle ($\alpha$) and radius ($R$, in \SI{}{\micro m}), calculated from eq. for a wavelength of $\lambda = \SI{632.8}{nm}$. The graphs in the right column (2,4,6) overlay the angles of SPP excitation as calculated from the model (eq. \refeq{eq-model-plasmon-incangles}, fig. \ref{graphic_spp-r-alpha}) over each of the left graphs for comparison. Only $Q_\text{abs}$ for p-polarization is shown, as the s-polarization does not exhibit a resonance structure.\label{graphic_scattering-colorplots-air}}
\end{figure} 

    For this analysis again it is important how the material properties are described. \citet{martinos-economou} point out that modelling the excitation of surface plasmons in scattering theory makes it necessary to explicitly introduce damping of the electromagnetic waves inside the material. This is especially the case as many theoretical treatments of the problem omit losses and additionally assume a dielectric permittivity of the metal of $\varepsilon_\text{metal} = -1$ resulting in strongest possible SPP resonance, although it is not experimentally applicable.

    The application of only the real part of $\varepsilon(\omega)$, thus omitting losses, in fact simplifies the numerical solution of the scattering problem significantly: Complex values as arguments of the Bessel functions, Hankel functions and their derivatives in the equation system \eqref{eq-scattering-coefficientmatrix} result in magnitudes for physically real parameters lead to problems in the numerical representation of 32 bit computer programs\footnote{An overview over the Bessel functions for complex arguments is compiled in appendix \ref{ch_bessel-functions}}. This was an even more significant challenge in earlier studies, as described by \citet{shah}.

    In the described calculations the wavelength dependent intrinsic dissipative properties of the material are entirely included by applying complex dielectric permittivities $\varepsilon(\omega)$.

    The first step to solve this problem in this work was to derive a linear equation system and the corresponding solution matrix \eqref{eq-scattering-coefficientmatrix}. This matrix is numerically solved instead of computing the coefficients $a_{n\text{s}},a_{n\text{p}},b_{n\text{s}},b_{n\text{p}}$\footnote{To achieve the scattering efficiencies, obviously only the coefficients $a,b$ for the fields outside the cylinder are necessary. $f$ and $g$ for the expansion of the fields inside the cylinder are only used for the computation of the field distribution as shown in section \ref{chapter_theory-fielddistrib}.} from equations \eqref{eq_scattering-an-bn}, which is a much more efficient approach and provides a higher precision of computation.

    The wire radius $R$ enters the arguments of the Bessel functions $\eta, \xi$ (\refeq{eq_scattering-eta},\refeq{eq_scattering-xi}), respectively of the whole expansion. It is shown that the higher $R$ is, the more orders $n$ of the expansion are necessary, to include all nonzero coefficients. The expansion is shown to converge for orders $m \gtrsim 15$ for small radii ($R \lesssim \SI{3}{\micro m}$) respectively small angles . For a computation of all incident angles $\ang{0}<\alpha<\ang{90}$, (e.g. in fig.s \ref{graphic_scattering-colorplots-air},\ref{graphic_scattering-graph}), the order $m$ of the expansion has to be increased to $m \leq 220$. A routine is implemented in the program to determine the necessary number of orders and to indicate results that cause erroneous results.

    The range of radii for certain angles can due to this method be enlarged, compared to a computation from equations \eqref{eq_scattering-an-bn}. It is therefore possible to calculate the scattering efficiencies for radii up $R=\SI{20}{\micro m}$ e.g. for silver and gold wires with surrounding air.

    Applying this method, extinction and scattering efficiencies $Q_\text{ext}$ and $Q_\text{sca}$ are calculated from eq. \eqref{eq_scattering-efficiencies}. Eventually the absorption efficiency is obtained from eq. \refeq{eq_scattering-abs}.

    The curves of the absorption efficiencies $Q_\text{abs}$ (fig. \ref{graphic_scattering-graph}) in the case of s-polarization have substantially different shapes than those for p-polarization:

\begin{description}
  \item[For s-polarization] $Q_\text{abs}$ increases with $\alpha$ (fig \ref{graphic_scattering-graph}). For the wavelength that is also experimentally applied (middle gray, dashed, \SI{632.8}{nm}), the curve increases monotonic and no resonance dips are observed. For shorter frequencies, this behaviour stays the same.
  \item[For p-polarization] the slope of $Q_\text{abs}$ remains positive. Different to s-polarization, a modulation of the absorption efficiency over $\alpha$ is clearly indicated.
\end{description}

    For longer wavelengths (black, \SI{1200}{nm}) the absorption efficiency shows anomalous behaviour. For both polarizations the curves are modulated by an additional fluctuation. Particularly in s-polarization this results in a modulation significantly visible in the region of small incident angles. It is not assumed that this superimposed effect is caused by the interband transitions of gold, as (see fig. \ref{graphic_au-palik}) these lie at much shorter wavelengths. A precise examination of the effect could reveal interesting results but is postponed as the focus of this work is the examination at $\lambda=\SI{632.8}{nm}$.

    The left plot in figure \ref{graphic_scattering-colorplots-air} displays the absorption efficiency $Q_\text{abs}$ for a silver wire in air ($\varepsilon_\text{o}=1$)\footnote{It shall be particularly mentioned that the blue ($=0$) area for radii $R>\SI{17}{\micro m}$ at small angles is a numerical artefact. The coefficients for these radii cannot be obtained from the applied theory, for the aforementioned numerical reasons and are therefore set to zero.}.

    Comparable plots of the computed scattering cross section over the wavelength $\lambda$ and a variation of the refractive index for coated cylinders with perpendicular incidence with the objective to examine finite wavelength cloaking have been recently published \citet{nicorovici}.

    \citet{martinos-economou} demonstrated that the imaginary part of the eigenfrequencies of the surface plasmon modes increases for higher mode order $n$, which causes the sharpness of the peaks in the absorption coefficients $Q_\text{abs}$ to decrease with $n$. This theoretical prediction can also be clearly seen in the graphs for p-polarization in figure \ref{graphic_scattering-graph}: For increasing $\alpha$, thus higher mode orders, the dips smooth out. \\

    \subsection{Inherence of SPP resonances \label{ch_scattering-resonances}}

    It can be shown that the denominators of equations (\refeq{eq_scattering-an-bn-p},~\refeq{eq_scattering-an-bn-p}) coincide morphologically with the resonance angles $\alpha$ calculated from the exact solution of the dispersion relation of SPP modes on the cylinder \citep{schmidt-theory} (as presented in sec. \ref{ch_theoresults-exact}).

    This observation suggests clearly that the derived scattering model accords to the description of SPP modes on the cylinder surface. The same is therefore true for the approximated model in its region of coincidence with the exact solution.

    \citet{englmann} already discussed a theoretical comparison of an exact solution for the dispersion of modes on ionic crystals of cylindrical shape to the scattering patterns calculated analogue to $Q_\text{abs}$. Although the size of their particles lies clearly in the regime much smaller than $\lambda$, in the limiting case their result also shows a coincidence of the resonances in the scattering theory and those from an exact solution of the mode dispersion.

    Recently \citet{she-2008} have reported a theoretical comparison of exact classical scattering theory to analytic formulae for plasmonic resonances on small wires ($D\ll \lambda$) with low dissipation. Their results also indicate an interconnection as the one proposed here, although their calculations were again only computed for ideal $\varepsilon_\text{i} = -1 + \i \delta$ with very small dissipation, which is then in detail transferred to very thin, coated Ag cylinders. The detailed results therefore cannot be compared. However, the resulting resonance behaviour can be compared to the one achieved in this work and shows coincidence.

    As the comparison of an exact solution for modes of low order (sec.~\ref{ch_theoresults-exact}) indicated a rather good agreement, at least for sufficiently large $R$ and $\alpha$, the calculated $R$ over $\alpha$ for wires of gold (2,4) and silver (6) are overlayed to the amplitude of $Q_\text{abs}$ (right plots (2,4,6) in fig.\;\ref{graphic_scattering-colorplots-air}). Particularly for large angles of incidence $\alpha$ a matching of both theories can be observed. However, for incident angles $\alpha < \ang{20}$, the calculated modes do not coincide with the absorption efficiencies. As the range of $\alpha$ in the experiments is particularly chosen in this range, therefore it is decided to compare the measured results primarily with the computed results for $Q_\text{abs}$.

    \section{Numerical computation of the scattered field distribution \label{chapter_theory-fielddistrib}\label{ch_scattering-s}}

     \begin{figure}
        \centering

\psfrag{alpha45}[c][c][0.9]{$\alpha=\ang{45}$}%
\psfrag{alpha90}[c][c][0.9]{$\alpha=\ang{90}$}%

\psfrag{x}[c][c][0.7]{$x$}%
\psfrag{y}[c][c][0.7]{$y$}%
\psfrag{z}[c][c][0.7]{$z$}%
\psfrag{E}[c][c][0.7]{$|E|$}%

\psfrag{yzs}[B][B][0.7]{$yz$ s-pol.}%
\psfrag{xzs}[B][B][0.7]{$xz$ s-pol.}%
\psfrag{xys}[Br][Br][0.7]{$xy$ s-pol.}%
\psfrag{yzp}[B][B][0.7]{$yz$ p-pol.}%
\psfrag{xzp}[B][B][0.7]{$xz$ p-pol.}%
\psfrag{xyp}[Br][Br][0.7]{$xy$ p-pol.}%


\psfrag{v01}[B][B][0.7]{0.2}%
\psfrag{v02}[B][B][0.7]{0.4}%
\psfrag{v03}[B][B][0.7]{0.6}%
\psfrag{v04}[B][B][0.7]{0.8}%
\psfrag{v05}[B][B][0.7]{1}%
\psfrag{v06}[B][B][0.7]{1.2}%
\psfrag{v07}[B][B][0.7]{1.4}%
\psfrag{v08}[B][B][0.7]{1.6}%
\psfrag{v09}[B][B][0.7]{-5}%
\psfrag{v10}[B][B][0.7]{-3}%
\psfrag{vx}[B][B][0.7]{-1}%
\psfrag{v12}[B][B][0.7]{1}%
\psfrag{v13}[B][B][0.7]{3}%
\psfrag{v14}[B][B][0.7]{5}%

\psfrag{v9001}[B][B][0.7]{0}%
\psfrag{v9002}[B][B][0.7]{0.2}%
\psfrag{v9003}[B][B][0.7]{0.4}%
\psfrag{v904}[B][B][0.7]{0.6}%
\psfrag{v9005}[B][B][0.7]{0.8}%
\psfrag{v9006}[B][B][0.7]{1}%
\psfrag{v907}[B][B][0.7]{1.2}%
\psfrag{v9008}[B][B][0.7]{1.4}%
\psfrag{v9009}[B][B][0.7]{1.6}%

\psfrag{x01}[t][t][0.7]{-5}%
\psfrag{x02}[t][t][0.7]{-3}%
\psfrag{x03}[t][t][0.7]{-1}%
\psfrag{x04}[t][t][0.7]{1}%
\psfrag{x05}[t][t][0.7]{3}%
\psfrag{x06}[t][t][0.7]{5}%

\psfrag{xi01}[B][B][0.5]{-1}%
\psfrag{xi02}[B][B][0.5]{-0.3}%
\psfrag{xi03}[B][B][0.5]{0.3}%
\psfrag{xi04}[B][B][0.5]{1}%
\psfrag{vi10}[l][l][0.5]{-1}%
\psfrag{vix}[l][l][0.5]{-0.3}%
\psfrag{vi12}[l][l][0.5]{0.3}%
\psfrag{vi13}[l][l][0.5]{1}%

            \includegraphics[width=\textwidth]{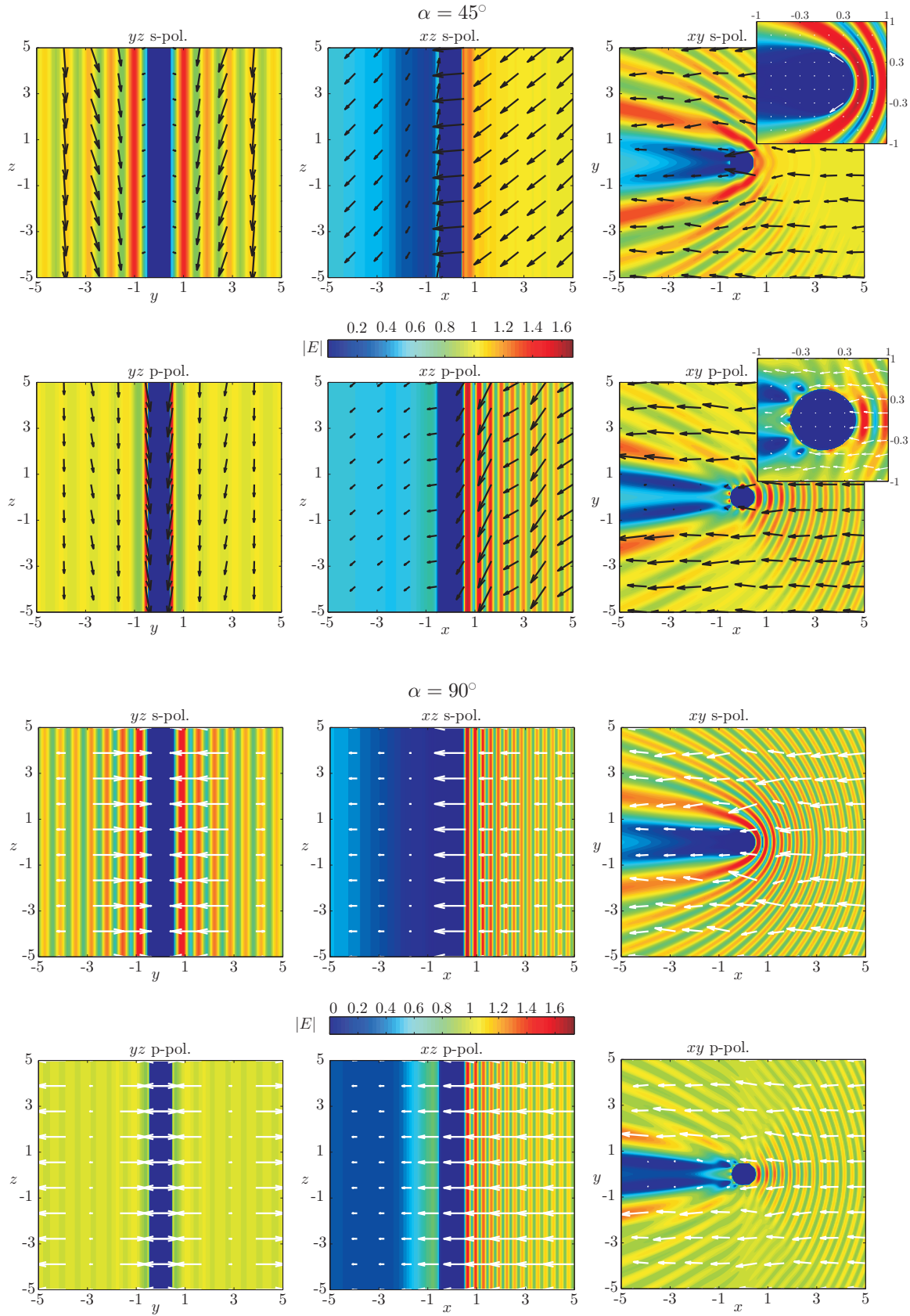}

        \caption[Field distribution and energy flux, calculated for oblique incidence on a single gold wire $D=\SI{1}{\micro m}$ in air.]{Field distribution and energy flux, calculated for $\alpha = \ang{45}$ and $\alpha = \ang{90}$. Plane waves with $\lambda=\SI{632.8}{nm}$ strike a single gold wire $D=\SI{1}{\micro m}$ in air. The absolute value of the electric field vector ($|\vc E|$) and the values and directions of the Poynting vector ($\vc S$) are displayed. The spatial coordinates are given in \SI{}{\micro m}.\label{sc-s-au-air-0.5}}
\end{figure} 

    Although the scattering coefficients (sec.\;\ref{chapter_theory-scatterincoeff}) allow a good insight into the angle dependency of the total scattering properties of the cylinder, a more detailed analysis of the scattered field distributions is worth being done. The field distributions and the energy flux, represented by the Poynting vector offer near field information about the scattering processes in contrast to the far field scattering properties of the scattering efficiency (as introduced in section \ref{ch_theory-scatteringefficiency}).

    Scattered field distributions, the Poynting vector field distribution and Poynting vector field bifurcations for the simplified case of normal incidence ($\alpha=\pi/2$) on small ($D \ll \lambda$) coated cylinders were recently studied in detail by \citet{lukyanchuk}.

    The field distribution $(\vc E (\vc r))_{r,\varphi,z}$ is calculated for all three cylindrical coordinates by solving matrix \eqref{eq-scattering-coefficientmatrix} analogue to the algorithm for the computation of the absorption coefficients $Q_\text{abs}$. For the field distribution inside the cylinder $\vc E_\text{i}$, not only the scattering coefficients $a_n,b_n$ but also $f_n, g_n$ are determined.

    After a transformation from cylindrical coordinates to carthesian coordinates, the field distribution is plotted for an area chosen to show the wire surface with high resolution.

    Figure \ref{sc-s-au-air-0.5} depicts the absolute field distributions ($|\vc{E}|$) exemplarily for incident plane waves with $\alpha = \ang{45}$ (top six pictures) and for normal incidence ($\alpha = \ang{90}$, bottom six pictures) as well as the projections of the Poynting vector. Each row of pictures shows the three perpendicular spatial planes in cartesian coordinates with respect to the cylinder and the plane of rotation, as illustrated in fig. \ref{graphic_scattering-drawing}.

    In all pictures, the plane waves approach the cylinder from the right side.

\begin{figure}
        \centering

\psfrag{x}[c][c][0.7]{$x$}%
\psfrag{y}[c][c][0.7]{$y$}%
\psfrag{E}[c][c][0.7]{$|E|$}%
\psfrag{Q}[c][c][0.9]{$Q_\text{abs}$}%
\psfrag{angle}[c][c][0.9]{incident angle $\alpha$ in $\degree$}%


\psfrag{1:7degs}[B][B][0.9]{$\alpha_1 = \ang{7}$ s-pol.}%
\psfrag{2:9degs}[B][B][0.9]{$\alpha_2 = \ang{9.5}$ s-pol.}%
\psfrag{3:18degs}[B][B][0.9]{$\alpha_3 = \ang{18.1}$ s-pol.}%
\psfrag{1:7degp}[t][t][0.9]{$\alpha_1 = \ang{7}$ p-pol.}%
\psfrag{2:9degp}[t][t][0.9]{$\alpha_2 = \ang{9.5}$ p-pol.}%
\psfrag{3:18degp}[t][t][0.9]{$\alpha_3 = \ang{18.1}$ p-pol.}%

\psfrag{1s}[l][l][0.5]{s-polarization \SI{400}{nm} }%
\psfrag{1p}[l][l][0.5]{p-polarization \SI{400}{nm} }%
\psfrag{2s}[l][l][0.5]{s-polarization \SI{632.8}{nm}}%
\psfrag{2p}[l][l][0.5]{p-polarization \SI{632.8}{nm}}%
\psfrag{3s}[l][l][0.5]{s-polarization \SI{1200}{nm}}%
\psfrag{3p}[l][l][0.5]{p-polarization \SI{1200}{nm}}%


\psfrag{0}[r][r][0.7]{$0$}%
\psfrag{0.5}[r][r][0.7]{$0.5$}%
\psfrag{1}[r][r][0.7]{$1$}%
\psfrag{1.5}[r][r][0.7]{$1.5$}%
\psfrag{2}[r][r][0.7]{$2$}%
\psfrag{2.5}[r][r][0.7]{$2.5$}%
\psfrag{3}[r][r][0.7]{$3$}%
\psfrag{3.5}[r][r][0.7]{$3.5$}%
\psfrag{4}[r][r][0.7]{$4$}%
\psfrag{4.5}[r][r][0.7]{$4.5$}%


\psfrag{h0}[t][t][0.7]{$0$}%
\psfrag{10}[t][t][0.7]{$10$}%
\psfrag{20}[t][t][0.7]{$20$}%
\psfrag{30}[t][t][0.7]{$30$}%
\psfrag{40}[t][t][0.7]{$40$}%
\psfrag{50}[t][t][0.7]{$50$}%
\psfrag{60}[t][t][0.7]{$60$}%
\psfrag{70}[t][t][0.7]{$70$}%
\psfrag{80}[t][t][0.7]{$80$}%
\psfrag{90}[t][t][0.7]{$90$}%

\psfrag{h1}[t][t][0.5]{$-5$}%
\psfrag{h2}[t][t][0.5]{$-3$}%
\psfrag{h3}[t][t][0.5]{$-1$}%
\psfrag{h4}[t][t][0.5]{$1$}%
\psfrag{h5}[t][t][0.5]{$3$}%
\psfrag{h6}[t][t][0.5]{$5$}%
\psfrag{v1}[r][r][0.5]{$-5$}%
\psfrag{v2}[r][r][0.5]{$-3$}%
\psfrag{v3}[r][r][0.5]{$-1$}%
\psfrag{v4}[r][r][0.5]{$1$}%
\psfrag{v5}[r][r][0.5]{$3$}%
\psfrag{v6}[r][r][0.5]{$5$}%

\psfrag{xi01}[B][B][0.5]{-1}%
\psfrag{xi02}[B][B][0.5]{-0.3}%
\psfrag{xi03}[B][B][0.5]{0.3}%
\psfrag{xi04}[B][B][0.5]{1}%
\psfrag{vi10}[l][l][0.5]{-1}%
\psfrag{vix}[l][l][0.5]{-0.3}%
\psfrag{vi12}[l][l][0.5]{0.3}%
\psfrag{vi13}[l][l][0.5]{1}%

            \includegraphics[width=\textwidth]{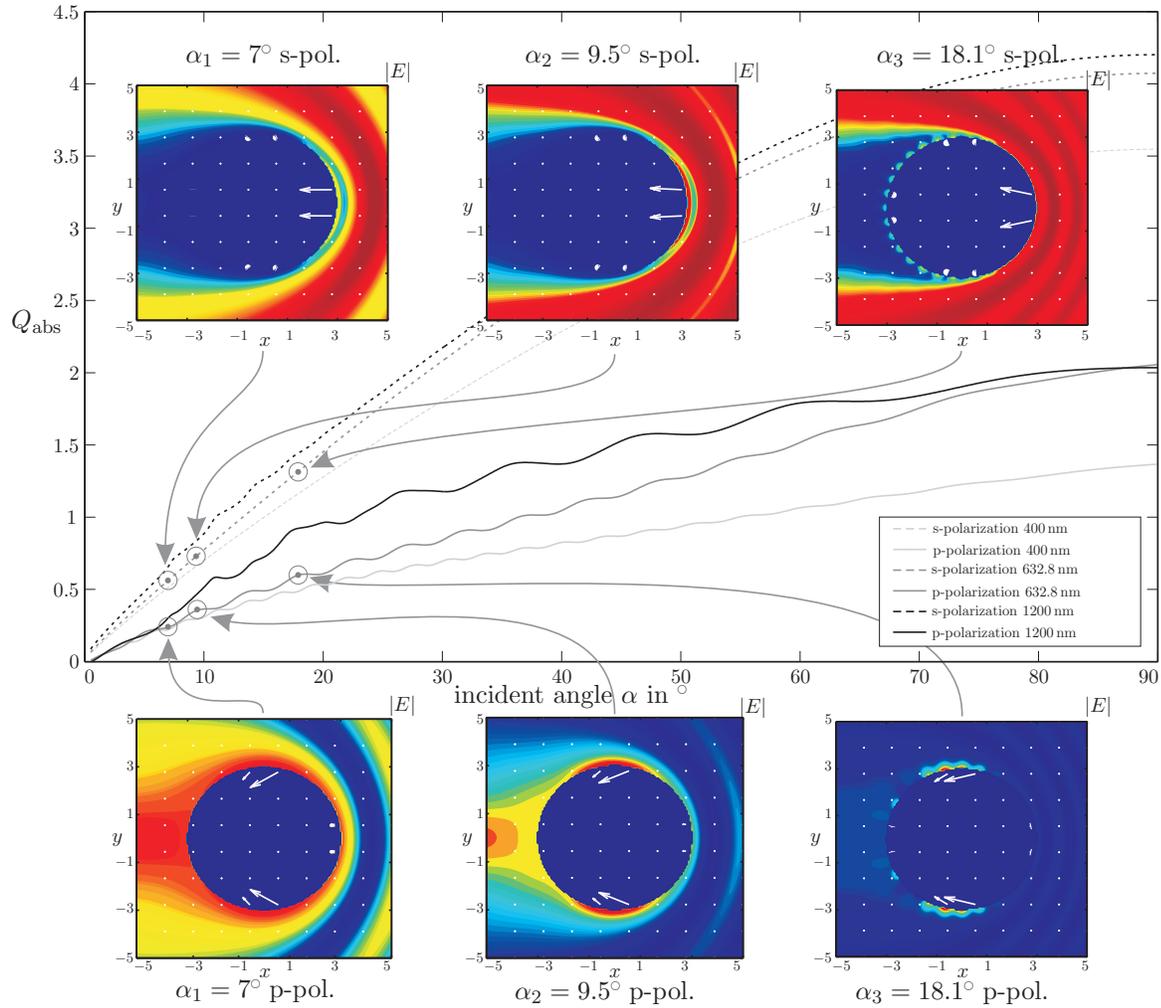}

        \caption[Absorption efficiency of a $R = \SI{3}{\micro m}$ gold wire. Plot for s- and p-polarization at the frequencies $\lambda = \SI{400}{nm},\SI{632.8}{nm},\SI{1200}{nm}$ and resonance field and Poynting vector distributions.]{Absorption efficiency $Q_\text{abs}$ of a gold wire in silica ($R = \SI{3}{\micro m}$). Plot for s- and p-polarization at the frequencies $\lambda = \SI{400}{nm}$ (light gray), $\lambda = \SI{632.8}{nm}$ (middle gray) and $\lambda = \SI{1200}{nm}$ (black) and resonance field ($|\vc E|$, arbitrary units) and Poynting vector distributions for \SI{632.8}{nm} at angles of \ang{7} (1), \ang{9} (2), \ang{18} (3) for p- (top) and s-polarization (bottom). Spatial units in \SI{}{\micro m}.\label{graphic_scattering-graph}}
\end{figure}

    Examining the field distributions over a variation of the incident angle reveals the formation of spots of field enhancement on the wire surface. The insets in figure \ref{graphic_scattering-graph} show that for both polarizations such field enhancement effects occur. It is assured by several checks of order of the field expansion and from checks of the continuity of the tangential field components that the observed effects are not computing artefacts.

    An striking effect is that also in the case of s-polarization, for increasing $\alpha$, on the non illuminated side of the wire for a particular choice of parameters, spots occur. These can be seen in figure \ref{graphic_scattering-graph}, whilst for the parameters in figure \ref{sc-s-au-air-0.5} the effect cannot be observed. It seems that this effect for the s-polarization is caused by localized surface plasmon (LSP) excitation, which cannot yet be proved due to restricted computation time.

    Particularly for p-polarization, for which from the model SPP excitation is expected, not only with increasing $\alpha$ discrete field spots on the surface occur. From a fine resolved simulation of field plots over $\alpha$, a periodic fluctuation of those is observed, which can be correlated to the modulation of the absorption efficiencies as depicted in figure \ref{graphic_scattering-graph}.

    The Poynting vectors for p-polarization indicate an energy flow along the surface of the wire. Different to the simplified schematic in figure \ref{graphic_model-drawing}, the beam, in the theoretical calculation the plane wave, extends not only over one point on the top or bottom half of a horizontal cylinder but hits its full lateral extension. This means that if the resonance condition is fulfilled, SPP modes into both, counter-propagating directions are expected to be excited.

    An outstanding feature of a comparison of the field distributions is the formation of a ``shadow'' behind the wire in the case of s-polarization, but the emission of radiation from the wire in p-polarization that causes no ``shadow'' to arise.

    Comparing the depicted resonance behaviour to recent publications \citep{she-2008,lukyanchuk}, analogies in the field distribution are observed. For these treatments of small cylinders with a very thin coating, as in the case described here, with higher mode order that is achieved by a variation of the frequency of incident light, discrete modes arise, which are represented by additional nodes of field accumulations on the first and second surface (outer coating interface and inner coating interface).

    The pattern of field distributions, observed in this work is a bit more complicated, although the complification of a surrounding thin layer is not treated. Additional modes are observed to occur at increasing $\alpha$ and a reorganization of the existing modes is seen in the final field plots. New modes arise at the upper and lower edge of the wire in a perpendicular cut as depicted in the right field plots and the spacing of the existing mode spots decreases.

    Due to the much higher range if $k_\text{inc}$ that is accessible by a variation of $\alpha$ compared with the reported, common variation of $\omega$ in the mentioned publications, a much higher mode order is reached already for relatively low incident angles.

\chapter{Experimental results and discussion \label{ch_results-discussion}}

\begin{flushright}\parbox[t]{\textwidth/2}{%
        \small{``Nur Beharrung f\"{u}hrt zum Ziel,

        Nur die F\"{u}lle f\"{u}hrt zur Klarheit,

        und im Abgrund liegt die Wahrheit.''\\ \slshape{(Friedrich Schiller)}}}
\end{flushright}

    Using the experimental setup as described in ch.\;\ref{ch_experimental-setup} series of measurements of scattering amplitudes are accomplished on wires (ch.\;\ref{ch_sample-fabrication}) of gold and silver. These two materials are chosen due to their good processing properties (sec.\;\ref{ch_material-properties}) and as for these two materials the filling method for fibers in PCF and single hole capillary (sec.\;\ref{ch_sample-silica}) shows the best results.

    An additional major advantage of wires of gold, fabricated by the sputtering technique is its extraordinary durability against surface degradation. In fact, a comparison of the results confirms that the measurement of SPP resonances on silver wires becomes increasingly difficult much faster than for gold wires.

    A sufficient number of free-standing sputtered samples shows distinct, reproducible dips for the theoretically predicted angles of SPP excitation. Therefore a concise comparison of the acquired data with theory is presented in sec.\;\ref{ch_result-measured-angles}. Good accordance between both is indicated. Therefore the first proof of radiative SPP on wires for nonperpendicular incidence is reported to be accomplished successfully.\\

    Also samples of small silver and gold wires and wire arrays in specially drawn silica fiber (sec.\;\ref{ch_sample-silica}) were fabricated successfully that are by electric conductivity and optical lateral examination proved to be continuous. The measurements of these samples with both methods (sec.s\;\ref{ch_setup-freestanding}, \ref{ch_setup-liquid}) turned out to be practically much more difficult. Scattering on single wires was first tried. The effective cross section of these samples was measured to be too small to obtain results with dips in the scattered intensity. Possible effects from SPP resonances could not be separated from measurement noise.

    Therefore the scattering properties of samples of wire arrays in PCF were measured. The complicated resulting patterns could not yet be correlated to theoretical effects. In this case coupling between the modes on adjacent wires occurs, causing a modification of the resonance properties. A further optimization of the setup to allow a successfull measurement and the derivation of a corresponding theory (which could base on the propositions of \citet{pitarke-cylinders}) were not accomplished in the course of this thesis.

\section{Observed scattering pattern}

    In accordance to the theoretical scattering pattern (sec.~\ref{ch_scattering-pattern}), the experimental setup exhibits circular rings in the experimentally accessed range of incident angles ($\ang{0}<\alpha<\ang{20}$). This observation is in accordance to the low eccentricity of the theoretically predicted ellipses due to low $\alpha$.
    
    \begin{figure}
        \centering

            \psfrag{al}[r][r][0.9]{\textcolor{white}{$\alpha$\hphantom{\ang{0}}}}
            \psfrag{2d}[r][r][0.9]{ \textcolor{white}{\ang{2} }}
            \psfrag{10d}[r][r][0.9]{\textcolor{white}{\ang{10} }}
            \psfrag{20d}[r][r][0.9]{\textcolor{white}{\ang{20} }}
            \psfrag{30d}[r][r][0.9]{\textcolor{white}{\ang{30} }}

            \includegraphics[width=0.5\textwidth]{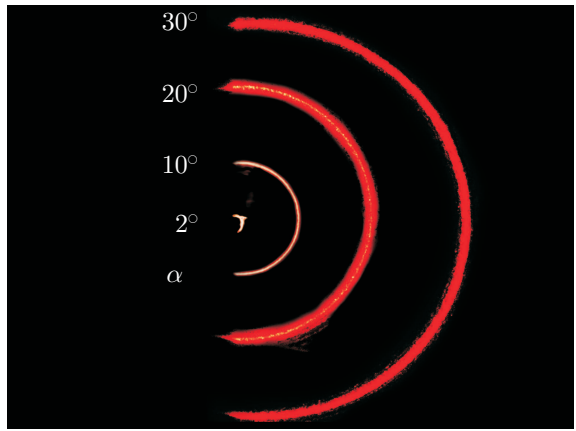}

        \caption[Scattering pattern of a single, free-standing silver wire.]{Scattering pattern of a single, free-standing silver wire with a diameter of $D=\SI{20}{\micro m}$. The displayed plane is the plane of the sensor (PD2), in which a ground glass screen is positioned for these pictures. The patterns are acquired under a variation of the incident angle $\alpha$ by photography. \label{graphic_scatteringpattern}}

\end{figure} 

    Displayed (fig.\;\ref{graphic_scatteringpattern}) are the observed half rings, as the samples are mounted on a holder (see sec.~\ref{ch_setup-freestanding}) that intentionally blocks the incident laser beam. This causes also blocking of the left half of the scattering patterns.

    The observed intensity distribution suggests that the main portion of light reflected by the cylinder, is radiated into a relatively small total solid angle and that it is strongly peaked on the surface of the scattered ellipses. This supports a measurement of scattering with a collimated beam, which is the best real approximation of plane waves: the relatively small portion of the beam that is scattered provides only a low intensity. This, on the other hand can be exploited well. Only small intensity is lost by scattering into the whole sphere surrounding the sample.

    For a variation of $\alpha$ with the detector (PD2) (sec. \ref{chapter_measurement}) at a fixed angle of observation $\gamma$ (fig. \ref{graphic_setup-scheme}) therefore, the maximum intensity step by step passes the aperture of the sensor.

    Best results are achieved with a choice of $\ang{10}<\gamma<\ang{20}$ for measurements with incident angles $\ang{0}<\alpha<\ang{10}$. As PD2 is placed very near to the sample and offers a comparably large aperture, followed by the focussing optics, it is possible to catch the scattered maximum already for low $\alpha$ and track it for the mentioned angle range with high sensitivity.

\section{Measurement of resonance angles \label{ch_result-measured-angles}}

\begin{figure}
        \centering

        \psfrag{f}[c][c][1][0]{a}
        \psfrag{b}[c][c][1][0]{b}
        \psfrag{e}[c][c][1][0]{c}
        \psfrag{d}[c][c][1][0]{d}
        \psfrag{c}[c][c][1][0]{e}
        \psfrag{a}[c][c][1][0]{f}

        \psfrag{angle}[c][c][0.9]{incident angle $\alpha$ in $\degree$}%
        \psfrag{amplitude}[c][c][0.9]{amplitude in arbitrary normalized units}%
        \psfrag{ratio}[l][l][0.9]{ratio p/s}%
        \psfrag{p}[l][l][0.9]{p-polarization}%
        \psfrag{s}[l][l][0.9]{s-polarization}%

\psfrag{v00}[t][t][0.7]{$0.00$}%
\psfrag{v25}[t][t][0.7]{$0.25$}%
\psfrag{v50}[t][t][0.7]{$0.50$}%
\psfrag{v75}[t][t][0.7]{$0.75$}%
\psfrag{v10}[t][t][0.7]{$1.00$}%
\psfrag{h4}[t][t][0.7]{$4$}%
\psfrag{h5}[t][t][0.7]{$5$}%
\psfrag{h6}[t][t][0.7]{$6$}%
\psfrag{h7}[t][t][0.7]{$7$}%
\psfrag{h8}[t][t][0.7]{$8$}%

\psfrag{vr00}[l][l][0.7]{$0.00$}%
\psfrag{vr025}[l][l][0.7]{$0.25$}%
\psfrag{vr050}[l][l][0.7]{$0.50$}%
\psfrag{vr075}[l][l][0.7]{$0.75$}%
\psfrag{vr100}[l][l][0.7]{$1.00$}%
\psfrag{vr125}[l][l][0.7]{$1.25$}%
\psfrag{vr150}[l][l][0.7]{$1.50$}%

        \includegraphics[width=0.8\textwidth]{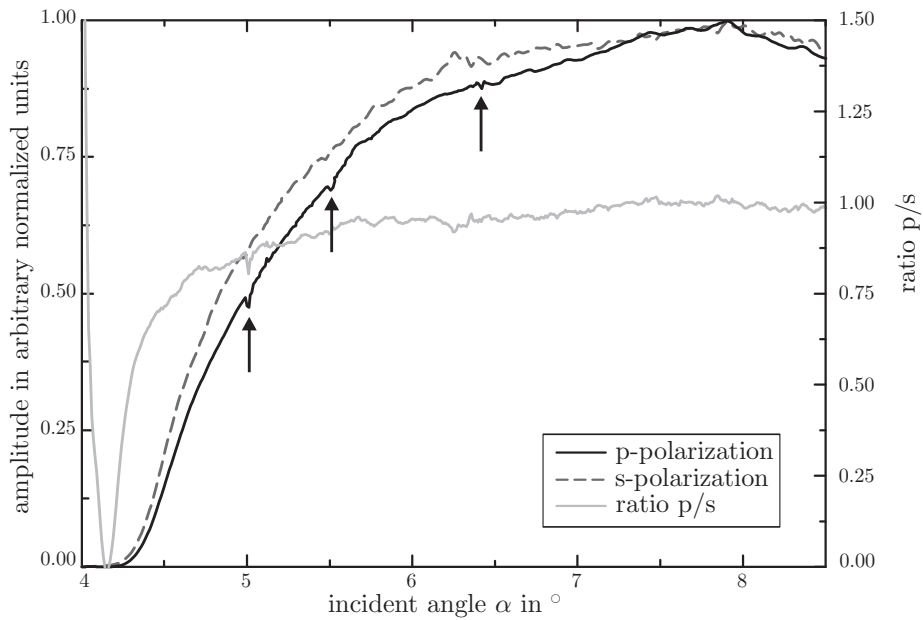}

        \caption[Measured intensity profile for scattering on a gold wire with $R=\SI{20.3}{\micro m}$ in s- and p-polarization.]{Example for a typical intensity profile for scattering on a gold wire with $R=\SI{20.3}{\micro m}$ in s- and p-polarization. This measurement is labelled as (f) in the summary and in fig. \ref{graphic_au-dips-color}.\label{graphic_au-graph}}
\end{figure} 

\subsection*{Measurement of the scattered intensity distribution}

    For the reported measurements continuously improved techniques are applied to produce the sputtered gold wires. A significant enhancement is observed by an optimization of the sputtering process (sec.\,\ref{ch_sample-freestanding}).


    For each of the samples a number of measurements are carried out. The measurement parameters are chosen for a range of $\ang{0}<\alpha<\ang{10}$. The strongly peaked intensity maximum that can be clearly seen in figure \ref{graphic_scatteringpattern} for each measurement travels through the aperture of the sensor, leading to a maximum at $\gamma = 2 \alpha$ (see fig \ref{graphic_setup-scheme}). This intensity profile is modulated by the absorption profile of the sample, which is to observe.

    All measurements are carried out with an angle resolution of at least $\Delta \alpha \leq \ang{0.01}$ with an optimized resolution of the acquired signal amplitudes. An increase of the angle resolution always causes an increase in the total time necessary for a complete measurement.

    Depending on the measurement conditions, a typical automated measurement takes two to four hours. Smaller diameters of the wire samples reduce the intensity of scattered light as the effective scattering cross section decreases. The consequently reduced signal-to-noise-level requires longer integration times (sec.\;\ref{ch_setup-lockindetection}). For very small wire diameters nevertheless the noise on the resulting measurement curves cannot be avoided to increase. 
    
    For a reliable identification of the small dips caused by SPP excitation, at least a ratio of the SPP dip-strength to noise of $5$ is found to be required.
    
\begin{figure}
        \centering

        \psfrag{f}[r][r][1][0]{a}
        \psfrag{b}[r][r][1][0]{b}
        \psfrag{e}[r][r][1][0]{c}
        \psfrag{d}[r][r][1][0]{d}
        \psfrag{c}[r][r][1][0]{e}
        \psfrag{a}[r][r][1][0]{f}

        \psfrag{angle}[c][c][0.9]{incident angle $\alpha$ in $\degree$}%
        \psfrag{radius}[c][c][0.9]{radius in \SI{}{\micro m}}%
        \psfrag{arbitrary units}[c][c][0.9]{arbitrary units}%
        \psfrag{Q_abs}[c][c][0.9]{$Q_\text{abs}$}%

\psfrag{a00}[t][t][0.7]{$0$}%
\psfrag{a01}[t][t][0.7]{$1$}%
\psfrag{a02}[t][t][0.7]{$2$}%
\psfrag{a03}[t][t][0.7]{$3$}%
\psfrag{a04}[t][t][0.7]{$4$}%
\psfrag{a05}[t][t][0.7]{$5$}%
\psfrag{a06}[t][t][0.7]{$6$}%
\psfrag{a07}[t][t][0.7]{$7$}%
\psfrag{a08}[t][t][0.7]{$8$}%
\psfrag{a09}[t][t][0.7]{$9$}%
\psfrag{a10}[t][t][0.7]{$10$}%
\psfrag{a11}[t][t][0.7]{$11$}%
\psfrag{a12}[t][t][0.7]{$12$}%
\psfrag{a13}[t][t][0.7]{$13$}%
\psfrag{a14}[t][t][0.7]{$14$}%
\psfrag{a15}[t][t][0.7]{$15$}%
\psfrag{a16}[t][t][0.7]{$16$}%
\psfrag{a17}[t][t][0.7]{$17$}%
\psfrag{a18}[t][t][0.7]{$18$}%
\psfrag{a19}[t][t][0.7]{$19$}%
\psfrag{a20}[t][t][0.7]{$20$}%
\psfrag{a21}[t][t][0.7]{$21$}%
\psfrag{a22}[t][t][0.7]{$22$}%
\psfrag{a30}[t][t][0.7]{$30$}%
\psfrag{a40}[t][t][0.7]{$40$}%
\psfrag{a50}[t][t][0.7]{$50$}%
\psfrag{a60}[t][t][0.7]{$60$}%
\psfrag{a70}[t][t][0.7]{$70$}%
\psfrag{a80}[t][t][0.7]{$80$}%

\psfrag{r01}[c][c][0.7]{$1$}%
\psfrag{r02}[c][c][0.7]{$2$}%
\psfrag{r03}[c][c][0.7]{$3$}%
\psfrag{r04}[c][c][0.7]{$4$}%
\psfrag{r05}[c][c][0.7]{$5$}%
\psfrag{r06}[c][c][0.7]{$6$}%
\psfrag{r07}[c][c][0.7]{$7$}%
\psfrag{r08}[c][c][0.7]{$8$}%
\psfrag{r09}[c][c][0.7]{$9$}%
\psfrag{r10}[c][c][0.7]{$10$}%
\psfrag{r11}[c][c][0.7]{$11$}%
\psfrag{r12}[c][c][0.7]{$12$}%
\psfrag{r13}[c][c][0.7]{$13$}%
\psfrag{r14}[c][c][0.7]{$14$}%
\psfrag{r15}[c][c][0.7]{$15$}%
\psfrag{r16}[c][c][0.7]{$16$}%
\psfrag{r17}[c][c][0.7]{$17$}%
\psfrag{r18}[c][c][0.7]{$18$}%
\psfrag{r19}[c][c][0.7]{$19$}%
\psfrag{r20}[c][c][0.7]{$20$}%
\psfrag{r21}[c][c][0.7]{$21$}%
\psfrag{r22}[c][c][0.7]{$22$}%
\psfrag{r30}[c][c][0.7]{$30$}%
\psfrag{r40}[c][c][0.7]{$40$}%
\psfrag{r50}[c][c][0.7]{$50$}%
\psfrag{r60}[c][c][0.7]{$60$}%
\psfrag{r70}[c][c][0.7]{$70$}%
\psfrag{r80}[c][c][0.7]{$80$}%

\psfrag{sv00}[l][l][0.7]{ $ 0$}%
\psfrag{sv0.1}[l][l][0.7]{$ 0.1$}%
\psfrag{sv0.2}[l][l][0.7]{$ 0.2$}%
\psfrag{sv0.3}[l][l][0.7]{$ 0.3$}%
\psfrag{sv0.4}[l][l][0.7]{$ 0.4$}%
\psfrag{sv0.5}[l][l][0.7]{$ 0.5$}%
\psfrag{sv0.6}[l][l][0.7]{$ 0.6$}%
\psfrag{sv0.8}[l][l][0.7]{$ 0.8$}%
\psfrag{sv01}[l][l][0.7]{$ 1$}%
\psfrag{sv1.2}[l][l][0.7]{$ 1.2$}%
\psfrag{sv1.4}[l][l][0.7]{$ 1.4$}%
\psfrag{sc1.6}[l][l][0.7]{$ 1.6$}%
\psfrag{sv1.8}[l][l][0.7]{$ 1.8$}%
\psfrag{sv02}[l][l][0.7]{$ 2$}%

\psfrag{sh00}[t][t][0.7]{$0$}%
\psfrag{sh0.1}[t][t][0.7]{$0.1$}%
\psfrag{sh0.2}[t][t][0.7]{$0.2$}%
\psfrag{sh0.3}[t][t][0.7]{$0.3$}%
\psfrag{sh0.4}[t][t][0.7]{$0.4$}%
\psfrag{sh0.5}[t][t][0.7]{$0.5$}%
\psfrag{sh0.6}[t][t][0.7]{$0.6$}%
\psfrag{sh0.8}[t][t][0.7]{$0.8$}%
\psfrag{sh01}[t][t][0.7]{$1$}%
\psfrag{sh1.2}[t][t][0.7]{$1.2$}%
\psfrag{sh1.4}[t][t][0.7]{$1.4$}%
\psfrag{sc1.6}[t][t][0.7]{$1.6$}%
\psfrag{sh1.8}[t][t][0.7]{$1.8$}%
\psfrag{sh02}[t][t][0.7]{$2$}%

        \includegraphics[width=0.75\textwidth]{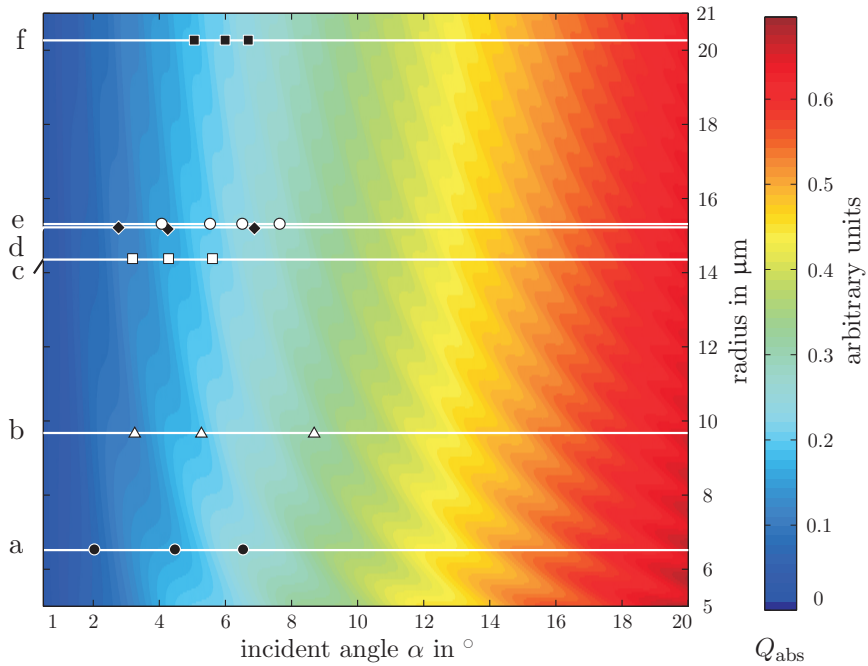}

        \caption[Graph of the scattering results for gold wires of different radii in air.]{Graph of the scattering results for gold wires of different radii in air. The horizontal lines indicate the radius for each measured wire. The symbols indicate the mean of the reproducibly measured dip positions. The data is overlayed over simulated absorption efficiencies from fig.\,\ref{graphic_scattering-colorplots-air}\label{graphic_au-dips-color}}
\end{figure} 

    The referenced signals for both incident polarizations (s and p) acquired simultaneously (sec.\;\ref{chapter_measurement}) are compared for each measurement.

    An exemplary curve of the reflected intensity over a variation of the incident angle $\alpha$ is depicted in figure \ref{graphic_au-graph}. The angle of observation for this measurement is chosen as $\gamma = \ang{16}$, thus at an incident angle of $\alpha = \ang{8}$ the maximum intensity (see fig.\;\ref{graphic_scatteringpattern}) passes the centre of the sensor aperture.

    The referenced intensity of s- and p-polarization in normalized, arbitrary units is displayed together with the ratio of p- over s- polarization. The ratio reveals the first dip (indicated with an arrow) to be measured only in p-polarization. The dips at higher $\alpha$ are measured in both polarizations but reproducible for repeated measurements.\\

\subsection*{Analysis and reproducibility}

\begin{figure}
        \centering

        \psfrag{d}[c][c][1][0]{a}
        \psfrag{a}[c][c][1][0]{b}
        \psfrag{b}[c][c][1][0]{c}
        \psfrag{c}[c][c][1][0]{d}

        \psfrag{angle}[c][c][0.9]{incident angle $\alpha$ in $\degree$}%
        \psfrag{radius}[c][c][0.9]{radius in \SI{}{\micro m}}%
        \psfrag{arbitrary units}[c][c][0.9]{arbitrary units}%
        \psfrag{Q_abs}[c][c][0.9]{$Q_\text{abs}$}%

\psfrag{a00}[t][t][0.7]{$0$}%
\psfrag{a01}[t][t][0.7]{$1$}%
\psfrag{a02}[t][t][0.7]{$2$}%
\psfrag{a03}[t][t][0.7]{$3$}%
\psfrag{a04}[t][t][0.7]{$4$}%
\psfrag{a05}[t][t][0.7]{$5$}%
\psfrag{a06}[t][t][0.7]{$6$}%
\psfrag{a07}[t][t][0.7]{$7$}%
\psfrag{a08}[t][t][0.7]{$8$}%
\psfrag{a09}[t][t][0.7]{$9$}%
\psfrag{a10}[t][t][0.7]{$10$}%
\psfrag{a11}[t][t][0.7]{$11$}%
\psfrag{a12}[t][t][0.7]{$12$}%
\psfrag{a13}[t][t][0.7]{$13$}%
\psfrag{a14}[t][t][0.7]{$14$}%
\psfrag{a15}[t][t][0.7]{$15$}%
\psfrag{a16}[t][t][0.7]{$16$}%
\psfrag{a17}[t][t][0.7]{$17$}%
\psfrag{a18}[t][t][0.7]{$18$}%
\psfrag{a19}[t][t][0.7]{$19$}%
\psfrag{a20}[t][t][0.7]{$20$}%
\psfrag{a21}[t][t][0.7]{$21$}%
\psfrag{a22}[t][t][0.7]{$22$}%
\psfrag{a30}[t][t][0.7]{$30$}%
\psfrag{a40}[t][t][0.7]{$40$}%
\psfrag{a50}[t][t][0.7]{$50$}%
\psfrag{a60}[t][t][0.7]{$60$}%
\psfrag{a70}[t][t][0.7]{$70$}%
\psfrag{a80}[t][t][0.7]{$80$}%

\psfrag{r01}[c][c][0.7]{$1$}%
\psfrag{r02}[c][c][0.7]{$2$}%
\psfrag{r03}[c][c][0.7]{$3$}%
\psfrag{r04}[c][c][0.7]{$4$}%
\psfrag{r05}[c][c][0.7]{$5$}%
\psfrag{r06}[c][c][0.7]{$6$}%
\psfrag{r07}[c][c][0.7]{$7$}%
\psfrag{r08}[c][c][0.7]{$8$}%
\psfrag{r09}[c][c][0.7]{$9$}%
\psfrag{r10}[c][c][0.7]{$10$}%
\psfrag{r11}[c][c][0.7]{$11$}%
\psfrag{r12}[c][c][0.7]{$12$}%
\psfrag{r13}[c][c][0.7]{$13$}%
\psfrag{r14}[c][c][0.7]{$14$}%
\psfrag{r15}[c][c][0.7]{$15$}%
\psfrag{r16}[c][c][0.7]{$16$}%
\psfrag{r17}[c][c][0.7]{$17$}%
\psfrag{r18}[c][c][0.7]{$18$}%
\psfrag{r19}[c][c][0.7]{$19$}%
\psfrag{r20}[c][c][0.7]{$20$}%
\psfrag{r21}[c][c][0.7]{$21$}%
\psfrag{r22}[c][c][0.7]{$22$}%
\psfrag{r30}[c][c][0.7]{$30$}%
\psfrag{r40}[c][c][0.7]{$40$}%
\psfrag{r50}[c][c][0.7]{$50$}%
\psfrag{r60}[c][c][0.7]{$60$}%
\psfrag{r70}[c][c][0.7]{$70$}%
\psfrag{r80}[c][c][0.7]{$80$}%

\psfrag{sv00}[l][l][0.7]{ $ 0$}%
\psfrag{sv0.1}[l][l][0.7]{$ 0.1$}%
\psfrag{sv0.2}[l][l][0.7]{$ 0.2$}%
\psfrag{sv0.3}[l][l][0.7]{$ 0.3$}%
\psfrag{sv0.4}[l][l][0.7]{$ 0.4$}%
\psfrag{sv0.5}[l][l][0.7]{$ 0.5$}%
\psfrag{sv0.6}[l][l][0.7]{$ 0.6$}%
\psfrag{sv0.8}[l][l][0.7]{$ 0.8$}%
\psfrag{sv01}[l][l][0.7]{$ 1$}%
\psfrag{sv1.2}[l][l][0.7]{$ 1.2$}%
\psfrag{sv1.4}[l][l][0.7]{$ 1.4$}%
\psfrag{sc1.6}[l][l][0.7]{$ 1.6$}%
\psfrag{sv1.8}[l][l][0.7]{$ 1.8$}%
\psfrag{sv02}[l][l][0.7]{$ 2$}%

\psfrag{sh00}[t][t][0.7]{$0$}%
\psfrag{sh0.1}[t][t][0.7]{$0.1$}%
\psfrag{sh0.2}[t][t][0.7]{$0.2$}%
\psfrag{sh0.3}[t][t][0.7]{$0.3$}%
\psfrag{sh0.4}[t][t][0.7]{$0.4$}%
\psfrag{sh0.5}[t][t][0.7]{$0.5$}%
\psfrag{sh0.6}[t][t][0.7]{$0.6$}%
\psfrag{sh0.8}[t][t][0.7]{$0.8$}%
\psfrag{sh01}[t][t][0.7]{$1$}%
\psfrag{sh1.2}[t][t][0.7]{$1.2$}%
\psfrag{sh1.4}[t][t][0.7]{$1.4$}%
\psfrag{sc1.6}[t][t][0.7]{$1.6$}%
\psfrag{sh1.8}[t][t][0.7]{$1.8$}%
\psfrag{sh02}[t][t][0.7]{$2$}%

        \includegraphics[width=0.75\textwidth]{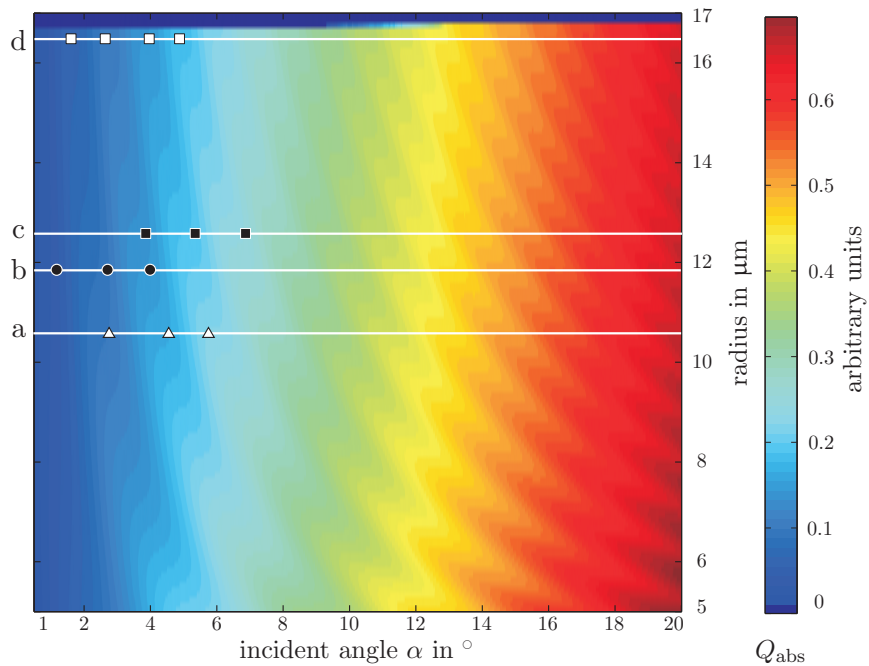}

        \caption[Graph of the scattering results for silver wires of different radii in air.]{Graph of the scattering results for silver wires of different radii in air. The horizontal lines indicate the radius for each measured wire. The symbols indicate the mean of the reproducibly measured dip positions. The data is overlayed over simulated absorption efficiencies from fig.\,\ref{graphic_scattering-colorplots-air}\label{graphic_ag-dips-color}.}
\end{figure} 

\begin{table}
        \centering
        \begin{tabular}{cccc}\toprule
          sample   &  R in \SI{}{\micro m}   &   identified dips & std. dev. $|\sigma|$ (in \textdegree)\\\midrule
          a &  \SI{6.5}{} & $3$ & 0.12\\\
          b & \SI{9.7}{} & $3$  & 0.05\\
          c & \SI{14.3}{} & $3$ & 0.08\\
          d & \SI{15.2}{} & $4$ & 0.15\\
          e & \SI{15.3}{} & $3$ & 0.21\\
          f & \SI{20.3}{} & $3$ & 0.20\\\bottomrule
        \end{tabular}
        \caption[Measured results for samples of sputtered gold wire of different radii.]{Measured results for samples of sputtered gold wire of different radii. For each sample the measurement was repeated at least twice. The measured number of dips and the mean standard deviation for all dips over the measurements are indicated and plotted in fig. \ref{graphic_au-dips-color}.\label{table_au-dips}}
\end{table}

    As it is shown in fig. \ref{graphic_scattering-colorplots-air} a theoretical comparison unveils that the SPP model is not a good approximation particularly for low incident angles. Therefore for figures (\ref{graphic_au-dips-color},\ref{graphic_ag-dips-color}) it is decided to compare the experimental data directly to the results of the scattering theory. This is legitimate for a proof of SPP as it is shown in sec.\;\ref{ch_scattering-resonances} that the scattering model inherently comprises the resonance condition for SPP modes.

    Each measurement of a sample showing resonance dips is repeated at least twice to reproduce the observed resonance angles. After at least three coinciding repeated measurements of at least three dips the sample is included in the overview of results for gold (tab.\;\ref{table_au-dips}) respectively for silver (tab.\;\ref{table_ag-dips}). For a comparison of the resonance dips to the calculated absorption efficiencies of gold (fig.\;\ref{graphic_au-dips-color}), respectively silver (fig.\;\ref{graphic_ag-dips-color}) the mean value of the angles of these reproduced dips is calculated and the standard deviation is indicated.
    
    In figures (\ref{graphic_au-dips-color}, \ref{graphic_ag-dips-color}) it can be seen that not for every probe subsequent dips are indicated. This is due to a strict evaluation if a resonance dip is at least reproduced for three times.

    Both graphs indicate that at least no resonances were measured in the maxima of $Q_\text{abs}$, but even the dip spacing corresponds well to the calculated distribution.

\subsection*{Limiting conditions}

\begin{table}
        \centering
        \begin{tabular}{cccc}\toprule
          sample   &  R in \SI{}{\micro m}   &   identified dips & std. dev. $|\sigma|$ (in \textdegree)\\\midrule
          a &  \SI{10.5}{} & $3$ & 0.23\\\
          b & \SI{11.9}{} & $3$  & 0.15\\
          c & \SI{12.6}{} & $3$ & 0.20\\
          d & \SI{16.15}{} & $3$ & 0.09\\\bottomrule
        \end{tabular}
        \caption[Measured results for samples of sputtered silver wire of different radii.]{Measured results for samples of sputtered silver wire of different radii. For each sample the measurement was repeated at least twice. The measured number of dips and the mean standard deviation for all dips over the measurements are indicated and plotted in fig. \ref{graphic_au-dips-color}.\label{table_ag-dips}}
\end{table}

    In the experimental data not only dips at the predicted angles are observed but also a fine superstructure occurs.

    However, it is not possible to exclude an additional source for absorption experimentally that might explain those: Coupling caused by surface roughness. In experiments to probe planar SPP excitations the systematic creation of such surface roughness in the size of \SI{10}{nm} to \SI{100}{nm} is applied to obtain coupling conditions.

    The surfaces of the probes, applied in this work are intended to be as smooth as possible. By inspectation with atomic force microscopy (AFM) during the optimization process for sample fabrication (sec.\;\ref{ch_sample-fabrication}) a surface roughness of \SI{5}{nm} for the desired thickness of the layer was verified for the probes applied.

    This technique should minimize effects from coupling due to surface roughness but cannot cancel them. It is difficult to predict the resulting structure of excitations by random surface structuring but it can be assumed that a portion of the observed fluctuations, at least of those that are observed repeatedly by a repetition for one probe, is caused by this effect.
    
    Another particularly observed effect is degradation of measurement features during the measurements. It is not expected due to the exceptionally low coupling efficiency to the SPP modes that this effect is caused by damage from energy accumulation in resonance as it happens for in-fiber excitation experiments on nanowires.
    
    A much simpler reason is revealed by a comparison of samples of gold and silver. For the measurements of silver wires it is observed that the distinction of the resonance dips decreases more rapidly with time after fabrication, instead the number and density of noise dips increases. This effect is attributed to the formation of an oxide layer on the silver surface. 
    
    This does not happen for gold samples due to its significantly lower chemical reactivity.
    Actually, for gold samples measurements over a longer period still exhibit the observed features much better. 
    
    The only possible way to prevent this effect for free-standing samples of metal is to encase the setup by a vacuum chamber. Another possibility, extending applying the proposed fabrication technique could be to use the new possibility of sputtering deposition of silica glass. A thin layer of silica, surrounding the sputtered gold wire would prevent the metal layer from any degradation effects caused by oxidization. Of course, as a side effect it would significantly change the properties of the SPP, but not in a way that would not be explainable by the theory that is presented in this work.
    
    \noindent However, embedded metal nanowires in silica fiber intrinsically do not suffer this limitation.\\
    
    Additional to the difference, observed between samples of gold and silver, also for gold a degradation of the resonance dips over time is observed. As a reason for that the unavoidable adsorption of dust can be identified which is in most cases not removable without destroying the sample.

    Fluctuations between curves obtained for a repeated measurement for one sample can be caused by the limitations of the measurement itself as well as by the assembly of small particles on the surface of the wires over time.

\subsection*{Comparison to theory}

    In figures \ref{graphic_au-dips-color} and \ref{graphic_ag-dips-color} for both materials good accordance of the measured resonances to the theoretical absorption efficiency $Q_\text{abs}$ is clearly indicated.

    From theory (sec.\;\ref{ch_theory}) it is anticipated that only for one polarization (p, TM, which means perpendicular to the plane of rotation of the wire) SPP excitation occurs. Nevertheless the measured data reveals dips in the scattered intensity in both polarizations for most of the samples. For no sample an undisturbed curve for s-polarization is obtained.

    It was experimentally checked that this effect is no measurement artefact as the separation of both simultaneously measured signals is of the magnitude of $1:200$. Mixing and interference between the two polarization channels can therefore not be the reason. The resonance dips physically occur for the indicated polarizations, contradictory to theory.\\

    A comparison to data, obtained by \citet{miziumski-ex},\citep{miziumski-theory},\citep{miziumski-proceeding} who showed a proof of concept of a scattering setup for the excitation of surface plasmons on fairly large ($\varnothing \approx \SI{50}{\micro m}$)\footnote{A measurement for smaller wire diameters was proposed, which were not feasible at that time. In those experiments no polarization dependency was measured. } tapered glass cylinders with a smooth evaporated surface coating of aluminium and silver of unknown but sufficient thickness, reveals that his measurement for low angles matches the resonance angles from the derived model for the planar SPP approximation.
    
    But in fact, the comparison of this model to the exact solution and to the theoretical absorption efficiencies, as shown in sec. \ref{chapter_theory-scatterincoeff} do explicitly not coincide with the model for the range of low incident angles that is also probed by Miziumski.

\chapter{Conclusion and outlook \label{ch_cnclusion}}

    This thesis shows a coherent treatise of the excitation of surface plasmon polaritons on small wires in theory and in experiment. The measured experimental results predominantly agree with the modelled results from theory. The results of this work are summarized by formulating answers to the main questions that are established in chapter~\ref{ch_introduction}.

\begin{flushleft}\parbox[t]{0.75\textwidth}{%
        ``What is the range of validity of the proposed model for the propagation of radiative SPPs on metal wires?''}
\end{flushleft}

    The model of spiralling surface plasmon polaritons~(sec.\;\ref{chapter_theory-model})~\citep{schmidt-prb-arrays-08,miziumski-theory,miziumski-proceeding,pfeiffer-theory} bases on the assumption that plasmons on a cylinder can be approximated by planar SPP~(sec.\;\ref{ch_theory-planar-spp}). The approximation appears intuitively to apply better for large wire diameters. This assumption is proved and compared to an exact solution of Maxwell's equations for SPPs on cylinders (sec.\;\ref{ch_theoresults-exact}). Both models show good agreement for sufficiently large incident angles ($\alpha > \ang{5}$). For small incident angles it is not valid as can be clearly seen~(fig.\;\ref{graphic_exact-vs-model}).

    Quantitatively the model matches the exact solution better with increasing radii ($R$), higher order modes ($m$) and larger angle of incidence.

\begin{flushleft}\parbox[t]{0.75\textwidth}{%
    ``Does a description of the scattering of light on metal wires comprise SPP resonances?''}
\end{flushleft}

    For comparison, a theoretical description of the scattering process was derived from classical Mie theory. The inherence of SPP resonances in this theory is proved~(sec.\;\ref{ch_scattering-resonances}). Comparing the angles of SPP excitation to calculated absorption efficiencies $Q_\text{abs}$ from this theory~(sec.\;\ref{ch_scattering-s}), good agreement for large angles of incidence but decreasing agreement in the range of $\alpha<\ang{20}$ is observed~(sec.\;\ref{ch_scattering-resonances}).

\begin{flushleft}\parbox[t]{0.75\textwidth}{%
    ``Does an experimental scattering examination of exciting SPPs on small metal wires reveal resonances that match the predicted dispersion relation in a range of wave vectors that is not feasible by in-fiber coupling?''}
\end{flushleft}

    An experimental setup was developed for the polarization dependent examination of scattering~(sec.\;\ref{ch_experimental-setup}) on samples of small wires of gold and silver~(sec.\;\ref{ch_sample-fabrication}).

    A low signal-to-noise-ratio was achieved by applying sophisticated noise reduction techniques~(sec.\;\ref{ch_setup-lockindetection}) allowing the observation even of small SPP resonance effects.

    Measurements of the scattering amplitude over a variation of the incident angle exhibit reproducible dips at those angles, predicted from Mie scattering theory and SPP dispersion theory. A comparison of the experimental results for scattering on single wires of silver and gold ($\SI{6.5}{\micro m}<R<\SI{20.3}{\micro m}$) shows good agreement to the calculated absorption efficiencies~(sec.\;\ref{ch_result-measured-angles}). A polarization dependency of the observed resonance dips could not be consistently observed. In most measurements the resonances that were identified occurred in both polarizations.

\begin{flushleft}\parbox[t]{0.75\textwidth}{%
``Is it possible to relate the observations from scattering to a near field description?''}
\end{flushleft}

    For an in depth understanding of the occurring processes of SPP excitation by scattering, the theoretical approach from Mie theory for far field scattering efficiencies was extended~(sec.\;\ref{ch_scattering-solution-fields}). With the obtained theory the near field distribution for s- and p-po\-la\-ri\-za\-tion was examined for the occurence of resonance features~(sec.\;\ref{chapter_theory-fielddistrib}).

    The prediction from the SPP model, that analogue to planar SPPs, resonances only occur for p-po\-la\-ri\-za\-tion~(TM) is proved by effects in the near field pattern and in the far field absorption efficiency.
    Arising field enhancements in p-polarization are observed in the vicinity of resonance angles in the near field distribution, which are identified as SPPs. The amplitude of these fluctuates smoothly with resonant versus not resonant incident angles.

    Additionally field enhancements in s-polarization on the surface of the wire can be seen, being probably identified as nonretarded localized surface plasmons~(LSP). In contrast the amplitude of SPP, no oscillating fluctuation in the strength of these field enhancements can be observed.

    In the near field distribution for both different effects (s-po\-la\-ri\-za\-tion, LSPs, p-po\-la\-ri\-za\-tion, SPPs) a remarkable difference was observed which accords to the general expectation for LSPs and radiative SPPs. In the vicinity of some $\SI{}{\micro m}$ behind the irradiated wire, in s-po\-la\-ri\-za\-tion a ``shadow'' is observed, in p-polarization this is not seen, which matches the interpretation as radiative SPP intrinsically radiate their energy away into the surrounding dielectric medium.\\

    A further examination of the excitation of SPPs on small metal wires still promises interesting insights. Especially as the theoretical treatment in this area of physics currently is ahead its experimental examination and most theoretical studies do not match the experimental requirements of materials and technical feasibility. Additionally some observations during the work on this thesis could not yet be explored in depth and some developed techniques promise further, exciting applications and further developments:

    The calculated near field patterns could be observed by extending the developed experimental setup by a scanning near field microscope (SNOM). A comparable approach for a planar glass-gold-air system was recently reported~ \citep{jose}. Particularly the detailed examination of the observed shadowing effect and differences in s- and p-polarization that could not be measured by the far field measurement technique, applied in this work, could reveal interesting conclusions.

    Not presented in this work are the obtained computational results for perpendicular incidence on a single nanowire with a variation of the wavelength instead of the incident angle. These computational results promise interesting experimental results. An experiment would be the complementary to the experiments carried out in this thesis. A possible realization of the setup could utilize a supercontinuum light source and a monochromator for scanning the frequency and measuring the absorption. The simulated results propose an interesting characteristic not only for the reflected but also for the transmitted light from a collimated beam.
    An experimental observation of the scattering effects on large ($\varnothing > \SI{100}{\micro m}$) dielectric cylinders for perpendicular light incidence and varied angle of observation and wavelength was reported by \citet{abushagur}. A comparable experiment on large diameter ($\varnothing \approx \SI{120}{\micro m}$) copper and brass cylinders in the infrared range was performed by \citet{cohen-experiment}. These results could be easily achieved and by far extended by the proposed setup.

    The techniques that were applied in this work offer the possibility of the fabrication of very small wires as well as dielectric wires with a very thin coating for the experimental testing of recently published theoretical studies. This makes it possible to probe cylindrical glass-metal-air configurations if a further optimization of the sputtering parameters could be achieved.

    Measuring experimentally the scattering properties of arrays of gold nanowires during the works for this thesis revealed more complicated resonance behaviour. This has to be theoretically and experimentally examined in detail. Particularly it is expected that in this case coupling between the wires occurs, changing the total dispersion relation. This cannot simply be modelled by classical scattering theory any more. An approach to solve this problem could follow the propositions of \citet{pitarke-cylinders}.

    The fabrication technique of the samples in this works could be extended to create wires with a nonuniform diameter over the length. Coupling light into the metal surface at a point of smaller diameter, followed by a cavity of large diameter and subsequently a decrease in diameter could result in a cavity resonance and allow an insight into the absorption of travelling SPP. A modification of the surface by application of focussed ion beam (FIB) or other microstructuring techniques also promises an interesting modification of the SPP guiding properties of the wires.

    During the works on this thesis a proof of concept for a technique of side polishing the fiber embedded wire samples was demonstrated. This approach could allow coupling to these very thin metal wires without disturbance due to the bent surface of the surrounding fiber. Additionally ATR (attenuated total reflection) coupling using prism coupling setups could provide access to the guided SPP modes on the wires that were not possible to probe by the techniques applied in this thesis.

\SaveTocDepth{-1}
\cleardoublepage
\appendix


\chapter{Mathematical and electromagnetic foundations }

\section{The Bessel differential equation and its solutions \label{ch_bessel-functions}}


Due to the special importance of Bessel functions for optics in general and for the derivations of section \ref{chapter_theory-scatterincoeff} in specific, some important properties of the Bessel functions shall be assembled here. A more in depth information about the Bessel functions and differential equation can be found in \citet{abramowitz,bateman}.\\

The Bessel differential equation is the solution of the scalar wave equation \ref{eq_scalar-wave-equation} in cylindrical coordinates~(eq.\;\ref{eq_scalar-wave-equation-cyli}). It can be expressed as:

\begin{equation}
x^2 \frac{\den^2}{\den x^2} Z + x \frac{\den}{\den x} Z + (x^2-n^2) Z = 0
\end{equation}

\begin{figure}[tbp]
        \centering

            \psfrag{J_0(x)}[c][c][0.8][0]{$J_0(x)$}
            \psfrag{J_1(x)}[l][l][0.8][0]{$J_1(x)$}
            \psfrag{J_2(x)}[l][l][0.8][0]{$J_2(x)$}
            \psfrag{J_3(x)}[l][l][0.8][0]{$J_3(x)$}
            \psfrag{J_4(x)}[l][l][0.8][0]{$J_4(x)$}
            \psfrag{J_5(x)}[l][l][0.8][0]{$J_5(x)$}
            \psfrag{Y_0(x)}[c][c][0.8][0]{$Y_0(x)$}
            \psfrag{Y_1(x)}[l][l][0.8][0]{$Y_1(x)$}
            \psfrag{Y_2(x)}[l][l][0.8][0]{$Y_2(x)$}
            \psfrag{Y_3(x)}[l][l][0.8][0]{$Y_3(x)$}
            \psfrag{Y_4(x)}[l][l][0.8][0]{$Y_4(x)$}
            \psfrag{Y_5(x)}[l][l][0.8][0]{$Y_5(x)$}

            \includegraphics[width=\textwidth]{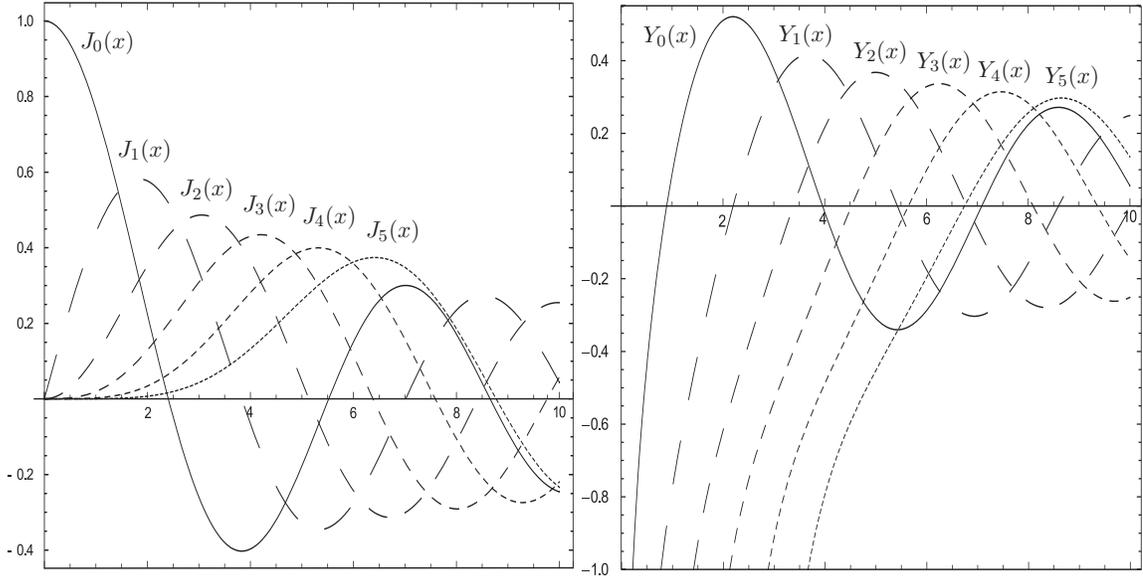}

        \caption[Plot of the Bessel functions of first $J_n(x)$ and second kind $Y_n(x)$ for real arguments and different orders.]{Plot of the Bessel functions of first $J_n(x)$ and second kind $Y_n(x)$ for real arguments $x$ and different orders $n=0\ldots5$.\label{graphic-bessel-jy}}

\end{figure}

Where $Z$ are Bessel functions. The Bessel functions of first $Z^{(1)}_n(x)=J_n(x)$ and second kind $Z^{(2)}_n(x)=Y_n(x)$ (the latter also called Neumann function) are the two linearly independent solutions to the Bessel differential equation. For integral orders $n$ these two solutions are depicted in figure \ref{graphic-bessel-jy} over real arguments.\\
\nomenclature[mjn]{$J_n(x)$}{Bessel function of type 1 and order $n$ of the argument $x$.}
\nomenclature[myn]{$Y_n(x)$}{Bessel function of type 2 (Neumann function) and order $n$ of the argument $x$.}

The first derivatives of the Bessel functions of type $\nu$, $Z^{(\nu)}_n(x)$ are given by the relation\footnote{See \citet{abramowitz} for further details.}:

\begin{equation}
 {Z^{(\nu)}_n}' = \frac{1}{2} (Z^{(\nu)}_{n-1}(x) - Z^{(\nu)}_{n+1}(x) ) \label{eq-bessel-derivative}
\end{equation}%
\nomenclature[mz]{$Z^{(\nu)}_n(x)$}{Bessel function of type $\nu$ and order $n$ of the argument $x$.}

A third and a fourth solution of the Bessel differential equation can be constructed from $J_n(x)$ and $Y_n (x)$, which are called Bessel functions of third and fourth kind or Hankel functions of first $H_n^{(1)}(x)$  and second kind $H_n^{(2)}(x)$:

\begin{align}
 H_n^{(1)}(x) &= J_n(x) + \i Y_n (x)     &     H_n^{(2)}(x) &= J_n(x) - \i Y_n (x)
\end{align}%
\nomenclature[mh]{$H_n^{(1)}(x), H_n^{(2)}(x)$}{Hankel function of type 1, respectively type 2 and order $n$ of the argument $x$.}

For the calculation of the coefficients as it is accomplished in sections \ref{chapter_theory-scatterincoeff} and \ref{chapter_theory-fielddistrib} for the expansion of the fields in vector cylindrical harmonics (equations \ref{eq-expansion-all} and \ref{eq-expansion-e_i2}) it is important to estimate the number of orders of Bessel functions that is necessary to achieve correct results.\\

Analytically a summation of $\inf$ orders is possible, in numerics the number of orders is crucial as the values of the Bessel functions have to be calculated for each order thus the time for computation scales linearly with the number of orders.\\

\begin{figure}[btp]
        \centering

            \psfrag{Re(H1_1)}[c][c][0.8][0]{$\text{Re}(H^{(1)}_1(x))$}
            \psfrag{Im(H1_1)}[c][c][0.8][0]{$\text{Im}(H^{(1)}_1(x))$}
            \psfrag{Re(H1_n)}[c][c][0.8][0]{$\text{Re}(H^{(1)}_n(x))$}
            \psfrag{Im(H1_n)}[c][c][0.8][0]{$\text{Im}(H^{(1)}_n(x))$}
            \psfrag{x=}[l][l][0.8][0]{$x=0.248+0.6477 \i$}
            \includegraphics[width=\textwidth]{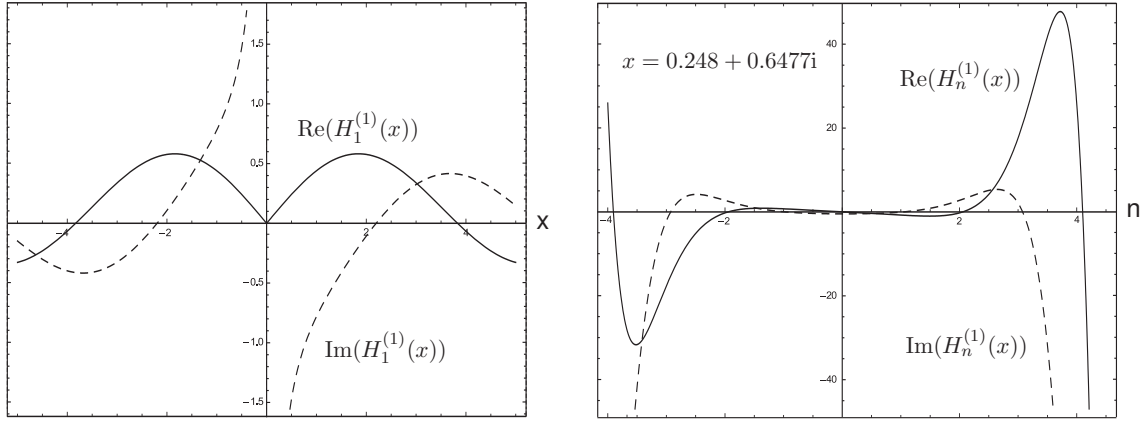}

        \caption[Plot of the Hankel function of first kind  and order $H_1(x)$ for real arguments and a plot over integral and fractional orders $n=-4 \ldots 4$ for a typical argument $H_n(x=0.248+0.6477 \i)$]{Plot of the Hankel function of first kind and order $H_1(x)$ for real arguments and a plot over integral and fractional orders $n=-4 \cdots 4$ for a typical argument $H_n(x=\xi=0.248+0.6477 \i)$ as it occurs for a $R=\SI{20}{\mu m}$ diameter Au wire in silica with an incident angle of $\ang{45}$.\label{graphic-bessel-h-over-n_x}}

\end{figure}

For typical arguments of the Bessel function $H_n(x)$ of $x=\eta=0.248+0.6477 \i$ (for a gold wire of diameter $R=\SI{20}{\micro m}$ with surrounding silica at an incident angle of $\alpha=\ang{45}$) a plot over the orders $n=-4 \ldots 4$ is depicted in fig.\;\ref{graphic-bessel-h-over-n_x}.

\section{The Poynting vector of plane waves \label{ch_plane-waves}}

    The intensity mathematically is derived from the nonobservable Poynting vector $\vc S$ of an electromagnetic wave.

    \begin{align}
     I(\vc r) &= |\vc{S} (\vc r)|      &        \vc S (\vc r)= \nicefrac{1}{2} \, \Re \left( \vc E (\vc r)\times \vc H (\vc r)^{\ast} \right)
    \end{align}

    The Poynting vector includes additional information about the direction of energy flux. In general $\vc S$ is a complex vector, whose real and imaginary parts represent the physical energy flux respectively an analogon to the idle power. In this thesis the Poynting vector is defined as only the real part.

    For a plane wave (see section \ref{ch_plane-waves-sol}, p. \pageref{ch_plane-waves-sol}), whose propagation direction is $\parallel \bvc{e}$ the Poynting vector $\vc{S}$ can be written in terms of only the electric or the magnetic field. Therefore it is possible \citep[p.7 ff]{lindlein} to describe the electric field respectively the magnetic field as:
    \nomenclature{$\vc{S}$}{The Poynting vector $\vc{S}= \nicefrac{1}{2}\Re(\vc{E} \times \vc{H^\ast})$ has the physical unit of an intensity $[S] = \SI{1}{V A / m^2} = \SI{1}{W/m^2}$ and describes the direction and magnitude of the electromagnetic energy flux. For plane waves in appendix \ref{ch_plane-waves} the relation is derived to reduce to $|\vc{S}|=\sqrt{\varepsilon_0 \varepsilon / (\mu_0\mu)} |\vc{E}|^2 = \sqrt{\varepsilon_0 \varepsilon/(\mu_0\mu)} |\vc{H}|^2$.}

        \begin{align}
        \vc{E} &= \mp \sqrt{\frac{\mu_0\mu}{\varepsilon_0 \varepsilon} } \bvc{e} \times \vc{H}  &
        \vc{H} &= \pm \sqrt{\frac{\varepsilon_0 \varepsilon}{\mu_0\mu} } \bvc{e} \times \vc{E} \label{eq-planewave-h}\\
        \end{align}

    Thus $\vc{E}$, $\vc{H}$ and $\bvc{e}$ form an orthogonal trihedron, describing a wave which propagates in a homogeneous medium. The Poynting vector of a plane wave can then be written as:

        \begin{align}
        \vc{S}  &= \vc{E} \times \vc{H^\ast}\\
                &= \left( \mp \sqrt{\frac{\mu_0\mu}{\varepsilon_0 \varepsilon}} \bvc{e} \times \vc{H} \right) \times \left( \mp \sqrt{\frac{\varepsilon_0 \varepsilon}{\mu_0\mu}} \bvc{e} \times \vc{E} \right)^\ast \\
                &= - \left( (\bvc{e} \times \vc{H}) \cdot \vc{E^\ast} \right) \bvc{e} + \left( (\bvc{e} \times \vc{H}) \cdot \bvc{e} \right) \vc{E^\ast}
        \end{align}

    The second term is zero. The first term remains and can be converted with equation \refeq{eq-planewave-h} to:

        \begin{align}
        \vc{S} &= \pm \sqrt{\frac{\varepsilon_0 \varepsilon}{\mu_0\mu}} |\vc{E}|^2 \bvc{e} \label{eq-s-e}\\
        \vc{S} &= \pm \sqrt{\frac{\mu_0\mu}{\varepsilon_0 \varepsilon}} |\vc{H}|^2 \bvc{e} \label{eq-s-h}
        \end{align}

    Equation \ref{eq-s-e} is also true for $\vc{H}$ (equation \ref{eq-s-h}) \footnote{ As $(\vc{a} \times \vc{b}) \cdot \vc{c} = - (\vc{a} \times \vc{c})  \cdot \vc{b}$ } with equation \ref{eq-planewave-h}. Additionally for a plane wave we achieve the basic correlation:

        \begin{equation}
        \mu_0\mu |\vc{H}|^2 = \varepsilon_0\varepsilon |\vc{E}|^2
        \end{equation}

    The intensity of an electromagnetic wave (electromagnetic power per area), incident on a surface perpendicular to $\vc{S}$ is defined as:

    \begin{equation}
    I = |\vc{S}|
    \end{equation}
    \nomenclature{$I$}{The intensity is defined by $I=|\vc{S}|$ (see also $\vc{S}$).} 
    \newpage
    \chapter{Conventions}

\section{Mathematical and physical conventions \label{ch_appendix-conventions} }

Throughout this thesis attention was paid to consistent mathematical, physical and experimental conventions. These are listed in the following lines.

\begin{multicols}{2}
\begin{description}
    \item[Complex relative dielectric permittivity] In this work $\varepsilon=\varepsilon_\text{r}+\i \varepsilon_\text{i}$ is defined dimensionless. This nomenclature was consistently chosen for reasons of mathematical convenience. It shall be pointed out that this is not the convention of some theoretical textbooks, including \citet{born-wolf} and \citet{jackson}.

        The conversion between both systems is simply $\varepsilon=\varepsilon_\text{abs}/\varepsilon_0$.

        Also \citet{bohren-huffman} in the beginning of his derivations applies their absolute dielectric permittivity nomenclature, whereas he later omits this completely as he expresses all relations in terms of $k$ which is independent of the nomenclature of $\varepsilon$. Therefore the later derivations of the scattering theory in \citet{bohren-huffman} are nevertheless directly comparable.
    \item[Complex quantities:] The real and imaginary of a complex quantity are written in the most widespread notation as $x = \Re(x) + \i \Im(x)$.
    \item[Constants], physical as well as mathematical are not typeset italic but upright, including $\pi$.
    \item[Equalities] are distinguished into nondirectional $=$, logically directional definitions $:=$ and $=:$, defining the quantity on the $:$ marked sinde. Approximate equality is indicated with $\approx$, combinations are $\leqslant,\geqslant,\lesssim,\gtrsim$. $\lesseqqgtr$ is only used to emphasize a relation in contrast to another relation.
    \item[Labelling] of equations is limited to those, which are referred to in later text or equations.
    \item[Vectors] are typeset boldface italic $\vc{a}$, latin letters as well as greek letters, following the angloamerican standard print notation. Unit vectors are typeset $\bvc{e}$.
    \item[Vector products] are typeset $\vc{a} \cdot \vc{\beta}$.
    \item[Vector differential operators] are noted as $\nabla$ which is a vector whose entries are the first derivatives into each spatial dimension. The Laplace differential operator, $\triangle \vc{A}=\nabla\cdot(\nabla \cdot \vc{A})$ which is built on that convention, shall not be confused with $(\nabla \cdot \nabla) \vc{A}$.
\end{description}
\end{multicols}

\section{Nomenclature \label{ch_nomenclature} }

The applied nomenclature in the theoretical derivations and the abbreviations in the experimental sections are chosen consistently. The following overview lists all those, mentioning the pages in parentheses where they are first mentioned. The list starts with mathematical and physical conventions and constants, that are mainly used in formulae and texts, that describe derivations. It continues with a list of the abbreviations, mainly used for the explanation of the experimental setup.

\begin{multicols}{2}
\printnomenclature
\end{multicols}

    \newpage
    \chapter{Applied instrumentation \label{ch_appendix-instrumentation} }

\section*{Sample manufacturing and preparation}
    \begin{description}
        \item[Tapering rig] Explained schematically in section \ref{ch_tapering}. For tapering silica glass fibers to the desired diameters.
        \item[Sputtering system] Emtech K575X turbo single magnetron sputter coater, peltier coold. For coating tapered fibers with thin layers of metals.
        \item[Sputtering system] AJA, ATC orion series uhv, water cooled multiple magnetron and RF sputter coater. For coating tapered fibers with thin layers of metals.
        \item[Optical microscope] Nicon Eclipse LV100 optical bright field/dark field digital microscope with a maximum magnification of $100 \times$.
        \item[Scanning electron microscope] Hitachi S-4800 field emission sem.
        \item[Scanning electron microscope / focused ion beam system] Zeiss field emission SEM/FIB Gemini Nvision 40 CrossBeam.
        \item[Atomic force microscope]
        \item[Fiber drawing tower] Explained schematically in section \ref{ch_fiber-drawing}. For the fabrication of special small hole capillary fibers and customized photonic crystal fibers for metal filling.
    \end{description}

\section*{Optics}
    \begin{description}
        \item[Laser] Thorlabs HeNe laser HRP 120. Specifications: 12 mW, $\lambda=\SI{632.8}{nm}$, linear polarization $>500:1$, mode $\text{TEM}_{00}>99\%$, $1/\e^2$ beam diameter \SI{0.88}{mm}, beam divergence \SI{0.92}{mrad}
        \item[Polarization maintaining fiber] Fibercore limited HB600, bowtie structure, single mode for $\lambda=\SI{633}{nm}$, effective core diameter \SI{3.2}{\micro m}, numeric aperture $\text{na}=0.14-0.18$
        \item[Photo detector](PD1) Thorlabs DET36A High speed si pin photo detector, range $\lambda=\SI{350}{}-\SI{1100}{nm}$, peak $\lambda=\SI{970}{nm}$, active area $3.6 \times \SI{3.6}{mm}$, diode capacitance $C_\text{d}=\SI{40}{pF}$, responsitivity $\mathcal{R}(\SI{633}{nm})=\SI{0.4}{A/W}$, bias voltage and max. output voltage $U=\SI{10}{V}$.
        \item[Amplified photo detector] (PD2) Thorlabs PDA100A switchable gain, transimpendance amplified si pin photo detector, reverse biased, range $\lambda=\SI{400}{}-\SI{1100}{nm}$, peak $\lambda=\SI{970}{nm}$, active area $\varnothing \SI{9.8}{mm}$, responsitivity $\mathcal{R}(\SI{633}{nm})=\SI{0.35}{A/W}$, adjustable gain of \SI{0}{dB}, \SI{10}{dB}, \SI{20}{dB}, \SI{30}{dB}, \SI{40}{dB}, \SI{50}{dB}, \SI{60}{dB}, \SI{70}{dB} with the bandwidth decreasing for increasing gain from \SI{1.5}{MHz} to \SI{2}{kHz}.
    \end{description}

\section*{Measurement instruments}
    \begin{description}
        \item[Lock-in amplifier] Stanford Research Systems RS830 DSP digital lockin amplifier, external inputs with a/d converters, GPIB controlled.
        \item[Lock-in amplifier module 1] Femto LIA-MV-200 single phase digital lockin module, analog, magnitude signal output.
        \item[Lock-in amplifier module 2] Femto LIA-MVD-200 dual phase digital lockin module, analog magnitude, phase and quadrature signal output.
        \item[Digital multimeter] Agilent 4110A digital multimeter, GPIB controlled and synchronized to the RS830.
        \item[Motorized precision rotation stage] Physik Instrumente (PI) M-037.DG servo motorized , referenced precision rotation stage with PI mercury servo motor controller, usb programmable.
    \end{description} 
\listoffigures
    \cleardoublepage

\nocite{gouesbet,jones}
\bibliographystyle{abbrvnat}  
\bibliography{cited/thesis-literature}
    \cleardoublepage
    \newpage
    
    \addchap*{Acknowledgements}

    I would like to thank Prof. Philip Russell for offering me the opportunity to accomplish my diploma thesis in his group. It was an honour for me to work together and learn from him and and the great people who have gathered in division three of the Max-Planck-Research-Group.

    Special thanks go to my supervisor Dr. Markus Schmidt. He teached me working in the laboratory, not believing everything that is reported in literature, not giving up, repeating and rethinking when experimental results, as usual, are difficult to obtain and gave me numerous advice during the year of this work.

    A nice scientific working environment and a lot of scientific as well as technical advice was always guaranteed by the people of the nanowire-group, Hemant Tyagi, Luis Prill-Sempere, Howard Lee and Jerry Chen.

    For helping me with the fabrication of fibers with the sputtering machine and for introducing me into the miracles of the focussed ion beam system and the scanning electron microscope, I have to say special thanks to Helga Hussy and Daniel Plo{\ss}. For aiding me in fabricating great fibers with our fiber drawing tower Silke Rammler and Michael Scharrer were responsible: Thank you two.

    Providing me with helpful advice for writing this thesis I thank especially Andre Brenn, Philipp H\"{o}lzer, Amir Abdolvand, Johannes Nold, Leyun Zang, Anna Butsch, Christine Kreuzer, Sebastian Stark and Martin Garbos.

    Last but not least, in the end of my undergraduate studies I thank my family and Sarina. Thank you for supporting me for the whole time. The thesis and my studies would not have been accomplished without you. \\

    The development of acknowledgements over time has recently been subject of study \citep{ben-ari} that indicates a clear interconnection between selection of the colleagues who are thanked for support and elementary interests of the author.

    As epistemically a minimum amount of subjectivity can neither way be excluded, the well-disposed reader is invited to read my acknowledgements as they are: Some subjective last sentences to thank those without whose support and inspiring interchange of ideas the presented work would not have been finished as it is. 
    \newpage
\addchap*{Eidesstattliche Erklärung}
    \vspace{5cm}
Ich versichere, dass ich die Diplomarbeit (im Folgenden Arbeit genannt) ohne fremde Hilfe und ohne Benutzung anderer als der angegebenen Quellen angefertigt habe und dass die Arbeit in gleicher oder \"{a}hnlicher Form noch keiner anderen Pr\"{u}fungsbeh\"{o}rde vorgelegen hat und von dieser als Teil einer Pr\"{u}fungsleistung angenommen wurde. Alle Ausf\"{u}hrungen, die w\"{o}rtlich oder sinngem\"{a}{\ss} \"{u}bernommen wurden, sind als solche ge\-kenn\-zeich\-net.

Mir ist ferner bekannt, dass die Friedrich-Alexander-Universit\"{a}t Erlangen-N\"{u}rnberg aufgrund der pr\"{u}fungsrechtlichen Vorschriften einen Anspruch auf das Original der Arbeit hat. Die erforderlichen Dateien werde ich dem Lehrstuhl zur Verf\"{u}gung stellen. Dieser Anspruch bezieht sich jedoch nur auf das k\"{o}rperliche Eigentum an der Arbeit als solches und auf deren Verwendung zu den in der Pr\"{u}fungsordnung festgelegten Zwecken.

\vspace{3cm}
Erlangen, den \rule{3cm}{1pt} \ \ \ \ \ \ \ \ \ \ \  \rule{6cm}{1pt} 
\RestoreTocDepth


\includepdf{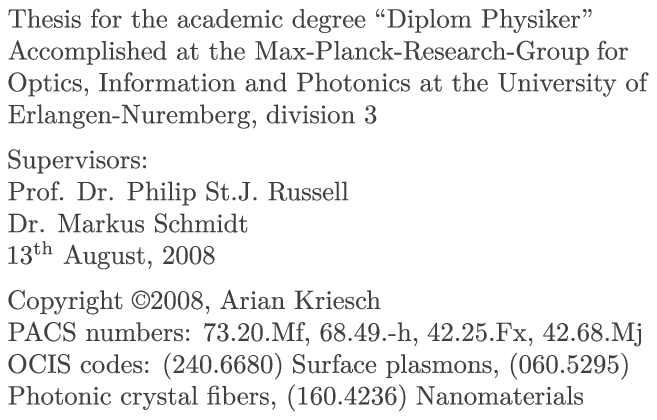}

\end{document}